\documentstyle[epsf,12pt]{book}
\begin{document}
\setlength{\baselineskip}{18pt}

\jot=8pt
\def\ove{\overline} 
\gdef\journal#1,#2,#3,#4.{{\it #1~}{\bf #2} (#4) #3}
\def\prd{\journal Phys. Rev. D,}
\def\prl{\journal Phys. Rev. Lett.,}
\def\npb{\journal Nucl. Phys. B,}
\def\plb{\journal Phys. Lett. B,}
\def\apj{\journal Ap. J.,}
\def\apjl{\journal Ap. J. Lett.,}
\def\MNRAS{\journal MNRAS,}
\def\be{\begin{equation}}  \def\bea{\begin{eqnarray}}  \def\beaa{\begin{eqnarray*}}
\def\ee{\end{equation}}     \def\eea{\end{eqnarray}}     \def\eeaa{\end{eqnarray*}}
\def\double{\baselineskip 24pt \lineskip 10pt}
\def\re#1{{[\ref{#1}]}}
\def\fun#1#2{\lower3.6pt\vbox{\baselineskip0pt\lineskip.9pt
        \ialign{$\mathsurround=0pt#1\hfill##\hfil$\crcr#2\crcr\sim\crcr}}}
\def\half{{\textstyle{ 1\over 2}}}
\def\frac#1#2{{\textstyle{#1\over #2}}}
\def\gsim{\mathrel{\raise.3ex\hbox{$>$\kern-.75em\lower1ex\hbox{$\sim$}}}}
\def\lsim{\mathrel{\raise.3ex\hbox{$<$\kern-.75em\lower1ex\hbox{$\sim$}}}}
\def\la{\bigl\langle} \def\ra{\bigr\rangle}
\def\cd{\!\cdot\!}
\def\a{\hat a}      \def\b{\hat b}      \def\c{\hat c}
\def\ab{\a\cd\b}    \def\ac{\a\cd\c}    \def\bc{\b\cd\c}
\def\cg{\cos\gamma} \def\ca{\cos\alpha} \def\cb{\cos\beta}
\def\gt{\hat\gamma_2}  \def\gth{\hat\gamma_3}
\def\gf{\hat\gamma_4}
\def\got{ \hat\gamma_1\cd\hat\gamma_2} \def\ggo{ \hat\gamma\cd\hat\gamma_1}
\def\goth{\hat\gamma_1\cd\hat\gamma_3} \def\ggt{ \hat\gamma\cd\hat\gamma_2}
\def\gtth{\hat\gamma_2\cd\hat\gamma_3} \def\ggth{\hat\gamma\cd\hat\gamma_3}

\def\n{\hat n}       \def\no{\hat n_1}   \def\nt{\hat n_2}  \def\nth{\hat n_3}
\def\nont{\no\cd\nt} \def\nonth{\no\cd\nth} \def\ntnth{\nt\cd\nth}

\def\nogo{\no\cd\hat\gamma_1} \def\nogt{\no\cd\hat\gamma_2}
\def\nogth{\no\cd\hat\gamma_3}
\def\ntgo{\nt\cd\hat\gamma_1} \def\ntgt{\nt\cd\hat\gamma_2}
\def\ntgth{\nt\cd\hat\gamma_3}
\def\nthgo{\nth\cd\hat\gamma_1} \def\nthgt{\nth\cd\hat\gamma_2}
\def\nthgth{\nth\cd\hat\gamma_3}

\def\D{ {\Delta T \over T} }   \def\dO{d\Omega}
\def\d{ {\delta T \over T} }
\def\DN{ {\Delta N_\nu} }
\def\bg{\bar{g}_\ast}

\def\etal{{\sl et al.}}
\def\eg{{\sl e.g.}}
\def\ie{{\sl i.e.}}

\def\simlt{\stackrel{<}{{}_\sim}}
\def\simgt{\stackrel{>}{{}_\sim}}
\def\lesssim{\mathrel{\mathop
  {\hbox{\lower0.5ex\hbox{$\sim$}\kern-0.8em\lower-0.7ex\hbox{$<$}}}}}
\def\greatsim{\mathrel{\mathop
  {\hbox{\lower0.5ex\hbox{$\sim$}\kern-0.8em\lower-0.7ex\hbox{$>$}}}}}

\newcommand{\passo}{\addvspace{1.2cm}}

\newcommand{\gi}{g_i}
\newcommand{\ei}{E_i}
\newcommand{\mi}{m_i}
\newcommand{\nni}{n_i}
\newcommand{\bvec}{\mathbf}
\newcommand{\rr}{\rho}
\newcommand{\ri}{\rr_i}
\newcommand{\p}{{\mathrm p}}
\newcommand{\pri}{\p_i}
\newcommand{\ba}{\begin{array}}
\newcommand{\ea}{\end{array}}
\newcommand{\sgi}{S_i}
\newcommand{\si}{s_i}
\newcommand{\gs}{g_\ast}
\newcommand{\gss}{g_{\ast s}}
\newcommand{\td}{T^i_D}
\newcommand{\gapproxeq}{\lower .7ex\hbox{$\;\stackrel{\textstyle >}{\sim}\;$}}
\newcommand{\lapproxeq}{\lower .7ex\hbox{$\;\stackrel{\textstyle <}{\sim}\;$}}
\newcommand{\ov}{\overline}
\newcommand{\rt}{\rightarrow}
\def\nm{\hbox{$\nu_\mu \!$ }}

\def\epr{E^\prime}
\def\eps{\varepsilon}
\def\ep{\epsilon}
\def\al{\alpha}
\def\ga{\gamma}
\def\Ga{\Gamma}
\def\om{\omega}
\def\OM{\Omega}  
\def\la{\lambda}
\def\La{\Lambda} 
\def\al{\alpha}
\def\lpr{l^\prime}
\def\phpr{\phi^\prime}
\def\hpr{h^\prime}
\def\Apr{A^\prime}
\def\Cpr{C^\prime}
\def\hprdag{h^{\prime\dagger}}
\def\hdag{h^{\dagger}}
\def\barl{\bar{l}} 
\def\barphi{\bar{\phi}} 
\def\barlpr{\bar{l}^\prime} 
\def\barphpr{\bar{\phi}^\prime} 
\newcommand{\mat}[4]{\left(\begin{array}{cc}{#1}&{#2}\\{#3}&{#4}
\end{array}\right)}
\newcommand{\matr}[9]{\left(\begin{array}{ccc}{#1}&{#2}&{#3}\\
{#4}&{#5}&{#6}\\{#7}&{#8}&{#9}\end{array}\right)}
\newcommand{\matrr}[6]{\left(\begin{array}{cc}{#1}&{#2}\\
{#3}&{#4}\\{#5}&{#6}\end{array}\right)}



\thispagestyle{empty}

\vspace*{-2.5 cm}
\begin{center}
\begin{tabular}{c}
\centerline{\bf \Huge University of Rome ``Tor Vergata''} \\ \\ \hline \\
\centerline{\bf \Large Physics Department} \\
\end{tabular}
\end{center}

\vspace{1.4 cm}
\centerline{\large {\bf Ph.D. Thesis in Astronomy} }

\vspace{3.2 cm}
\begin{center}
{\Huge {\bf Cosmology of the \\[.2truecm] 
                     Mirror Universe }}

\end{center}
\vspace{0.8cm}
\centerline{\Large {\bf Paolo Ciarcelluti} }

\vspace{1.2cm}

\centerline{\large{April 2003}}
\vspace{4.2cm}

\begin{flushleft}
  {} \hfill {Thesis supervisor:}\\ [.4truecm]
  {} \hfill {\large \bf Prof. Zurab Berezhiani}\\
\end{flushleft}

\begin{center}

\vspace{2.8cm}

\begin{tabular}{c}
\hline \\
\centerline{\large Academic year 2001-2002}
\end{tabular}
\end{center}




\newpage

\thispagestyle{empty}
\mbox{}

\vspace{2cm}

\begin{flushright} 

{\bf {To my little dog Luna.}}

\vspace{2.2cm}

{\bf To all those who dream a better world...}

\vspace{.2cm}

{\bf To all those who fight for a better world.}

\end{flushright}




\newpage
\pagestyle{empty}

\noindent {\huge {\bf Acknowledgements}}
\addcontentsline{toc}{chapter}{{Acknowledgments}}

\vspace{2cm}

\noindent 
I would like to express my gratitude to prof. Roberto Buonanno and prof. Nicola Vittorio for helping me to follow my scientific interests and giving me useful advices.

I'm specially thankful to my supervisor prof. Zurab Berezhiani, for his 
invaluable help, patience and encouragement during my Ph.D. thesis work. 

I would like to thank my other collaborators during the PhD: Santi Cassisi, Denis Comelli and Francesco Villante. 
I'm also indebted with the professors Silvio Bonometto, Stefano Borgani, Alfonso Cavaliere and Andrei Doroshkevic for illuminating discussions, and Bruno Caccin for his encouragement.

I'm grateful to my friend Angelo for patiently improving my poor english, and my colleagues Valentina and Fabrizio for the enjoying times spent together in Rome.

Who missed the most by my endeavours is my little dog Luna: she always waited for my return and always was close to me during these years. Thanks for your fidelity and love, you are an example for all humans.

And last but not least, I would like to thank my family and friends in Pescara for standing by me in these years, with a special thank to Rita, who was always close to me when I needed.



\pagenumbering{roman}

\tableofcontents
\label{contents}
\markboth{CONTENTS}
                    {CONTENTS}



\newpage

\begin{center}
\large \bf Units and Conventions
\end{center}
\addcontentsline{toc}{chapter}{Units and Conventions}

\markboth{Units and Conventions}
                    {Units and Conventions}

In this thesis we study the Universe starting from its early phase and following its expansion till cosmological scales. Thus, discussions will span a wide variety of length scales, ranging from the Planck length 
up to the size of the Universe as a whole.

In any branch of physics the chosen units and conventions are almost always dictated by the problem at hand.
It is not hard to imagine then that also here disparate units come into the game. Astronomers will not always feel comfortable with those units employed by particle physicists, and viceversa, and so it is worthwhile to spend some words in order to fix notation.

We will employ the so-called {\sl natural units}, namely $\hbar = c = k_B = 1$, unless otherwise indicated. $\hbar$ is Planck's constant divided by $2\pi$, $c$ is the speed of light, and $k_B$ is Boltzmann's constant. Thus, all dimensions can be expressed in terms of one energy-unit, which is usually chosen as GeV $=10^9$eV, and so $$[{\rm Energy}] = [{\rm Mass}] = [{\rm Temperature}] = [{\rm Length}]^{-1}=[{\rm Time}]^{-1} ~.$$ Some conversion factors that will be useful in what follows are

\vspace*{0.5truecm}
\begin{tabular}{llll}
{}~~~~~~~~~~~~
&1 {\rm GeV} & = 1.60$\times 10^{-3}$ {\rm erg}       & ({\rm Energy}) \\
{}~~~~~~~~~~~~
&            &                                        &                \\
{}~~~~~~~~~~~~
&1 {\rm GeV} & = 1.16$\times 10^{13}$ {\rm K}    & ({\rm Temperature}) \\
{}~~~~~~~~~~~~
&            &                                        &                \\
{}~~~~~~~~~~~~
&1 {\rm GeV} & = 1.78$\times 10^{-24}$ {\rm g}        & ({\rm Mass})   \\
{}~~~~~~~~~~~~
&            &                                        &                \\
{}~~~~~~~~~~~~
&1 {\rm GeV}$^{-1}$ & = 1.97$\times 10^{-14}$ {\rm cm}  & ({\rm Length}) \\
{}~~~~~~~~~~~~
&                 &                                   &                \\
{}~~~~~~~~~~~~
&1 {\rm GeV}$^{-1}$ & = 6.58$\times 10^{-25}$ {\rm s} & ({\rm Time})   \\
\end{tabular}
\vspace*{0.5truecm}

The Planck mass $m_P$ and associated quantities are given (both in natural and cgs units) by
\bea
m_P &=& 1.22 \times  10^{19} {\rm GeV} = 2.18 \times 10^{-5 } {\rm g} \nonumber \\
l_P &=& 8.2  \times 10^{-20} {\rm GeV}^{-1} = 1.62 \times 10^{-33} {\rm cm} \nonumber \\
t_P &=& 5.39 \times 10^{-44}{\rm s} \nonumber 
\eea
In these units Newton's constant is given by
$G = 6.67\times 10^{-8} {\rm cm}^3 {\rm g}^{-1} {\rm sec}^{-2} = m_P^{-2}$.

These fundamental units will be replaced by other, more suitable, ones when studying issues on large-scale structure formation. In that case it is better to employ astronomical units, which for historical reasons are based on solar system quantities. Thus, we use the {\sl parsec} (the distance at which the Earth-Sun mean distance subtends one second of arc), or better the {\sl megaparsec}, which is more useful when dealing with structures on cosmological scales. Regarding masses, the standard unit is given by the solar mass $M_{\odot}$.
\bea
1 {\rm pc} &=& 3.26 ~{\rm light ~years} = 3.1\times 10^{18} {\rm cm} \nonumber \\
1 M_{\odot} &=& 1.99\times 10^{33} {\rm g} \nonumber \\
1 {\rm Mpc} &=& 3.1\times 10^{24} {\rm cm} \nonumber
\eea

It is worthwhile also recalling some cosmological parameters that will be used in this thesis, like 
the present Hubble time ($H_0^{-1}$) and distance ($c H_0^{-1}$), and the critical density $\rho_c$.
\bea
H_0^{-1} &=& 3.09\times 10^{17} h^{-1} {\rm ~s} \nonumber \\
c H_0^{-1} &=& 2997.9 h^{-1} {\rm ~Mpc} \nonumber \\
\rho_c &=& 3 H_0^2 / (8\pi G) = 1.88 h^2\times 10^{-29} {\rm ~g ~cm}^{-3} = 1.05 h^2 \times 10^4 {\rm ~eV ~cm}^{-3} \nonumber
\eea

We adopt the following conventions: Greek letters denote spacetime indices and run from 0 to 3; spatial indices run from 1 to 3 and are given from Latin letters. 

For any given quantity, the subscript zero ($_0$) obviously indicates the present value, while the prime indicates that it is referred to the mirror sector instead of the ordinary one (example: $\Omega_b$ is the density parameter of ordinary baryons, $\Omega'_b$ is the one of mirror baryons).

Below we list some symbols and abbreviations used in the text.
  
\vspace*{0.2truecm}
\vspace*{0.2truecm}
\begin{tabular}{lllll}
$\Omega$ &=& total ~density ~parameter \\
$\Omega_\lambda$ &=& {\rm density ~parameter ~of ~vacuum} \\
$\Omega_m$ &=& {\rm density ~parameter ~of ~matter} \\
$\Omega_b$ &=& {\rm density ~parameter ~of ~baryons} \\
$\omega_b$ &=& $\Omega_b ~h^2$ \\
\end{tabular}

\vspace*{0.2truecm}

\begin{tabular}{lllll}
O &=& ordinary \\
M &=& mirror \\
\end{tabular}

\vspace*{0.2truecm}

\begin{tabular}{lllll}
CMB &=& cosmic microwave background \\
LSS &=& large scale structure \\
CDM &=& cold dark matter \\
WDM &=& warm dark matter \\
HDM &=& hot dark matter \\
SIDM &=& self-interacting dark matter \\
MBDM &=& mirror baryon dark matter \\
\end{tabular}

\newpage

\vspace*{0.2truecm}

\begin{tabular}{lllll}
FRW &=& Friedmann--Robertson--Walker \\
BA &=& baryon asymmetry \\
BBN &=& Bib Bang nucleosynthesis \\
GUT &=& grand unified theory \\
EW &=& electroweak \\
SM &=& standard model \\
MSSM &=& minimal supersymmetric standard model \\
RD &=& radiation dominated \\
MD &=& matter dominated \\
MRE &=& matter--radiation equality \\
MRD &=& matter--radiation decoupling \\
\end{tabular}
\vspace*{0.5truecm}


\newpage

\clearpage
\pagenumbering{arabic}
\setcounter{page}{1}
\setcounter{chapter}{0}
\pagestyle{myheadings}



\def \chap-exp-univ{Standard cosmological paradigm}
\chapter{\chap-exp-univ}
\label{chap-exp-univ}
\markboth{Chapter \ref{chap-exp-univ}. ~ \chap-exp-univ}
                    {Chapter \ref{chap-exp-univ}. ~ \chap-exp-univ}


\passo
\def \sec-RW-metric{The expanding Universe}
\section{\sec-RW-metric}
\label{sec-RW-metric}
\markboth{Chapter \ref{chap-exp-univ}. ~ \chap-exp-univ}
                    {\S \ref{sec-RW-metric} ~ \sec-RW-metric}

On large scales the Universe is homogeneous and isotropic (to a good approximation), then it does not possess any privileged positions or directions: this idea is stated as the {\sl Cosmological Principle} (also known as the {\sl Copernican Principle}: no observer occupies a preferred position in the Universe).\footnote{There is a stronger formulation called the {\sl Perfect Cosmological Principle}, in which the Universe is the same not only in space, but also in time; this idea led to develop the {\sl Steady State} cosmology, theory that implies the continuous creation of matter to keep the density of the expanding Universe constant. This theory was abandoned for the difficulties encountered to explain the properties of the cosmic microwave background, radio sources and the primordial nucleosyntesis.} In effect, we know only the observable Universe (our present Hubble volume), but for purposes of description we may assume the entire Universe is homogeneous and isotropic.

The most general space-time metric describing a Universe in which the cosmological principle is obeyed is the maximally-symmetric {\sl Friedmann-Robertson-Walker metric} (FRW), written as
\be
\label{frw}
ds^2 = dt^2 - a^2(t) 
  \left[ {dr^2\over 1 - kr^2} + r^2 d\theta^2 + r^2 \sin^2\theta ~ d\phi^2 \right] \,,
\ee
\noindent where ($t$, $r$, $\theta$, $\phi$) are {\sl comoving} coordinates\footnote{ An important concept is that of the ``comoving observer''. Loosely speaking, a comoving observer follows the expansion of the Universe including the effects of any inhomogeneities that may be present.  }, $a(t)$ is the {\sl cosmic scale factor}, and, with an appropriate rescaling of the coordinates, {\sl k} can be chosen to be +1, $-$1, or 0 for spaces of constant positive, negative, or zero spatial curvature, respectively. The coordinate $r$ is dimensionless, while $a(t)$ has dimensions of length.

The time coordinate in eq.~(\ref{frw}) is just the {\sl proper} (or clock) time measured by an observer at rest in the comoving frame, i.e.~$(r, \theta, \phi)=$ const.; this coordinate system is called the {\sl synchronous gauge} (see appendix \ref{mabert3}). 

It is often convenient to express the metric in terms of {\sl conformal time} $\tau$, defined by $d\tau  = dt / a(t)$
\be
\label{conformaltime}
ds^2 = a^2(\tau) \left[ 
  d\tau^2 - {dr^2\over 1 - kr^2} - r^2 d\theta^2 - r^2 \sin^2\theta  ~d\phi^2 \right] \,.
\ee 


\def \sec-z-H{Redshift and Hubble law}
\subsubsection{\sec-z-H}
Consider the equation of geodesic motion (\ref{geodesic}) of a particle that is not necessarily massless. The four-velocity $u^\alpha $ of a particle with respect to the comoving frame is referred to as its {\sl peculiar} velocity. If we choose the $\alpha  = 0$ component of the geodesic equation, in the particular case of the FRW metric, we find that $u \propto a^{-1}$, and the magnitude of the momentum of a freely-propagating particle ``red shifts'' as: $|{\bvec p}| \propto a^{-1}$.

In this way we can see why the comoving frame is the natural frame. Consider an observer initially (at time $t$) moving non-relativistically with respect to the comoving frame with physical three velocity of magnitude $v$. At a later time $t_0$, the magnitude of the observer's physical three velocity $v_0$, will be
\be
\label{v_redsh}
v_0 = v {a(t) \over a(t_0)} \;.
\ee
\noindent Then, in an expanding Universe the free-falling observer is destined to come to rest in the comoving frame even if he has some initial velocity with respect to it (in this respect the term {\sl comoving} is well chosen). 

The light emitted by a distant object can be viewed quantum mechanically as freely-propagating photons. Since the wavelength of a photon is inversely proportional to its momentum ($\lambda=h/p$), that changes in proportion to $a^{-1}$, the wavelength at time $t_0$, denoted as $\lambda_0$, will differ from that at time $t$, denoted as $\lambda$, by
\be
\label{l_redsh}
{\lambda \over \lambda_0} = {a(t) \over a(t_0)} \;.
\ee
\noindent As the Universe expands, the wavelength of a freely-propagating photon increases, just as all physical distances increase with the expansion.

Hence, we introduce a new variable related to the scale factor $a$ which is more directly observable: the {\bf redshift} of an object, $z$, defined in terms of the ratio of the detected wavelength $\lambda _0$ to the emitted wavelength $\lambda$
\be
\label{redshift}
z = {{\lambda_0 -\lambda} \over \lambda} ~~~~~~~~~~~~~ 1+z = {a(t_0) \over a(t)} \;.
\ee
\noindent Any increase (decrease) in $a(t)$ leads to a red shift (blue shift) of the light from distant sources. Since today astronomers observe distant galaxies to have red shifted spectra, we can conclude that the Universe is expanding.

{\bf Hubble's law}, the linear relationship between the distance to an object and its observed red shift, can be expressed (assuming negligible peculiar velocity) as 
\be
z \approx Hd \approx 10^{-28} {\rm cm}^{-1} \times d \;,
\ee
\noindent where $d$ is the {\sl proper distance} \footnote{The proper distance of a point $P$ from another point $P_0$ is the distance measured by a chain of rulers held by observers which connect $P$ to $P_0$.} of a source, and $H$ is the {\sl Hubble constant} or, more accurately, the {\sl Hubble parameter} (because it is not constant in time, and in general varies as $t^{-1}$), defined by
\be
H = {\dot a(t) \over a(t)} \;.
\ee

Now, we can expand $a(t)$ in a Taylor series about the present epoch (\ie, for times $t$ close to $t_0$):
\be
\label{taylor1}
a(t) = a(t_0)\left[ 1 + H_0 (t-t_0) - {1 \over 2} ~ q_0 H_0^2 (t-t_0)^2 + ... \right] \,,
\ee
\noindent where
\be
\label{h0q0}
H_0 = {\dot a(t_0) \over a(t_0)} ~~~~ , ~~~~~~~~~ q_0 = {-\ddot a(t_0) \over \dot a^2(t_0)} ~ a(t_0) = {-\ddot a(t_0) \over a(t_0) H_0^2}
\ee
\noindent are the present values of the Hubble constant and the so-called {\sl deceleration parameter}. 

At present time the Hubble parameter $H_0$ is not known with great accuracy, so it is indicated by
\be
\label{consthubble}
H_0 = 100 ~h ~ {\rm km ~ s^{-1}Mpc^{-1}} ~~~~~~~~~~~~~ 0.5 \leq h \leq 0.8 \;,
\ee
\noindent where the dimensionless parameter $h$ contains all our ignorance on $H_0$.

The present age and the local spatial scale for the Universe are set by the {\sl 
Hubble time}\footnote{Note that earlier than some time, say $t_X$, or better for $a$ less than some $a_X$, our knowledge of the Universe is uncertain, so that the time elapsed from $a = 0$ to $a = a_X$ cannot be reliably calculated. However, this contribution to the age of the Universe is very small, and most of the time elapsed since $a = 0$ accumulated during the most recent few Hubble times.} and {\sl radius}
\be
H_0^{-1} \simeq 9.778 \times 10^9 h ^{-1} ~{\rm yr} \simeq 3000 ~h ^{-1} ~{\rm Mpc} \;. 
\ee


\passo
\def \sec-Friedmann{The Friedmann equations and the equation of state}
\subsubsection{\sec-Friedmann}
\label{sec-Friedmann}

The expansion of the Universe is determined by the Einstein equations
\be
G_{\alpha \beta } \equiv  R_{\alpha \beta } - {1\over 2} R g_{\alpha \beta } = {8\pi G \over c^4} T_{\alpha \beta } + \Lambda g_{\alpha \beta } \;,
\ee
where $R_{\alpha \beta}$ is the {\sl Ricci tensor}, $R$ is the {\sl Ricci scalar}, $g_{\alpha \beta}$ is the {\sl metric tensor}, $T_{\alpha \beta}$ is the {\sl energy-momentum tensor}, and $\Lambda$ is the {\sl cosmological constant}\footnote{The cosmological constant $\Lambda $ was introduced by Einstein for the need to have a static Universe. 
However, now we know that the Universe is expanding for sure. In fact, one of the biggest theoretical problems of the modern physics is to explain why the cosmological term is small and not order $ M_P^2 $.}
For the FRW metric (\ref{frw}), they 
are reduced to the form (see appendix \ref{gen_rel})
\be
\label{friedmann1}
{{\ddot a}\over a} = -{{4 \pi}\over 3}G {\left (\rho + 3p \right)}
\ee
for the time-time component, and
\be
\label{friedmann2}
{{\ddot a}\over a} + 2{\left( {\dot a} \over a \right)}^2 + 2{k \over a^2} = 4 \pi G {\left( \rho - p \right)}
\ee
for the space-space components, where, if the cosmological constant is present, $ p $ and $ \rho $ are modified according to eq.~(\ref{p_rho_wl}). From eqs.~(\ref{friedmann1}) and (\ref{friedmann2}) we obtain also
\be
\label{friedmann3}
{\left( {\dot a} \over a \right)}^2 + {k \over {a^2}} = {{8 \pi}\over 3}G \rho \;.
\ee
The equations (\ref{friedmann1}) and (\ref{friedmann3}) are the {\bf Friedmann equations}; they are not independent, because the second can be recovered from the first if one takes the adiabatic expansion of the Universe into account.

From eq.~(\ref{friedmann1}), models of the Universe made from fluids with $-1/3 < w < 1$ have $\ddot a$ always negative; then, because today $\dot a \geq  0$, they possess a point in time where $a$ vanishes and the density diverges; this instant is called the {\sl Big Bang singularity}. Note that the expansion of the Universe described in the Big Bang model is not due in any way to the effect of pressure, which always acts to decelerate the expansion, but is a result of the initial conditions describing a homogeneous and isotropic Universe. 

The Friedmann equation (\ref{friedmann3}) can be recast as
\be
\label{friedmann4}
{k \over {H^2 a^2}} = {\rho \over {3 H^2} / 8 \pi G} -1 \equiv  \Omega -1 \;,
\ee
where $\Omega $ is the ratio of the density to the {\sl critical density} $\rho _c$ necessary for closing the Universe \footnote{The present value of the critical density is $\rho_{0c} = 1.88 ~h^2 \times 10^{-29}$ g cm$^{-3}$, and taking into account the range of permitted values for $h$, this is $\sim (3 - 12) \times 10^{-27}$ kg m$^{-3}$, which in either case corresponds to a few H atoms/m$^3$. Just to compare, a `really good' vacuum (in the laboratory) of $10^{-9}$ N$/$m$^{2}$ at 300 K contains $\sim 2\times 10^{11}$ molecules/m$^3$. The Universe seems to be empty indeed!} 
\be
\label{omega}
\Omega \equiv  {\rho \over \rho _c} ~~~~~~,~~~~~~~~~~ \rho _c \equiv {{3 H^2} \over {8 \pi G}} \;.
\ee
Since $ H^2 a^2 \geq 0 $, there is a correspondence between the sign of $ k $, and the sign of ($ \Omega - 1 $)

  \begin{center}
CLOSED $\;\;\;\;\;\;\;\;\;\;\;\; k = +1\;\;\;\;\;\;\;\; \Longrightarrow \;\;\;\;\;\; \Omega > 1 $

FLAT $\;\;\;\;\;\;\;\;\;\;\;\;\;\;\;\;\; k = 0 \;\;\;\;\;\;\;\;\;\;\; \Longrightarrow \;\;\;\;\;\; \Omega = 1 $ 

OPEN $\;\;\;\;\;\;\;\;\;\;\;\;\;\;\;\; k = -1 \;\;\;\;\;\;\;\;\, \Longrightarrow \;\;\;\;\;\; \Omega < 1 $ 
  \end{center}

From equation (\ref{friedmann4}) we find for the scale factor today
\be
\label{a0}
a_0 \equiv H_0^{-1} \left( k \over {\Omega _0 -1} \right)^{1/2} \approx {{3000 ~h^{-1}~{\rm Mpc}}\over |\:{\Omega _0 -1}|^{1/2}} \;,
\ee
which can be interpreted as the current radius of curvature of the Universe. If $\Omega _0 = 1$, then $a_0$ has no physical meaning and can be chosen arbitrarily (it will always cancel out when some physical quantity is computed).

In order to derive the dynamical evolution of the scale factor $a(t)$, it is necessary to specify the {\bf equation of state} for the fluid, $p = p (\rho)$. It is standard to assume the form
\be
\label{eos}
p = w \rho 
\ee
\noindent and consider different types of components by choosing different values for $w$ (assumed to be constant).

If the Universe is filled with pressureless {\sl non--relativistic matter} (dust), we are in the case where $p \ll \rho$ and thus $w = 0$.
Instead, for radiation, the ideal {\sl relativistic gas} equation of state $p = 1/3 \rho$ is the most adapt, and therefore $w = 1/3$.
Another interesting equation of state is $p = - \rho$, corresponding to $w = -1$.
This is the case of {\sl vacuum energy} \footnote{
There is also a widely popular idea that the ``dark energy'' can be variable in time, related to the energy density of very slowly rolling scalar field (so-called ``quintessence''). This situation would imply $w$ in general different from -1, and it can be tested by high precision data on CMB and matter power spectra (see e.g refs.~\cite{bacmatvit2002,balmatvit2003}). However, in this thesis we do not consider this possibility and in our calculations assume the dark energy to be constant in time.
}, which will be the relevant form of energy during the so--called inflationary epoch; we will give a brief review of inflation in \S\ref{sec-inflat}.

The $\alpha = 0$ component of the conservation equation for the energy--momentum tensor, $T^{\alpha\beta}_{~~;\beta} = 0$, gives the 1st law of thermodynamics
\be
\label{FistLaw2}
\dot\rho + 3 H {\left( \rho + p \right)} = 0 \;,
\ee
where the second term corresponds to the dilution of $\rho$ due to the expansion ($H = \dot a / a$) and the third stands for the work done by the pressure of the fluid.

From equations (\ref{eos}) and (\ref{FistLaw2}) we can obtain the relation
\be
\label{conserv}
\rho a^{3(1+w)} = const \;.
\ee
In particular we have, for a {\sl dust} or {\sl matter Universe}
\be
\label{dustuniverse}
w = 0 ~~ \longrightarrow ~~ \rho_m a^3 = const ~~ \longrightarrow ~~  \rho_m \propto \rho_{0m}(1+z)^3 \;,
\ee
for a {\sl radiative Universe} \footnote{Consider for example photons: not only their density diminishes due to the growth of the volume ($\propto a^{-3}$), but the expansion also {\sl stretches} their wavelength {\sl out}, which corresponds to lowering their frequency, \ie, they {\sl redshift} (hence the additional factor $\propto a^{-1}$).}

\be
\label{radiativeuniverse}
w = 1/3 ~~ \longrightarrow ~~ \rho_r a^4 = const
 ~~ \longrightarrow ~~ \rho_r \propto \rho_{0r}(1+z)^4 \;,
\ee
and for a {\sl vacuum Universe} 
\be
\label{vacuumuniverse}
w = -1 ~~ \longrightarrow ~~ \rho_\lambda = \rho_{0 \lambda} = const \;.
\ee

Recalling the definition of the deceleration parameter $q$ from eq.~(\ref{h0q0}) and using eqs.~(\ref{friedmann1}) and (\ref{friedmann3}), we may express it as follows
\be
q = {3\over 2} \left({1\over 3} + w \right) \Omega \;.
\ee
This shows that $w = - {1 / 3}$ is a critical value, separating qualitatively different models. A period of evolution such that $w < - {1 / 3}$ ($q<0$)
is called {\sl inflationary}.
For the present value $q_0$, we find for a matter dominated (MD) model $q_0 = \Omega _0 / 2$, for a radiation dominated (RD) model $q_0 = \Omega _0$, and for a vacuum dominated model $q_0 = -\Omega _0$ (the expansion is accelerating, $\ddot a > 0$).

If we want to calculate the density parameter $\Omega $ at an arbitrary redshift $z$ as a function of the present density parameter $\Omega _0$, we can use the expressions (\ref{omega}) 
together with the equations (\ref{conserv}) and (\ref{friedmann3}) to obtain the relation
\be
\label{omegaz}
\Omega (z) = {{\Omega _0 (1+z)^{1+3w}} \over {(1 - \Omega _0) + \Omega _0 (1+z)^{1+3w}}} \;.
\ee
Note that if $\Omega _0 > 1$, then $\Omega (z) > 1$ for all $z$; likewise, if $\Omega _0 < 1$, then $\Omega (z) < 1$ for all $z$; on the other hand, if $\Omega _0 = 1$, then $\Omega (z) = 1$ for all time. The reason for this is clear: the expansion cannot change the sign of the curvature parameter $k$. It is also worth noting that, as $z \rightarrow \infty $, i.e., as we move closer and closer to the Big Bang, $\Omega (z)$ always tends towards unity: any Universe with $\Omega _0 \not= 1$ behaves like a flat model in the vicinity of the Big Bang. We shall come back to this later when we discuss the {\sl flatness problem} in \S\ref{sec-inflat}.


\passo
\def \sec-flatmod{Flat models}
\section{\sec-flatmod}
\label{sec-flatmod}
\markboth{Chapter \ref{chap-exp-univ}. ~ \chap-exp-univ}
                    {\S \ref{sec-flatmod} ~ \sec-flatmod}

From equations (\ref{friedmann3}) and (\ref{conserv}) we can obtain for $ k = 0 $ 
\be
\label{friedmann5}
{\left( \dot a \over a_0 \right)}^2 = H_0^2{\left[ \Omega_0 {\left( a_0 \over a \right)}^{1+3w} + {\left( 1 - \Omega_0 \right) } \right]} \;.
\ee
Now we shall find the solution to this equation appropriate to a {\sl flat Universe}. For $\Omega_0 = 1$, integrating equation (\ref{friedmann5}) one can obtain
\be
\label{aflat}
a(t) = a_0 {\left( t \over t_0 \right)}^{2/3 \left( 1+w \right)} \;,
\ee
which shows that the expansion of a flat Universe lasts an indefinite time into the future; equation (\ref{aflat}) is equivalent to the relation between cosmic time $t$ and redshift 
\be
\label{tflat}
t = t_0 {\left( 1+z \right)}^{-3 \left( 1+w \right)/2} \;.
\ee
From equations (\ref{aflat}), (\ref{tflat}) and (\ref{conserv}), we can derive
\bea
\label{Hflat}
H & \equiv & {\dot a \over a} = {2 \over 3 \left( 1+w \right)t} = H_0{t_0 \over t} = H_0 {\left( 1+z \right)}^{3 \left( 1+w \right)/2} ~  \;,\\
\label{qflat}
q & \equiv & -{{a \ddot a} \over \dot a^2} = {{1+3w} \over 2} = {\rm const.} = q_0 ~  \;,\\
\label{t0flat}
t_0 & = & {2 \over {3(1+w) H_0}} ~  \;,\\
\label{rhoflat}
\rho & = & \rho_0 { \left( t \over t_0 \right) }^{-2} = { 1 \over {6 (1+w)^2 \pi G t^2}} ~ \;.
\eea

\addvspace{0.8cm}
\noindent In appendix \ref{app-flat} we report the above relations for the special cases of a Universe dominated by matter or radiation. 
A general property of flat Universe models is that the scale factor $a$ grows indefinitely with time, with constant deceleration parameter $q_0$. The role of pressure can be illustrated again by the fact that increasing $w$ and, therefore, increasing the pressure causes the deceleration parameter also to increase.

A cosmological model in which the Universe is empty of matter and has a positive cosmological constant is called the {\sl de Sitter Universe}. From equations (\ref{p_rho_wl}) 
and (\ref{friedmann3}) we obtain
\be
\left( {\dot a} \over a \right)^2 = {8 \pi G \over 3} \rho_\Lambda \;, 
\ee
which for positive $\Lambda $ implies the exponentially fast expansion
\be
\label{a_exp}
a(t) = a_0 \, e^{H (t - t_0)} ~,~~~~~~~~~~~ H = {8 \pi G \over 3} \rho_\Lambda = {\Lambda \over 3} \;,
\ee
corresponding to a Hubble parameter constant in time. In the de Sitter vacuum Universe test particles move away from each other because of the repulsive gravitational effect of the positive cosmological constant. 

Finally, the age of flat Universe that contains both matter and positive vacuum energy ($\Omega = \Omega_{\rm m} + \Omega_\Lambda = 1$) is
\be
\label{t0_flat_vacuum}
t_0 = {2 \over 3 H_0} {1 \over \Omega _{\Lambda}^{1/2}} \ln \left[{1 + \Omega_{\Lambda}^{1/2}} \over { \left( 1 -\Omega_{\Lambda} \right)^{1/2}} \right] ~,~~~~~~ \Omega_\Lambda = {\rho_\Lambda \over \rho_{\rm c}}\;.
\ee

It is interesting to note that, unlike previous models, a model Universe with $\Omega _{\Lambda} \geq 0.74$ is older than $H_0^{-1}$; this occurs because the expansion rate is accelerating. In the limit $\Omega _{\Lambda} \rightarrow 1$, $t_0 \rightarrow \infty $. For this reason the problem of reconciling a young expansion age with other independent age determinations (like for example the globular clusters) has at several times led cosmologists to invoke a cosmological constant.


\passo
\def \sec-inflat{Inflation}
\section{\sec-inflat}
\label{sec-inflat}
\markboth{Chapter \ref{chap-exp-univ}. ~ \chap-exp-univ}
                    {\S \ref{sec-inflat} ~ \sec-inflat}

The {\sl inflation} is a microphysical mechanism which operates at very early times ($t \sim 10^{-34}$ s, well above the Planck time\footnote{The {\sl Planck time} $t_P \sim 10^{-43} s$ is the time after the Big Bang for which quantum fluctuations are no longer negligible and the theory of General Relativity should be modified in order to take account of quantum effects. One can then define {\sl Planck mass} $M_P \simeq \rho_P \; l_P^3 \simeq 1.22 \times 10^{19}$ GeV, where $l_P \simeq c \, t_P$ is the Planck length and $\rho_P \simeq (G \, t_P^2)^{-1}$ is the Planck density.}), according to which the Universe, at that stage dominated by vacuum energy and filled with the potential energy of a scalar field $\phi$ (called {\sl inflaton}), underwent a brief epoch of exponential expansion of the type (\ref{a_exp}). This field is initially displaced from the minimum of its potential, 
so it decays from the false vacuum to the true vacuum (lowest energy) state, and the system has to tunnel across the bump and then it slowly rolls down the potential. After reaching the minimum of the potential, it executes damped oscillations, during which energy is thermalised and entropy is increased ({\sl reheating}). From that point onwards the Universe is well described by the hot Big Bang model.

Due to the exponential expansion, primordial ripples in all forms of matter--energy perturbations at that time were enormously amplified and stretched to cosmological sizes and, after the action of gravity, became the large-scale structure that we see today, as firstly argued by Guth in 1981 \cite{guth23}. 

In a Universe dominated by a homogeneous scalar field $\phi$ with potential $V(\phi )$, minimally coupled to gravity,  the equation of motion is given by
\be
\label{LL52}
\ddot \phi + 3 H \dot\phi + { \partial V \over \partial\phi } = 0 \;.
\ee

The criterion that the potential must be sufficiently flat is quantified by the {\sl slow-rolling} conditions:
\begin{eqnarray}
\epsilon(\phi) &=& {{1}\over{16\pi G}} \left({{V'}\over{V}} \right)^2 \ll 1 ~\;,\\
|\eta(\phi)| &=& \left| {{1}\over{8\pi G}} {{V''}\over{V}} \right| \ll 1 ~\;,
\end{eqnarray}
where primes denote differentiation with respect to $\phi$.  

Due to the period of accelerated expansion of the Universe, all those perturbations present before inflation were rendered irrelevant for galaxy formation, since inflation washes out all initial inhomogeneities; otherwise, the same mechanism stretched the quantum mechanical fluctuations enough to produce {\sl scalar} (density) and {\sl tensor} (gravitational waves) perturbations on cosmologically interesting scales.

An important feature of the inflation is the ability to generate power spectrum of scalar and tensor perturbations as power laws with spectral index respectively \cite{lidd291,kolb50,stew302}
\be
\label{LL541}
n \equiv n_s = 1 + 2\eta - 6\epsilon ~~~~~~~~~~~~~~~~ n_t = - 2 \epsilon ~\;,
\ee
where $n=1$ represents scale invariance. Thus, we see that, according to the slow-rolling conditions, {\sl the inflation naturally produces flat spectra}, with $n=1$ and $n_t=0$.

The great success of the inflation is to solve some shortcomings present in Big Bang cosmology, namely the flatness, the horizon, and the unwanted relics problems.

{\bf The horizon problem.}
The CMB radiation is known to be isotropic with a very high precision. In fact, two microwave antennas pointed in opposite directions in the sky do collect thermal radiation with $\Delta T/T\leq 10^{-5}$, $T$ being the black body temperature. The problem is that these two regions from which the radiation of strikingly uniform temperature is emitted cannot have been in causal contact at the time of last scattering \cite{horizon}, since the causal horizon at that time subtends an apparent angle of order only 2 degrees today. How then causally disconnected spots of the sky got in agreement to produce the same temperature anisotropies?

The inflation can solve this problem: in fact, two regions initially causally connected are moved away one from another by the exponential inflationary expansion on the Universe itself and exit from their horizons, so that now we see regions of the sky disconnected at decoupling, but that were connected before inflation.

{\bf The flatness problem.}
From equation (\ref{omegaz}) we learned that the density parameter $\Omega$ rapidly evolves away from 1 while the Universe expands. Given that observations indicate that $\Omega$ is very close to 1 today, we conclude that it should have been much closer to 1 in the past. Going to the Planck time we get $|\Omega (10^{-43} {\rm s}) - 1| \sim {\cal O} (10^{-57})$, and even at the time of nucleosynthesis we get $|\Omega (1 {\rm s}) - 1| \sim {\cal O} (10^{-16})$. This means, for example, that if $\Omega $ at the Planck time was slightly greater than 1, say $\Omega (10^{-43} {\rm s}) = 1 + 10^{-55}$, the Universe would have collapsed millions of years ago. How the Universe was created with such fine-tuned closeness to $\Omega =1$?

Inflation can avoid this fine-tuning of initial conditions, because, whatever was the curvature of the local space-time, the huge inflationary expansion stretches all length, included the curvature radius, so that it becomes large enough that at the end of the inflation the Universe is as flat as we want.

{\bf The unwanted relics problem.}
Arising from phase transitions in the early Universe, there is an overproduction of unwanted relics. As an example, defects (e.g., magnetic monopoles) are produced at an abundance of about 1 per horizon volume (Kibble mechanism), yielding a present $\Omega_{\rm monop}$ far in excess of 1, clearly cosmologically intolerable. How can the cosmology rid the Universe of these overproduced relics?

Even now, the key is the huge expansion produced by an early inflationary era, which dilutes the abundance of (the otherwise overproduced) magnetic monopoles or any other cosmological ``pollutant'' relic.\footnote{This also implies, however, that any primordial baryon number density will be diluted away and therefore a sufficient temperature from reheating	as well as baryon--number and CP-violating interactions will be required after inflation (see Kolb \& Turner, 1990 \cite{kolbbookeu}).} 


\def \sec-thermodyn{Thermodynamics of the Universe}
\section{\sec-thermodyn}
\label{sec-thermodyn}
\markboth{Chapter \ref{chap-exp-univ}. ~ \chap-exp-univ}
                    {\S \ref{sec-thermodyn} ~ \sec-thermodyn}

In this section we will study the properties of the Universe considered as a thermodynamic system composed by different species (electrons, photons, neutrinos, nucleons, etc.) which, in the early phases, were to a good approximation in thermodynamic equilibrium, established through rapid interactions. Obviously, coming back to the past, decreasing the cosmic scale factor we have an increase of the temperature. 

Let us evaluate the total energy density $\rho$ and pressure $p$ of the cosmological fluid, considering all the particles in thermal equilibrium in the Universe, and express them in terms of photon temperature $T_\gamma \equiv T$ using the eqs.~(\ref{rho_equil2}) and (\ref{p_equil2})
\bea
  \rho & = & T^4 \, \sum_i \, \left( {T_i \over T} \right)^4 \, {g_i \over 2 \pi^2} \, \int_{x_i}^{\infty} \, 
{{(u_i^2 - x_i^2)^{1/2} \, u_i^2 \, d u_i} \over {\exp \left\{ u_i - y_i \right\} \, {\pm} \, 1}} ~~~,   \\
  p & = & T^4 \, \sum_i \, \left( {T_i \over T} \right)^4 \, {g_i \over 6 \pi^2} \, \int_{x_i}^{\infty} \, {{(u_i^2 - x_i^2)^{3/2} \, d u_i} \over {\exp \left\{ u_i - y_i \right\} \, {\pm} \, 1}} ~~~, 
\eea
where the sums run over all species $i$ (which may have a thermal distribution, but with a different temperature than that of the photons) and $u_i = \ei / T_i$, $x_i = \mi / T_i$, $y_i = \mu_i / T_i$. Note that, from eqs.~(\ref{n_nonrel}) and (\ref{rho_nonrel}), non relativistic particles contribute negligibly to the energy density in the radiation dominated era, since their energy density is exponentially suppressed with respect to the case of relativistic particles; thus we can neglect the contribution of non relativistic species in the sums above. Assuming all species non degenerate, we then get
\bea
 \rho & \simeq & {{\pi^2} \over {30}} \, \gs \, T^4  ~~~, 
\label{rho_univ} \\
 p & \simeq & {1 \over 3} \, \rr \; = \; {\pi^2 \over 90} \, \gs \, T^4  ~~~,
\label{p_univ}
\eea
where in $\gs$ contribute only the relativistic degrees of freedom
\begin{equation}\label{degfree}
  \gs \; \simeq \; \sum_B \, \gi \left( {T_i \over T} \right)^4 \; + \; {7 \over 8} \, \sum_F \, \gi \left( {T_i \over T} \right)^4
\end{equation}
($B$ = bosons, $F$ = fermions). Note that $\gs$ is in general a function of $T$, since the number of degrees of freedom becoming relativistic at a given temperature depends on $T$ itself. Moreover, at a given time, not all (relativistic) particles in the bath are in equilibrium at a common temperature $T$.\footnote{Some example: for $T \ll$ 1 MeV (the only relativistic species are the three neutrino species and the photon) $g_\ast = 3.36$; for $ T \gsim$ 1 MeV, up to 100 MeV, $g_\ast = 10.75$ (the electron and positron are additional relativistic degrees of freedom); for $T > 100$ GeV (all the species in the standard model -- 8 gluons, $W^\pm$, $Z$, $\gamma$, 3 generations of quarks and leptons, and 1 complex Higgs doublet -- should have been relativistic) $g_\ast = 106.75$.} A particle will be in kinetic equilibrium with the background plasma (that is $T_i = T$) only as long as its interaction with the plasma is fast enough; although the conditions for this to occur will be discussed in \S \ref{sec-therm-evol}, it is obvious that these involve a comparison between the particle interaction and the expansion rate $H$. 

Using definition (\ref{entropy_dens}) and eqs.~(\ref{rho_univ}), (\ref{p_univ}) we can evaluate the total entropy density in the comoving volume, obtaining
\begin{equation}\label{s_com}
  s \; = \; {2 \pi^2 \over 45} \, \gss \, T^3 \;,
\end{equation}
where
\begin{equation}\label{degfrees}
  \gss \; \simeq \; \sum_B \, \gi \left( T_i \over T \right)^3 \; + \; {7 \over 8} \, \sum_F \, \gi \left( T_i \over T \right)^3 ~~~
\end{equation}
and the sums again run only over the relativistic degrees of freedom in equilibrium (in the considered approximation). 
For most of the history of the Universe all particle species had a common temperature, and $g_\ast$ can be replaced by $g_{\ast s}$. Note that $s$ in eq. (\ref{s_com}) can be parametrized in terms of the photon number density, using the eq.~(\ref{n_rel_nondeg}), as follows
\begin{equation}\label{s_com_fot}
  s \; = \;  {{\pi^4} \over {45 \zeta(3)}} \, \gss \, n_\gamma \; \simeq \; 1.80 \, \gss \, n_\gamma \;,
\end{equation}
where $\zeta(x)$ is the Riemann zeta-function and $\zeta(3) \simeq 1.2021$. From entropy conservation we can now obtain the scaling law relating the cosmic scale factor with the temperature; in fact, from eqs.~(\ref{S_conserv}) and
(\ref{s_com}) we get
\begin{equation}\label{T-a_equil}
  T \; \sim \; \gss^{- \, 1/3} \, a^{-1} \;.
\end{equation}
We stress that only if $\gss$ is constant we can obtain the familiar result $T \propto a^{-1}$, valid for pure expansion. 


\subsubsection{Time - temperature relationship}

The useful relationship between the time $t$ and the background (photon) temperature $T$ in the radiation dominated era can be obtained straightforwardly by integrating the Friedmann equations by means of entropy conservation. In fact, we have
\begin{equation}
  t \; = \; \int_0^{a(t)} {1 \over H} \, {da \over a} \;.
\end{equation}
By using the equations (\ref{friedmann3}) (neglecting the curvature term in the radiation dominated era) and (\ref{rho_univ}), considering that $G \sim M_P^{-2}$, we obtain 
\begin{equation}\label{H_Mp}
  H \; \simeq \; \sqrt{{4 \pi^3} \over {45 M_P^2}} \, \gs^{1/2} \, T^2 ~ \simeq ~ 1.66 \; g_\ast^{1/2} {T^2 \over M_P} \;,
\end{equation}
from which, considering the entropy conservation, and during the periods when both $\gs$ and $\gss$ are approximately constant (i.e.~away from phase transitions and mass thresholds where relativistic degrees of freedom change), we find
\begin{equation}
  t \; \simeq \; 0.301 \; \gs^{- 1/2} \; {M_P \over T^2} \; \; \longrightarrow \; \; t({\rm sec}) \sim T^{-2}({\rm MeV}) \;.
\end{equation}


\passo
\def \sec-therm-evol{Thermal evolution and nucleosynthesis}
\section{\sec-therm-evol}
\label{sec-therm-evol}
\markboth{Chapter \ref{chap-exp-univ}. ~ \chap-exp-univ}
                    {\S \ref{sec-therm-evol} ~ \sec-therm-evol}

Sufficiently away from the Big Bang event, we can approximate the evolution of the Universe as made of several subsequent phases of (different) thermal equilibrium with temperature $T$ mainly varying as $a^{-1}$. Thermal equilibrium is realized if the reactions between the particles in the heat bath take place very rapidly compared with the expansion rate, set by $H$; so the key to understanding the thermal history of the Universe is the comparison of the particle interaction rates and the expansion rate. 

Denoting with $\Gamma = n \langle \sigma v \rangle$ the thermal average of the given scattering reaction rate ($n$ being the number density of target particles, $\sigma$ and $v$ the cross section and the particle velocity, respectively), the (approximate) condition for having equilibrium is  $\Gamma \; \gsim \; H$, and the decoupling temperature $T_D$ is defined by \footnote{The correct way of proceeding is to solve the Boltzmann transport equations for the given species, but here we are interested in a semi-quantitative discussion, and this criteria fit well.}
\begin{equation}\label{T_D}
  \Gamma(T_D) \; = \; H(T_D) \;.
\end{equation}
If we ignore the temperature variation of $g_\ast$, in the RD era the expansion rate is approximately, from eq.~(\ref{H_Mp})
\begin{equation}\label{H_Mp2}
  H \; \sim \; {{T^2} \over {M_P}} \;,
\end{equation}
while the scattering rate depends on the particular interactions experienced by the particles.


\vspace{0.2cm}
\subsubsection{Neutrino decoupling}

Massless neutrinos are maintained in equilibrium by reactions such as $\nu \ov{\nu} \leftrightarrow e^+ e^-$, $\nu e \leftrightarrow \nu e$ and so on, whose 
reaction rates are given by
\begin{equation}\label{gam_mass}
  \Gamma \; \sim \; G_F^2 T^5 \;, 
\end{equation}
where $G_F \simeq 1.17 \times 10^{-5}$ GeV$^{-2}$ is the Fermi coupling constant. From eqs.~(\ref{T_D}), (\ref{H_Mp2}), (\ref{gam_mass}) we then find the neutrino decoupling temperature 
\begin{equation}
  T_D^\nu \; \sim \; 1 \, {\rm MeV} \;.
\end{equation}
Thus, above $\sim 1 \, {\rm MeV}$ neutrinos are in equilibrium with the plasma of photons with $T_\nu = T$ and from eq.~(\ref{n_rel_nondeg}) $n_\nu = (3/4) n_\gamma$, afterwards they decouple from the plasma and their temperature $T_\nu$ scales approximately as $a^{-1}$. Subsequently, as the temperature drops below $\sim 0.5 \, {\rm MeV}$, $e^+$ and $e^-$ annihilate, heating the photons but not the decoupled neutrinos. 
Thus, from eq.~(\ref{T-a_equil}) with $\gs = \gss$ we find that $\gs (aT)^3 = const$, and so across $e^{\pm}$ annihilation $(aT_\nu)$ remains constant, while the $e^\pm$ entropy transfer increases $(aT_\gamma)$ by a factor 
\begin{equation}\label{T-Tnu}
  {T \over T_\nu} \; \simeq \; \left[ {{\gs(m_e \lapproxeq T \lapproxeq T_D^\nu)} \over {\gs(T \lapproxeq m_e)}} \right]^{{1 \over 3}} \;\simeq \;  \left( {11 \over 4} \right)^{{1 \over 3}} \; \simeq \; 1.40 \;.
\end{equation}

After $e^{\pm}$ annihilation there are no other relativistic species that can become non relativistic (altering the effective degrees of freedom), so that eq.~(\ref{T-Tnu}) is just the relation between the present values of  $\gamma$ and $\nu$ temperature (given $T = T_\gamma \simeq 2.73 \:{\rm K}$, $T_\nu \simeq 1.96 \:{\rm K}$).


\vspace{0.4cm}
\subsubsection{Matter - radiation equality}
\label{matrad_eq}

Remembering the equations (\ref{dustuniverse}) and (\ref{radiativeuniverse}), which relates matter and radiation densities to the redshift, we note that they scale with $a(t)$ in different ways, so that there must be a moment in the history of the expanding Universe when pressureless matter will dominate (if very early on it was radiation the dominant component, as is thought to be the case within the hot Big Bang). In fact, $\rho_{mat} / \rho_{rad} \propto a(t)$. This moment, known as the {\sl matter-radiation equality time} (MRE), is an important time scale in the thermal history of the Universe, and is usually denoted as $t_{eq}$
\be \label{matradt}
\rho_r (t_{eq}) = \rho_m (t_{eq}) \;.
\ee
Then it follows that the redshift, temperature and time  of equal matter and radiation energy densities are given by
\bea
 1 + z_{eq} & \equiv & { a_0 \over a_{eq} } = a_{eq}^{-1} 
  = {\Omega_{\rm m} \over \Omega_{\rm r}} = 2.32 \times 10^4 \, \Omega_{\rm m} h^2 \label{mreq} ~~,\\
 T_{eq} & = & T_0(1 + z_{eq}) = 5.50 \, \Omega_{\rm m} h^2 \; {\rm eV} ~~,\\
 t_{eq} & \simeq & {2 \over 3} H_0^{-1} \Omega_0^{-1/2} (1 + z_{eq})^{-3/2} \simeq \nonumber \\
    & \simeq & 1.4 \times 10^3 (\Omega_0 h^2)^{-1/2} (\Omega_{\rm m} h^2)^{-3/2} \; {\rm yr} ~~.
\eea


\vspace{0.4cm}
\subsubsection{Photon decoupling and recombination}

In the early Universe matter and radiation were in good thermal contact, because of rapid interactions between the photons and electrons. Since the photons continue to meet electrons, they do not propagate in straight paths for very long distances; in other words, the Universe is opaque to electromagnetic radiation.
However, when the electrons become slow enough, they are captured by the protons, forming stable hydrogen atoms. When this process, called {\sl recombination} \footnote{The term {\sl re}combination is slightly misleading, because the electrons have never been combined into atoms before.}, is completed, the photons find no more free electrons to scatter against, and they do not scatter against  the neutral hydrogen atoms; thus the matter becomes transparent to radiation, and the decoupled photons continue moving along in straight lines (or, more precisely, geodesics) from the {\sl last scattering surface}\footnote{The last scattering did not occur to all photons at the same time, so this surface is really a shell of thickness $\Delta z \simeq 0.07 z$.} until today. It is these photons, redshifted by the expansion of the Universe,  that are known as the {\sl cosmic microwave background radiation} (CMB) (for more details see chapter \ref{chap-cmbslow}).

This occurs when the interaction rate of the photons $\Gamma _\gamma$ 
becomes equal to the Hubble rate $H$, with
\be
\Gamma _\gamma = n_e \, \sigma _T \;,
\ee
\noindent where $n_e$ is the number density of free electrons, and $\sigma _T = 6.652 \cdot 10^{-25} \;{\rm cm^{2}}$ is the Thomson cross section.

For simplicity we assume all the baryons in the form of protons. The charge neutrality of the Universe implies $n_p = n_e$, and baryons conservation implies $n_b = n_p + n_H$. In thermal equilibrium, at temperatures less than $m_i$, where $i = e, p, H$, the equation (\ref{n_nonrel}) is valid for every species $i$, and in chemical equilibrium the process $p + e \rightarrow H + \gamma $ guarantees that $\mu _p + \mu _e = \mu _H$. By introducing the fractional ionization $X_e \equiv {n_p / n_b}$, 
we can obtain the {\sl Saha equation} for the equilibrium ionization fraction $X_e^{eq}$
\be
{{1 - X_e^{eq} \over (X_e^{eq})^2}} = {4 \sqrt{2} \zeta (3) \over \sqrt{\pi}} \; \eta {\left( T \over m_e \right)}^{3/2} \exp(B/T) \;,
\ee 
where $B \equiv m_p + m_e - m_H \simeq 13.6$ eV is the binding energy of hydrogen and $\eta = n_{\rm b} / n_\gamma $ is the {\sl baryon to photon ratio}. 

If we define recombination as the point when 90\% of the electrons have combined with protons, we find that it occurred at a redshift $z_{rec} \sim 1100$, with temperature \footnote{Note that recombination occurs at a temperature of about 0.3 eV, not at $T \sim B \sim 13.6$ eV, because the released binding energy reheats the remaining electrons, and the large amount of entropy in the Universe favours free photons and electrons.}
\be
T_{rec} = T_0 (1 + z_{rec}) \simeq 3000 \; {\rm K} \simeq 0.26 \; {\rm eV} \;.
\ee
If the Universe was matter dominated at decoupling, its age was
\be
t_{\rm dec} \simeq t_{\rm rec} = {2 \over 3} H_0^{-1} \; \Omega _0^{-1/2} \; (1+z_{rec})^{-3/2} \simeq 180000  \; (\Omega _0 \, h^2)^{-1/2} \; {\rm yr} \;. 
\ee


\vspace{0.4cm}
\begin{flushleft}
{\bf Primordial nucleosynthesis (BBN)} \footnote{Here we describe only few generalities on this phenomenon, referring the reader to \S \ref {nucleosyn} for a more quantitative discussion.}
\end{flushleft}

According to this theory, going back in time, we reach densities and temperatures high enough for the synthesis of the lightest elements: when the age of Universe was  between 0.01 sec and 3 minutes and its temperature was around 10 -- 0.1 MeV the synthesis of D, $^3$He, $^4$He, $^7$Li took place. The nuclear processes lead primarily to $^4$He, with a primordial mass fraction of about 24\%, while lesser amounts of the other elements are produced: about $10^{-5}$ of D and $^3$He, and about $10^{-10}$ of $^7$Li by number relative to H. The prediction of the cosmological abundance of these elements is one of the most useful probes of the standard hot Big Bang model (and certainly the earliest probe we can attain) \cite{paddybook,weinbookgc}.

The outcome of primordial nucleosynthesis is very sensitive to the baryon to photon ratio $\eta = n_b / n_{\gamma}$ and the number of very light particle species, usually quantified as the equivalent number of light neutrino species $N_{\nu }$ (=3 in the Standard Model physics). Then, the nucleosynthesis predictions can be compared with observational determinations\footnote{As the ejected remains of stellar nucleosynthesis alter the light element abundances from their primordial values, but also produce heavy elements (``metals''), one seeks astrophysical sites with low metal abundances, in order to measure light element abundances very close to primordial.} of the light elements abundances, obtaining
\begin{equation}
\label{etarange}
2.6 \times 10^{-10} \leq \eta \leq 6.2 \times 10^{-10} \;.
\end{equation}
This parameter provides a measure of the baryon content of the Universe. In fact, fixed $n_\gamma$ by the present CMB temperature (from eq.~(\ref{n_rel_nondeg}) $ n^0_\gamma = (2 \zeta(3) / {\pi^2}) \, T^3 \; \simeq \; 422 \, {\rm cm^{-3}} $), we find
\begin{equation}
\Omega _b h^2 \simeq 3.66 \times 10^7 \eta \;.
\end{equation}

The presence of additional neutrino flavors (or of any other relativistic species) at $T_W \sim 1$ MeV (the ``freeze-out'' temperature of weak interactions) increases $g_*$, hence the expansion rate, the value of $T_W$, the neutron to proton ratio, and therefore $Y_{BBN}$. Otherwise, the limits on $N_\nu$ can be translated into limits on other types of particles that would affect the Hubble expansion rate during BBN. In chapter \ref{chap-mirror_univ_1} we will use these concepts in the presence of a mirror sector.


\newpage

\passo
\def \sec-baryogen{Baryogenesis}
\section{\sec-baryogen}
\label{sec-baryogen}
\markboth{Chapter \ref{chap-exp-univ}. ~ \chap-exp-univ}
                    {\S \ref{sec-baryogen} ~ \sec-baryogen}

Today the Universe seems to be populated exclusively by matter rather than antimatter. There is in fact strong evidence against primary forms of antimatter in the Universe. Furthermore, as seen in the previous section, the density of baryons compared to the density of photons is extremely small, $\eta \sim 10^{-10}$.

It is well known that a non-zero baryon asymmetry (BA) can be produced in the initially baryon symmetric Universe if three following conditions are fulfilled: 
(i) B violation, (ii) C and CP violation, and (iii) departure from the thermal equilibrium \cite{sakha67}. 
The first two of these ingredients are expected to be contained in grand unified theories (GUT) as well as in the non-perturbative sector of the standard model.

Generally speaking, the baryogenesis scenarios can be divided in two categories, in which the out of equilibrium conditions are provided (a) by the Universe expansion itself, or (b) by fast phase transition and bubble nucleation.  In particular, the latter concerns the {\sl electroweak baryogenesis} schemes, while the former is typical for a {\sl GUT type baryogenesis or leptogenesis}. In \S \ref{baryogen} we will treat in more detail these two scenarios in both the standard and mirror sectors.

At present it is not possible to say which of the known mechanisms is responsible for the observed BA.  We only know that the baryon to photon number density ratio $\eta= n_{\rm b}/n_\gamma$ is restricted by the BBN constraint given by eq.~(\ref{etarange}). 


\passo
\def \sec-now-cosm{Present status of cosmology}
\section{\sec-now-cosm}
\label{sec-now-cosm}
\markboth{Chapter \ref{chap-exp-univ}. ~ \chap-exp-univ}
                    {\S \ref{sec-now-cosm} ~ \sec-now-cosm}

In the previous sections we gave a brief review of the so-called ``standard cosmological paradigm''. As we saw, there is some free parameter which describes the state and the evolution of the Universe; in this section we try to describe the present status in the determination of these parameters.

First of all, it is worthwhile to remember that during the last years cosmology seems to have reached a so-called {\sl concordance} \cite{tegmark1}, given that almost all parameters, as measured by sometimes very different methods, are converging to the same values. This is very important, as it hardly suggests that we are working in the right direction, but, at the same time, it confirms the existence of some trouble in the today's picture of the Universe, with a strong evidence of the need of new physics or new forms of matter, till unknown. In particular, the estimates of the cosmological parameters renews one of the most exciting problem in astrophysics, the {\sl dark matter} (see next chapter).

A powerful instrument to estimate {\sl all} the ``dynamical'' cosmological parameters is the joint analysis of CMB and LSS power spectra (see chapter \ref{chap-cmbslow}), as made for example by Percival et al.~\cite{perc327,percetal337}, Wang et al.~\cite{wang}, and Sievers et al.~\cite{siev}, all assuming that the initial seed fluctuations were adiabatic, Gaussian, and well described by power law spectra (for more details see next chapters). We will study the CMB and LSS for a mirror Universe in chapter \ref{chap-mirror_univ_3}.

\vspace{.2cm} 
\noindent {\bf Total density of the Universe.} 
From the first results of BOOMERANG experiment \cite{dber}, it seemed immediately clear that the Universe is approximately spatially {\sl flat}, as showed by the CMB anisotropies. Subsequently, other experiments (MAXIMA \cite{leea}, DASI \cite{halv568,pryk568}, VSA \cite{scot}, CBI \cite{maso0205384,pear0205388}, and again BOOMERANG \cite{nett}) confirmed and are still confirming this result, which, very interestingly, is in agreement with the inflationary prediction (see \S\ref{sec-inflat}). Using the result of ref. \cite{siev}, we have
\be \label{omega0}
\Omega _0 = 1.00 \pm ^{0.03}_{0.02} ~.
\ee

\vspace{.2cm} 
\noindent {\bf Baryonic density.} According to the standard BBN scenario, its value can be obtained in direct measurements of the abundances of the lights elements (see \S \ref{sec-therm-evol}), and its strongest constraint comes from the primordial deuterium abundance observed in Lyman-$\alpha $ feature in the absorption spectra of high redshift quasars, which in a recent analysis \cite{burl} gives
\be
\omega _{\rm b} (^2{\rm H}) = 0.020 \pm 0.001 ~,
\ee
while, from the cited indirect observations provided by the CMB and LSS analysis, we obtain
\be
\omega _{\rm b} ({\rm CMB}) = 0.022 \pm 0.002 \pm 0.001 ~, 
\ee
values which are in excellent agreement.

\vspace{.2cm} 
\noindent {\bf Matter density.} Combining the results for $\Omega_{\rm m}$ of the three cited joint compilations of CMB and LSS power spectra, we find
\be \label{omegam}
\Omega_{\rm m} = 0.29 \pm 0.05 \pm 0.04 ~.
\ee
This is in agreement with the constraint on the total density of clustered matter in the Universe which comes from the combination of {\sl x}-ray measurements of galaxy clusters with large hydrodynamic simulations.\footnote{
Using measurements of both the temperature and luminosity of the {\sl x}-rays coming from hot gas which dominates the baryon fraction in clusters, under the assumption of hydrostatic equilibrium,
one obtains the gravitational potential, and in particular the ratio of baryon to total mass of these systems. Employing the constraint on the cosmic baryon density coming from BBN, and assuming that galaxy clusters provide a good estimate of the total clustered mass in the Universe, one can then arrive at a range for the total mass density in the Universe \cite{erdogdu1,evrard,krauss2,white}.}  

\vspace{.2cm} 
\noindent {\bf Vacuum density.} Whatever it is, cosmological constant or quintessence,
the vacuum energy is today the dominant form of energy density of the Universe. In fact, given $\Omega _\Lambda = \Omega _0 - \Omega _{\rm m}$, combining the results (\ref{omega0}) and (\ref{omegam}), we obtain
\be
\Omega_\Lambda  = 0.71 \pm 0.07 ~.
\ee
Furthermore, a non-zero cosmological constant is favoured by the type-Ia supernovae measurements (Perlmutter et al, (1998) \cite{perl391}; Schmidt et al (1998) \cite{schm507}). 

\vspace{.2cm} 
\noindent {\bf Spectral index.} In the usual assumption of scalar adiabatic perturbations (see next chapter), which provide good fits with experimental data and are exactly what inflation predicts, tensorial spectral index is obviously $n_t = 0$, while for the scalar spectral index we obtain, from the cited joint compilations,
\be
n_s = 1.02 \pm 0.06 \pm 0.05 ~,
\ee
again in agreement with the inflationary theory, which predicts a scale invariant spectrum.

\vspace{.2cm} 
\noindent {\bf Hubble parameter.} The situation for the determination of $H_0$ is slightly different from the previous ones, as in this case the concordance of different methods is less evident. A great effort in this direction was made by the Hubble Space Telescope (HST) Key Project \cite{free}, which, using a calibration based on revised Cepheid distances, found the value of $H_0$ for five different methods (type Ia supernovae, Tully-Fisher relation, surface brightness fluctuations, type II supernovae, and fundamental plane), with a weighed average given by
\be
H_0 = 72 \pm 8 ~{\rm km}~{\rm s^{-1}Mpc^{-1}} ~.
\ee

However, significantly smaller values of $H_0$ are obtained by other methods, such as measurements of time delays in gravitationally lensed systems, or the Sunyaev-Zeldovich (SZE) effect in X-ray emitting galaxy clusters, which bypass the traditional ``distance ladder'' and probe far deeper distances than the objects used by the Key Project. Recently, Kochaneck used five well-constrained gravitational lenses, obtaining
\be
H_0 = 62 \pm 7 ~{\rm km}~{\rm s^{-1}Mpc^{-1}} ~.
\ee
Measurements of the SZE in 14 clusters also indicate a value of $H_0\sim60~{\rm km}~{\rm s}^{-1}~{\rm Mpc}^{-1}$ with presently a large ($\sim30\%$) systematic uncertainty \cite{sze}. Rowan-Robinson has argued that the Key Project data need to be corrected for local peculiar motions using a more sophisticated flow model than was actually used, and also for metallicity effects on the Cepheid calibration, then lowering the value of $H_0$ inferred from the same dataset to $63\pm6~{\rm km}~{\rm s}^{-1}~{\rm Mpc}^{-1}$ \cite{row}. Furthermore, it is worth noting that all joint analysys of CMB and LSS indicate a value of $H_0$ closer to the lower estimation.

\vspace{.8 cm}
The figure below shows an histogram which compares the amounts of the different components of the Universe. The second component from left is the dark matter (this argument will be treated in the next chapter), which together with the third component, the baryonic matter, constitutes the total matter content of the Universe. We can also note that the vacuum energy density is today comparable to (or better higher than) matter density.

At this point, the raw conclusion is very simple: the Universe is spatially flat, some 70\%  of the total energy content is dark, possibly in the form of a cosmological constant, and some 25\%  of gravitating matter is dark and unknown!

\begin{figure}[h]
  \begin{center}
    \leavevmode
    \epsfxsize = 8cm
    \epsffile{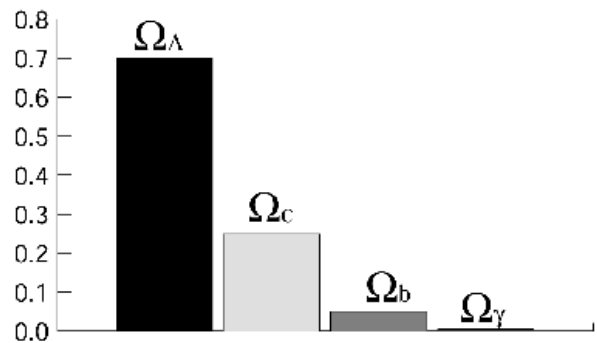}
  \end{center}
    \caption{\small Histogram of the composition of the Universe, given, from left to right, by: energy of the vacuum ($70\% \pm 10\%$), 
dark matter ($26\% \pm 10\%$), ordinary baryonic matter ($4\% \pm 1\%$), and      negligible energy density from photons.
About 13\% of the matter in the Universe is baryonic (${\Omega_{\rm b}}/{\Omega_{\rm m}} = {0.04}/{0.3} \simeq 0.13$). 
The baryons can be further divided into $3\%$ warm invisible gas, $0.5\%$ optically visible stars and $0.5\%$ hot gas visible in the {\sl x}-rays \cite{Fukugita}.}
\label{enerdens}
\end{figure}



\def \chap-lin-form{Cosmological structure formation and dark matter}
\chapter{\chap-lin-form}
\label{chap-lin-form}
\markboth{Chapter \ref{chap-lin-form}. ~ \chap-lin-form}
                    {Chapter \ref{chap-lin-form}. ~ \chap-lin-form}


\passo
\def \dark-matt{Dark matter}
\section{\dark-matt}
\label{dark-matt}
\markboth{Chapter \ref{chap-lin-form}. ~ \chap-lin-form}
                    {\S \ref{dark-matt} ~ \dark-matt}

With the term {\sl dark matter} cosmologists indicate an hypothetic material component of the Universe which does not emit directly electromagnetic radiation (unless it decays in particles having this property).

Historically, the first evidences of the existence of some form of dark matter comes from measurements of velocity dispersion of galaxies in clusters (firstly obtained by Zwicky in 1933 \cite{zwicky}) and the surprisingly flat rotation curves of spiral galaxies\footnote{Rotation curves of galaxies are characterized by a peak reached at distances of some Kpc and a behaviour typically flat for the regions at distance larger than that of the peak. A peculiarity is that the expected Keplerian fall is not observed. These rotation curves are consistent with extended halos containing masses till 10 times the galactic mass observed in the optical \cite{albada1}.}. What we deduce is that on galactic scales and above the mass density associated with luminous matter (stars, hydrogen, clouds, {\sl x}-ray gas in clusters, etc.) cannot account with the observed dynamics on those scales.

From figure \ref{enerdens} and from the parameter estimation of the \S\ref{sec-now-cosm} we clearly see the discrepancy existent between the total matter density and the baryon density. The conclusion is that most of the matter in the Universe is dominated by an unknown form of matter or by an unfamiliar class of dark astrophysical objects \cite{kormendy}.

There are essentially two ways in which matter in the Universe can be revealed: by means of radiation, by itself emitted, or by means of its gravitational interaction with baryonic matter which gives rise to cosmic structures. In the first case, electromagnetic radiation permits to reveal only baryonic matter. In the second case, we can only tell that we are in presence of matter that interacts by means of gravitation with the luminous mass in the Universe. 

The most widespread hypothesis is that dark matter is in form of collisionless particles, but we still do not exactly know neither the nature nor the masses of them. High energy physics theories, however, provide us with a whole ``zoo'' of candidates which are known as {\sl cosmic relics} or {\sl relic WIMPs} (Weakly Interacting Massive Particles). Typically, we distinguish between ``thermal'' and ``non-thermal'' relics. The former are kept in thermal equilibrium with the rest of the Universe until they decouple (a characteristic example is the massless neutrino), while the latter (such as axions, magnetic monopoles and cosmic strings) have been out of equilibrium throughout their lifetime. 
Thermal relics are further subdivided into three families.

\begin{itemize}
\item {\sl Hot Dark Matter} (HDM): ``hot'' relics which are still relativistic when they decouple. A typical hot thermal relic is a light neutrino with $m_{\nu}\simeq10~{\rm eV}$. 

\item {\sl Cold Dark Matter} (CDM): ``cold'' relics which go non-relativistic before decoupling.\footnote{In general non-thermal relics are also indicated as CDM.} The best motivated cold relic is the lightest supersymmetric partner of the standard model particles\footnote{ Supersymmetric particles were postulated in order to solve the strong CP problem in nuclear physics. This problem arises from the fact that some interactions violate the parity P, time inversion T, and CP.  If these are not eliminated, they give rise to a dipole momentum for the neutron which is in excess of ten order of magnitude with respect to experimental limits (Kolb and Turner, 1990 \cite{kolbbookeu}).}, identified in the ``neutralino'', with $m \geq 1~{\rm GeV}$.

\item {\sl Warm Dark Matter} (WDM): the intermediate case, given by thermal relics with masses around $1$~keV. Right-handed neutrinos, axinos and gravitinos have all been suggested as potential warm relic candidates.

\end{itemize}

The study of dark matter has as its finality the explanation of formation of galaxies and in general of cosmic structures. For this reason, in the last decades, the origin of cosmic structures has been ``framed" in models in which dark matter constitutes the skeleton of cosmic structures and supply the most part of the mass of which the same is made. In this chapter, after a brief review of linear perturbation theory (for more details see appendix \ref{app-strucform}), we will see some features of the most commonly used dark matter scenarios.


\passo
\def \intro_struct_form{Introduction to structure formation}
\section{\intro_struct_form}
\label{intro_struct_form}
\markboth{Chapter \ref{chap-lin-form}. ~ \chap-lin-form}
                    {\S \ref{intro_struct_form} ~ \intro_struct_form}

We know that the Universe was very smooth at early times (as recorded by the CMB) and it is very lumpy now (as observed locally). Cosmologists believe that the reason is {\sl gravitational instability}: small fluctuations in the density of the primeval cosmic fluid grew gravitationally into the galaxies, the clusters and the voids we observe today. However, there is the lack of the exact scenario, which is to say the lack of the exact composition of the Universe (``dark matter'' and ``dark energy'' problems). 

The idea of gravitational instability was first introduced in the early 1900s by Jeans, who showed that a homogeneous and isotropic fluid is unstable to small perturbations in its density~\cite{jean199,jeanbookac}. What Jeans demonstrated was that density inhomogeneities grow in time when the pressure support is weak compared to the gravitational pull.

The Newtonian theory,
the extension of Jeans theory to an expanding Universe, is only applicable to scales well within the Hubble radius, where the relativistic effects
are negligible, and, even in this context, one can only analyze density perturbations in the non-relativistic component. Perturbations on scales over the Hubble radius, or in the relativistic matter at all scales, require the full relativistic theory. 

Using these theories, if we want to construct a detailed scenario of structure formation, we need to know:
(i) the composition of the Universe; (ii) the contribution of its various components to the total density 
(namely $\Omega_{\rm b}$ from ordinary baryons, $\Omega_{\rm WIMP}$ from relic WIMPs, $\Omega_{\gamma}$ from relativistic particles, $\Omega_{\Lambda}$ from a potential cosmological constant, etc.); (iii) the spectrum and the type (adiabatic or isocurvature) of the primeval density perturbations.

\vspace{0.2 cm}
\noindent {\bf Adiabatic and isocurvature perturbations.} In general, when dealing with the pre-recombination plasma, we distinguish between two types of perturbations, namely between ``isoentropic" ({\sl adiabatic}) and ``entropic" ({\sl isocurvature} or {\sl isothermal}) modes (Zeldovich, 1967) \cite{Z}. 
Before recombination, a generic perturbation can be decomposed into a superposition of independently propagating adiabatic and entropic modes, but, after matter and radiation have decoupled, perturbations evolve in the same way regardless of their original nature. 

``Adiabatic'' modes contain fluctuations both in the matter and the radiation components, while keeping the entropy per baryon conserved. So, if we remember eq. (\ref{entrradbar}), defining, as usual, $\delta \equiv \left( {\delta\rho} / \rho \right)$, we find the condition for adiabaticity
\begin{equation}
{{\delta{\rm S}} \over {\rm S}} = 0 \,\Rightarrow\,\,\, 
  \left({3 \over 4} {{\delta\rho_r} \over {\rho_r}} - {{\delta\rho_{\rm b}} \over {\rho_{\rm b}}}\right) = 0 \,\Rightarrow\,\,\,
  \delta_{\rm b} = {3\over4}\delta_r \; . \label{adm}
\end{equation}
These perturbations are naturally generated in the simplest inflationary models through the vacuum fluctuation of the inflaton field (see Liddle \& Lyth (2000)~\cite{liddbookcilss}).

In the ``isocurvature'' modes we have 
\begin{equation}
\delta\rho=0 \,\Rightarrow\,\,\, 
  \rho_{\rm b}\delta_{\rm b} + \rho_{\gamma}\delta_{\gamma}=0 \,\Rightarrow\,\,\,
  {{\delta_{\gamma}} \over {\delta_{\rm b}}} = 
    - {{\rho_{\rm b}} \over {\rho_{\gamma}}} \;.  \label{iscm}
\end{equation}
This implies that the geometry of the 3-dimensional spatial hypersurfaces remains unaffected, hence the name isocurvature. 
Unlike adiabatic disturbances, isocurvature perturbations are usually absent from the simplest models of inflation. 

In this thesis we study only adiabatic perturbations, which are today the preferred perturbation modes, and we leave out the isocurvature modes, which could also have a contribution, but certainly  cannot be the dominant component \cite{enqvd62,enqvd65}.


\passo
\def \lin_newt_pert{Linear evolution of perturbations}
\section{\lin_newt_pert}
\label{lin_newt_pert}
\markboth{Chapter \ref{chap-lin-form}. ~ \chap-lin-form}
                    {\S \ref{lin_newt_pert} ~ \lin_newt_pert}

As long as the inhomogeneities are small, they can be studied by the linear perturbation theory. A great advantage of the linear regime is that the different perturbative modes evolve independently and therefore can be treated separately. In this respect, it is natural to divide the analysis of cosmological perturbations into two regimes. The early phase, when the perturbation is still outside the horizon, and the late time regime, when the mode is inside the Hubble radius. In the first case microphysical processes, such as pressure effects for example, are negligible and the evolution of the perturbation is basically kinematic. After the mode has entered the horizon, however, one can no longer disregard microphysics and damping effects.

The Newtonian treatment suffices on sub-horizon scales and as long as we deal with fluctuations in the non-relativistic component, while general relativity is necessary on scales outside the horizon and also when studying perturbations in the relativistic component. Here we remember only the principal results of these theories and some important concept, as the Jeans length, referring to appendix \ref{app-strucform} for a more quantitative discussion.

The complete set of the relativistic equations reveals three types of perturbations: {\sl tensor, vector and scalar modes}. Tensor perturbations correspond to the traceless, transverse part of $\delta g_{\alpha \beta}$. They describe gravitational waves and have no Newtonian analogue. Vector and scalar perturbations, on the other hand, have Newtonian counterparts. Vector modes correspond to rotational perturbations of the velocity field, while scalar modes are associated with longitudinal density fluctuations. We will only consider the latter type of perturbations.

As explained in appendix \ref{app-strucform}, the key scales for the study of the perturbation evolution are the {\bf Jeans length} $\lambda_{\rm J}$ and the related {\bf Jeans mass} $M_{\rm J}$
\begin{equation}
\lambda_{\rm J}=v_{\rm s}\sqrt{{\pi} \over {G\rho_0}} ~~~~~~~~~~~~~~~~
M_{\rm J}={4\over3}\pi\rho\left({\lambda_{\rm J}} \over {2}\right)^3 \;,
\end{equation}
where $v_{\rm s}$ is the {\sl adiabatic sound speed} defined in eq.~(\ref{soundspeed}), $\rho_0$ is the unperturbed density and $\rho$ is the density of the perturbed component. The Jeans scale constitutes a characteristic feature of the perturbation. It separates the gravitationally stable modes from the unstable ones: fluctuations on scales well beyond $\lambda_{\rm J}$ grow via gravitational instability, while modes with $\lambda\ll\lambda_{\rm J}$ are stabilized by pressure.

\begin{table}
\caption{Growth of perturbations.}
\begin{center}
\begin{tabular}{|c|c|c|c|} \hline \hline
  Epoch & $\delta _{\rm r}$ & $\delta_{\rm dm}$ & $\delta _{\rm b}$ \\
  \hline \hline
  $a < a_{\rm enter} < a_{\rm eq}$ ~~ ($ \lambda > \lambda _{\rm H}) $ & grows as $a^2$ & grows as $a^2$ & grows as $a^2$ \\
  \hline
  $a_{\rm enter} < a < a_{\rm eq}$ ~~ ($ \lambda < \lambda _{\rm H}) $ & oscillates & grows as ln a & oscillates\\
  \hline
  $a_{\rm eq} < a < a_{\rm dec}$ ~~~~ ($ \lambda < \lambda _{\rm H} $) & oscillates & grows as a & oscillates \\
  \hline
  $a_{\rm dec} < a$ ~~~~~~~~~~~~ ($ \lambda < \lambda _{\rm H} $) & oscillates & grows as a & grows as a \\
  \hline \hline
\end{tabular}
\end{center}
\label{tb2}
 \end{table}

\begin{figure}[h]
  \begin{center}
    \leavevmode
    \epsfxsize = 9cm
    \epsfysize = 6cm
      \epsffile{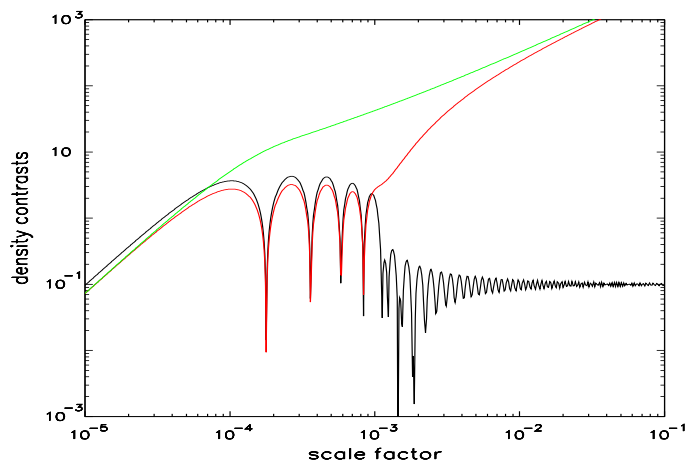}
  \end{center}
\caption{\small Evolution of the perturbation amplitudes on a scale $M \simeq 10^{15}~{\rm M}_{\odot}$ for the dark matter component (green line), baryons (red) and radiation (black).} 
\label{evol-std-1}
\end{figure}

Starting from the results obtained in appendix \ref{app-strucform}, we summarize the trends of density contrasts for radiation, dark matter and baryons in table \ref{tb2}, where $a_{\rm enter}$ indicates the epoch at which a given perturbation enters the Hubble radius $\lambda _{\rm H} $. In figure \ref{evol-std-1} we depict the perturbation evolution of the same components. 

Before the horizon entry fluctuations in all components grows as $a^2$. After this epoch baryons and photons enter in causal contact and, being tightly coupled, begin to oscillate until the moment of decoupling. Meanwhile, given that WIMPs interact with matter by means of gravity only, and do not feel the effect of pressure forces due to interaction with radiation, dark matter perturbations continue to grow, but whit a slower rate, as $ {\rm ln}\, a $ between horizon entry and matter-radiation equivalence and as $a$ after equivalence. Finally, after baryon to photon decoupling, photons continue to oscillate, while perturbations in the baryonic distribution will be driven by the gravitational potential of the collisionless species, and grow rapidly to soon equalize with those in the dark matter ({\sl dark matter boost}). Subsequently, perturbations in both components grow proportionally to the scale factor. Note that the presence of dark matter is necessary, because without it baryonic fluctuations do not have enough time to produce the observed structures compatibly with the extreme smoothness of the CMB temperature (see next chapter). In this context the key role played by dark matter is to start growing perturbations in its density earlier than those in the baryonic component; in chapter \ref{chap-mirror_univ_2} we will see that the same result could also be obtained with a particular self-collisional form of dark matter. 


\passo
\def \non_bar_struct_form{Non-baryonic structure formation}
\section{\non_bar_struct_form}
\label{non_bar_struct_form}
\markboth{Chapter \ref{chap-lin-form}. ~ \chap-lin-form}
                    {\S \ref{non_bar_struct_form} ~ \non_bar_struct_form}


\subsection{Evolution of the Jeans mass}

If the dark component is made up of weakly interacting species, the particles do not feel each other's presence via collisions. Each particle moves along a spacetime geodesic, while perturbations modify these geodesic orbits. 
We can treat the collisionless species as an ideal fluid, so the associated Jeans length is obtained similarly to the baryonic one. When dealing with a collisionless species, however, one needs to replace eqs.~(\ref{Neqc1}), (\ref{Neqc2}) with the Liouville equation (see Coles \& Lucchin (1995)~\cite{coles95}). Then,
\begin{equation}
\lambda_{\rm J}=v_{\rm dm}\sqrt{{{\pi} \over {G\rho}}} \;\;, \label{nbJl}
\end{equation}
where now $v_{\rm dm}$ is the velocity dispersion of the dark matter component. The corresponding Jeans mass is
\begin{equation}
M_{\rm J}={4\over3}\pi\rho_{\rm dm}\left({{\lambda_{\rm J}} \over {2}}\right)^3 \;,  \label{nbJm}
\end{equation}
where $\rho_{\rm dm}$ is the density of the non-baryonic matter.


\vspace{0.2 cm}
\noindent {\bf Hot thermal relics.} 
They decouple while they are still relativistic, that is, defining $t_{\rm nr}$ as the time when the particles become non-relativistic, $t_{\rm dec}<t_{\rm nr}$, where we assume that $t_{\rm nr}<t_{\rm eq}$. Throughout the relativistic regime $v_{\rm dm}\sim1$, $\rho\simeq\rho_{\gamma}\propto a^{-4}$ and $\rho_{\rm dm}\propto a^{-4}$, implying that $\lambda_{\rm J}^{({\rm h})}\propto a^2$ and $M_{\rm J}^{({\rm h})}\propto a^2$. Once the species have become non-relativistic and until matter-radiation equality, $v_{\rm dm}\propto a^{-1}$ (due to the redshifting of the momentum $p \propto a^{-1}$ when the particles are non-relativistic), $\rho\simeq\rho_{\gamma}\propto a^{-4}$ and $\rho_{\rm dm}\propto a^{-3}$. Recall that the particles have already decoupled, which means that $T_{\rm dm}\neq T_{\gamma}$. Consequently, $\lambda_{\rm J}^{({\rm h})}\propto a$ and $M_{\rm J}^{({\rm h})}={\rm const}$. After equipartition $v_{\rm dm}\propto a^{-1}$ and $\rho\simeq\rho_{\rm dm}\propto a^{-3}$, which translates into $\lambda_{\rm J}^{({\rm h})}\propto a^{1/2}$ and $M_{\rm J}^{({\rm h})}\propto a^{-3/2}$. Overall, the Jeans mass of hot thermal relics evolves as
\begin{equation}
M_{\rm J}^{({\rm h})}\propto\left\{\begin{array}{l}
a^2 \hspace{21,5mm} a<a_{\rm nr} \;,\\ 
{\rm constant} \hspace{10mm} a_{\rm nr}<a<a_{\rm eq} \;,\\
a^{-3/2} \hspace{16mm} a_{\rm eq}<a\;.
\end{array}\right.  \label{hJm}
\end{equation}
Clearly, $M_{\rm J}^{({\rm h})}$ reaches its maximum at $a_{\rm nr}$. In fact, the highest possible value corresponds to particles with $a_{\rm nr} = a_{\rm eq}$ such as neutrinos with $m_{\nu}\simeq10~{\rm eV}$. In this case $(M_{\rm
J}^{({\nu})})_{\rm max}\simeq3.5\times10^{15}(\Omega_{\nu}h^2)^{-2}~{\rm M}_{\odot}$ (see Coles \& Lucchin (1995)~\cite{coles95}). For a typical hot thermal relic $(M_{\rm J}^{({\rm h})})_{\rm max}\sim10^{12}-10^{14}~{\rm M}_{\odot}$.


\vspace{0.2 cm}
\noindent {\bf Cold thermal relics.} 
Cold thermal relics decouple when they are already non-relativistic (i.e.~$t_{\rm nr}<t_{\rm dec}<t_{\rm eq}$). Thus, for $t<t_{\rm nr}$ we have $v_{\rm dm}\sim1$, $\rho_{\rm dm}\propto a^{-4}$ and $\rho\simeq\rho_{\gamma}\propto a^{-4}$,
implying that $\lambda_{\rm J}^{({\rm c})}\propto a^2$ and $M_{\rm J}^{({\rm c})}\propto a^2$. In the interval between $t_{\rm nr}$ and $t_{\rm dec}$ the key variables evolve as $v_{\rm dm}\propto a^{-1/2}$ (recall that $T_{\rm dm}\simeq T_{\gamma}$ until $t_{\rm dec}$), $\rho_{\rm dm}\propto a^{-3}$ and
$\rho\simeq\rho_{\gamma}\propto a^{-4}$. As a result, $\lambda_{\rm J}^{({\rm c})}\propto a^{3/2}$ and $M_{\rm J}^{({\rm c})}\propto a^{3/2}$. After the particles have decoupled $T_{\rm dm}\neq T_{\gamma}$, which means that $v_{\rm dm}\propto a^{-1}$. At the same time $\rho_{\rm dm}\propto a^{-3}$ and $\rho\simeq\rho_{\gamma}\propto a^{-4}$, ensuring that $\lambda_{\rm J}^{({\rm c})}\propto a$ and $M_{\rm J}^{({\rm c})}={\rm constant}$. After equality $v_{\rm dm}\propto a^{-1}$ and $\rho\propto\rho_{\rm dm}\propto a^{-3}$, implying that $\lambda_{\rm J}^{({\rm c})}\propto a^{1/2}$ and $M_{\rm J}^{({\rm c})}\propto a^{-3/2}$. In short, the Jeans mass of cold thermal relics evolves as
\begin{equation}
M_{\rm J}^{({\rm c})}\propto\left\{\begin{array}{l}
a^2\hspace{21,5mm} z>z_{\rm nr} \;,\\
a^{3/2}\hspace{18,5mm}z_{\rm nr}>z>z_{\rm dec} \;,\\ 
{\rm constant}\hspace{10mm}z_{\rm dec}>z>z_{\rm eq} \;,\\
a^{-3/2}\hspace{16mm}z_{\rm eq}>z \;.
\end{array}\right.  \label{cJm}
\end{equation}
Accordingly, the maximum value for $M_{\rm J}^{({\rm c})}$ corresponds to species with $t_{\rm dec}=t_{\rm eq}$. In other words, the sooner the particles cease being relativistic and decouple, the smaller the associated maximum Jeans mass. Typically $(M_{\rm J}^{({\rm c})})_{\rm max}\ll10^{12}~{\rm M}_{\odot}$.


\subsection{Dissipative effects}
\label{diseffsec}

The ideal fluid approximation for collisionless species holds on sufficiently large scales only. On small scales, the free geodesic motion of the particles will wipe out any structure. This phenomenon, known as ``Landau damping" or ``free streaming", consists in the smoothing of inhomogeneities in the primordial Universe because of the motion of collisionless particles from overdense to underdense regions. 

Consider the coordinate (comoving) distance traveled by a free streaming particle
\begin{equation}
x_{\rm FS}=\int_0^t{{v_{\rm dm}} \over {a}}{\rm d}t \;, \label{fss}
\end{equation}
where $\lambda _{\rm FS}=ax_{\rm FS}$ is the corresponding physical (i.e.~proper) distance. Clearly, perturbations in the dark matter component on scales smaller than $\lambda _{\rm FS}$ will be wiped out by free streaming. Integrating the above during the three intervals $t<t_{\rm nr}$~, $t_{\rm nr}<t<t_{\rm eq}$ and $t>t_{\rm eq}$~, we find the total physical free-streaming scale 
\begin{equation}
\lambda _{\rm FS}=\left\{{{2t_{\rm nr}} \over {a_{\rm nr}}}\left[1 + \ln\left({{a_{\rm eq}} \over {a_{\rm nr}}}\right)\right] + {{3t_{\rm nr}} \over {a_{\rm nr}}}\left[1-\left({{a_{\rm eq}} \over {a}}\right)^{1/2}\right]\right\}a \;.  \label{pfss}
\end{equation}
At late times, when $a\gg a_{\rm eq}$, the above approaches its maximum value
\begin{equation}
\lambda _{\rm FS}\rightarrow \left(\lambda _{\rm FS}\right)_{\rm max}=
{{t_{\rm nr}} \over {a_{\rm nr}}}\left[5+2\ln\left({{a_{\rm eq}} \over {a_{\rm nr}}}\right)\right] \;.  \label{ltpfss}
\end{equation}
To obtain numerical estimates we need to identify the epoch the species become non-relativistic. Assuming that the transition takes place when $T\sim m_{\rm dm}/3$, we find (see Padmanabhan (1993)~\cite{padmbooksfu})
\begin{equation}
\left(\lambda _{\rm FS}\right)_{\rm max}\simeq0.5\left({{m_{\rm dm}} \over {1~{\rm keV}}}\right)^{-4/3}\left(\Omega_{\rm dm}h^2\right)^{1/3}~{\rm Mpc} \;, \label{mfss}
\end{equation}
where $m_{\rm dm}$ is the mass of the collisionless particles.
Accordingly, the minimum scale that survives collisionless dissipation depends crucially on the mass of the dark matter species. For neutrinos with $m_{\rm \nu}\simeq30$~eV we find $(\lambda _{\rm FS})_{\rm max} \simeq 28$~Mpc
and a corresponding mass scale $(M_{\rm FS})_{\rm max}\sim10^{15}~{\rm M}_{\odot}$.
For a much heavier candidate, say $m_{\rm dm}\simeq1$~keV, we find $(\lambda _{\rm FS})_{\rm max}\sim0.5$~Mpc and $(M_{\rm FS})_{\rm max}\sim10^9~{\rm M}_{\odot}$. {\sl In general, the lighter the dark matter species less power survives on small scales.}

Cold thermal relics and non-thermal relics have very small dispersion velocities. As a result, the maximum values of the Jeans mass and of the free streaming mass are very low. In this case, perturbations on all scales of cosmological interest grow unimpeded by damping processes, although they suffer stagnation due to the Meszaros effect until matter-radiation equality. After
recombination the potential wells of the collisionless species can boost the growth of perturbations in the baryonic component on scales of the order of 
$(M_{\rm J})_{\rm rec}\sim10^5~{\rm M}_{\odot}$.


\subsection{Evolution of the Hubble mass}
\label{subsec_evol_MH1}

We can build a useful quantity analogous to the Jeans mass by means of the Hubble radius $\lambda _{\rm H}$, and call it the {\sl Hubble mass}
\begin{equation}
M_{\rm H} = {4\over3}\pi\rho\left({{\lambda_{\rm H}} \over {2}}\right)^3\,, \label{nbHm}
\end{equation}
where $\rho$ is the energy density of the perturbed species. This is an additional important scale for structure formation; $\lambda_{\rm H}$ and $M_{\rm H}$ define the scale over which the different parts of a perturbation are in causal contact. Note that a mass scale $M$ is said to be entering the Hubble radius when $M=M_{H}$. 

For $t<t_{\rm nr}$ we have $\rho\propto a^{-4}$, $\lambda_{\rm H}\propto t\propto a^2$. During the interval $t_{\rm nr}<t<t_{\rm eq}$ we have $\rho\propto a^{-3}$ and $\lambda_{\rm H}\propto t\propto a^2$. Finally, when $t_{\rm eq}<t$, $\rho\propto a^{-3}$ and $\lambda_{\rm H}\propto t\propto a^{3/2}$. Overall, taking in mind that $a_{\rm nr}$ and $a_{\rm eq}$ depend on the species, the Hubble mass of every component evolves as
\begin{equation}
M_{\rm H}\propto\left\{\begin{array}{l}
a^2\hspace{17mm} a<a_{\rm nr} \;,\\
a^3\hspace{17mm}a_{\rm nr}<a<a_{\rm eq} \;,\\
a^{3/2}\hspace{14mm}a_{\rm eq}<a \;.
\end{array}\right.  \label{nbHme}
\end{equation}
Following definitions (\ref{nbJl}), (\ref{nbJm}) and (\ref{nbHm}), one can easily verify that the Jeans mass and the Hubble mass are effectively identical as long as the relic species are relativistic, namely $a<a_{\rm nr}$.


\subsection{Scenarios, successes and shortcomings}
\label{cdmscen}

The study of origin and formation of structures in the Universe has been historically fundamentally framed into the two HDM and CDM scenarios, according to what is the dominant form of dark matter. Although both theories have the same starting points (flat Universe fundamentally constituted by dark matter; small baryonic contribution to the mass of the Universe; primordial fluctuations adiabatic, scale invariant and Gaussian), structure formation in these scenarios is completely different since hot and cold relics are subject to different physical phenomena.

\begin{figure}[h]
  \begin{center}
    \leavevmode
    \epsfxsize = 8cm
    \epsfysize = 5.4cm
      \epsffile{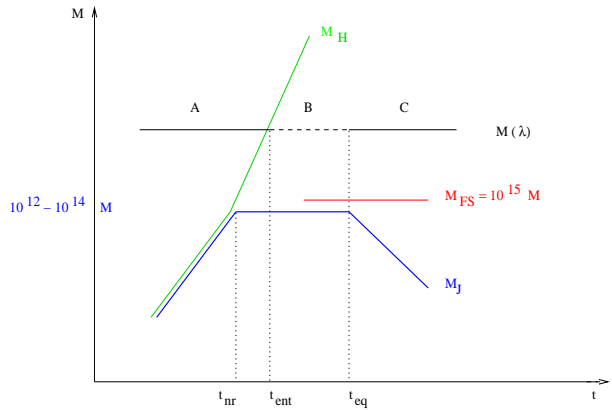}
  \end{center}
\caption{\small Evolution of the Hubble and Jeans mass for dark matter, with indicated a perturbed scale $M=M(\lambda)$, the three stages of the text and the neutrino free streaming mass.}
\label{fig2tsagas}
\end{figure}

As shown in figure \ref{fig2tsagas}, there are essentially three stages in the evolution of a mode which enters the Hubble radius between $a_{\rm nr}$ and $a_{\rm eq}$: (A) ($t<t_{\rm ent}<t_{\rm eq}$) the wavelength of the perturbation is bigger than the Hubble radius, and the mode grows as $a^2$; (B) ($t_{\rm ent}<t<t_{\rm eq}$) the wavelength is inside the Hubble radius and is bigger than $\lambda _{\rm J}$, so pressure support cannot stop the collapse, but the perturbation is frozen in due to the Meszaros effect; (C) ($t_{\rm eq} < t$) the wavelength is inside the Hubble radius and is bigger than $\lambda _{\rm J}$, the mode becomes unstable again and grows as $a$. Note that fluctuations with size smaller than $M_{\rm FS}$ are wiped out by neutrino free streaming.

\vspace{0.2 cm}
\noindent {\bf Hot dark matter models.} 
Keeping a light neutrino species with $m_{\nu}\sim10$~eV (which implies $a_{\rm nr} \simeq a_{\rm eq}$), the key feature of the perturbation spectrum (see chapter \ref{chap-cmbslow}) is the cutoff at $\lambda _{\rm FS}\simeq40$~Mpc due to the neutrino free streaming. Because of this, the first structures to form have sizes of approximately $10^{15}\,{\rm M}_{\odot}$, which corresponds to a supercluster of galaxies (see figure \ref{fig2tsagas}). Moreover, because the scale is very large, collapse must have occurred at relatively recent times (i.e.~at $z<3$). Thus, in a Universe dominated by hot thermal relics, structure formation proceeds in a ``top-down'' fashion. 
Perturbations on scales as large as $10^{15}\,{\rm M}_{\odot}$ go no linear in a highly non-spherical way. As a result, they collapse to one dimensional objects called ``pancakes'' (Zeldovich (1970)~\cite{Ze}). Once the pancake forms and goes non-linear in one of its dimensions, the baryons within start colliding with each other and dissipate their energy. Thereby, the baryonic component fragments and condenses into smaller galaxy-sized objects. 

These models have problems reproducing the small scale clustering properties of galaxies. In particular, the HDM simulations can agree with the observed galaxy-galaxy correlation function (see chapter \ref{chap-cmbslow}) only if the epoch of pancaking takes place at $z\simeq1$ or less. This seems too late to account for the existence of galaxies with redshift $z>1$ and of quasars with $z\simeq5$. 


\vspace{0.2 cm}
\noindent {\bf Cold dark matter models.} 
For such particles the maximum damping scale is too small ($\ll1$~Mpc) to be of any cosmological relevance, so that the CDM model has a spectrum without a cut-off at short wave-length (at least till scales much smaller than the galactic ones). Another important feature is the weak growth experienced by perturbations between horizon crossing and equipartition. It means that the density contrast increases as we move to smaller scales, or that the perturbation spectrum has more small-scale power. 
Thus, in standard CDM scenarios the first objects to break away from the background expansion have sub-galactic sizes ($<10^6~{\rm M}_{\odot}$). These structures virialize through violent relaxation (Lynden-Bell (1967)~\cite{lynd136}; Shu (1978)~\cite{shu225}) into gravitationally bound configurations that resemble galactic halos. At the same time the baryons can dissipate their energy and condense further into the cores of these objects. As larger and larger scales go non-linear, bigger structures form through tidal interactions and mergers. Hence, according to the CDM scenario structures form in a ``bottom-up'' or ``hierarchical'' fashion.


\vspace{0.2 cm}
\noindent {\bf Dark matter model problems.} 
Today the standard model consists of an Universe in which all the dark matter is made of CDM, and the remaining energy density needed to reach the critical density is provided by the dark energy: this is the so-called $\Lambda$CDM scenario. This model obtained noteworthy successes in the description of the characteristics of the Universe (clustering statistics of galaxies, peculiar velocities, CMB fluctuations) from the galactic scale on 
\cite{bar,blu,efs,fre,peeb263,whi}, 
but it has shown various discrepancies, when compared with data on smaller (subgalactic) scales, 
at least when any bias of the distribution of galaxies relative to the mass (see chapter \ref{chap-cmbslow}) is constant with scale \cite{babul1,bower1,popolo1,popolo2}. Some of these problems come from the following.

\begin{enumerate}  

\item \label{prob1} $N$-body CDM simulations give cuspy halos with divergent profiles toward the galactic center~\cite{NFW}, in 
disagreement with the galaxy rotation curves~\cite{binney} and with observations from gravitational lensing~\cite{flores}. In fact, in CDM halos we find $\rho(r) \sim r^{-\alpha}$ with $\alpha \sim 1$, so that the rotation velocity at the centers of galaxies should increase as $r^{1/2}$, but the data, especially that on dark-matter-dominated dwarf galaxies, instead showed a linear increase with radius, corresponding to roughly constant density in the centers of galaxies \cite{moore94}.

\item \label{prob3} One of the most striking features of halos in high resolution CDM simulations~\cite{Klypin:1999uc} is that they are heavily populated with small subhalos or subclumps, but their number is $1$--$2$ orders of magnitude more than what is observed~\cite{Moore:1999wf}, leading to an evident excess  of small scale structures~\cite{ss1985}. Even if there are mechanisms which inhibit star formation in small clumps and thereby leave them dark, a large population of clumps may heat the galactic disk and even endanger its stability \cite{MooreSub,TothOstriker,Weinberg}.

\item \label{prob4} The distribution of specific angular momentum in dark matter halos has a universal profile~\cite{bullock00}.  But if the baryons have the same angular momentum distribution as the dark matter, this implies that there is too much baryonic material with low angular momentum to form the observed rotationally supported exponential disks~\cite{bullock00,vdbosch01}.  

\item \label{prob2} Bar stability in high surface brightness spiral galaxies also demands low-density cores~\cite{debattista}. 

\end{enumerate}  

It is conceivable that these discrepancies are due to problems with current simulations, the quality of present observations or the omission of important astrophysical processes in the models. However, if any one of them persists, they may be an indication that the dark matter is not collisionless. 

Moreover, in the last years studies on predicted $\gamma$-ray emission from CDM annihilation in clumps are severely constraining the CDM particles parameters in the light of recent (EGRET) and future (GLAST) experiments, giving important indication on the reliability of present CDM candidates \cite{blasi1, taylor1}.


\vspace{0.2 cm}
\noindent {\bf Alternative options.} 
Issues of the $\Lambda$CDM model that have arisen on small scales
have prompted people to propose alternatives to CDM, such as ``warm dark matter'' (WDM) \cite{bode} and ``self-interacting dark matter'' (SIDM) \cite{sidm}. 

Warm dark matter particles have relatively high thermal velocities, 
so free streaming can suppress the formation of structure on small scales, while on larger scales the spectrum is the same as for the CDM case. 
Simulations \cite{colin00,bode} show that there are far fewer small satellite halos with $\Lambda$WDM than $\Lambda$CDM.  
Dark matter halos nevertheless have density profiles much like those in CDM \cite{huss,moore99,navstein01}. 
Hydrodynamical simulations also indicate that the disk angular momentum problem may be resolved with this suppression of small scale power \cite{sldolgov}.

While CDM assumes that the dark matter particles have only weak interactions with each other and with other particles, SIDM assumes that they have strong elastic scattering cross sections, but negligible annihilation or dissipation. In this way SIDM might suppress the formation of the dense central regions of dark matter halos.

Finally, another dark matter candidate is the so-called {\bf mirror dark matter}, which in principle could contain the virtues of all the previous CDM, WDM and SIDM. His study is just the aim of this thesis!



\def \chap-cmbslow{CMB, LSS and evolutionary equations}
\chapter{\chap-cmbslow}
\label{chap-cmbslow}
\markboth{Chapter \ref{chap-cmbslow}. ~ \chap-cmbslow}
                    {Chapter \ref{chap-cmbslow}. ~ \chap-cmbslow}


\def \cmb1{The cosmic microwave background (CMB)}
\section{\cmb1}
\label{cmb1}
\markboth{Chapter \ref{chap-cmbslow}. ~ \chap-cmbslow}
                    {\S \ref{cmb1} ~ \cmb1}

As explained in \S \ref{sec-therm-evol}, at recombination the photons decoupled from the baryons, and began to travel nearly unperturbed from the {\sl last scattering surface}\footnote{The sphere surrounding us at $z\simeq 1100$, which represents the position at which the CMB photons seen today last interacted directly with matter, is called the ``last scattering surface''.}
until the present day, 
where we observe them peaked in the microwave region of the spectrum as the {\sl cosmic microwave background radiation} (CMB). Thus, they provide a fossil record of our observable Universe when it was roughly $10^5$ years old and about $10^3$ times smaller than today.

The existence of this background of cold photons (with a present number density $n_{\gamma} \simeq 420~{\rm cm^{-3}}$), discovered 
in 1965 by Penzias and Wilson \cite{penwil65},  was predicted several years before by Gamow, Alpher and Herman \cite{alpher48,gamow48} as a consequence of the hot Big Bang theory.

The cosmic background radiation is the most perfect blackbody ever seen, according to the FIRAS (Far InfraRed Absolute Spectrometer) instrument \cite{fixsen96} of the Cosmic Background Explorer (COBE) satellite, which measured in the early 90's a temperature of $T_0 = 2.725 \pm 0.002$ K \cite{mather1}.

In addition to the FIRAS results, the other major result of COBE comes from the DMR (Differential Microwave Radiometer) instrument, which measured a dipole ($\Delta T_{\rm dip} = 3.353 \pm 0.024$~mK) associated with the 300 km/s flow of the Earth in the CMB frame, an intrinsic quadrupole ($\ell = 2$) amplitude $\Delta T_2 = 10.0^{+3.8}_{-2.8}~\mu$K, and provided the first unambiguous detection of anisotropies at a level $\Delta T/T \approx 10^{-5}$ on large angular scales  ($> 7^\circ$) \cite{smoot92}. This breakthrough immediately stimulated the realization of many new experiments aiming at measuring the CMB angular distribution with increasing resolution and sensitivity, revealing temperature anisotropies on smaller angular scales, which correspond to the physical scale of observed structures such as galaxies and clusters of galaxies.


\def \cmb4{Anisotropies and power spectrum}
\subsection{\cmb4}
\label{cmb4}
\markboth{Chapter \ref{chap-cmbslow}. ~ \chap-cmbslow}
                    {\S \ref{cmb4} ~ \cmb4}

Theories of the formation of large-scale structure predict the existence of slight inhomogeneities in the distribution of matter in the early Universe, which eventually underwent gravitational collapse to form galaxies, galaxy clusters, and superclusters.  These density inhomogeneities lead to temperature anisotropies in the CMB, because the radiation leaving a dense area of the last scattering surface is gravitationally redshifted to a lower apparent temperature and viceversa for an underdense region. 

Mapping the temperature on the celestial sphere, 
the temperature anisotropy at a point on the sky $(\theta,\phi)$ can be  expressed in the basis of spherical harmonics as 
\begin{equation}
{{\Delta T}\over{T}} (\theta, \phi) = \sum_{l=0}^{\infty} \sum_{m=-l}^{m=+l} a_{lm} Y_{lm} (\theta,\phi) \;,
\end{equation}
\noindent where $\ell \sim 180^\circ/\theta$, and $a_{\ell m}$ represent the multipole moments that, in the case of a Gaussian distribution of density perturbations (as suggested by inflationary models), should be characterized \cite{coles95} by
zero mean, $\langle a_{\ell m}\rangle=0$, and non-zero variance 
\be
C_\ell\equiv\left\langle |a_{\ell m}|^2\right\rangle
\ee 
(the angle brackets indicate an average over all observers in the Universe; the absence of a preferred direction implies that $\langle |a_{\ell m}|^2\rangle$ should be independent of $m$). The set of $C_\ell$ is known as the {\sl angular power spectrum}.
A cosmological model predicts the amplitude of the $a_{\ell m}$ coefficients,
which can be expressed in terms of $C_{\ell}$ alone in the hypothesis of gaussianity.

Temperature fluctuations in the CMB around a mean temperature in a direction $\alpha$ on the sky can be analyzed in terms of the {\sl correlation function} $C(\theta)$
\be
C(\theta)=\left\langle{\delta T\over T}(\alpha){\delta T\over T}
(\alpha+\theta)\right\rangle\ \;,
\ee
\noindent which measures the average product of temperatures in two directions separated by an angle $\theta$. For small angles $\theta$ the temperature correlation function can be expressed as a sum of Legendre polynomials $P_{\ell}(\theta)$ of order $\ell$, 
with coefficients or powers $a_{\ell}^2$,
\be
C(\theta)={1 \over 4\pi}\sum_{\ell=2}^{\infty}a_{\ell}^2(2\ell+1)P_{\ell}(\cos\theta)\ \;.
\ee
\noindent All analyses start with the quadrupole mode $\ell=2$ because the $\ell=0$ monopole mode is just the mean temperature over the observed part of the sky, and the $\ell=1$ mode is the dipole anisotropy due to the motion of Earth relative to the CMB. The higher the angular resolution, the more terms of high $\ell$ must be included.

Anisotropies on scales $0.1^\circ\lsim\theta\lsim 2^\circ$ are related to causal processes occurring in the photon-baryon fluid until recombination. Photons and baryons are in fact tightly coupled and behave like a single fluid. In the presence of gravitational potential forced acoustic oscillations in the fluid arise: they can be described by an harmonic oscillator where the driving forces are due to gravity, inertia of baryons and pressure from photons. 
Recombination is a nearly instantaneous process and modes of acoustic oscillations with different wavelength are ``frozen" at different phases of oscillation. The first (so-called Doppler) peak at degree scale in the power spectrum is therefore due to a wave that has a density maximum just at the time of last scattering; the secondary peaks at higher multipoles are higher harmonics of the principal oscillations and have oscillated more than once. 
A precise measurement of the acoustic peaks can reveal information on the cosmological parameters.

In the short but finite time taken for the Universe to recombine, the photons can diffuse a certain distance. Anisotropies on scales smaller than this mean free path will be erased by diffusion, leading to the quasi-exponential damping \cite{silk68,huwh97} in the spectrum at large $\ell$'s. This is called ``Silk damping'' (see chapter \ref{chap-mirror_univ_2}) and becomes quite effective at $\ell\gsim 1000$, corresponding to angular scales $\theta\lsim10'$. 

In all models the power spectrum contains the bulk of the statistical content in CMB maps. A statistically isotropic sky means that all the $(2 \ell + 1)$ $m$'s are equivalent and the power at each $\ell$ can be written as $ (2 \ell + 1) C_\ell / (4\pi) $. For an idealized full-sky observation, the variance of each measured $C_\ell$ is $ [2 / (2 \ell + 1)] C_\ell^2 $. This sampling variance (known as {\sl cosmic variance}), which comes about because each $C_\ell$ is $\chi ^2$ distributed with $(2 \ell + 1)$ degrees of freedom for our observable volume of the Universe \cite{whitess94}, sets the ultimate limit on the accuracy of our estimates of the power spectrum.\footnote{Of course similar considerations to those discussed above apply when an only patch of the sky is available, so temperature maps may have sampling variance too.}


\def \lss1{The large scale structure (LSS)}
\section{\lss1}
\label{lss1}
\markboth{Chapter \ref{chap-cmbslow}. ~ \chap-cmbslow}
                    {\S \ref{lss1} ~ \lss1}


\subsubsection{Fourier description and power spectrum}

Let us now turn to the distribution of matter in the Universe, where
naturally appear random fluctuations around the mean density $\bar{\rho}(t)$, manifested by compressions in some regions and rarefactions in other regions. 

It is generally assumed that this distribution is given by the superposition of plane waves, each with its characteristic wavelength $\lambda$ or comoving wave number $k$ and its amplitude $\delta_k$, that evolve independently, at least until they are in the linear regime (see \S \ref{lin_newt_pert}). Let we divide the Universe in cells of volume $V_{\rm U}$ (for example a cube of side $L$) and impose periodic conditions on the surfaces. 
We define, as usual, the density contrast as 
\be
\delta({\bf x}) = { \rho({\bf x}) - \bar{\rho} \over \bar{\rho} } \;,
\ee
and we assume this to be expressible as a Fourier expansion
\be
\delta({\bf x})=\sum_{{\bf k}}\ \delta_{{\bf k}}\ \exp (i{\bf  k} \cdot {\bf x})=
\sum_{\bf k}\ \delta^*_{\bf k}\ \exp(-i{\bf k}\cdot {\bf  x}) \;,
\label{fourierseries} 
\ee 
where $ k_{x,y,z} = {2 \pi n_{x,y,z}} / L $ and, for the periodicity condition, $ \delta(x,y,L) = \delta(x,y,0)$ (and similar conditions for the other components). The inverse relationship of the (\ref{fourierseries}) gives the Fourier coefficients $ \delta_{{\bf k}}$ 
\be  
\delta_{\bf k} = {1 \over V_{\rm U}}\int_{V_{\rm U}} \delta({\bf x}) \exp(-i{\bf k}\cdot {\bf x})d{\bf x} \;,
\ee 
which are complex quantities
\be 
\delta_{\bf k} = {\vert \delta_{\bf k} \vert} \exp{(i\phi_{\bf k})} \;.
\ee
The assumption of periodic boundaries results in a discrete ${\bf k}$-space representation.
Conservation of mass in $V_{\rm U}$ implies $\delta_{{\bf k}=0}=0$ and the reality of $\delta({\bf x})$ requires $\delta_{\bf k}^* = \delta_{-{\bf k}}$.

If we consider $n$ volumes $V_{\rm U}$, we have the problem of determining the distribution of Fourier coefficients $ \delta_{{\bf k}} $ and that of  $ \left|\delta\right| $. 
If we suppose that phases $\phi_{\bf k}$ are random, in the limit $ V_{\rm U} \rightarrow \infty $ it is possible to show that  we get $ \left|\delta\right|^{2} =\sum_{{\bf k}} \left|\delta_{{\bf k}}\right|^{2} $. The Central Limit theorem leads us to conclude that the distribution for $ \delta $ is Gaussian 
\begin{equation}
P ( \delta ) \propto \exp \left( -{{\delta^{2}} \over {2 \sigma^{2}}} \right) \;.
\label{eq:gau}
\end{equation}
The quantity $\sigma$ present in eq.~(\ref{eq:gau}) is the {\sl variance} of the density field, defined as
\begin{equation}
\sigma^{2} = \langle \delta ^{2} \rangle = \sum_{{\bf k}} \langle \left|\delta_{{\bf k}}\right|^{2} \rangle = {{1}\over{V_{u}}}\sum_{{\bf k}}  \delta_{k}^{2} \;.
\end{equation}
This quantity characterizes the amplitude of the inhomogeneity of the density field. If $ V_{\rm U} \rightarrow \infty$, we obtain the more usual relation
\begin{equation}
\sigma^{2} = {{1}\over{\left(2 \pi\right)^3}} \int P(k) d^{3} k = {{1}\over{2 \pi^{2}}} \int P(k) k^{2} d k \;.
\end{equation}
The term $ P( k ) = \langle \left|\delta_k\right|^{2} \rangle $ is called {\sl power spectrum} of perturbations. It is function only of $k$ because the ensemble average in an isotropic Universe depends only on $x$. 

With the lack of more accurate knowledge, one assumes for simplicity that the power spectrum of primordial gravitational fluctuations is specified by a power law
\be
P(k) = A\,k^{n_{\rm s}} \;,
\ee
\noindent where $n_s$ is the spectral index of scalar fluctuations and $A$ is the amplitude, which is expected to be equal on all scales. Inflationary models also predict that the power spectrum of matter fluctuations is almost scale-invariant as the fluctuations cross the Hubble radius. This is the Harrison--Zeldovich spectrum, for which $n_{\rm s}=1$.  


\subsubsection{Correlation function}

An important quantity connected with the spectrum is the {\sl two-points correlation function} $ \xi(r) $, which describes the real-space statistical properties of spatial density perturbations. It can be defined as the joint probability of finding an overdensity $\delta$ in two distinct points of space
\be
  \xi({\bf r})={\langle[\rho({\bf x})-\bar{\rho}] [\rho({\bf x}+{\bf r})-\bar{\rho}]\rangle\over \bar{\rho}^2}=\langle\delta({\bf x})\delta ({\bf x}+{\bf r})\rangle \;, 
\label{2_pt_def} 
\ee 
where averages are averages on an ensemble obtained from several realizations of Universe. Correlation function can be expressed as the joint probability of finding a galaxy in a volume $ \delta V_{1} $ and another in a volume $ \delta V_{2} $ separated by a distance $ r_{12} $
\begin{equation}
\delta ^{2} P = n_{V}^{2} [1+ \xi(r_{12} )] \delta V_{1} \delta V_{2} \,,
\end{equation}
where $ n_{V} $ is the average number of galaxies per unit volume. The concept of correlation function, given in this terms, can be enlarged to the case of three or more points. 

Correlation functions have a fundamental role in the study of clustering of matter. In order to show the relation between perturbation spectrum and two-points correlation function, we apply the Fourier machinery to equation (\ref{2_pt_def}) and arrive at the relation

\be
\xi({\bf r})= \sum_{\bf k} \langle\vert \delta_{\bf k}\vert^2\rangle \exp(-i{\bf k}\cdot {\bf r}) \;, 
\label{weiner_finite} 
\ee
which, in passing to the limit $V_{\rm U}\rightarrow\infty$, becomes 
\be
\xi({\bf r})={1 \over (2\pi)^3}\int P(k) \exp(-i{\bf k}\cdot {\bf r}) d{\bf k} \;. 
\label{weiner_cont} 
\ee 
This result shows that {\sl the two-point correlation function is the Fourier transform of the power spectrum} (this is the so-called {\it Wiener--Khintchin theorem}). This is similar to the situation in the context of CMB anisotropies, where the waves represented temperature fluctuations on the surface of the surrounding sky, and the powers $a_{\ell}^2$ were coefficients in the Legendre polynomial expansion.

In an isotropic Universe, it is $ |{\bf r}| = r $ and then $ | {\bf k} | =k $, and the spectrum can be obtained from an integral on $ |{\bf k}| = k $, so that correlation function may be written as
\begin{equation}
\xi( r ) = {{1}\over{2 \pi^{2}}} \int k^{ 2} P (k ) {{\sin( k r )}\over{k r}} d k \;.
\end{equation}
Averaging equation (\ref{weiner_finite}) over ${\bf r}$ gives
\be
\langle \xi({\bf r}) \rangle _{{\bf r}} = {1 \over V_{\rm U}} \sum_{\bf k} \langle\vert \delta_{\bf k}\vert ^2\rangle \int \exp(-i{\bf k}\cdot {\bf r}) d {\bf r} = 0 \;.
\ee

Phenomenological models of density fluctuations can be specified by the amplitudes $\delta_k$ of the correlation function $\xi(r)$. In particular, if the fluctuations are Gaussian, they are completely specified by the power spectrum $P(k)$. The models can then be compared to the real distribution of galaxies and galaxy clusters, and the phenomenological parameters determined.


\subsubsection{Transfer function}

During the evolution of the Universe and after perturbations enter the horizon, the spectrum is subject to modulations because of physical processes characteristic of the model itself (free streaming for acollisional components, Silk damping for collisional particles, etc.; see previous and next chapters). Amplification, stagnation or damping cause an evolution of the fluctuations different from one spatial scale to another. The combined effect of the various processes involved in changing the shape of the original power spectrum can be summarized in a single quantity, the {\sl transfer function} $ T(k;t) $,
which connects the primordial spectrum $ P( k; t_{p} )$ at time $t_p$ to the final spectrum at time $t_f$
\begin{equation}
P(k;t_{f}) = \left[{{b(t_{f})}\over{b(t_{p})}}\right]^{2}T^{2}(k;t_{f}) P( k;t_{p}) \;,
\end{equation}
where $b(t)$ is the law of growth of perturbations in the linear regime. In the absence of other physical effects, this spectrum would simply scale with time in accord with the linear growth law for each perturbation mode.

This processing of the primordial spectrum happens in a way that depends on cosmological parameters and the form of any non-baryonic dark matter, so that this function can be calculated for every model.


\subsubsection{Biasing factor}

Attempts to confront theories of cosmological structure formation with observations of galaxy clustering are complicated by the uncertain and possibly biased relationship between galaxies and the distribution of gravitating matter. 

The problem is that one observes the distribution of galaxies, that is of luminous objects, which are luminous in a very specific way\footnote{We might imagine that galaxies should form not randomly sprinkled around according to the local density of matter, but at specific locations where collapse, cooling and star formation can occur.}, rather than the distribution of the total matter, while there is no {\sl a priori} reason why the galaxy distribution should be a good tracer of the mass distribution of the Universe. Indeed, observations show that it definitely cannot be; the correlation functions for, to give an example, galaxies selected optically and galaxies selected in the infrared are different and hence clearly cannot both trace the mass distribution accurately. This effect is known as {\sl bias} in the galaxy distribution, and it seriously impairs our abilities to use it to constrain the matter spectrum \cite{dekels86,rees85,silk85}.

The literature contains at least three different definitions of the {\sl biasing factor} or {\sl bias parameter}, $b$, which are not equivalent to one another, but all represent the possible difference between mass statistics and the statistics of galaxy clustering. 
They are:
\be
  b^2 \equiv {(\sigma_8^2)_{galaxy} \over (\sigma_8^2)_{mass}} \;\;\;\;\; ; \;\;\;\;\;  
  \left({\delta \rho \over \rho} \right)_{galaxy} = b \left({\delta \rho \over \rho} \right)_{mass} \;\;\;\;\; ; \;\;\;\;\;
  \xi_{galaxy} = b^2 \xi_{mass} \;,
\ee
where $\sigma_8^2$ represents the dimensionless variance in either galaxy counts or mass in spheres of radius $8 h^{-1} {\rm Mpc}$.\footnote{This choice is motivated by the observational result that the variance of counts of galaxies in spheres of this size is of order unity, so that $b \simeq 1/\sigma_8$(mass).}
In general $b$ need not be a constant. However, it may well be adequately represented by a constant across some range of scales, and indeed there is observational evidence supporting this as long as we look to large enough scales, which in practice more or less means scales in the linear regime \cite{peacockd1}.

In practice, $b$ parametrizes our ignorance of galaxy formation in the same way as the mixing-length parameter does in the theory of stellar convection. To understand how this occurs we need to understand not only gravitational clustering, but also star formation and gas dynamics. All this complicated physics is supposed to be contained in the free parameter $b$.


\def \cmb6{Evolutionary equations}
\section{\cmb6}
\label{cmb6}
\markboth{Chapter \ref{chap-cmbslow}. ~ \chap-cmbslow}
                    {\S \ref{cmb6} ~ \cmb6}

To predict CMB anisotropies one has to solve the equations for the evolution of all particle species present (see e.g. \cite{bond84,peebles70, vittorio84}). 
In this thesis we base on the work done by Ma \& Bertschinger in 1995 \cite{mabert1}, where they studied evolutionary equations in two different gauges (syncronous and Newtonian conformal). In particular, we focus on the syncronous gauge, briefly describing in the following sections the equations used for our work.

Although the process of galaxy formation in recent epochs is well described by Newtonian gravity (and other physical processes such as hydrodynamics), a general relativistic treatment is required for perturbations on scales larger than the horizon size or before the horizon crossing time (see chapter \ref{chap-lin-form}).  The use of general relativity brought in the issue of gauge freedom, which has caused some confusion over the years. Lifshitz (1946) \cite{lifshitz1} adopted the ``synchronous gauge'' for his coordinate system, which has since become the most commonly used gauge for cosmological perturbation theories.\footnote{However, some complications associated with this gauge, such as the appearance of coordinate singularities and spurious gauge modes (see appendix \ref{mabert3}), prompted Bardeen (1980) \cite{bardeen1} and others (e.g. Kodama \& Sasaki, 1984 \cite{kosa1}) to formulate alternative approaches that deal only with gauge-invariant quantities.}

We consider only spatially flat background spacetimes with isentropic scalar metric perturbations (see \S \ref{intro_struct_form} and \S \ref{lin_newt_pert}).  
Here we give a complete discussion of CDM, baryons, photons and massless neutrinos in flat models and present the coupled, linearized Einstein, Boltzmann, and fluid equations for the metric and density perturbations.  

The CDM and the baryon components behave like collisionless and collisional fluids, respectively, while the photons and the neutrinos require a phase-space description governed by the Boltzmann transport equation. We also derive analytically the time dependence of the perturbations on scales larger than the horizon.  This information is needed in the initial conditions for the numerical integration of the evolution equations.

The photon and neutrino distribution functions are expanded in Legendre polynomials, reducing the linearized Boltzmann equation to a set of coupled ordinary differential equations for the expansion modes.


\subsection{Phase space and the Boltzmann equation}

A phase space is described by six variables: three positions $x^i$ and
their conjugate momenta $P_i$.\footnote{The conjugate momentum is related to the proper momentum $p^i$ measured by an observer at a fixed spatial coordinate value by $P_i = a(\delta_{ij}+\frac{1}{2}h_{ij}) p^j$.}
The phase space distribution of the particles gives their number in a differential volume $dx^1 dx^2 dx^3 dP_1 dP_2 dP_3$ in phase space
\begin{equation}
	f(x^i,P_j,\tau)dx^1 dx^2 dx^3 dP_1 dP_2 dP_3 = dN \; .
\end{equation}
Importantly, $f$ is a scalar and is invariant under canonical transformations.  The zeroth-order phase space distribution is the Fermi-Dirac distribution for fermions ($+$ sign) and the Bose-Einstein distribution for bosons ($-$ sign)
\begin{equation}
\label{equil-dist}
   f_0=f_0(\epsilon) = {g_s\over h_{\rm P}^3}{1\over e^{\epsilon/
     k_{\rm B} T_0} \pm 1} \;,
\end{equation}
where $\epsilon = a(p^2+m^2)^{1/2} =  (P^2 + a^2m^2)^{1/2}\,$, $T_0 = a T$
denotes the temperature of the particles today, the factor $g_s$ is the
number of spin degrees of freedom, and $h_{\rm P}$ and $k_{\rm B}$ are
the Planck and the Boltzmann constants.

Following common practice (e.g., Bond \& Szalay 1983 \cite{bondsza1}) we shall find it convenient to replace $P_j$ by $q_j\equiv ap_j$ in order to eliminate the metric perturbations from the definition of the momenta.  Moreover, we shall write the comoving 3-momentum $q_j$ in terms of its magnitude and direction: $q_j=qn_j$ where $n^in_i=\delta_{ij}n^in^j=1$.  Thus, we change our phase space variables, replacing $f(x^i,P_j,\tau)$ by $f(x^i,q,n_j,\tau)$. While this is not a canonical transformation (i.e., $q_i$ is not the momentum conjugate to $x^i$), it is perfectly valid provided that we correctly transform the momenta in Hamilton's equations.  Note that we do not transform $f$.  Because $q_j$ are not the conjugate momenta, $d^3xd^3q$ is not the phase space volume element, and $fd^3xd^3q$ is not the particle number.

In the perturbed case we shall continue to define $\epsilon$ as $a(\tau)$
times the proper energy measured by a comoving observer,  $\epsilon=
(q^2+a^2m^2)^{1/2}$.  
For the models we are interested in, the photons and the massless neutrinos at the time of neutrino decoupling are all ultra-relativistic particles, so $\epsilon$ in the unperturbed Fermi-Dirac and Bose-Einstein distributions can be simply replaced by the new variable $q$.

The general expression for the energy-momentum tensor written in terms of the distribution function and the 4-momentum components is given by
\begin{equation}
\label{tmunu}
	T_{\alpha\beta}= \int dP_1 dP_2 dP_3\,(-g)^{-1/2}\,
	{P_\alpha P_\beta\over P^0} f(x^i,P_j,\tau) \;,
\end{equation}
where $g$ denotes the determinant of $g_{\alpha\beta}$. It is convenient to write the phase space distribution as a zeroth-order distribution plus a perturbed piece in the new variables $q$ and $n_j$  
\begin{equation}
\label{f-pert}
	f(x^i,P_j,\tau) = f_0(q) \left[ 1 + \Psi(x^i,q,n_j,\tau) \right] \,.
\end{equation}

\noindent In the synchronous gauge, from equation (\ref{tmunu}) we obtain
to linear order in the perturbations
\begin{eqnarray}
\label{tmunu2}
	T^0{}_{\!0} &=& - a^{-4} \int q^2dq\,d\Omega\,
		\sqrt{q^2+m^2a^2}\,f_0(q)\,(1+\Psi) \;,\nonumber\\
	T^0{}_{\!i} &=& a^{-4} \int q^2dq\,d\Omega\,
		q\,n_i\,f_0(q)\,\Psi \;,\\
	T^i{}_{\!j} &=& a^{-4} \int q^2dqd\Omega
		\,{q^2 n_i n_j\over \sqrt{q^2+m^2a^2}}\,f_0(q)\,(1+\Psi) \;.
		\nonumber
\end{eqnarray}
The phase space distribution evolves according to the Boltzmann equation, which can be written to first order in $k$-space as
\begin{equation}
\label{bolt-syn}
  {\partial \Psi \over \partial \tau} + i\,{q\over\epsilon}
	({\bf k}\cdot \hat{n})\Psi +
     {d\ln f_0 \over d\ln q}\, \left[\dot{\eta} - {\dot{h}+6\dot{\eta}
       \over 2}(\hat{k}\cdot\hat{n})^2 \right] =
     {1\over f_0}\,\left( {\partial f \over \partial\tau} \right)_C \;,
\end{equation}
where 
\be
{Df \over d\tau} = {\partial f \over \partial \tau} + {dx^i \over d\tau}{\partial f\over \partial x^i} + {dq \over d\tau}{\partial f\over \partial q} + {dn_i \over d\tau}{\partial f\over \partial n_i} = \left( {\partial f \over \partial\tau} \right)_C \;.
\ee

The terms in the Boltzmann equation depend on the direction of the
momentum $\hat{n}$ only through its angle with ${\bf k}$.  
Therefore, we shall assume that the initial momentum-dependence is axially symmetric, so that $\Psi$ depends on ${\bf q}=q\hat{n}$ only through $q$ and $\hat{k}\cdot\hat{n}$. 


\subsection{Cold dark matter}

CDM interacts with other particles only through gravity and can be
treated as a pressureless perfect fluid.  The CDM particles can be
used to define the synchronous coordinates and therefore have zero
peculiar velocities in this gauge.  Setting $\theta=\sigma=0$ and
$w=\dot{w}=0$ in equations (\ref{fluid}) leads to
\begin{equation}
\label{cdm}
	\dot{\delta_c} = -{1 \over 2}\,\dot{h} ~~,~~~~~~ \dot{\theta_c} = 0 \;.
\end{equation}


\subsection{Massless neutrinos}

\label{sec:lessnu}
The energy density and the pressure for massless neutrinos 
are $\rho_\nu = 3p_\nu = -T^0{}_{\!0} = T^i{}_{\!i}$.  From equations (\ref{tmunu2}) the unperturbed energy density ${\rho}_\nu$ and pressure $p_\nu$ are given by
\begin{equation}
	 {\rho}_\nu = 3 \p_\nu = a^{-4}
	\int q^2dq d\Omega\,q f_0(q) \;,
\end{equation}
and the perturbations of energy density $\delta\rho_\nu$ and pressure $\delta
p_\nu$
are 
\begin{eqnarray}
     \delta\rho_\nu &=& 3 \delta p_\nu = a^{-4}
        \int q^2dq d\Omega\,q f_0(q) \Psi 
	\;.
\end{eqnarray}

The Boltzmann equation simplifies for massless particles, for which
$\epsilon=q$.  To reduce the number of variables we integrate out the
$q$-dependence in the neutrino distribution function and expand the
angular dependence of the perturbation in a series of Legendre polynomials
$P_l(\hat{k}\cdot\hat{n})$:
\begin{equation}
\label{fsubl}
      F_\nu({\bf k},\hat{n},\tau) \equiv {\int q^2 dq\,q f_0(q)\Psi
	\over \int q^2 dq\,q f_0(q)} \equiv \sum_{l=0}^\infty(-i)^l
	(2l+1)F_{\nu\,l}({\bf k},\tau)P_l(\hat{k}\cdot\hat{n})\,.
\end{equation}

In terms of the new variable $F_\nu({\bf k},\hat n,\tau)$ and its harmonic
expansion coefficients, the perturbations $\delta_\nu$, $\theta_\nu$, and $\sigma_\nu$, defined in eq.~(\ref{theta}), take the form
\begin{eqnarray}
\label{fsubnu}
      \delta_\nu &=& {1\over 4\pi} \int d\Omega
		F_\nu({\bf k},\hat{n},\tau)=F_{\nu\,0} \;,\nonumber\\
      \theta_\nu &=& {3i \over 16\pi} \int d\Omega\,
		({\bf k}\cdot\hat{n}) F_\nu({\bf k},\hat{n},\tau)
		={3\over 4} k F_{\nu\,1} \;, \\
      \sigma_\nu &=& -{3 \over 16\pi} \int d\Omega
	\left[(\hat{k}\cdot\hat{n})^2 - {1\over 3}\right]
	 F_\nu({\bf k},\hat{n},\tau)={1\over 2} F_{\nu\,2}\;.\nonumber
\end{eqnarray}

Integrating equation (\ref{bolt-syn}) over $q^2dq\,q f_0(q)$ and dividing it by $\int q^2dq\,q f_0(q)$, the Boltzmann equation for massless neutrinos becomes
\be\label{bolmn}
  {\partial F_\nu\over\partial\tau}+ik\mu F_\nu =
    -{2 \over 3}\dot h -{4 \over 3}(\dot h+6\dot\eta)P_2(\mu) \;,
\ee
where $\mu\equiv\hat k\cdot\hat n$ and $P_2(\mu)=(1/2)(3\mu^2-1)$ is the Legendre polynomial of degree 2.  Substituting the Legendre expansion for $F_\nu$ and using the orthonormality of the Legendre polynomials and the recursion relation $(l+1)P_{l+1}(\mu) = (2l+1) \mu P_l(\mu) -l P_{l-1}(\mu)$, we obtain the following equations:
\begin{eqnarray}
\label{massless}
	\dot{\delta}_\nu &=& -{4\over 3}\theta_\nu
		-{2\over 3}\dot{h} \;,\nonumber\\
	\dot{\theta}_\nu &=& k^2 \left(\frac{1}{4}\delta_\nu
		- \sigma_\nu \right) \;,\nonumber\\
	\dot{F}_{\nu\,2} &=& 2\dot\sigma_\nu = {8\over15}\theta_\nu
	    	- {3\over 5} k F_{\nu\,3} + {4\over15}\dot{h}
	  	+ {8\over5} \dot{\eta} \;,\nonumber\\
	\dot{F}_{\nu\,l} &=& {k\over2l+1}\left[ l
	 	F_{\nu\,(l-1)} - (l+1)
		F_{\nu\,(l+1)} \right]\,, \quad l \geq 3 \;.
\end{eqnarray}
This set of equations governs the evolution of the phase space distribution of massless neutrinos.  Note that a given mode $F_l$ is coupled only to the $(l-1)$ and $(l+1)$ neighboring modes. In this way the Boltzmann equation (\ref{bolmn}) has been transformed into an infinite hierarchy of moment equations that must be truncated at some maximum multipole order $l_{\rm max}$ (for more details see \S~\ref{mirror_mod}).


\subsection{Photons}

Photons evolve differently before and after recombination.  Before recombination, photons and baryons are tightly coupled, interacting mainly via Thomson scattering (and the electrostatic coupling of electrons and ions).  
After recombination, the Universe gradually becomes transparent to radiation and photons travel almost freely, although Thomson scattering continues to transfer energy and momentum between the photons and the matter.

The evolution of the photon distribution function can be treated in a similar way as the massless neutrinos, with the exception that the collisional terms on the right-hand side of the Boltzmann equation are now present and they depend on polarization.  
We shall track both the sum (total intensity) and difference (Stokes parameter $Q$) of the phase space densities in the two polarization states, denoted respectively by $F_\gamma({\bf k},\hat{n},\tau)$, defined as in equation (\ref{fsubl}), and $G_\gamma({\bf k},\hat{n},\tau)$.

Expanding $F_\gamma ({\bf k},\hat{n},\tau)$ and $G_\gamma({\bf k}, \hat{n},\tau)$ in Legendre series as in equation (\ref{fsubl}),
the collision operators can be rewritten as
\begin{equation}
\label{thom1}
	\left( {\partial F_\gamma \over \partial\tau} \right)_C
	    = a n_e \sigma_T \left[ {4i\over k}
	    (\theta_\gamma-\theta_b)P_1 + \left(9\sigma_\gamma - {1\over 2}
          G_{\gamma\,0} - {1\over 2} G_{\gamma\,2}\right) P_2
	    - \sum_{l\ge 3}^\infty (-i)^l(2l+1) F_{\gamma\,l} P_l \right] \;, \\
\end{equation}
\begin{equation}
\label{thom2}
	\left( {\partial G_\gamma \over \partial\tau} \right)_C
	    = a n_e \sigma_T \left[ {1\over 2} \left(
	    F_{\gamma\,2} + G_{\gamma\,0}+ G_{\gamma\,2}\right)(1-P_2)
	    - \sum_{l\ge 0 }^\infty (-i)^l(2l+1) G_{\gamma\,l} P_l \right] \;,
\end{equation}
where $n_e$ is the proper mean density of the electrons. The left-hand-side of the Boltzmann equation for $F_{\gamma}$ and $G_{\gamma}$ remain the same as for the massless neutrinos, so we obtain
\bea 
\label{photon}
     \dot{\delta}_\gamma &=& -{4\over 3}\theta_\gamma
	-{2\over 3}\dot{h} \;,\nonumber\\
     \dot{\theta}_\gamma &=& k^2 \left({1\over 4}\delta_\gamma
	- \sigma_\gamma\right)
	+ a n_e \sigma_T (\theta_b - \theta_\gamma) \;,\nonumber\\
     \dot{F}_{\gamma\,2} &=& 2\dot\sigma_\gamma={8\over15}\theta_\gamma
       - {3\over 5} k F_{\gamma\,3} + {4\over15}\dot{h}
       + {8\over5} \dot{\eta} - {9\over5} an_e \sigma_T \sigma_\gamma
       + {1\over10} an_e \sigma_T \left(G_{\gamma\,0}+G_{\gamma\,2}\right)
	\;, \nonumber\\
     \dot{F}_{\gamma\,l} &=& {k\over2l+1}\left[ l
	F_{\gamma\,(l-1)} - (l+1) F_{\gamma\,(l+1)}\right]
	- an_e \sigma_T F_{\gamma\,l} \;,\quad l \geq 3 \;,\nonumber\\
     \dot{G}_{\gamma\,l} &=& {k\over2l+1}\left[ l
        G_{\gamma\,(l-1)} - (l+1) G_{\gamma\,(l+1)}\right] + \nonumber\\
        &&+ an_e \sigma_T \left[ -G_{\gamma\,l} + {1\over2} \left(
	F_{\gamma\,2}+G_{\gamma\,0}+G_{\gamma\,2}\right)\left(
	\delta_{l0}+{\delta_{l2}\over5}\right)\right] \;.
\eea 

We truncate the photon Boltzmann equations in a manner similar to
massless neutrinos, 
except that Thomson opacity terms must be added.  For $l=l_{\rm max}$ we replace equations (\ref{photon}) by
\bea 
\label{truncphot}
     \dot{F}_{\gamma\,l} &=& k F_{\gamma\,(l-1)} - {l+1\over\tau}
        F_{\gamma\,l} - an_e \sigma_T F_{\gamma\,l} \;,\nonumber\\
     \dot{G}_{\gamma\,l} &=& k G_{\gamma\,(l-1)} - {l+1\over\tau}
        G_{\gamma\,l} - an_e \sigma_T G_{\gamma\,l} \;.
\eea 


\subsection{Baryons}

The baryons (and electrons) behave like a non-relativistic fluid described, in the absence of coupling to radiation, by the energy-momentum conservation equation (\ref{fluid}) with $\delta p_b/\delta\rho_b = c_s^2=w\ll1$ and $\sigma=0$.  
Before recombination, however, the coupling of the baryons and the photons causes a transfer of momentum and energy between the two components.

The momentum transfer into the photon component is represented by $an_e\sigma_T(\theta_b - \theta_\gamma)$ of equation (\ref{photon}). Momentum conservation in Thomson scattering then implies that a term $(4\rho_\gamma/3 \rho_b)\,an_e\sigma_T(\theta_\gamma - \theta_b)$ has to be added to the equation for $\dot{\theta}_b\,$ (where we have used ${p}_b \ll {\rho}_b$), so equation (\ref{fluid}) is modified to become 
\begin{eqnarray}
\label{baryon}
	\dot{\delta}_b &=& -\theta_b - {1\over 2}\dot{h} \;, \nonumber\\
	\dot{\theta}_b &=& -{\dot{a}\over a}\theta_b
	+ c_s^2 k^2\delta_b + {4\bar\rho_\gamma \over 3\bar\rho_b}
	 an_e\sigma_T (\theta_\gamma-\theta_b) \;.
\end{eqnarray}
The square of the baryon sound speed is evaluated from
\begin{equation}
\label{soundsp}
        c_s^2={\dot p_b\over\dot\rho_b}={k_{\rm B}T_b\over\mu}
	  \left(1-{1\over3}{d\ln T_b\over d\ln a}\right)\,,
\end{equation}
where $\mu$ is the mean molecular weight (including free electrons and all ions of H and He) and, in the second equality, we have neglected the slow time variation of $\mu$.\footnote{This approximation is adequate because even during recombination, when $\dot\mu$ is largest, the baryons contribute very little to the pressure of the photon-baryon fluid.}


\def \mabert6{Super-horizon-sized perturbations and initial conditions}
\subsection{\mabert6}
\label{mabert6}

The evolution equations derived in the previous sections can be solved numerically once the initial perturbations are specified.  We start the integration at early times when a given $k$-mode is still outside the horizon, i.e., $k\tau \ll 1$, where $k\tau$ is dimensionless. 
The behaviour of the density fluctuations on scales larger than the horizon is gauge-dependent. The fluctuations can appear as growing modes in one coordinate system and as constant modes in another. 

We are concerned only with the radiation-dominated era since the numerical integration for all the $k$-modes of interest will start in this era.  At this early time
the CDM and the baryons make a negligible contribution to the total energy density of the Universe: ${\rho}_{\rm total} = {\rho}_\nu + {\rho}_{\gamma}$.  The expansion rate is $\dot{a}/a=\tau^{-1}$. We can analytically extract the time-dependence of the metric and density perturbations $h$, $\eta$, $\delta$, and $\theta$ on super-horizon scales ($k\tau\ll1$) from equations (\ref{massless}) and (\ref{photon}).  The large Thomson damping terms in equations (\ref{photon}) drive the $l\ge 2$ moments of the photon distribution function $F_{\gamma\,l}$ and the polarization function $G_{\gamma\,l}$ to zero.  Similarly, $F_{\nu\,l}$ for $l\ge3$ can be ignored because they are smaller than $F_{\nu\,2}$ by successive powers of $k\tau$.  Equations (\ref{ein-syna}), (\ref{ein-sync}), (\ref{massless}), and (\ref{photon}) then give
\begin{eqnarray}
\label{press}
  && \tau^2\ddot{h} + \tau\dot{h} + 6 [(1-R_\nu)\delta_\gamma
	+R_\nu\delta_\nu] = 0 \;, \nonumber\\
  && \dot{\delta}_\gamma + {4\over 3}\theta_\gamma + {2\over 3}\dot{h}=0\,,
	\qquad \dot{\theta}_\gamma - {1\over 4}k^2 \delta_\gamma = 0
	\;,\nonumber\\
  && \dot{\delta}_\nu + {4\over 3}\theta_\nu + {2\over 3}\dot{h}=0\,,
   	\qquad \dot{\theta}_\nu - {1\over 4}k^2
	(\delta_\nu-4\sigma_\nu) = 0 \;, 
\end{eqnarray}
where we have defined $R_\nu \equiv \rho_\nu / (\rho_\gamma +\rho_\nu)$.  For $N_\nu$ flavors of neutrinos ($N_\nu=3$ in the standard model), after electron-positron pair annihilation and before the massive neutrinos become non relativistic, $\rho_\nu / \rho_\gamma=(7N_\nu/8)(4/11)^{4/3}$ is a constant.

To lowest order in $k\tau$, the terms $\propto k^2$ in equations (\ref{press}) can be dropped, and we have $\dot{\theta}_\nu=\dot {\theta}_\gamma =0$. Then these equations can be combined into a single fourth-order equation for $h$
\begin{equation}
   \tau {d^4h\over d\tau^4} + 5{d^3h\over d\tau^3}=0 \;,
\end{equation}
whose four solutions are power laws: $h \propto \tau^n$ with $n=0$, 1, 2, and $-2$.  From equations (\ref{press}) we also obtain
\begin{eqnarray}
\label{modes}
	h &=& A+B(k\tau)^{-2}+C(k\tau)^2 + D(k\tau) \;, \nonumber\\
	\delta\equiv(1-R_\nu)\delta_\gamma+R_\nu\delta_\nu &=&
		-{2\over 3}B(k\tau)^{-2} - {2\over 3}C(k\tau)^2
		- {1\over 6}D(k\tau) \;, \nonumber\\
	\theta\equiv(1-R_\nu)\theta_\gamma+R_\nu\theta_\nu &=&
		-{3\over 8}Dk \;,
\end{eqnarray}
and $A$, $B$, $C$, and $D$ are arbitrary dimensionless constants. The other metric perturbation $\eta$ can be found from equation (\ref{ein-syna})
\begin{equation}
	\eta = 2C + {3\over 4}D(k\tau)^{-1} \;.
\end{equation}

Press \& Vishniac (1980) \cite{pres239} derived a general expression for the time dependence of the four eigenmodes.  They showed that of these four modes, the first two (proportional to $A$ and $B$) are gauge modes that can be eliminated by a suitable coordinate transformation.  The latter two modes (proportional to $C$ and $D$) correspond to physical modes of density perturbations on scales larger than the Hubble distance in the radiation-dominated era.  Both physical modes appear as growing modes in the synchronous gauge, but the $C(k\tau)^2$ mode dominates at later times.  In fact, the mode proportional to $D$ in the radiation-dominated era decays in the matter-dominated era (Ratra 1988 \cite{ratra1}). We choose our initial conditions so that only the fastest-growing physical mode is present (this is appropriate for perturbations created in the early Universe), in which case $\theta_\gamma=\theta_\nu = \dot\eta = 0$ to lowest order in $k\tau$.  To get nonzero starting values we must use the full equations (\ref{press}) to obtain higher order terms for these variables.  To get the perturbations in the baryons we impose the condition of constant entropy per baryon. Using all of these inputs, we obtain the leading-order behaviour of super-horizon-sized perturbations in the synchronous gauge
\begin{eqnarray}
\label{super}
	&&\delta_\gamma = -{2\over 3}C (k\tau)^2 \;, \qquad
	  \delta_c = \delta_b = {3\over 4}\delta_\nu =
		{3\over 4}\delta_\gamma \;, \nonumber\\
	&&\theta_c=0 \;, \quad
	\theta_\gamma = \theta_b = -{1\over 18}C(k^4 \tau^3) \;,
	\quad \theta_\nu={23+4R_\nu \over 15+4R_\nu}
	\theta_\gamma \;, \nonumber\\
	&&\sigma_\nu= {4C \over 3(15+4R_\nu)} (k\tau)^2 \;, \nonumber\\
	&&h = C(k\tau)^2 \;, \qquad
	\eta = 2C - {5+4R_\nu\over 6(15+4R_\nu)}C(k\tau)^2 \;.
\end{eqnarray}



\def \mir_univ_1{The Mirror Universe}
\chapter{\mir_univ_1}
\label{chap-mirror_univ_1}
\markboth{Chapter \ref{chap-mirror_univ_1}. ~ \mir_univ_1}
                    {Chapter \ref{chap-mirror_univ_1}. ~ \mir_univ_1}


\passo
\def \intro_mirror{Introduction to the mirror world}
\section{\intro_mirror}
\label{intro_mirror}
\markboth{Chapter \ref{chap-mirror_univ_1}. ~ \mir_univ_1}
                    {\S \ref{intro_mirror} ~ \intro_mirror}

In 1956 Li and Yang \cite{li104} proposed that the interactions of the fundamental particles could be non invariant under mirror reflection of the coordinate system (Parity transformations). 
Subsequently experiments confirmed that the weak interactions indeed violate the Parity. In particular they are left-chiral, i.e. they have V$-$A form.

From the modern point of view, fundamental interactions (strong, weak and electromagnetic) are described by the Standard Model (SM) based on the gauge symmetry $ SU(3) \times SU(2) \times U(1) $. 
The fermion fields quarks and leptons are ascribed to the certain representation of this symmetry. 
The electroweak symmetry $ SU(2) \times U(1) $ is spontaneously broken down to electromagnetic $ U(1) _{\rm em} $ at the scale $ v \sim 100 $ GeV, and as a result, the gauge bosons $ W^{\pm} $ and $ Z $ acquire masses $\sim$ 100 GeV.
Parity violation in weak interactions is related to the fact that the left-handed components of quarks $ q_i = (u,d)_{iL} $ and leptons $ l_i = (\nu,e)_{iL} $ transform as doublets of $ SU(2) \times U(1) $, while the right-handed components $ u_{iR} $, $ d_{iR} $ and $ e_{iR} $ are the singlets ($ i = 1, 2, 3 $ is the family index). 
Thus, for particles one observes that their weak interactions have V$-$A form.

However, the fact that the observable matter is constituted by quarks and leptons and not by their antiparticles is a consequence of the baryon asymmetry of our Universe.
If instead the baryon asymmetry would have the opposite sign, then the observed Universe would be made of antiparticles, namely anti-quarks $ \bar q_i = C \ove {q_i}^T $ and anti-leptons $ \bar l_i = C \ove {l_i}^T $, which are now right-handed, 
and so we would observe the V$+$A form in their weak interactions, i.e. our world would be the right-handed.\footnote{
In fact, one cannot exclude the possibility that in some patch of the Universe, very far from us, the baryon asymmetry has the opposite sign and thus antiparticles rather than particles are dominant.}

The fact that we have particle excess over antiparticles is probably related to the CP violating features of the gauge and particle sector of our world. So, in the context of the full theory (Grand Unification or seesaw model of neutrino masses or Supersymmetric Standard Model) the sign of the baryon asymmetry of the Universe is defined by the CP violation phases in the corresponding baryogenesis mechanism (respectively GUT baryogenesis, leptogenesis or electroweak baryogenesis).

In short, for creating the matter in the Universe, one should have to violate not only C and P, but also the CP symmetry, which actually should be identified as the particle-antiparticle symmetry.

Today, it is widely believed that {\sl mirror symmetry} is in fact violated in nature.
The weak nuclear interaction is the culprit, with the asymmetry being particularly striking for the weakly interacting neutrinos. For example, today we know that neutrinos only spin with one orientation. Nobody has ever seen a right-handed neutrino.

If one wants to retain the mirror symmetry, the only possibility is a complete doubling of the number of particles. 
There is an old idea suggested by Li and Yang \cite{li104} that there can exist a hidden mirror sector of particles and interactions which is the exact duplicate of our visible world, i.e. with the Lagrangian exactly identical to the Lagrangian of the ordinary particles, however with right-handed 
interactions.
In other words, for each type of particle, such as electron, proton and photon, there is a mirror twin, so that the ordinary particles favor the left hand, the mirror particles favor the right hand. If such particles exist in nature, then mirror symmetry would be exactly conserved.
In fact one could have the Parity symmetry as an exact symmetry of exchange of the ordinary and mirror particles.

{\em The mirror particles can exist without violating any known experiment.}  Thus, the correct statement is that the experiments 
have only shown that the interactions of the {\it known} particles are not mirror symmetric, they have not demonstrated that mirror symmetry is broken in nature.

What really needs to be done is to understand the experimental implications of the existence of mirror particles and find out whether such things could describe our Universe. In particular, it is exactly the aim of this thesis to investigate the cosmological implications of the mirror world.

The hypothesis of the mirror sector has attracted a significant interest over last years, in particular being motivated by the problems of neutrino physics \cite{akhm69,bere52,foota7,footd52,volk58}, gravitational microlensing  
\cite{bere27,bere375,blin9801015,foot452,moha478}, gamma ray bursts \cite{blin9902305,volk13}, ultra-high energy cosmic rays \cite{Venya}, flavour and CP violation \cite{bere417,bere0009290,ruba65}, etc. The basic concept is to have a theory given by the product $G\times G'$ of two identical gauge factors with the identical particle contents, which could naturally emerge, e.g., in the context of $E_8\times E'_8$ superstring. Two sectors communicate through gravity and perhaps also via some other messengers.  A discrete symmetry $P(G\leftrightarrow G')$ interchanging corresponding fields of $G$ and $G'$, so called {\sl mirror parity},  implies that both particle sectors are described by the same Lagrangians.\footnote{In the brane world picture, the M sector can be the same O world realized on a parallel brane, $G'=G$ \cite{Manyfold}.}

In particular, one can consider a minimal symmetry $G_{SM}\times G'_{SM}$, where $G_{SM}=SU(3)\times  SU(2)\times U(1)$ stands for the standard model of observable particles: three families of quarks and leptons $q_i, ~\bar u_i, ~\bar d_i; ~l_i, ~\bar e_i$ ($i=1,2,3$) and the Higgs doublet $\phi$, while $G'_{SM}=[SU(3)\times SU(2)\times U(1)]'$ is its mirror gauge counterpart with analogous particle content: fermions $q'_i, ~\bar u'_i, ~\bar d'_i; ~l'_i, ~\bar e'_i$  and the Higgs $\phi'$. (From now on all fields and quantities of the mirror (M) sector will have an apex to distinguish from the ones belonging to the observable or ordinary (O) world.) The mirror parity 
implies that all coupling constants (gauge, Yukawa, Higgs) have the same pattern in both sectors and thus their microphysics is the same.\footnote{ The mirror parity could be spontaneously broken and the weak interaction scales
$\langle \phi \rangle =v$ and $\langle \phi' \rangle =v'$ could be different, which leads to somewhat different particle physics in the mirror sector \cite{akhm69,bere27,bere375,bere52,moha478,moha462,moha0001362}. In this thesis we treat only the simplest case,  $ v = v' $, in which the M sector has exactly the same physics as the O one. A possibility for further studies is to add another free parameter related to the difference of the weak scales in the two sectors, but at the moment it would be a useless complication.} 

According to this theory the mirror partners have the same mass as their ordinary counterparts, 
but the mirror particles interact with ordinary particles predominately by gravity only. The three non-gravitational forces act on ordinary and mirror particles completely separately (and with opposite handedness: where the ordinary particles are left-handed, the mirror particles are right-handed). For example, while ordinary photons interact with ordinary matter,
they {\it do not} interact with mirror matter\footnote{ On the quantum level, small new fundamental interactions connecting ordinary and mirror particles are possible. Various theoretical constraints suggest only a few possible types of interactions: neutrino-mirror neutrino mass mixing and photon-mirror photon kinetic mixing \cite{foot272,foota7,kobz3}. For the purposes of this thesis, this possibility is ignored. }. Similarly, the ``mirror image'' of this statement must also hold, that is, the mirror photon interacts with mirror matter but does not interact with ordinary matter.  The upshot is that we cannot see mirror photons because we are made of ordinary matter. The mirror  photons would simply pass right through us without interacting at all! 

The mirror symmetry requires that the mirror photons interact with mirror electrons and mirror protons in exactly the same way in which ordinary photons interact with ordinary electrons and ordinary protons. A direct consequence of this is that a mirror atom made of mirror electrons and a mirror nucleus, composed of mirror protons and mirror neutrons, can exist. In fact, mirror matter made of mirror atoms would also exist with exactly the same internal properties as ordinary matter, but would be completely invisible to us! 

Today the most important aspect of the mirror scenario is that it predicts the existence of dark matter in the Universe in a very natural manner. In fact, mirror matter would be invisible, making its presence felt by its gravitational effects, which is exactly the definition of ``dark matter''! 

One could naively think that due to mirror parity the O and M particles have not only the exactly identical particle physics (microphysics), but also the same cosmology at all stages of the Universe evolution. However, identical Lagrangians of two particle sectors do not necessarily imply that the ordinary and mirror worlds should be realized in the same initial states.\footnote{For analogy two oscillators with the same Lagrangians could naturally have different amplitudes.} In fact, should the O and M particles have the same cosmological densities, this would be in the immediate conflict with the Big Bang nucleosynthesis (BBN) bounds on  the effective number of extra light neutrinos, $\DN <1$ \cite{LSV}: the mirror photons, electrons and neutrinos would give a contribution to the Hubble expansion rate exactly equal to the contribution of the ordinary ones, equivalent to $\DN\simeq 6.14$. Therefore, to satisfy the nucleosynthesis limit, the M particles density in the early Universe should be appropriately reduced, i.e. at the BBN epoch they should have a temperature smaller than the ordinary one, $ T' < T $.
Namely, the conservative bound $\DN < 1$ implies that $T'/T < 0.64$. 

This situation is plausible if two following conditions are satisfied: 

A. At the initial moment two systems are born with different densities.   In particular, the inflationary reheating temperature in the M sector should be lower than in the visible one, $T'_R < T_R$, which can be achieved in certain models \cite{bere27,bere375,Venya,KST}.

B. The M and O particles interact very weakly, so that two systems do not come into the thermal equilibrium with each other in the early Universe.  This condition is automatically fulfilled  if two worlds communicate only via the gravity.\footnote{More generally, there could be other messengers like superheavy gauge singlet fields or light singlets of the moduli type. In either case, they should mediate the effective couplings between the O and M particles suppressed by a large mass factor $M\sim M_{P}$ or so.}

If two sectors have different reheating temperatures, and if they do not come into the thermal contact at later stages, then during the Universe expansion they evolve independently and approach the BBN epoch with different temperatures.  

Several possibilities were discussed in the literature when after inflation two particle sectors could be reheated at different temperatures. 
For example one can consider the chaotic inflation scenario with two inflaton fields, the ordinary one $ \eta $ and the mirror one $ \eta ' $, having identical harmonic potentials $ \mu^2 \eta^2 + \mu^2 \eta'^2 $ and interacting respectively with the O and M particles.
In this case one can show that different initial amplitudes of $ \eta $ and $ \eta' $ would result in the different temperatures of the O and M thermal bathes born after inflaton decay \cite{Venya,hodg47}. 
The thermal energies of two sectors simply reflect the difference of the initial energies of two oscillators $ \eta $ and $ \eta' $.
This naive scenario, however, is not valid for inflaton with unharmonic potentials $ \lambda \eta^4 + \lambda \eta'^4 $ \cite{Venya}.

An attractive realization of the inflationary paradigm is provided by supersymmetric models of hybrid inflation. 
The simplest model is based on the superpotential $W=\lambda S(\Phi^2 -\mu^2)$ containing the inflaton field $S$ and the additional ``orthogonal'' field $\Phi$, where $\lambda$ is order 1 coupling constant and $\mu$ is a dimensional parameter of the order of the GUT scale ($ \mu \sim 10^{15} $ GeV) \cite{Gia}. 
The supersymmetric vacuum is located at $S=0$, $\Phi = \mu$, while for the field values $\Phi=0$, $S > \mu$ the tree level potential has a flat valley with an energy density $V = \lambda^2 \mu^4$. Since the supersymmetry is broken by the non-vanishing $F$-term, $F_S=\lambda\mu^2$, the flat direction is lifted by radiative corrections and the potential of $S$ gets a slope which is appropriate for the slow roll conditions. The COBE results on the CMB anisotropy (see previous chapter) imply that $V^{1/4} \simeq \epsilon ^{1/4} \times 7\cdot 10^{16}$ GeV, where $\epsilon \ll 1$ is a slow-roll parameter. 

In ref. \cite{BCT} there was suggested to modify the superpotential to the following form
\be \label{W}
W_{\rm preheat} 
=\lambda S(\Phi^2 -\mu^2) + g\Phi(\Psi^2 + \Psi'^2) \;,
\ee 
where $ \psi $ and $ \psi' $ are some additional superfields.
Such a modification was motivated by the necessity to solve the initial conditions problem.
Fast damping of $\Phi$ is a pretext of inflationary stage which allows the inflaton energy density $\sim \mu^4$ to dominate and then $S$ can slowly roll to the origin. This function is carried by the second term in (\ref{W}) -- the oscillating orthogonal field $\Phi$ fastly decays into $\Psi$ and $\Psi'$ particles which have practically no contact to inflaton $S$. 
In addition, with vanishing $\Phi$, also effective mass terms of $\Psi$'s disappear and the latter fields start behaving as massless -- they stop oscillating and freeze. In general, oscillations $\Psi$ and $\Psi'$  freeze at different amplitudes, typically $\sim \mu$, at which they are catched by the moment when their mass $g\Phi$ drops below the Hubble parameter (see fig.~3 of ref.~\cite{BCT}). When slow-roll ends up, all fields start oscillating around their vacuum values, $S=0$, $\Phi=\mu$, $\Psi,\Psi'=0$, and reheat the Universe e.g. through the decays of $ \Psi \to \phi_1 \phi_2 $ and $ \Psi' \to \phi'_1 \phi'_2 $, due to superpotential terms $ \Psi \phi_1\phi_2 $ and $ \Psi' \phi'_1\phi'_2 $, where $\phi_{1,2}$ and $\phi'_{1,2}$ are the Higgs doublet superfields respectively for the O and M sectors.  Then different magnitudes of $\Psi$ and $\Psi'$ at the end of slow-roll phase should reflect into difference of reheating temperatures $T_R$ and $T'_R$ in two systems, simply because of the energy difference stored into $\Psi$ and $\Psi'$ oscillations which decay respectively into O and M Higgses. 
\footnote{For other scenarios of the asymmetric reheating see \cite{bere27,bere375,Venya,KST,moha478,moha462,moha0001362}.} 

We have also to make sure that after reheating two sectors do not come into the thermal equilibrium to each other, or in other words, that interactions between O and M particles are properly suppressed.
In particular, the quartic interaction between the O and M Higgs scalars $ \lambda (\phi^\dagger \phi) (\phi^{\prime\dagger} \phi^\prime) $ are dangerous, since they would bring to sectors in thermal equilibrium unless the coupling constant is very small, $ \lambda < 10^{-8} $ \cite{bere375}.
Another dangerous coupling can be presented by the kinetic mixing of the O and M photons, $ \alpha F^{\mu\nu} F'_{\mu\nu} $, which can be safe only if $ \alpha < 10^{-9} $.

The operator which can link O and M sectors, has a higher dimension 
\be \label{Planck} 
{D_{ij}\over{M}} (l_i\phi)(l'_{j} \phi') \; + \; {h.c.} \;,    
\ee
and is cutoff by a large mass factor $M\sim M_P$ or so, and thus are safe.
This operator creates mixing between the O and M neutrinos and so they could oscillate in each other.
From the point of view of the ordinary observer, the mirror neutrinos should appear as the sterile ones.

In a supersymmetric version of the theory the contact between two sectors is naturally weak. In particular, the dangerous mixed term $ \lambda (\phi^\dagger \phi) (\phi^{\prime\dagger} \phi^\prime) $ is not allowed anymore, and the lowest order operator between the Higgses of two sectors in the superpotential has a dimension 5
\be \label{Planck1} 
{{\beta}\over{M}} (\phi_1 \phi_2)(\phi'_1 \phi'_2) \;,    
\ee
which are suppressed by a large mass factor $M$ as the operators (\ref{Planck}), and thus are safe. The same holds true for soft supersymmetry breaking terms like $(m_{3/2}/{M}) (\phi_1 \phi_2)(\phi'_1 \phi'_2)$, etc., where $m_{3/2} \sim 100 $ GeV is the gravitino mass. 

As for the kynetic mixing term $F^{\mu\nu} F'_{\mu\nu}$ between the field-strength tensors of the gauge factors $U(1)$ and $U(1)'$, it can be forbidden by embedding $G_{SM}\times G'_{SM}$ in the grand unified group like $SU(5)\times SU(5)'$ or $SO(10)\times SO(10)'$.       


\passo
\def \term_mir_univ{Thermodynamics of the Mirror Universe}
\section{\term_mir_univ}
\label{term_mir_univ}
\markboth{Chapter \ref{chap-mirror_univ_1}. ~ \mir_univ_1}
                    {\S \ref{term_mir_univ} ~ \term_mir_univ}

Once the O and M systems are decoupled already after reheating, at later times $t$ they will have different temperatures $T(t)$ and $T'(t)$, and so  
different energy and entropy densities
\be \label{rho}
\rho(t) = {\pi^{2}\over 30} g_\ast(T) T^{4} \;, ~~~ 
\rho'(t) = {\pi^{2}\over 30} g'_\ast(T') T'^{4} \;,~  
\ee
\be \label{s}
s(t) = {2\pi^{2}\over 45} g_{s}(T) T^{3} \;, ~~~ 
s'(t) = {2\pi^{2}\over 45} g'_{s}(T') T'^{3} \;.~
\ee
The factors $g_{\ast}$, $g_{s}$ and $g'_{\ast}$, $g'_{s}$ account for the effective number of the degrees of freedom in two systems, and in general can be different from each other.   During the Universe expansion, the two sectors evolve with separately conserved entropies. Therefore, the ratio $x\equiv (s'/s)^{1/3}$ is time independent\footnote{ We assume that expansion goes adiabatically in both sectors and neglect the additional entropy production due to the possible weakly first order electroweak or QCD phase transitions.}, 
while the ratio of the temperatures in two sectors is simply given by
\be \label{t-ratio}
{{T'(t)} \over {T(t)}} = x \cdot 
\left[{g_{s}(T)} \over {g'_{s}(T')} \right] ^{1/3} ~.
\ee

The Hubble expansion rate is determined by the total energy density ${\bar{\rho}}=\rho+\rho'$, $H=\sqrt{(8\pi/3) G_N{\bar{\rho}}}$. Therefore, at a given time $t$ in a radiation dominated epoch we have 
\be \label{Hubble}
H(t) = {1\over 2t} = 1.66 \sqrt{\bg(T)} {{T^2} \over {M_{Pl}}} = 
1.66 \sqrt{\bg'(T')} {{T'^2} \over {M_{Pl}}} ~  
\ee
in terms of O and M temperatures $T(t)$ and $T'(t)$, where 
{\begin{eqnarray} \label{g-ast}} 
\bg(T) = g_\ast (T) (1 + ax^4) \;, ~~~~ 
\bg'(T') = g'_\ast (T')\left(1 + {{1}\over{ax^4}}\right) \;.  
{\end{eqnarray}
Here the factor $ a(T,T') = [g'_\ast (T') / g_\ast (T)] \cdot [g_{s}(T) / g'_{s}(T')]^{4/3} $ takes into account that for $T'\neq T$ the relativistic particle contents of the two worlds can be different. However, except for very small values of $x$, we have $a \sim 1$. So hereafter we always take $\bg(T) = g_\ast (T) (1 + x^4)$ and $\bg'(T') = g'_\ast (T')(1 + x^{-4})$. In particular, in the modern Universe we have $a(T_0,T'_0) =1$, $g_s(T_0) = g'_s(T'_0)=3.91$, and $x= T'_0/T_0$, where $T_0,T'_0$ are the present temperatures of the O and M relic photons.\footnote{
The frozen ratio of the neutrino and photon temperatures in the M sector $r'_0=T'_{\nu 0}/T'_0$ depends on the $\nu'$ decoupling temperature from the mirror plasma, which scales approximatively as $T'_D \sim x^{-2/3} T_D$,  where $T_D= 2-3$ MeV is the decoupling temperature of the usual neutrinos. Therefore, unless $x < 10^{-3}$, $r'_0$ has a standard value $r_0= T_{\nu 0}/T_0=(4/11)^{1/3}$.  
For $x < 10^{-3}$, $T'_D$ becomes larger than the QCD scale $\Lambda \simeq 200$ MeV, so that due to the mirror gluons and light quarks $u',d',s'$ contribution we would obtain $r'_0=(4/53)^{1/3}$, $g'_s(T'_0)=2.39$ and so $T'_0/T_0 \simeq 1.2 \: x$.  However, in the following such small values of $x$ are not of our interest.} 

It is useful to note that, due to the difference in the initial temperature conditions in the two sectors, reactions at the same temperature  $T_* = T_*'$ occur at different times $t_*' = x^2 t_*$, which implies different rates of the Hubble expansion (in particular $H(t_*') > H(t_*)$, given $x < 1$), while reactions at the same time $t_* = t_*'$ occur at different temperatures  $T_*' = x T_*$.

In fact, $x$ is the only free parameter in our model, and in general it could be constrained by the BBN bounds. 
The observed abundances of light elements are in good agreement with the standard nucleosynthesis predictions, when at $T\sim 1$ MeV we have $g_\ast=10.75$ as it is saturated by photons $\gamma$, electrons $e$ and three neutrino species $\nu_{e,\mu,\tau}$. The contribution of mirror particles ($\gamma'$, $e'$ and $\nu'_{e,\mu,\tau}$) would change it to $\bg =g_\ast (1 + x^4)$. Deviations from $g_\ast=10.75$ are usually parametrized in terms of the effective number of extra neutrino species, $\Delta g= \bar{g}_\ast -10.75=1.75\Delta N_\nu$.  Thus we have
\be \label{BBN}
\Delta N_\nu = 6.14\cdot x^4 ~. 
\ee
In view of the present observational situation, a reliable bound is $\Delta N_\nu < 1$ \cite{BCT}, which translates as $x < 0.64$. This limit very weakly depends on $\DN$; e.g., $\Delta N_{\nu} < 1.5$ implies $x < 0.70$.  

As far as $x^4\ll 1$, in a relativistic epoch the Hubble expansion rate (\ref{Hubble}) is dominated by the O matter density and the presence of M sector 
practically does not affect the standard cosmology of the early ordinary Universe. However, even if the two sectors have the same microphysics, the cosmology of the early mirror world can be very different from the standard one as far as the crucial epochs like baryogenesis, nucleosynthesis, etc. are concerned. Any of these epochs is related to an instant when the rate of the relevant particle process $\Gamma(T)$, which is generically a function of the temperature, becomes equal to the Hubble expansion rate $H(T)$ (see chapter \ref{chap-exp-univ}). Obviously, in the M sector these events take place earlier than in the O sector, 
and as a rule, the relevant processes in the former freeze out at larger temperatures than in the latter. 

In the matter domination epoch the situation becomes different. 
In particular, we know that ordinary baryons can provide only a minor fraction of the present cosmological density, 
whereas the observational data indicate the presence of dark matter. So, it is interesting to question whether the missing matter density of the Universe could be due to mirror baryons. 


\passo
\def \baryogen{Baryogenesis features in ordinary and mirror sectors}
\section{\baryogen}
\label{baryogen}
\markboth{Chapter \ref{chap-mirror_univ_1}. ~ \mir_univ_1}
                    {\S \ref{baryogen} ~ \baryogen}

As we expressed in \S \ref{sec-baryogen}, in general there are two kinds of baryogenesis scenarios: the electroweak baryogenesis and the GUT baryogenesis or leptogenesis.

Whatever the mechanism responsible for the observed baryon asymmetry (BA) in the O sector $\eta= n_{B}/n_\gamma$, it is most likely to think that the BA in the M sector $\eta'=n'_B/n'_\gamma$ is produced by the same mechanism, and moreover the rates of the $B$ and CP violation processes are parametrically the same in both cases.  However, the out of equilibrium conditions should be different since at relevant temperatures the Universe expansion is faster for the M sector.  Below we show that by this reason $\eta'$ typically emerges  larger than $\eta$ for either type (a) or (b) scenarios.  

The M baryons can be of the cosmological relevance if $\Omega'_{\rm b}$ exceeds $\Omega_{\rm b} = 3.66\times 10^7\eta h^{-2} =0.04-0.05$, whereas $\Omega'_{\rm b} > 1$ would overclose the Universe. So we are interested in a situation when the ratio \be \label{betadef}
\beta = { \Omega'_{\rm b} \over \Omega_{\rm b} }
\ee
falls in the range from 1 to few tens. Since $n'_\gamma = x^3 n_\gamma$, we obtain $\beta = x^3\eta'/\eta$.  Therefore, $\eta' > \eta$ does not a priori mean that $\beta > 1$, and in fact there is a lower limit $x>10^{-2}$ or so for the relevant parameter space. Indeed, it arises from $x^3 = \beta \eta/\eta'$ by recalling that $\eta \sim 10^{-9}$, while $\eta'$ can be taken at most $\sim 10^{-3}$, the biggest value which can be principally realized in any baryogenesis scheme under the realistic assumptions. 


\subsubsection{GUT Baryogenesis }

The GUT baryogenesis mechanism is typically based on a superheavy boson $X$ undergoing the B and CP violating decays into quarks and leptons. 
The following reaction rates are of relevance:
\par\noindent 
{\it Decay}:   
$\Gamma_D \sim \alpha_X M_X$ for $T \lesssim M_X$ or
$\Gamma_D \sim \alpha_X M_X^2/T$ for $T \greatsim M_X$, 
where $\alpha_X$ is the coupling strength of $X$ to fermions and $M_X$ is its mass; 
\par\noindent
{\it Inverse decay}: 
$\Gamma_{ID} \sim  \Gamma_D (M_X/T)^{3/2} \exp(-M_X/T)$ for $T \lesssim M_X$ or
$\Gamma_{ID} \sim \Gamma_D$ for $T \greatsim M_X$;
\par\noindent
{\it The $X$ boson mediated $2\leftrightarrow 2$ processes}:  
$\Gamma_S \sim n_X \sigma\sim  A \alpha_X^2T^5/(M_X^2+T^2)^{2}$, 
where the factor $A$ amounts for the possible reaction channels.

The final BA depends on a temperature at which $X$ bosons go out from equilibrium. 
One can introduce a parameter which measures the effectiveness of the decay at the epoch $T\sim M_X$ \cite{kolbbookeu}: $k=(\Gamma_D/2H)_{T=M_X}=0.3\bg^{-1/2}(\alpha_X M_{Pl}/M_X)$.  
The larger is $k$ the longer equilibrium is maintained and the freeze-out abundance of $X$ boson becomes smaller. Hence, the resulting baryon number to entropy ratio, $B=n_{B}/s\simeq 0.14\eta$ 
is a decreasing function of $k$.    It is approximately given as $B \simeq {({\epsilon}/{g_s})} F(k,k_c)$, where $\epsilon$ is the CP violating factor and 
\begin{equation}\label{Fkkc}
F(k,k_c)=\cases{1 & if $ k < 1$ \cr
             0.3 k^{-1}(\log k)^{-0.6} 
             & if $ 1< k <  k_c$ \cr
                 \sqrt{A\,\alpha_X\,k}\;
e^{-\frac{4}{3}(A\,\alpha_X\,k)^{1/4}} & if $ k > k_c$\cr}
\end{equation}
Here $k_c$ is a critical value defined by equation $k_c(\log k_c)^{-2.4}=300/(A \alpha_X)$. It distinguishes between the regimes $k < k_c$, in which inverse decay is relevant, and $k > k_c$, in which instead $2\leftrightarrow2$ 
processes are the dominant reason for baryon damping.   

In a general context, without referring to a particular model, it is difficult to decide which range of parameters $k$ and $k_c$  can be relevant for baryogenesis. One can impose only the most reasonable constraints  $g_s(T=M_X) \geq 100$ and $\epsilon \leq 10^{-2}$, and thus $\epsilon/g_s < 10^{-4}$ or so. For a given mechanism responsible for the observed baryon asymmetry $B \sim 10^{-10}$, this translates into a lower bound $F(k,k_c) > 10^{-6}$.

\begin{figure}[h]
  \begin{center}
    \leavevmode
      {\hbox 
      {\epsfxsize = 7cm \epsffile{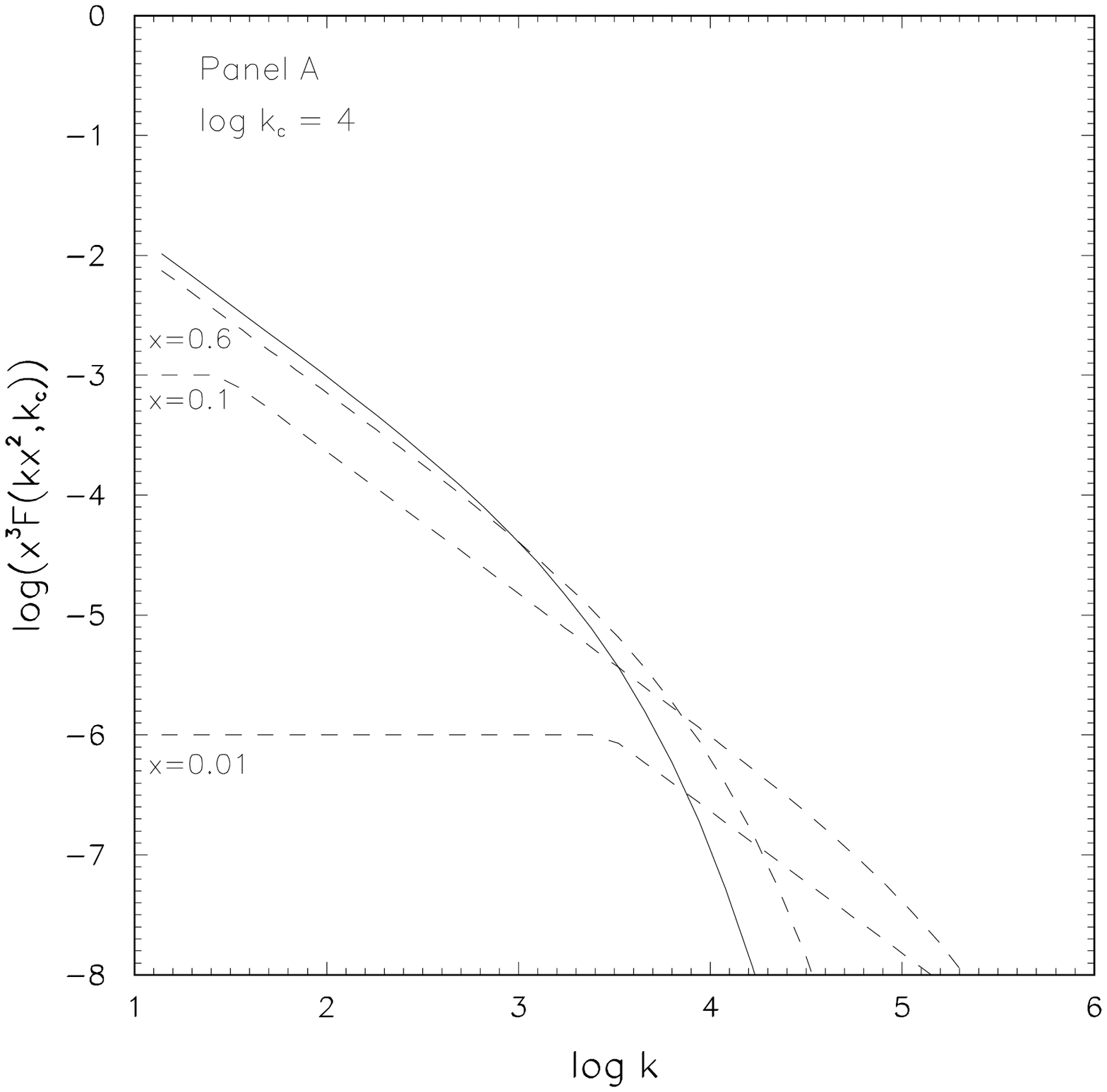} }
      {\epsfxsize = 7cm \epsffile{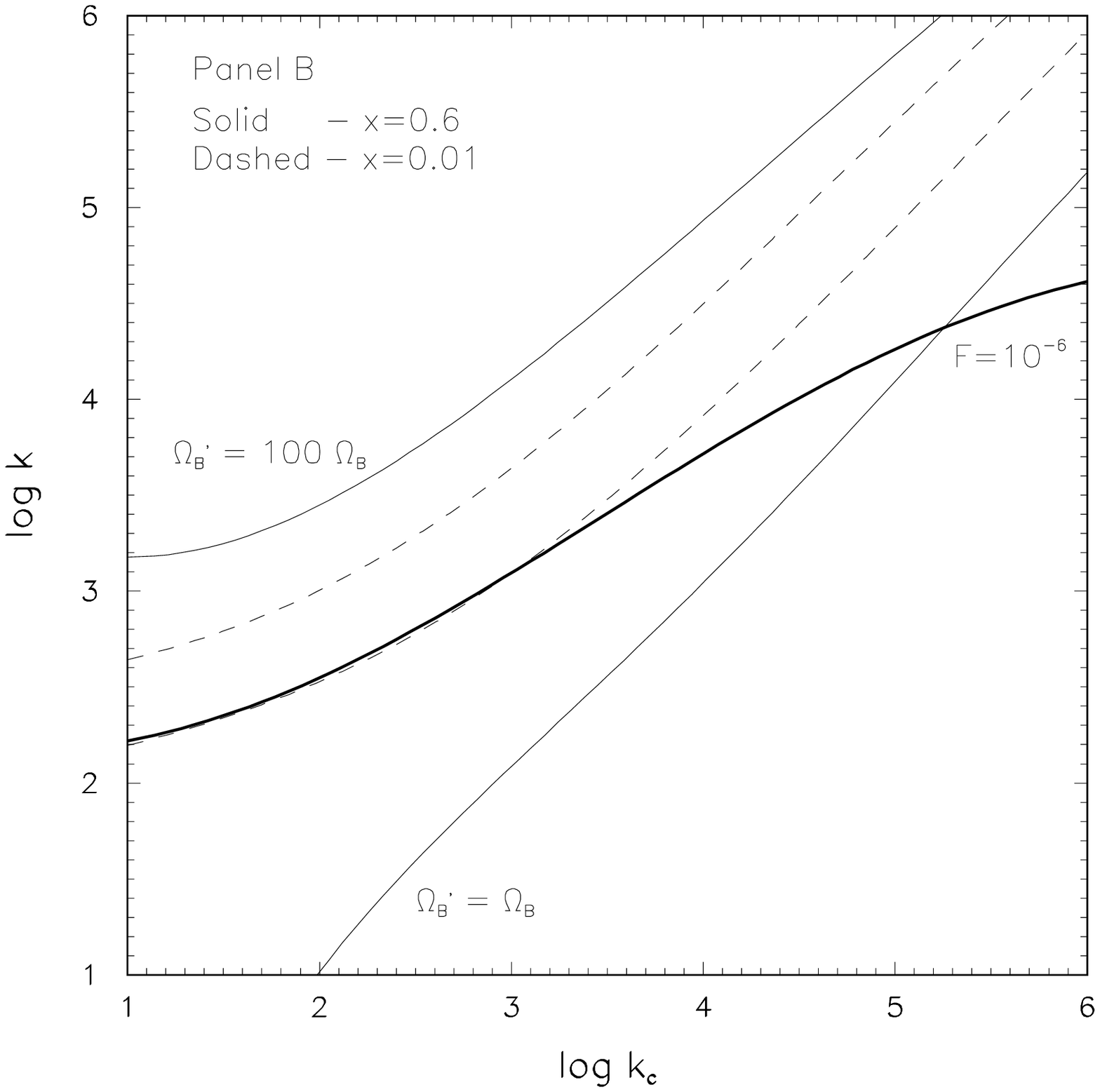} } }
  \end{center}
\caption{\small {\it Panel A}. The combination $x^3F(kx^2,k_c)$ as a function of $k$ for $k_c=10^4$ and $x=0.6,\;0.1,\;0.01$ (dash line). The solid curve corresponding to $x=1$ in fact measures the possible BA in the ordinary world, $F(k,k_c)=(g_s/\epsilon)B(k)$.  {\it Panel B}. The curves confining the parameter region in which $\beta=\Omega'_{\rm b}/\Omega_{\rm b}$ varies from 1 to 100, for $x=0.6$ (solid line) and for $x=0.01$ (dash). The parameter area above thick solid curve corresponds to $F(k,k_c) < 10^{-6}$ and it  is excluded by the observable value of $\eta$. } 
\label{fig_baryogen}
\end{figure}

The presence of the mirror sector practically does not alter the ordinary baryogenesis.  The effective particle number is $\bar g_\ast (T) = g_\ast(T)(1+x^4)$ and thus the contribution of M particles to the Hubble constant at $T\sim M_X$ is suppressed by a factor $x^4$ (which is very small, if we consider the BBN limits explained in \S ~\ref{term_mir_univ}). 

In the mirror sector everything should occur in a similar way, apart from the fact that now at $T'\sim M_X$ the Hubble constant is not dominated by the mirror species but by ordinary ones:   $\bar g'_\ast (T')\simeq g'_\ast (T')(1+ x^{-4})$.  As a consequence, we have $k' = (\Gamma'_D/2H)_{|T'=M_X} = k x^2$. Since the value of $k_c$ is the same in the two sectors, the mirror baryon asymmetry can be simply obtained by replacing $k\rightarrow k'=kx^2$ in eq.~(\ref{Fkkc}), i.e. $B'=n'_B/s' \simeq (\epsilon/g'_s) F(k'=kx^2,k_c)$. Since $F$ is a decreasing function of $k$, then for $x < 1$ we have $F(kx^2,k_c) > F(k,k_c)$ and thus we conclude that the mirror world always gets a {\it larger} BA than the visible one, $B' > B$. 

However, this does not a priori mean that $\Omega'_{\rm b}$ is always larger than $\Omega_{\rm b}$. Since the entropy densities are related as $s'/s=x^3$, for the ratio $\beta =\Omega'_{\rm b}/\Omega_{\rm b}$ we have
\be \label{B-ratio}
\beta (x) =  {{ n'_B} \over {n_{B}}} = {{B's'} \over {Bs}} = x^3 \; {{F(kx^2,k_c)} \over {F(k,k_c)}} \;.~
\ee
The behaviour of the factor $x^3 F(kx^2,k_c)$ as a function of $k$ for different values of the parameter $x$ is given in the fig.~\ref{fig_baryogen}A. Clearly, in order to have $\Omega'_{\rm b} > \Omega_{\rm b}$ the function $F(k,k_c)$ have to decrease faster than $k^{-3/2}$ between $k'=kx^2 $ and $k$. Closer inspection of the function (\ref{Fkkc}) reveals that the M baryons can be overproduced only if $k$ is order $k_c$ or larger.  In other words, the relevant interactions in the observable sector maintain equilibrium longer than in the mirror one, and thus ordinary BA can be suppressed by an exponential Boltzmann factor, while the mirror BA could be produced still in non-exponential regime $k' < k_c$.  

In fig.~\ref{fig_baryogen} B we show the parameter region in which $\beta =\Omega'_{\rm b}/\Omega_{\rm b}$ falls in the range $1-100$, in confront to the parameter area excluded by condition $F(k,k_c)>10^{-6}$.  We see that for $x=0.6$ there is an allowed parameter space in which $\beta$ can reach values up to 10, but $\beta=100$ is excluded. For a limiting case $x=10^{-2}$, as it was expected, the parameter space for $\beta > 1$ becomes incompatible with $F(k,k_c)> 10^{-6}$. For intermediate values of $x$, say $x\sim 0.1-0.3$, also the values $\beta \sim 100$ can be compatible.  
 
The above considerations can be applied also in the context of leptogenesis. One should remark, however, that potentially both the GUT baryogenesis or 
leptogenesis scenarios are in conflict with the supersymmetric inflation scenarios, because of the upper limit on the reheating temperatures about $T_R < 10^9$ GeV from the thermal production of gravitinos \cite{ellib118}.  Moreover, it was shown recently that the non-thermal gravitino production can impose much stronger limits, $T_R < 10^5$ GeV or so \cite{feld0002,giud9911}. This problem can be fully avoided in the electroweak baryogenesis scenario, which instead is definitely favoured by the supersymmetry.   


\subsubsection{Electroweak Baryogenesis }

The electroweak (EW) baryogenesis mechanism is based on the anomalous B-violating processes induced by the sphalerons, which are quite rapid at high temperatures, but become much slower when temperature drops below 100 GeV. A successfull scenario needs the first order EW phase transition and sufficient amount of CP violation, which conditions can be satisfied in the frames of the supersymmetric standard model, for certain parameter ranges \cite{bau}. 

The characteristic temperature scales of the electroweak phase transition are fixed entirely by the form of the finite temperature effective potential
\be \label{effective}
V(\phi,T)=D(T^2-T_0^2)\phi^2- E T\phi^3 + \lambda_T\phi^4 \;,
\ee 
where all parameters can be expressed in terms of the fundamental couplings in the Lagrangian.  For large temperatures, $T \gg 100$ GeV, the electroweak  
symmetry is restored and $V(\phi,T)$ has a minimum at $\phi=0$. 
With the Universe expansion the temperature drops, approaching the specific values which define the sequence of the phase transition. These are all in the 100 GeV range and ordered as $T_1> T_c > T_{\rm b} > T_0$. Namely, below $T=T_{1}$ the potential gets a second local minimum $\phi_{+}(T)$. At the critical temperature $T=T_c$ the latter becomes degenerate with the false vacuum $\phi=0$. 
At temperatures $T<T_c$ to the true vacuum state $\phi=\phi_+(T)$ becomes energetically favoured, and transition to this state can occur via thermal 
quantum tunneling, through the nucleation of the bubbles which then expand fastly, percolate and finally fill the whole space within a horizon by the true vacuum.  

The bubble production starts when the free energy barrier separating the two minima becomes small enough. 
The bubble nucleation temperature $T_{\rm b}$ is defined as a temperature at which the probability for a single bubble to be nucleated within a horizon volume becomes order 1
\be \label{nucl}
P(T_{\rm b})= \omega \left( {{T_c} \over {H(T_c)}} \right)^4 \left(1 - {{T_{\rm b}} \over {T_c}}\right) e^{ - {{F_c(T_{\rm b})} \over {T_{\rm b}}}} \sim 1 \;,
\ee
where $F_c(T)$ is the free energy and $\omega$ is an order 1 coefficient \cite{bau}. In particular, in the limit of thin wall approximation we have
\be \label{thin}
{{F_c(T)} \over {T}}= {{64 \pi} \over {81}} {{E} \over {(2 \lambda)^{3/2}}}
\left( {{T_c-T_0} \over {T_c-T}} \right)^2 \;. 
\ee
The condition (\ref{nucl}) results into large values of $F_c(T_{\rm b})/T_{\rm b}$, typically order $10^2$. 

Once the bubble nucleation and expansion rate is larger than the Hubble parameter, 
the out of equilibrium condition for anomalous B violating processes is provided by the fast phase transition itself.  The BA can be produced inside the bubbles due to the CP violation since the quarks and antiquarks have different 
reflection coefficients on the bubble wall. 
The baryogenesis rate is completely independent from the Universe expansion 
and it occurs practically in one instant as compared to the cosmological time scale of this epoch.

As for the mirror sector, it is described by the same Lagrangian as the ordinary one and so the effective thermal potential of the mirror Higgs $V(\phi',T')$ has the same form as (\ref{effective}). Then the temperature scales, which are defined entirely by the form of the effective potential, should be exactly the same for O and M sectors.  Namely, $T'_1=T_1$ and $T'_c=T_c$.

The equation (\ref{nucl}) is the same for the M sector apart from the fact that the corresponding Hubble constant is different: $H(T'=T_c)= x^{-2} H(T=T_c)$. Therefore, we obtain  
\be
{{F_c(T'_{\rm b})} \over {T'_{\rm b}}} = {{F_c(T_{\rm b})} \over {T_{\rm b}}} + 8 \log{x} \;,
\ee 
which in turn tells that the bubble nucleation temperatures in two sectors are practically equal: 
$T'_{\rm b}=T_{\rm b}(1+0.01 \log x)$. (Clearly, the phase transition in M sector occurs at earlier time than in O sector: $t'_{\rm b} \simeq x^{-2} t_{\rm b}$.) The reason is that between the temperature scales $T_c$ and $T_0$ the free energy $F_c(T)$ is a rapidly changing function but the change in temperature itself is insignificant. Hence, we expect that the initial BA's produced right at the phase transition to be the same in O and M sectors: $B(T=T_{\rm b})=B'(T'=T_{\rm b})$.

However, the instantly produced baryon number can be still washed out by the sphaleron interactions. The anomalous B violation rate $\Gamma(T) \sim \exp[-F(T)/T]$, where $F(T)$ is the sphaleron free energy at finite $T$, may be large enough inside the bubble as far as the temperature is large. But it quickly falls as the temperature decreases, and baryon number freezes out as soon as $\Gamma(t)$ drops below $H(t)$, given by eq. (\ref{Hubble}).   The wash-out equation $dB/dt=-\Gamma(t) B$ can be rewritten as 
\be \label{wah-out}
{{dB} \over {B}} = {{\Gamma(T)} \over {H T}}dT
\ee
and integrated. Then the final BA in the O and M sectors can be expressed respectively as 
\be \label{B-EW}
B=B(T_{\rm b}) D^{(1+x^4)^{-1/2}} \;, ~~~~~~
B'=B(T_{\rm b}) D^{x^2(1+x^4)^{-1/2}} \;, 
\ee
where $D<1$ is the baryon number depletion factor
\be \label{D}
D=\exp\left[- 0.6 g_{\ast}^{-1/2} M_{Pl} 
\int_{0}^{T_{\rm b}} dT \frac{\Gamma(T)}{T^3} \right] \;,~
\ee
and  $g_{\ast}\sim O(100)$ in the supersymmetric standard model. Thus we always have $B' > B$, while the M baryon mass density relative to O baryons reads
\be \label{baew}
\beta (x) = {{\Omega'_{\rm b}} \over {\Omega_{\rm b}}} = x^3 {{B'} \over {B}} = x^3 D^{-K(x)}\;, 
  ~~~~~ K(x) = {{1-x^2} \over {\sqrt{1+x^4}} } \;.~
\ee

\begin{figure}[h]
  \begin{center}
    \leavevmode
    \epsfxsize = 10cm
    \epsffile{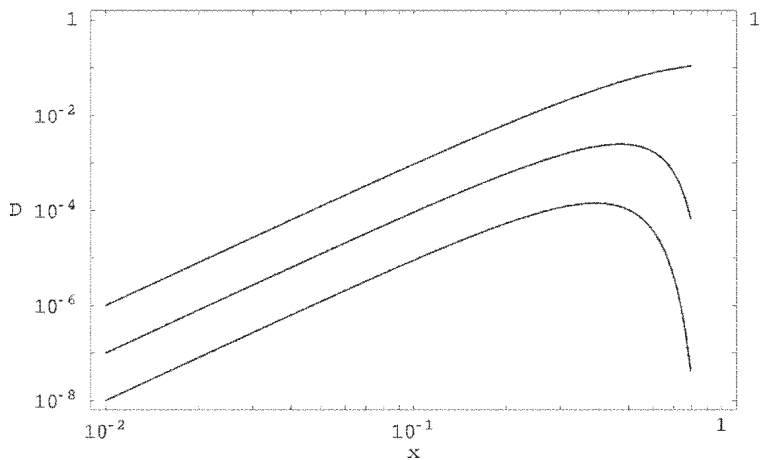}
  \end{center}
\caption{\small The contours of $\beta =\Omega_{\rm b}'/\Omega_{\rm b}$ in the plane of the parameters $x$ and $D$, corresponding to $\beta =1,10$ and 100 from top to bottom. }
\label{bb}
\end{figure}

Lacking a precise theory for non-perturbative sphaleron transitions in the broken phase, the exact value of $D$ cannot be calculated even in the context of concrete models. If $D \sim 1$, the wash-out is ineffective and practically all BA produced right at the bubble nucleation is conserved. In this case $\Omega'_{\rm b}$ should be smaller than $\Omega_{\rm b}$. However, if $D$ is enough small, one can achieve sufficiently large $\Omega'_{\rm b}$.  The contour plot for the  parameters $x$ and $D$ for which $\beta$ falls in the range $1-100$ is given in fig. (\ref{bb}). For small $x$ we have essentially $\beta \simeq x^3 D^{-1}$ and thus $100 > \beta > 1$ requires a depletion factor in the interval $D = (10^{-2} -1) x^3$. Once again, for $x \sim 10^{-2}$ one needs the marginal values $D\sim 10^{-8} - 10^{-6}$ below which the observable BA $B\sim 10^{-10}$ cannot be produced at all. 


\passo
\def \baryogen2{Baryogenesis via particle exchange between ordinary and mirror sectors}
\section{\baryogen2}
\label{baryogen2}
\markboth{Chapter \ref{chap-mirror_univ_1}. ~ \mir_univ_1}
                    {\S \ref{baryogen2} ~ \baryogen2}

Long time ago Sakharov \cite{sakha67} has suggested that a non-zero baryon asymmetry  can be produced in the initially baryon symmetric Universe if three conditions are fulfilled: B-violation, C- and CP-violation and departure from thermal equilibrium.  These conditions can be satisfied in the decays of heavy particles of grand unified theories. On the other hand, the non perturbative sphaleron processes, which violate $ B+L $ but conserve $ B-L $, are effective at temperatures from about $ 10^{12} $ GeV down to 100 GeV \cite{KRS155}. Thus, one actually needs to produce a non-zero $ B-L $ rather than just $ B $, a fact that disfavors the simplest baryogenesis picture based on grand unification models like $ SU(5) $. When sphalerons are in equilibrium, the baryon number and  $ B-L $ are related as $ B = a(B-L) $, where $ a $ is a model dependent order one coefficient \cite{bau}. Hence, the observed baryon to entropy density ratio, $ B = n_{B} / s = (0.6 - 1) \times 10^{-10} $, needs to produce $ B-L \sim 10^{-10} $.

The seesaw mechanism for neutrino masses offers an elegant possibility of generating non-zero $ B-L $ in CP-violating decays of heavy Majorana neutrinos $ N $ into leptons and Higgses, the so called {\sl leptogenesis} scenario \cite{FY86,FY90}. Namely, due to complex Yukawa constants, the decay rates $ \Gamma (N\to l\phi) $ and $ \Gamma (N \to \barl \barphi) $ can be different  from each other, so that leptons $ l $ and anti-leptons $ \barl $ are produced in different amounts.  

An alternative mechanism of leptogenesis based on the scattering processes rather than on decay was suggested in ref. \cite{benber87}. The main idea consists in the following. There exists some hidden (shadow) sector of new particles which are not in thermal equilibrium with the ordinary particle world as far as the two systems interact very weakly, e.g.~if they only communicate via gravity. However, other messengers may well exist, namely, superheavy gauge singlets like right-handed neutrinos which can mediate very weak effective interactions between the ordinary and hidden leptons. Then, a net $ B-L $ could emerge in the Universe as a result of CP-violating effects in the unbalanced production of hidden particles from ordinary particle collisions. 

Here we consider the case when the hidden sector is a mirror one. As far as the leptogenesis is concerned, we concentrate only on the lepton sector of both O and M worlds. Therefore we consider the standard model, containing among other particles species, the lepton doublets $ l_i = (\nu, e)_i $ ($ i=1,2,3 $ is the family index) and the Higgs doublet $ \phi $. Then the mirror standard model contains the lepton doublets $ l'_i = (\nu', e')_i $ and the Higgs doublet $ \phi' $, so that the products $ \l_i \phi $ and $ \l'_i\phi' $ are gauge invariant in both sectors. 

In the spirit of the seesaw mechanism, one can introduce some number of heavy fermions $ N_a $ ($ a=1,2,.. $), called heavy neutrinos, which are gauge singlets and thus can couple to $l,\phi$ as well as to $l^\prime,\phi^\prime$. In this way, they play the role of messengers between ordinary and mirror particles. The relevant Yukawa couplings have the form
\be
\label{Yuk} 
h_{ia} l_i N_a \phi + \hpr_{ka} \lpr_k N_a \phpr + 
  {1 \over 2} M_{ab} N_a N_b + {\rm h.c.}  
\end{equation} 
(charge-conjugation matrix $C$ is omitted); all fermion states $l,N,l^\prime$ are taken to be left-handed while their $C$-conjugate, right-handed anti-particles are denoted as $\bar{l},\bar{N},\bar{l}^\prime$. It is convenient to present the heavy neutrino mass matrix as $M_{ab} = g_{ab} M$, $M$ being the overall mass scale and $g_{ab}$ some typical Yukawa constants. (Without loss of generality, $g_{ab}$ can be taken diagonal and real.)  

Due to mirror Parity, both particle sectors are described by identical Lagrangians, that is, all coupling constants (gauge, Yukawa, Higgs) have the same pattern in both sectors and thus their microphysics is the same. Namely, the M Parity as a discrete symmetry under the exchange $\phi \to \phi^{\prime\dagger}$, $l \to \bar{l}^\prime$, etc., implies $h^\prime_{ia} =h^\ast_{ia}$. 

Integrating out the heavy states $N$ in the couplings (\ref{Yuk}), we get the effective operators 
\be
\label{op} 
{{A_{ij}} \over {2 M}} l_i l_j \phi \phi + 
  {{D_{ik}} \over {M}} l_i \lpr_k \phi \phpr + 
  {{\Apr_{kn}} \over {2 M}} \lpr_k \lpr_n \phpr \phpr  + {\rm h.c.} \;,
\end{equation}
with coupling constant matrices of the form $A = h g^{-1} h^T$, $\Apr = \hpr g^{-1} h^{\prime T}$ and $D = h g^{-1} h^{\prime T}$. Thus, the first operator in eq.~(\ref{op}), due to the ordinary Higgs vacuum expectation value (VEV) $\langle\phi \rangle = v\sim 100$ GeV, induces the small Majorana masses of the ordinary (active) neutrinos. In addition, if the mirror Higgs $\phi^\prime$ also has a non-zero VEV $\langle\phi^\prime \rangle = v^\prime \ll M$, then the third operator provides the masses of the mirror neutrinos contained in $\lpr$ (which in fact are sterile for the ordinary observer). And finally, as we see, the operator (\ref{Planck}) is also induced by the heavy neutrino exchanges (the second term in (\ref{op}), which gives rise to the mixing mass terms between the active and sterile neutrinos). The total mass matrix of neutrinos $\nu\subset l$ and $\nu^\prime\subset l^\prime$ reads as~\cite{bere52}
\begin{equation} \label{numass} 
M_\nu = \mat{m_\nu}{m_{\nu\nu^\prime}}{m_{\nu\nu^\prime}^T}
{m_{\nu^\prime}} = 
  {1 \over M} \mat{Av^2}{Dv v^\prime}{D^Tv v^\prime}
{A^\prime v^{\prime 2}} \, .
\end{equation}
Thus, this model provides a simple explanation of why sterile neutrinos could be light (on the same grounds as the active neutrinos) and could have significant mixing with the ordinary neutrinos. For example, if $v^\prime \sim 10^2 \, v$, then the mirror neutrinos $\nu^\prime$ with masses of keV order could provide the warm dark matter component in the Universe \cite{bere27,bere375,Venya}. Instead, if $\langle\phi^\prime\rangle =0$, the $\nu^\prime$ are massless and unmixed with the ordinary neutrinos. For our considerations with exact mirror Parity, we have $ v' = v $.

Let us discuss now the leptogenesis mechanism in our scenario. A crucial role in our considerations is played by the reheating temperature $T_R$, at which the inflaton decay and entropy production of the Universe is over, and after which the Universe is dominated by a relativistic plasma of ordinary particle species. As we discussed above, we assume that after the postinflationary reheating, different temperatures are established in the two sectors: $T'_R < T_R$,  i.e. the mirror sector is cooler than the visible one, or ultimately, even completely ``empty". This situation is motivated by the BBN constraints, and it can be achieved in the context of certain inflationary models \cite{bere27,bere503,bere375,Venya,KST}. In addition, the two particle systems should interact very weakly so that they do not come in thermal equilibrium with each other after reheating. The heavy neutrino masses are much larger than the reheating temperature $T_R$ and thus cannot be thermally produced. As a result, the usual leptogenesis mechanism via  $N\to l\phi$ decays is ineffective. 

Now, the important role is played by lepton number violating scatterings mediated by the heavy neutrinos $ N $. The ``cooler" mirror world starts being ``slowly" occupied due to the entropy transfer from the ordinary sector through the $ \Delta L=1 $ reactions $ l_i \phi \to \bar \lpr_k \barphpr $, $ \bar l_i \barphi \to \lpr_k \phpr $. In general these processes violate CP due to complex Yukawa couplings in eq.~(\ref{Yuk}), and so the cross-sections with leptons and anti-leptons in the initial state are different from each other. As a result, leptons leak to the mirror sector more (or less) effectively than antileptons and a non-zero $ B-L $ is produced in the Universe. It is essential that these processes stay out of equilibrium. In other words, their rate should be less than the Hubble parameter $ H = 1.66 \, g_\ast^{1/2} T^2 / M_{Pl} $ ($ g_\ast $ being the effective number of particle degrees of freedom) for temperatures $ T \leq T_R $. 

For the rate of $\Delta L=1$ reactions we have $\Gamma_1 = \sigma_1 n_{eq} $, where $n_{eq}\simeq (1.2/\pi^2)T^3$ is an equilibrium density per degree of freedom and $\sigma_1$ is the total cross section of  $l\phi \to \barlpr\barphpr$ scatterings
\begin{equation} 
\label{sigma}
\sigma_{1} = \sum \sigma (l\phi \to \barlpr\barphpr) = {{Q_1} \over {8\pi M^2}} \;. 
\end{equation}
The sum is taken over all flavor and isospin indices of initial and final states, and  $Q_1={\rm Tr}(D^\dagger D) = {\rm Tr}[(\hprdag\hpr) g^{-1}(h^\dagger h)^\ast g^{-1}]$. Hence, the out-of-equilibrium condition for this process reads as
\begin{equation} \label{K}
K_1 = \left({{\Gamma_1} \over {2H}}\right)_{R} \simeq 1.5 \times
10^{-3}\, {{Q_1 T_R M_{Pl}} \over {g_\ast^{1/2}M^2}} < 1 \, .  
\end{equation}

However, there are also scattering processes with $\Delta L =2$ like $l\phi \to \barl\barphi$ etc., which can wash out the produced $B-L$ unless they are out of equilibrium \cite{FY86,FY90}. Their total rate is given as $\Gamma_2 \simeq (3 Q_2/4 \pi M^2)n_{\rm eq}$ where $Q_2 = {\rm Tr}(A^\dagger A) = {\rm Tr}[(h^\dagger h) g^{-1} (h^\dagger h)^\ast g^{-1}]$. Therefore, for a given reheat temperature $T_R$, the eq. (\ref{K}) and the analogous condition $K_2=(\Gamma_2/2H)_{R} < 1$ translate into the lower limit on the heavy neutrino mass scale $M$
\begin{equation}  \label{M1}
M_{12} Q_1^{-1/2} > 4.2\, g_\ast^{-1/4} T_9^{1/2} , ~~~~ 
M_{12} Q_2^{-1/2} > 10.4\, g_\ast^{-1/4} T_9^{1/2} \, . 
\end{equation}
where $M_{12}\equiv (M/10^{12}~{\rm GeV})$, $T_{9}\equiv (T_R/10^{9}~{\rm GeV})$ and $g_\ast\approx 100$ in the standard model. Clearly, if the Yukawa constants $h_{ia}$ and $\hpr_{ia}$ are of the same order, the out-of-equilibrium conditions for $\Delta L=1$ and $\Delta L=2$ processes are nearly equivalent to each other.  

\begin{figure}[t]
  \begin{center}
    \leavevmode
    \epsfxsize = 7cm
    \epsffile{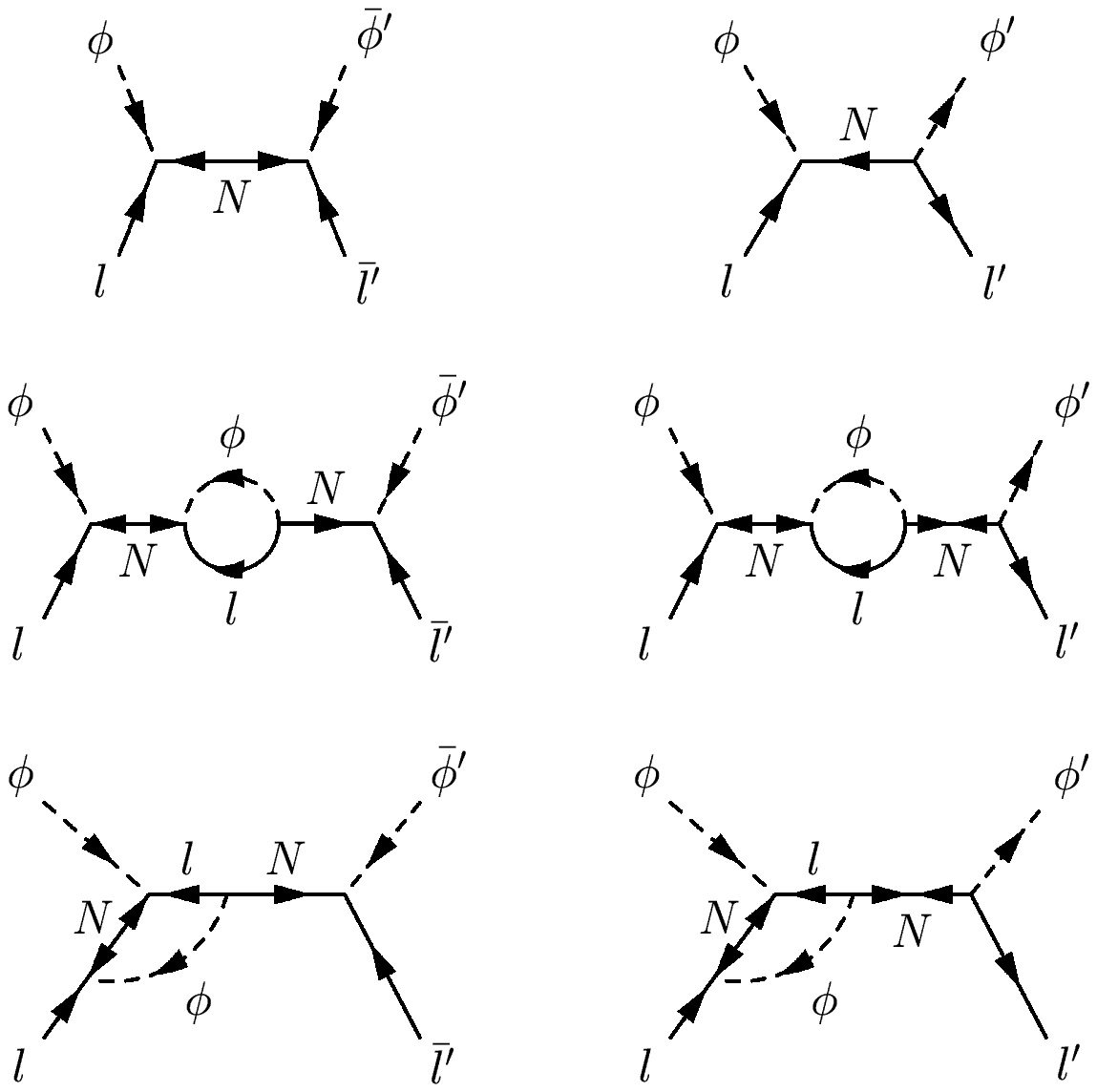}
  \end{center}
\caption{\small Tree-level and one-loop diagrams contributing to the CP-asymmetries in $l \phi \to \barlpr \barphpr$ (left column) and $l \phi \to \lpr \phpr$ (right column).}
\label{fig1}
\end{figure}

Let us turn now to CP-violation. In $\Delta L=1$ processes the CP-odd lepton number asymmetry emerges from the interference between the tree-level and one-loop diagrams of fig.~\ref{fig1}. The tree-level amplitude for the dominant channel $l\phi\to \barlpr\barphpr$ goes as $ 1/M$ and the radiative corrections as $ 1/M^3$. For the channel $l\phi\to \lpr\phpr$ instead, both tree-level and one-loop amplitudes go as $ 1/M^2$. As a result,  the cross section CP asymmetries are the same for both $l\phi\to \barlpr\barphpr$ and $l\phi\to \lpr\phpr$ channels (on the contrary, the diagrams with $\lpr\phpr$ inside the loops, not shown in fig.\ \ref{fig1}, yield asymmetries, $\pm \Delta \sigma'$, symmetric to each other). However, CP-violation takes also place in $\Delta L=2$ processes (see fig.\ \ref{fig2}). This is a consequence of the very existence of the mirror sector, namely the contribution of the mirror particles to the one-loop diagrams of fig.\ \ref{fig2}. The direct calculation gives
\begin{eqnarray}\label{CP}
&& 
\sigma (l\phi\to \barlpr\barphpr) -
\sigma(\barl\barphi \to \lpr\phpr) = 
(- \Delta\sigma  - \Delta\sigma' ) /2
\, ,  \nonumber \\
&&  
\sigma (l\phi\to \lpr\phpr) -
\sigma(\barl\barphi \to \barlpr\barphpr) = 
( -\Delta\sigma + \Delta\sigma' )/2
\, ,   \nonumber  \\
&&  
\sigma (l\phi\to \barl\barphi) -
\sigma(\barl\barphi \to l\phi) = \Delta\sigma \, ;  \\
&& 
\Delta\sigma = {{3J\, S} \over {32\pi^2 M^4}} \, ,  
\end{eqnarray}
where $J= {\rm Im\, Tr} [ (\hprdag\hpr) g^{-2}(h^\dagger h) g^{-1} (h^\dagger h)^\ast g^{-1}]$  is the CP-violation parameter and $S$ is the c.m.\ energy square ($\Delta\sigma'$ is obtained from $\Delta\sigma$ by exchanging $h$ with $h'$).  
   
This is in  perfect agreement with CPT invariance that requires that the total cross sections for particle and anti-particle scatterings are equal to each other: $\sigma(l\phi \to X) = \sigma(\barl\barphi \to X)$. Indeed, taking into account that $\sigma(l\phi \to l\phi) = \sigma(\barl\barphi \to \barl\barphi)$ by CPT, we see that CP asymmetries in the $\Delta L=1$ and $\Delta L=2$ processes should be related as  
\bea
\label{CPT}  
&&
\sigma(l\phi \to X^\prime) - \sigma(\barl\barphi \to X^\prime) =   
- [ \sigma (l\phi\to \barl\barphi) -
\sigma(\barl\barphi \to l\phi) ] = - \Delta \sigma   \;,
\eea
where $X^\prime$ are the mirror sector final states, $\barlpr\barphpr$ and $\lpr\phpr$. That is, the  $\Delta L=1$ and $\Delta L=2$ reactions have CP asymmetries with equal intensities but opposite signs. But, as $L$ varies in each case by a different amount, a net lepton number decrease is produced, or better, a net increase of $B-L$ $ \propto \Delta\sigma$ (recall that the lepton number $L$ is violated by the sphaleron processes, while $B-L$ is changed solely by the above processes). 

\begin{figure}[t]
  \begin{center}
    \leavevmode
    \epsfxsize = 7cm
    \epsffile{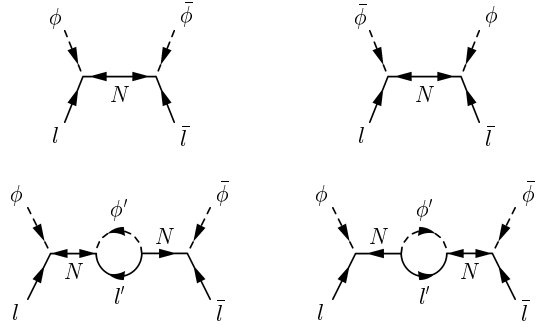}
  \end{center}
\caption{\small Tree-level and one-loop diagrams contributing to the CP-asymmetry of $l \phi \to \barl \barphi$. The external leg labels identify the initial and final state particles.}
\label{fig2}
\end{figure}

As far as we assume that the mirror sector is cooler and thus depleted of particles, the only relevant reactions are the ones with ordinary particles in the initial state. Hence,  the evolution of the $B-L$ number density is determined by the CP asymmetries shown in eqs.~(\ref{CP}) and obeys the equation 
\begin{equation} \label{L-eq}
{{ d n_{B-L} } \over {dt}} + 3H n_{B-L} = 
{3 \over 4} \Delta\sigma \, n_{\rm eq}^2   \; .
\end{equation} 
Since the CP asymmetric cross section $\Delta\sigma$ is proportional to the thermal average c.m.\ energy square $S \simeq 17\, T^2$ and  $H=1/2\, t \propto T^2$, one integrates the above equation from $T=T_R$ to the low temperature limit and obtains the final $B-L$ asymmetry of the Universe as
\begin{equation} \label{BL} 
B-L = {{ n_{B-L} } \over {s}} =
\left[{{\Delta\sigma\, n_{\rm eq}^2 } \over {4 H s}} \right]_{R} \, ,
\end{equation} 
where $s=(2\pi^2/45)g_\ast T^3$ is the entropy density.

The following remark is in order. In fact, the lepton number production starts as soon as the inflaton starts decaying and the particle thermal bath is produced  before the reheating temperature is established. (Recall that the maximal temperature at the reheating period is usually larger than $T_R$.) In this epoch the Universe is still dominated by the inflaton oscillations and therefore it expands as $ t^{2/3}$ while the entropy of the Universe grows as $t^{5/4}$. The integration of eq.~(\ref{L-eq}) from some higher temperatures down to $T_R$ gives an asymmetry 1.5 times larger than the estimation (\ref{BL}). Taking all these into account, the final result can be recasted as follows \footnote{ Observe that the magnitude of the produced $B-L$ strongly depends on the temperature, namely, larger $B-L$ should be produced in the patches where the plasma is hotter. In the cosmological context, this would lead to a situation where, apart from the adiabatic density/temperature perturbations, there also emerge correlated isocurvature fluctuations with variable $B$ and $L$ which could be tested with the future data on the CMB anisotropies and large scale structure.}
\begin{equation} \label{B-L}
B-L \approx 2 \times 10^{-3} \, {{J\, M_{Pl} T_R^3} \over {g_\ast^{3/2} M^4 }}
  \approx 2 \times 10^{-8} \, {{J\, T_9^3} \over {M_{12}^{4}}} \, ,
\end{equation}
where we have taken again $g_\ast \approx 100$. Taking also into account the lower limits (\ref{M1}), we obtain the upper limit on the produced $B-L$
\begin{equation} 
\label{upper}
B-L <   10^{-8}\, {{J\, T_9} \over {Q^2}} \, ; \quad
Q = \max\{Q_1,  6\, Q_2 \}  \, .
\end{equation} 
This shows that for Yukawa constants spread e.g.\ in the range $ 0.1-1 $, one can achieve $ B-L = {\cal O}(10^{-10}) $ for a reheating temperature as low as $ T_R\sim 10^9 $ GeV. Interestingly, this coincidence with the upper bound from the thermal gravitino production, $ T_R < 4\times 10^9 $ GeV or so \cite{ellib118}, indicates that our scenario could also work in the context of supersymmetric theories. 

As far as the mirror sector includes the gauge coupling constants, which are the same as the standard ones, the mirror particles should be thermalized at a temperature $ T^\prime $. Once $ K_1 < 1 $, $ T^\prime $ will be smaller than the parallel temperature of the ordinary system $ T $. Obviously, the presence of the out-of-equilibrium hidden sector does not affect much the Big Bang nucleosynthesis epoch. Indeed, if the two sectors do not come into full thermal equilibrium at temperatures $ T\sim T_R $, then they evolve independently during the Universe expansion and approach the nucleosynthesis era with different temperatures. For $ K_1 < 1 $, the energy density transferred to the mirror sector will be crudely $ \rho^\prime \approx (8 K_1/g_\ast)\rho $, where $ g_\ast(\approx 100) $ is attained to the leptogenesis epoch. Thus, assuming that at the BBN epoch the mirror sector is dominated by relativistic degrees of freedom, we obtain an effective number of extra light neutrinos $ \Delta N_\nu \approx K_1/2 $. 

Now it is important to stress that this mechanism would generate the baryon asymmetry not only in the observable sector, but also in the mirror sector. In fact, two sectors are completely similar, and have similar CP-violating properties. We have scattering processes which transform the ordinary particles into their mirror partners, and CP-violation effects in this scattering owing to the complex coupling constants. These exchange processes are active at some early epoch of the Universe, perhaps even enough  close to equilibrium. In this case, a hypothetical O observer could detect during the contact epoch that (i) matter slowly (in comparison to the Universe expansion rate) disappears from the thermal bath of our world, and, in addition, (ii) particles and antiparticles disappear with different rates, so that after the contact epoch ends up, he observes that his world is left with non-zero baryon number even if initially it was baryon symmetric. 

On the other hand, his mirror analogue, M observer, should see that (i) the matter creation takes place in his world, and (ii) particles and antiparticles emerge with different rates. Therefore, after the contact epoch, he also would observe the non-zero baryon number in his world.  

One would naively expect that in this case the baryon asymmetries in the O and M sectors should be literally equal, given that the CP-violating factors are the same for both sectors. However, we show in that in reality, the BA in the M sector, since it is colder, can be about an order of magnitude bigger than in O sector, as far as washing out effects are taken into account. Indeed, this effects should be more efficient for the hotter O sector while they can be negligible for colder M sector, which could provide reasonable differences between two worlds in case the exchange process is not too far from equilibrium. 
The possible marriage between dark matter and the leptobaryogenesis mechanism is certainly an attractive feature of our scheme.

Now, let us discuss how the mechanism considered above produces not only the lepton number in the ordinary sector, but also the lepton prime asymmetry in the mirror sector. The amount of this asymmetry will depend on the CP-violation parameter that replaces $ J $ in eqs.\ (\ref{CP}) and $ \Delta \sigma' $, namely $ J^\prime = {\rm Im\, Tr}[ (h^\dagger h) g^{-2} (h^{\prime\dagger} h^\prime) g^{-1} (h^{\prime\dagger} h^\prime)^\ast g^{-1}] $. The M parity under the exchange $ \phi \to \phi^{\prime\dagger} $, $ l \to \bar{l}^\prime $, etc., implies that the Yukawa couplings are essentially the same in both sectors, $ h^\prime_{ia} = h^\ast_{ia} $. Therefore, in this case also the CP-violation parameters are the same, $ J^\prime = J $.  

Therefore, one naively expects that $ n_{B-L} = n^\prime_{B-L} $ and the mirror baryon number density should be equal to the ordinary baryon density, $ \Omega'_{\rm b} = \Omega_{\rm b} $.\footnote{ The mirror parity could be also spontaneously broken by the difference in weak scales $ \langle \phi \rangle =v $ and $ \langle \phi' \rangle = v' $, which would lead to somewhat different particle physics in the mirror sector \cite{bere27,bere375,bere52,Venya}, e.g.\ the mirror leptons and baryons could be heavier than the ordinary ones. But, as the mechanism only depends on the Yukawa constant pattern in (\ref{Yuk}), one still has $ n_{B-L} = n^\prime_{B-L} $, while $ \Omega'_{\rm b} > \Omega_{\rm b} $. } 
However, now we show that if the $ \Delta L = 1 $ and $ \Delta L = 2 $ processes are not very far from equilibrium, i.e. $ K_{1,2} \sim 1 $, the mirror baryon density should be bigger than the ordinary one.

Here we should notice that eq.~(\ref{L-eq}) for $ n_{B-L} $ was valid if these processes were very far from equilibrium, $ K_{1,2} \ll 1 $. If instead  $ K_{1,2} \sim 1 $ then eq. (\ref{L-eq}) should be modified as 
\begin{equation} \label{L-eq-2}
{{ d n_{B-L} } \over {dt}} + 3H n_{B-L} + \Gamma n_{\rm B-L} = 
  {3 \over 4} \Delta\sigma \, n_{\rm eq}^2   \;,
\end{equation} 
where $ \Gamma = \Gamma_1 + \Gamma_2 = ( Q_1 + 6Q_2 ) n_{\rm eq} / 8\pi M^2 $ is the total rate of the $ \Delta L = 1 $ and $ \Delta L = 2 $ reactions.
Now, solving this equation we obtain the expression for $ B-L $ as
\be
B-L = (B-L)_0 \cdot D(2K) \;,
\ee
where $ (B-L)_0 $ is the solution of eq.~(\ref{L-eq}), given by expressions (\ref{BL}) or (\ref{B-L}), and the depletion factor $D(k)$ is given by
\be{Dk}
D(k) = \frac35\, e^{-k} F(k) + \frac25\, G(k) 
\ee
where 
\bea
&& F(k) = \frac{1}{4 k^4} \left[(2k -1)^3 + 6k-5 +6 e^{-2k}\right] , 
\nonumber \\
&& G(k) = \frac{3}{k^3} \left[ 2 -(k^2 + 2k +2) e^{-k} \right] . 
\eea
These two terms in $D(k)$ 
correspond to the integration of (\ref{L-eq-2}) respectively
in the epochs before and after reheating 
($T > T_R$ and $T < T_R$).
Obviously, for $ k \ll 1 $ the depletion factor 
$ D(k) \to 1 $ and thus we recover the result as in (\ref{BL}) or
(\ref{B-L}): $ B-L = (B-L)_0 $.
However, for large $k$ the depletion can be reasonable,
e.g. for $k=1,2$ we have respectively $D(k) = 0.34, 0.1$.

As far as the mirror sector is concerned, the evolution of the $B-L$  number density, $n' (B-L)$, obeys the equation
\begin{equation} \label{L-eq-3}
{{ d n'_{B-L} } \over {dt}} + 3H n'_{B-L} + \Gamma' n'_{\rm B-L} = 
  {3 \over 4} \Delta\sigma \, n_{\rm eq}^2   \; ,
\end{equation} 
where now $ \Gamma' = \Gamma'_1 + \Gamma'_2 = ( Q_1 + 6Q_2 ) n'_{\rm eq} / 8\pi M^2 $ is the total reaction rate of the $ \Delta L' = 1 $ and $ \Delta L' = 2 $ processes in the mirror sector, and $ n'_{\rm eq} = (1.2 / \pi^2) T^{\prime 3} = x^3 n_{\rm eq} $ is an equilibrium number density per a degree of freedom in a mirror sector. Therefore we have $ K' = \Gamma' / 2 H = x^3 K $, and so for $ K \sim 1 $ and $ x < 0.64 $ (BBN limit) $ K' \ll 1$.

For the mirror sector we have $ (B-L)' = (B-L)_0 \cdot D(2K') $, where the depletion can be irrelevant. Now taking into the account that in both sectors the $ B-L $ densities are reprocessed in the baryon number densities by the same sphaleron processes, we have $ B = a (B-L) $ and $ B' = a (B-L)' $, with coefficients $ a \approx 1/3 $ equal for both sectors. Therefore, we see that the cosmological densities of the ordinary and mirror baryons should be related as
\be
\label{omegabp}
\Omega'_{\rm b} = {D(2K') \over D(2K)} \: \Omega_{\rm b} \;.
\ee 
Therefore, for $K \ll 1$ and thus $K' = x^3 K \ll 1$, depletion factors in both sectors are $ D \approx D' \approx 1 $ and thus we have that the mirror and ordinary baryons have the same densities, $\Omega'_{\rm b} \approx \Omega_{\rm b}$. In this case mirror baryons are not enough to explain all dark matter and one has to invoke also some other kind of dark matter, presumably cold dark matter.

If instead $K \sim 1$, then we would have $ \Omega'_{\rm b} > \Omega_{\rm b} $, and thus all dark matter of the Universe could be in form of the mirror baryons. E.g., for $ K = 0.5 $ we have from eq.~(\ref{omegabp}) that $ \Omega'_{\rm b} \approx 5 \, \Omega_{\rm b} $, and hence for $ \Omega_{\rm b} \approx 0.05 $ we have $ \Omega'_{\rm b} \approx 0.25 $, which is exactly about the best fit value of the dark matter density.  


\passo
\def \nucleosyn{Primordial nucleosynthesis and mirror helium abundance}
\section{\nucleosyn}
\label{nucleosyn}
\markboth{Chapter \ref{chap-mirror_univ_1}. ~ \mir_univ_1}
                {\S \ref{nucleosyn} ~ \nucleosyn}

The time scales relevant for standard BBN are defined by the ``freeze-out'' temperature of weak interactions $T_{W}\simeq 0.8$ MeV ($t_W\sim 1$ s) and by the ``deuterium bottleneck'' temperature $T_{N}\simeq 0.07$ MeV ($t_N \sim 200$ s) \cite{kolbbookeu}. When $T > T_{W}$, weak interactions transform neutrons 
into protons and viceversa and keep them in chemical equilibrium. The neutron abundance $X_n = n_n/n_B$, defined as the ratio of neutron to baryon  densities, is given by $X_{n}(T)=[1+\exp(\Delta m/T)]^{-1}$,
where $\Delta m\simeq 1.29$ MeV is the neutron-proton mass difference.
For $T < T_W$ the weak reaction rate $\Gamma_W \simeq G_F^2 T^5$ drops below the Hubble expansion rate $H(T) \simeq 5.5 T^2/M_{Pl}$, the neutron abundance freezes out at the equilibrium value $X_{n}(T_{W})$ and it then evolves only due to the neutron decay: $X_{n}(t)=X_{n}(T_{W})\exp(-t/\tau)$, 
where $\tau=886.7$ s is the neutron lifetime. 

At temperatures $T > T_{N}$, the process $p+n \leftrightarrow d+\gamma$ is faster than the Universe expansion, and free nucleons and deuterium are in chemical equilibrium.  The light element nucleosynthesis essentially begins 
when the system cools down to the temperature 
\be \label{T_N}
T_{N} \simeq {{B_{d}} \over {-\ln(\eta)+1.5\ln(m_{N}/T_N)}} \simeq 0.07 ~ {\rm MeV} \;, 
\ee
where $B_{d}=2.22$ MeV is the deuterium binding energy, and $m_{N}$ is the nucleon mass. 
Thus, the primordial $^{4}$He mass fraction is
\be \label{helium}
Y_{4}\simeq2X_{n}(t_{N})=
  {{2\exp(-t_N/\tau)} \over {1+\exp(\Delta m/T_{W})}} \simeq 0.24 \;.
\ee

As we have already discussed (see \S ~\ref{term_mir_univ}), the presence of the mirror sector with a temperature $T' < T$ has practically no impact the standard BBN, in the limit $x < 0.64$, which in fact has been set by uncertainties of the present observational situation. In the mirror sector nucleosynthesis proceeds along the same lines. However, the impact of the O world for the mirror BBN is dramatic!

For any given temperature $T'$, using eqs. (\ref {Hubble}) and (\ref {g-ast}) now we have 
\be \label{hub_mir}
H(T') \simeq 5.5 (1+x^{-4})^{1/2} T'^2/M_{Pl}
\ee
for the Hubble expansion rate. Therefore, comparing $H(T')$ with the reaction rate $\Gamma (T') \propto {T'}^5$ (see eq.~(\ref {gam_mass})), we find a freeze-out temperature $T'_W=(1+x^{-4})^{1/6} T_W$, which is larger than $T_{W}$, whereas the time scales as $t'_W = t_W/(1+x^{-4})^{5/6} < t_W$ (obtained using eq.~(\ref {Hubble}) and the relation $t \propto H^{-1}$). In addition, $\eta'$ is different from $\eta\simeq 5 \times 10^{-10}$. However, since $T_N$ depends on baryon density only logarithmically (see eq.~(\ref{T_N})), the temperature $T'_N$ remains essentially the same as $T_N$, while the time $t'_N$ scales as $t'_N = t_N/(1+x^{-4})^{1/2}$. Thus, for the mirror $^4$He mass fraction we obtain:  
\be \label{m_helium}
Y'_{4}\simeq 2X'_n(t'_N)= 
{{ 2\exp[-t_N/\tau(1+x^{-4})^{1/2}] } \over
{1+\exp[\Delta m/T_W(1+x^{-4})^{1/6} ] }} ~. 
\ee
We see that $Y'_{4}$ is an increasing function of $x^{-1}$.  In particular, for $x\rightarrow 0$ one has $Y'_{4}\rightarrow 1$.   

\begin{figure}[h]
  \begin{center}
    \leavevmode
    \epsfxsize = 9cm
    \epsffile{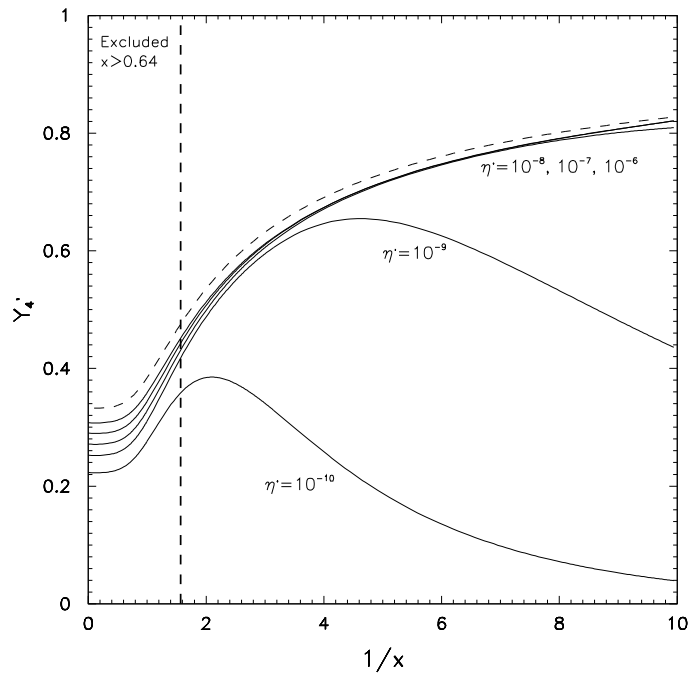}
  \end{center}
\caption{\small The primordial mirror $^4$He mass fraction as a function of $x$. The dashed curve represents the approximate result of eq. (\ref{helium}).
The solid curves obtained via exact numerical calculation correspond, from bottom to top, to $\eta'$ varying from $10^{-10}$ to $10^{-6}$. }
\label{bbn}
\end{figure}

In reality, eq.~(\ref{m_helium}) is not valid for small $x$, since in this case deuterium production through reaction $n+p\leftrightarrow d+\gamma$ can become ineffective. By a simple calculation one can make sure that for $x < 0.3 \cdot (\eta'\times 10^{10})^{-1/2}$, the rate at which neutrons are captured to form the deuterium nuclei, $\Gamma '_N = n'_p \sigma_N \sim \eta' n'_\gamma \sigma_N$, where $n'_\gamma \sim T'^3$ is the M photons density and $\sigma_N \simeq 4.5\cdot 10^{-20}$ cm$^3$ s$^{-1}$ 
is the thermal averaged cross section, becomes smaller than the Hubble rate $H(T')$ for temperatures $T' > T'_N$. In this case M nucleosynthesis is inhibited, because the neutron capture processes become ineffective before deuterium abundance grows enough to initiate the synthesis of the heavier elements. Therefore, for any given $\eta'$, $Y'_4$ first increases with increasing $1/x$, reaches a maximum and then starts decreasing. The true dependence of $Y'_4$ on the $x$ computed for different values of $\eta'$ by standard BBN code \cite{Kawano}, is presented in fig.~\ref{bbn}.  The Hubble expansion rate of the mirror world was implemented, for each value of $x$, by taking an effective number of extra neutrinos as $\Delta N_\nu = 6.14\cdot x^{-4}$. 

We have to remark, however, that in the most interesting situation when 
$\beta =\Omega'_b/\Omega_b = x^3 \eta'/\eta > 1$, the condition $x < 0.3 \cdot (\eta'\times 10^{10})^{-1/2}$ is never fulfilled and the behaviour of $Y'_4$ is well described by the approximate formula (\ref{m_helium}). Hence, in this case $Y'_4$ is always bigger than $Y_4$.   In other words, if dark matter of the Universe is represented by the baryons of the mirror sector, it should contain considerably bigger fraction of primordial $^4$He than the ordinary world. 
In particular, the helium fraction of mirror matter is comprised between 20\% and 80\%, depending on the values of $x$ and $\eta '$. This is a very interesting feature, because it means that mirror sector can be a helium dominated world, with important consequences on star formation and evolution, and other related astrophysical aspects, as we will see in chapter \ref{chap-mirror_univ_4}.


\passo
\def \mirror_dm{Mirror baryons as dark matter: general discussion}
\section{\mirror_dm}
\label{mirror_dm}
\markboth{Chapter \ref{chap-mirror_univ_1}. ~ \mir_univ_1}
                    {\S \ref{mirror_dm} ~ \mirror_dm}

As we explained in \S~\ref{intro_mirror}, mirror matter is an {\sl ideal candidate} for the inferred dark matter in the Universe, being naturally stable and ``dark'', interacting only gravitationally with the ordinary matter. 

At present time various observations indicate that the amount of dark matter is order 10 times the amount of ordinary matter in the Universe (for a picture of the present observational status see \S~\ref{sec-now-cosm}). This is not a problem for the mirror matter theory, because a mirror symmetric microscopic theory does not actually imply equal numbers of ordinary and mirror atoms in the Universe, as someone could erroneously think. The point is that, according to what we said in the previous sections, the initial conditions need not be mirror symmetric, and the Universe could have been created with more mirror matter than ordinary matter.

In the most general context, the present energy density contains relativistic (radiation) component $\Omega_r$, non-relativistic (matter) component $\Omega_m$ and the vacuum energy density $\Omega_\Lambda$ (cosmological term). According to the inflationary paradigm the Universe should be almost flat,  
$\Omega_0=\Omega_m + \Omega_r + \Omega_\Lambda \approx 1$, which well agrees with the recent results on the CMB anisotropy \cite{balb0005124,lang0005004}. 
For redshifts of the cosmological relevance, $1+z = T/T_0 \gg 1$, the Hubble parameter is expressed by (using eqs. in appendix ~\ref{app-flat})
\be \label{H}
H(z)= H_0 \left[\Omega_{r}(1+z)^4 + \Omega_{m} (1+z)^3 \right]^{1/2} \;,
\ee
where now $\Omega_{r}$ and $\Omega_{m}$ represent the total amount of radiation and matter of both ordinary and mirror sectors. In the context of our model, the relativistic fraction is represented by the ordinary and mirror photons and neutrinos, and, using eq.~(\ref{g-ast}) and the value of the observable radiation energy density $ \Omega_r \, h^2 \simeq 4.2 {\times} 10^{-5} $, it is given by 
\be
\Omega_r = 4.2 \times 10^{-5}\,h^{-2}\,(1+x^4) \simeq 4.2 \times 10^{-5}\,h^{-2} \;,
\ee
where the contribution of the mirror species is negligible in view of the BBN constraint $x< 0.64$. As for the non-relativistic component, it contains the O baryon fraction $\Omega_b$ and
the M baryon fraction $\Omega'_b = \beta\Omega_b$, while the other types of dark matter
could also present.
Obviously, since mirror parity doubles {\sl all} the ordinary particles, even if they are ``dark'' (i.e., we are not able to detect them now), whatever the form of dark matter made by some exotic ordinary particles, there will exist a mirror partner made by the mirror counterpart of these particles. 
In the context of supersymmetry, the CDM component could exist in the form of the lightest supersymmetric particle (LSP).  It is interesting to remark that the mass fractions of the ordinary and mirror LSP are related as $\Omega'_{\rm LSP} \simeq x\Omega_{\rm LSP}$. In addition, a significant HDM component $\Omega_\nu$ could be due to neutrinos with order eV mass. The contribution of the mirror massive neutrinos scales as $\Omega'_\nu = x^3 \Omega_\nu$ and thus it is irrelevant. In any case, considering the only CDM component, which is now the preferred candidate, we can combine both the ordinary and mirror components, given that their physical effects are exactly the same. So, in our work we consider a matter composition of the Universe made in general by
\be
\Omega_m=\Omega_b+\Omega'_b+\Omega_{\rm CDM} \;.
\ee
  
The important moments for the structure formation are related to the matter-radiation equality (MRE) epoch and to the plasma recombination and matter-radiation decoupling (MRD) epochs (see \S~\ref{sec-therm-evol}). The MRE occurs at the redshift 
\be \label{z-eq} 
1+z_{\rm eq}= {{\Omega_m} \over {\Omega_r}} \approx 
 2.4\cdot 10^4 {{\Omega_{m}h^2} \over {1+x^4}} \;,~~
\ee
which is {\sl always smaller than the value obtained for an ordinary Universe}, but approximates it for low $x$ (see fig.~\ref{fig3}). If we consider only ordinary and mirror baryons and photons, we find
\be \label{z-eq_2} 
1+z_{\rm eq} = {{\bar \rho_{b0}} \over {\bar \rho_{\gamma 0}}} =
 {{\rho_{b0} + \rho_{b0}'} \over {\rho_{\gamma 0} + \rho_{\gamma 0}'}} = 
 {{\rho_b \left( 1+\beta \right)} \over {\rho_{\gamma } \left( 1+x^4 \right)}} 
 {\left( 1+z \right)} =
 {{\rho_b' \left( 1+\beta^{-1} \right)} \over {\rho_{\gamma }' \left( 1+x^{-4} \right)}} 
 {\left( 1+z \right)} \;,~~
\ee
where $\bar \rho_b$ and $\bar \rho_\gamma$ indicate respectively the sums of baryons and photons of the two sectors. 
So, in the presence of a mirror sector the matter - radiation equality epoch shifts as 
\be \label{shifteq} 
1+z_{\rm eq} ~~ \longrightarrow ~~ { {\left( 1+\beta \right)} \over {\left( 1+x^4 \right)} } ~ (1+z_{\rm eq}) \;.
\ee

The MRD, instead, takes place in every sector only after the most of electrons and protons recombine into neutral hydrogen and the free electron number density  $n_{e}$ diminishes, so that the photon scattering rate $\Gamma_\gamma=n_{e}\sigma_{T}=X_{e}\eta n_{\gamma} \sigma_{T}$ drops below the Hubble expansion rate $H(T)$, where $\sigma_T=6.65\cdot 10^{-25}$ cm$^{2}$ is the Thomson cross section. In condition of chemical equilibrium, the fractional ionization $X_e=n_e/n_b$ is given by the Saha equation, which for $X_e \ll 1$ reads
\be \label{Saha} 
X_e \approx (1-Y_4)^{1/2}\; {{0.51} \over {\eta^{1/2}}} 
\left({T} \over {m_e}\right)^{-3/4} e^{-B/2T} \;,
\ee 
where $B=13.6$ eV is the hydrogen binding energy.  Thus we obtain the familiar result that in our Universe the MRD takes place in the matter domination period, at the temperature $T_{\rm dec} \simeq 0.26$ eV which 
corresponds to redshift $1+z_{\rm dec}=T_{\rm dec}/T_0 \simeq 1100$.

\vskip .4cm

\begin{figure}[h]
  \begin{center}
    \leavevmode
    \epsfxsize = 10cm
    \epsffile{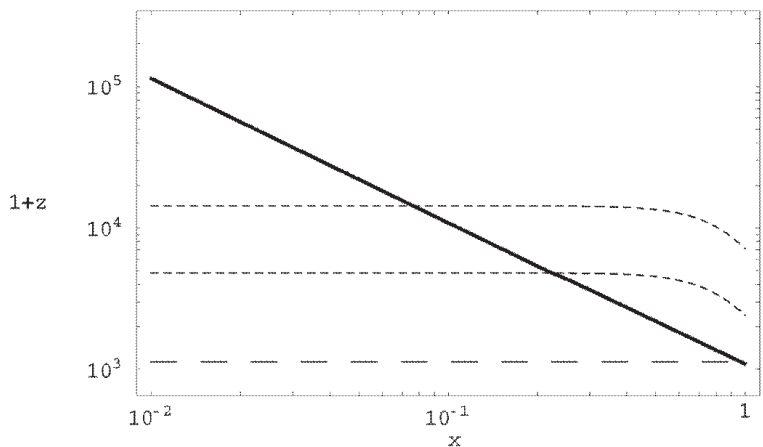}
  \end{center}
\caption{\small The M photon decoupling redshift $1+z_{\rm dec}'$ as a function of $x$ (solid line).  The long-dash line marks the ordinary decoupling redshift $1+z_{\rm dec} = 1100$. We also show the matter-radiation equality redshift $1+z_{\rm eq}$ for the cases $\Omega_m h^2 =0.2$ and $\Omega_m h^2=0.6$ (respectively lower and upper dash). }
\label{fig3}
\end{figure}

\vskip .4cm

The MRD temperature in the M sector $T'_{\rm dec}$ can be calculated following the same lines as in the ordinary one. Due to the fact that in either case the photon decoupling occurs when the exponential factor in eq.~(\ref{Saha}) becomes very small, we have $T'_{\rm dec} \simeq T_{\rm dec}$, 
up to small 
corrections related to $\eta'$, $Y'_4$ different from $\eta$, $Y_4$. Hence, considering that $T' = x \cdot T$ (see \S ~\ref{term_mir_univ}), we obtain
\be \label{z'_dec}
1+z'_{\rm dec} \simeq x^{-1} (1+z_{\rm dec}) 
\simeq 1.1\cdot 10^3 x^{-1} \;, ~~
\ee
so that {\sl the MRD in the M sector occurs earlier than in the ordinary one}. Moreover, comparing eqs.~(\ref{z-eq}) and (\ref{z'_dec}), which have different trends with $x$, we find that,  for $x$ less than a value $x_{\rm eq}$ given by
\be
x_{\rm eq} \approx 0.046(\Omega_m h^2)^{-1} \;, 
\ee  
the mirror photons would decouple yet during the radiation dominated period  
(see fig.~\ref{fig3}). Assuming, e.g., the ``standard'' values $\Omega_m = 0.3$ and $h = 0.65$, we obtain the approximate value $x_{\rm eq} = 0.36$, which indicates that below about this value the mirror decoupling happens in the radiation dominated period, whith consequences on structure formation (see next chapter).

We have shown that mirror baryons could provide a significant contribution to the energy density of the Universe and thus they could constitute a relevant component of dark matter. Immediate question arises: how the mirror baryonic dark matter (MBDM) behaves and what are the differences from the more familiar dark matter candidates as the cold dark matter (CDM), the hot dark matter (HDM), etc.
In next chapters  we discuss the problem of the cosmological structure formation in the presence of M baryons as a 
dark matter component. Namely, M baryons being a sort of self-interacting dark matter could provide interesting signatures on the CMB anisotropy, the large scale structure of the Universe, the form of the galactic halos, microlensing, etc. 



\def \mir_univ_2{Structure formation for a Mirror Universe}
\chapter{\mir_univ_2}
\label{chap-mirror_univ_2}
\markboth{Chapter \ref{chap-mirror_univ_2}. ~ \mir_univ_2}
                    {Chapter \ref{chap-mirror_univ_2}. ~ \mir_univ_2}


\passo
\def \intro_mu2{Introduction}
\section{\intro_mu2}
\label{intro_mu2}
\markboth{Chapter \ref{chap-mirror_univ_2}. ~ \mir_univ_2}
                    {\S \ref{intro_mu2} ~ \intro_mu2}

In chapter \ref{chap-lin-form} we presented the linear structure formation theory in a non-baryonic dark matter scenario (which is now considered the 
``standard'' paradigm). We introduced some fundamental concepts (like the 
Jeans length and the collisionless dissipation) and gave both the 
Newtonian and the general relativistic approaches (briefly reviewed in appendix \ref{app-strucform}). 
Here, we extend the theory to the case of dark matter with a non-negligible mirror baryon component.

In a Mirror Universe we assume that a mirror sector is present, so that the matter is made of ordinary baryons (the only certain component), non-baryonic (dark) matter, and mirror baryons. Thus, it is necessary to study the structure formation in all these three components. This requires essentially the study of the baryonic structure formation (given also that we already described the non-baryonic case in chapter \ref{chap-lin-form}). There are two reasons for that: first, the baryons are the only certain component of the universal matter (even if the lowest); second, as seen in the previous chapter, the physics of mirror baryons is the same as ordinary ones but with different initial conditions and time-shifted key epochs.

As for the non-baryonic dark matter, we continue to focus only on adiabatic perturbations, that we introduced in \S~\ref{intro_struct_form}. Here we remember only that an adiabatic perturbation satisfies the {\sl condition for adiabaticity}
\begin{equation}
\delta_{\rm m}={3\over4}\delta_{\rm r} \;, 
\label{adcond}
\end{equation}
which relates perturbations in matter ($ \delta_{\rm m} $) and radiation ($ \delta_{\rm r} $) components.


\passo
\def \baryo_struct_form{Baryonic structure formation (ordinary and mirror)}
\section{\baryo_struct_form}
\label{baryo_struct_form}
\markboth{Chapter \ref{chap-mirror_univ_2}. ~ \mir_univ_2}
                    {\S \ref{baryo_struct_form} ~ \baryo_struct_form}

We will now consider cosmological models where baryons, ordinary or mirror, are the dominant form of matter, but it should be made clear at the outset that purely ordinary baryonic models cannot successfully explain the origin of the observed structure. Nevertheless, it's very important to look at the details of these scenarios for two reasons. First, even if they are very few, ordinary baryons do exist in the Universe, and it's necessary to know as they behave. Second, mirror baryons, which follow the same physics of the ordinary ones, could be the dominant form of matter in the Universe. Thus, it is crucial to study the interaction between baryonic matter and radiation during the plasma epoch in both sectors, and the simplest way of doing it is by looking at models containing only these two matter components.

As explained in appendix \ref{app-strucform}, the relevant length scale for the gravitational instabilities is characterized by the Jeans scale (see \S~\ref{newth-app}), which now needs to be defined in both the ordinary and mirror sectors.


\subsection{Evolution of the adiabatic sound speed}
\label{evol_Vs}

Given the expression of the Jeans length (\ref{Jl}), it is clear that the key issue is the evolution of the sound speed, since it determines the scale of gravitational instability. If we remember eq.~(\ref{soundspeed}), we obtain for the two sectors

\be
  v_s = {\left({\partial p} \over {\partial \rho} \right)}_S^{1/2} = w^{1/2} ~~~~~~~~~~~~~~
  v_s '= {\left({\partial p'} \over {\partial \rho'} \right)}_S^{1/2} = (w')^{1/2} \;,
\ee
where $S$ denotes the total entropy and $w'$ is relative to the mirror equation of state $p' = w' \rho'$.


\subsubsection{The ordinary sound speed}

In a mixture of radiation and baryonic matter the total density and pressure are $\rho=\rho_{\gamma}+\rho_{\rm b}$ and $p\simeq p_{\gamma}=\rho_{\gamma}/3$ respectively (recall that 
$ p_{\rm b} \simeq 0 $). Hence, the adiabatic sound speed is given by
\begin{equation}
v_{\rm s} = \left({{\partial p} \over {\partial \rho}}\right)^{1/2}
\simeq {1\over\sqrt{3}}\left(1+{{3\rho_{\rm b}} \over {4\rho_{\gamma}}}\right)^{-{1/2}} \;, \label{vsord}
\end{equation}
where we have used the adiabatic condition (\ref{adcond}).
In particular, using eqs.~(\ref {dustuniverse}) and (\ref {radiativeuniverse}) together with the definition of matter - radiation equality (\ref {mreq}) (where now we consider that matter is only baryonic), we obtain
\be \label{ass2} 
v_{\rm s}(z) \simeq {1 \over {\sqrt3}}
 \left[ 1 +{3 \over 4} \left({{1+z_{\rm eq}} \over {1+z}}\right)\right]^{-1/2} \;. 
\ee 
In fact, the relation above is valid only in an ordinary Universe, and it is an approximation, for small values of $x$ and the mirror baryon density (remember that $\beta = {\Omega_b' / \Omega _b}$), of the more general equation for a Universe made of two sectors of baryons and photons, obtained using eqs.~(\ref{vsord}) and (\ref{z-eq_2}) and given by
\be \label{ass3} 
v_{\rm s}(z) \simeq {1 \over {\sqrt3}}
\left[ 1 +{3 \over 4}  \left({1+x^{4}} \over {1+\beta } \right) 
\left( {{1+z_{\rm eq}} \over {1+z}} \right) \right]^{-1/2} \;.
\ee 
In the most general case, the matter is made not only of ordinary and mirror baryons, but also of some other form of dark matter, so the factor $1+\beta $ is replaced by $1+\beta +\beta _{DM}$, where $\beta _{DM} = {\left(\Omega _m - \Omega _b - \Omega _b' \right) / \Omega _b}$. From eqs.~(\ref {ass2}) and (\ref {ass3}) we obtain that, given the conditions $ x < 1 $ (from the BBN bound presented in \S~\ref{term_mir_univ}) and $ \beta >1 $ (cosmologically interesting situation, i.e., significant mirror baryonic contribution to the dark matter), the ordinary sound speed in a Universe made of two sectors is always higher than that in an ordinary Universe. This is due to the presence of the term $ [(1 + x^4) / (1 + \beta) ] $ linked to the shift of matter-radiation equality epoch (\ref {shifteq}).

Now we define $a_{b\gamma }$ as the scale factor corresponding to the redshift 
\be
(1+z_{b\gamma }) = (a_{b\gamma})^{-1} = (\Omega _b / \Omega _\gamma ) = 3.9\cdot 10^4 (\Omega _b h^2) \;. 
\ee
Given that $1+z_{\rm rec} \simeq 1100$, baryon-photon equipartition occurs before recombination only if $\Omega_{\rm b}h^2>0.026$ (which seems unlikely, given the current estimates reported in \S~\ref{sec-now-cosm}). On the other hand, $a_{b\gamma}$ is always higher than $a_{eq}$, given that $\Omega_{\rm m} \geq \Omega_{\rm b}$. In the radiation era $\rho_{\gamma}\gg\rho_{\rm b}$, ensuring that $v_{\rm s} \simeq 1/\sqrt{3}$. In the interval between equipartition and decoupling, when $\rho_{\rm b}\gg\rho_{\gamma}$, eq.~(\ref{vsord}) gives $v_{\rm s} \simeq \sqrt{4\rho_{\gamma}/3\rho_{\rm b}}\propto a^{-1/2}$. After decoupling there is no more pressure equilibrium between baryons and photons, and $v_{\rm s}$ is just the velocity dispersion of a gas of hydrogen and helium, $v_{\rm s}\propto a^{-1}$. The situation whit $\Omega_{\rm b}h^2 > 0.026$ is resumed below:
\begin{equation}
v_{\rm s}(a) \propto\left\{\begin{array}{l}
 const \hspace{11mm} a<a_{b\gamma} \;,\\
 a^{-1/2} \hspace{12mm} a_{b\gamma}<a<a_{\rm dec} \;,\\
 a^{-1}  \hspace{15mm} a>a_{dec} \;.\\\end{array}\right.  \label{vsz}
\end{equation}
If $\Omega_{\rm b}h^2 < 0.026$, which is a more probable and simple case, $a_{b\gamma } > a_{dec}$, and the intermediate situation does not arise. It's very important to observe (see Padmanabhan (1993) \cite{padmbooksfu}) that at decoupling $v_{\rm s}^2$ drops from $(p_\gamma /\rho _b)$ to $(p_b /\rho _b)$. Since $p_\gamma \propto n_\gamma T$ while $p_b \propto n_b T$ with $(n_\gamma /n_b) \simeq 10^9 \gg 1$, this is a large drop in $v_{\rm s}$ and consequently in $\lambda _J$. More precisely, $v_{\rm s}^2$ drops from the value $(1/3)(\rho _\gamma / \rho _b) = (1/3)(\Omega _\gamma / \Omega _b)(1+z_{dec})$ to the value $(5/3) (T_{dec}/m_b) = (5/3) (T_0/m_b)(1+z_{dec})$, with a reduction factor 
\be
F_1 (\Omega_{\rm b}h^2 > 0.026) = { (v_{\rm s}^2)^{\rm (+)}_{\rm dec} \over (v_{\rm s}^2)^{\rm (-)}_{\rm dec} } = 6.63 \cdot 10^{-8} (\Omega _b h^2) \;,
\ee 
where $(v_{\rm s}^2)^{\rm (-)}_{\rm dec}$ and $(v_{\rm s}^2)^{\rm (+)}_{\rm dec}$ indicate the sound speed respectively just before and after decoupling.
In the case $\Omega_{\rm b}h^2 < 0.026$, $v_{\rm s}^2$ drops directly from $(1/3)$ to $(5/3) (T_{dec}/m_b) = (5/3) (T_0/m_b)(1+z_{dec})$ with a suppression
\be
F_2 (\Omega_{\rm b}h^2 < 0.026) = { (v_{\rm s}^2)^{\rm (+)}_{\rm dec} \over (v_{\rm s}^2)^{\rm (-)}_{\rm dec} } = 1.9 \cdot 10^{-9} \;.
\ee

In fig.~\ref{scalord1} we plot the previously discussed trends of the ordinary sound speed for a typical model with $\Omega_{\rm b}h^2 > 0.026$. If we reduce the value $ \Omega_{\rm b}h^2 $, $ a_{\rm b\gamma} $ goes toward higher values, while $ a_{\rm dec} $ remains fixed, so that for $\Omega_{\rm b}h^2 < 0.026$ decoupling happens before equipartition and the intermediate regime, where $ v_{\rm s} \propto a^{-1/2} $, disappears.

\begin{figure}[h]
  \begin{center}
    \leavevmode
    \epsfxsize = 9cm
    \epsffile{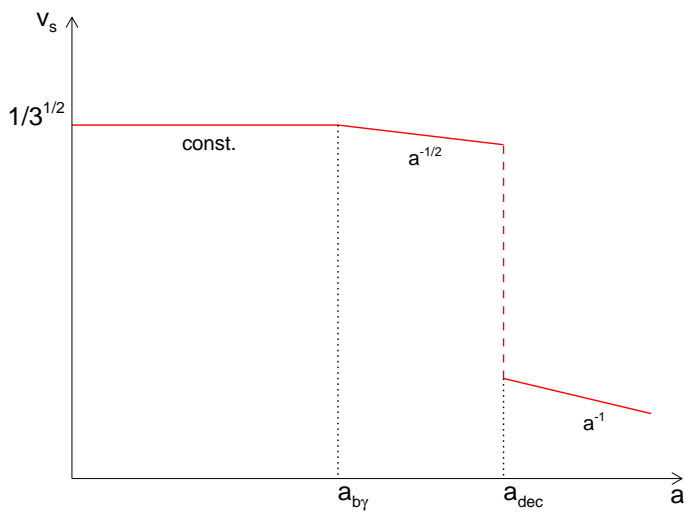}
  \end{center}
\caption{\small The trends of the ordinary sound speed as discussed in the text for the case $ \Omega_{\rm b}h^2 = 0.08 $; in this case there is also the intermediate regime $ v_{\rm s} \propto a^{-1/2} $. }
\label{scalord1}
\end{figure}


\subsubsection{The mirror sound speed}
\label{subsec_mirror_vs}

The mirror plasma contains more baryons and less photons than the ordinary one, $\rho'_b=\beta \rho_b$ and $\rho'_\gamma = x^4\rho_\gamma$.  Then, using eqs.~(\ref {vsord}) and (\ref {z-eq_2}), we have
\be \label{mirsound} 
v'_s(z) \simeq {1 \over {\sqrt3}} \left(1+ {{3\rho'_b} \over {4\rho'_\gamma}}\right)^{-1/2} 
\approx {1 \over {\sqrt3}} \left[ 1 +{3 \over 4} \left({{1+x^{-4}} \over {1+\beta ^{-1}}}\right) \left({{1+z_{\rm eq}} \over {1+z}}\right)\right]^{-1/2} \;. 
\ee 
Let us consider for simplicity the case when dark matter of the Universe is entirely due to M baryons, $\Omega_m\simeq\Omega'_b$ (i.e., $\beta \gg 1$). Hence, for the redshifts of cosmological relevance, $z\sim z_{\rm eq}$, we have $v'_s \sim 2x^2 /3$, which is always less than $v_s \sim 1/\sqrt{3}$  
(some example: if $x = 0.7$, $v_s' \approx 0.5 \cdot v_s$; if $x = 0.3$, $v_s' \approx 0.1 \cdot v_s$). In expression (\ref{mirsound}) it is crucial the presence of the factor $[(1+x^{-4}) / (1+\beta ^{-1})]$, which is always greater than 1, so that $v'_s < v_s$ during all the history of the Universe, and only in the limit $a \ll a_{\rm eq}$ we obtain $v'_s \simeq v_s \simeq 1/\sqrt{3}$. As we will see in the following, this has important consequences on structure formation scales.

According to what found in \S~\ref{mirror_dm}, in the mirror sector the scale of baryon - photon equality $a_{\rm b\gamma}'$ is dependent on $x$; if we remember the definition of the quantity $x_{\rm eq} \approx 0.046(\Omega_m h^2)^{-1}$, we find that for $x > x_{\rm eq}$ the decoupling occurs after the equipartition (as in the ordinary sector for $\Omega_{\rm b}h^2>0.026$), but for $x < x_{\rm eq}$ it occurs before (as for $\Omega_{\rm b}h^2<0.026$). It follows that, by taking care to interchange $a_{\rm b\gamma}$ with $a_{\rm b\gamma}'$ and $a_{\rm dec}$ with $a_{\rm dec}'$, we have for the sound speed the same trends with the scale factor in both sectors, as expressed in eq.~(\ref{vsz}), though with the aforementioned differences in the values. 

In fact, the matter-radiation equality for a single sector (ordinary) Universe, $ (a_{\rm eq})_{\rm ord} $, is always bigger than that for a two sectors (mirror) one, $ (a_{\rm eq})_{\rm mir} $, according to
\be \label{shiftaeq} 
(a_{\rm eq})_{\rm mir} = { {\left( 1+x^4 \right)} \over {\left( 1+\beta \right)} } ~ (a_{\rm eq})_{\rm ord} < (a_{\rm eq})_{\rm ord} \;,
\ee
while the baryon-photon equipartition transforms as
\be \label{shiftabg} 
a_{\rm b\gamma}' = { \Omega_{\gamma}' \over \Omega_b' } \simeq { \Omega_{\gamma} \, x^4 \over \Omega_b \, \beta} = a_{\rm b\gamma} { x^4 \over \beta } < a_{\rm b\gamma} \;.
\ee
Recalling our hypothesis $ x < 1 $ and $ \beta > 1 $, it is always verified that
\be \label{hiera1} 
a_{\rm b\gamma}' < a_{\rm eq} < a_{\rm b\gamma} \;.
\ee

If we consider now the drop in $(v_{\rm s}')^2$ at decoupling and call $F_1'$, $F_2'$ the factors of this drop in the cases when, respectively, $a_{b\gamma}'<a_{\rm dec}'$ or $a_{b\gamma}'>a_{\rm dec}'$, we find 
\be \label{f1pf2p}
F_1' = \beta x^{-3} F_1 ~~~~~~~~ {\rm and} ~~~~~~~~ F_2' = F_2 \;.
\ee
Some example: for $x = 0.7$, $F_1' \approx 2.9 \beta F_1$; for $x = 0.5$, $F_1' = 8 \beta F_1$; for $x = 0.3$, $F_1' \approx 37 \beta F_1$. We note that, if $\beta \geq 1$, $F_1'$ is at least an order of magnitude larger than $F_1$. In fact, after decoupling $(v_{\rm s}')^2 = (5/3) (T_{\rm dec}'/m_{\rm b}) = (5/3) (T_{\rm dec}/m_{\rm b}) = (v_{\rm s})^2$ (given $T_{\rm dec}' = T_{\rm dec}$), and between equipartition and recombination $(v_{\rm s}')^2 < (v_{\rm s})^2$. The relation above means that the drop is smaller in the mirror sector than in the ordinary one.
Obviously, before equipartition $(v_{\rm s}')^2 = (v_{\rm s})^2 = 1/3$, and for this reason the parameter $F_2$ is the same in both sectors.

In figure \ref{scalord2} we show the trends with scale factor of the mirror sound speed, in comparison with the ordinary one. The ordinary model is the same as in figure \ref{scalord1}, while the mirror model has $ x = 0.6 $ and $ \beta = 2 $ (this means that mirror baryons are twice the ordinary ones, but in these models we chose the latter four times their current estimation to better show the general behaviour). In the same figure are also indicated the aforementioned relative positions of the key epochs (photon-baryon equipartition and decoupling) for both sectors, together with the matter-radiation equality.

\begin{figure}[h]
  \begin{center}
    \leavevmode
    \epsfxsize = 12cm
    \epsffile{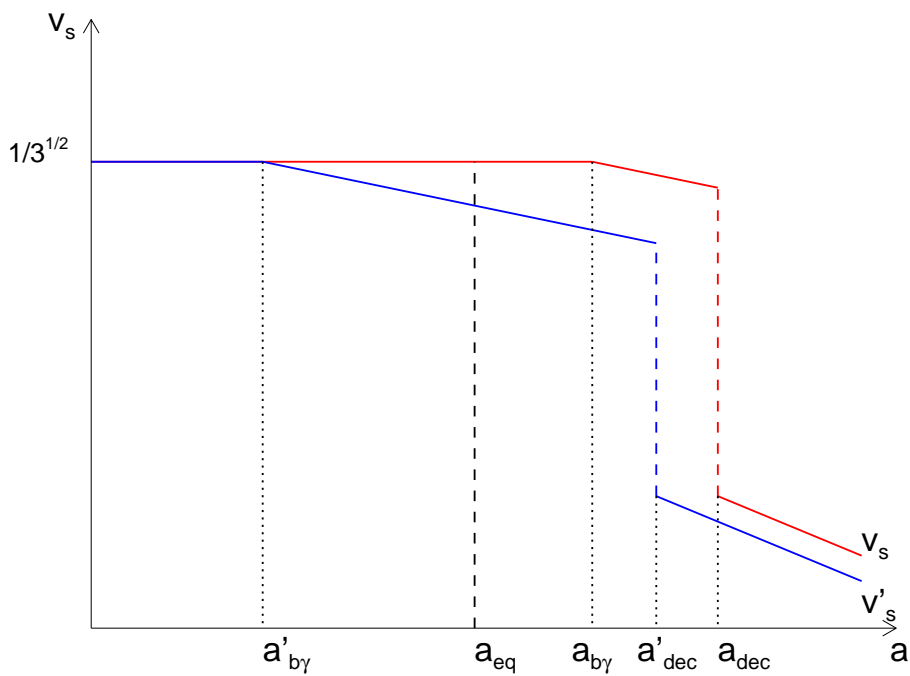}
  \end{center}
\caption{\small The trends of the mirror sound speed (blue line) as a function of the scale factor, compared with the ordinary sound speed (red line). The ordinary model has $\Omega_{\rm b}h^2 = 0.08 $ (as in figure \ref{scalord1}), while the mirror model has $ x = 0.6 $ and $ \beta = 2 $. Are also reported all the key epochs: photon-baryon equipartition and decoupling in both sectors, and the matter-radiation equality. }
\label{scalord2}
\end{figure}

In figure \ref{scalord3} the same ordinary and mirror sound speeds are plotted together with the velocity dispersion of a typical non baryonic cold dark matter candidate of mass $ \sim $ 1GeV. Note that the horizontal scale is expanded by some decade, because the key epochs for the CDM velocity evolution (when the particles become non relativistic, $ a_{\rm nr} $, and decoupling, $ a_{\rm d} $) occur at a much lower scale factor (for a brief review of the dark matter see chapter \ref{chap-lin-form}).

\begin{figure}[h]
  \begin{center}
    \leavevmode
    \epsfxsize = 12cm
    \epsffile{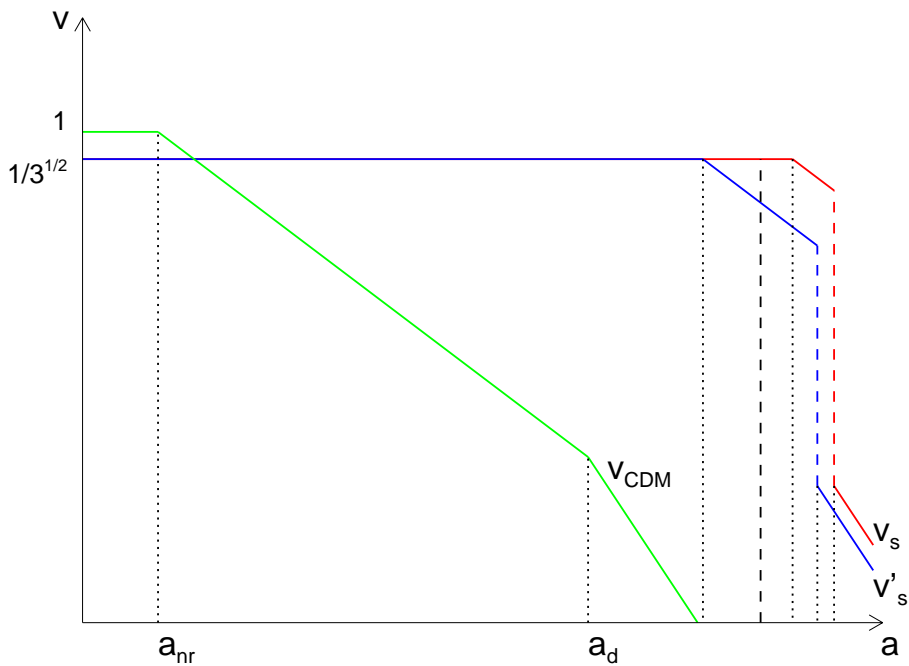}
  \end{center}
\caption{\small The trends of the mirror (blue line) and ordinary (red) sound speed compared with the velocity dispersion of a typical non baryonic cold dark matter candidate of mass $ \sim 1 \, {\rm GeV} $ (green); $ a_{\rm nr} $ and $ a_{\rm d} $ indicate the scale factors at which the dark matter particles become non relativistic or decouple. The models are the same as in figure \ref{scalord2}, but the horizontal scale is expanded by some decade to show the CDM velocity.}
\label{scalord3}
\end{figure}


\subsection{Evolution of the Jeans length and the Jeans mass}


\subsubsection{Ordinary sector}

Recalling what stated in \S~\ref{newth-app}, we have the Jeans length
\begin{equation}
\lambda_{\rm J}= 
  v_{\rm s}\sqrt{{\pi} \over {G\rho_{\rm dom}}} \;, \label{Jl1}
\end{equation}
where $\rho_{\rm dom}$ is the density of the dominant species, and the Jeans mass
\begin{equation}
M_{\rm J} = {4\over3}\pi\rho_{\rm b}\left({\lambda_{\rm J}} \over {2}\right)^3 
 = {\pi\over6}\rho_{\rm b}\left({\lambda_{\rm J}}\right)^3 \;, \label{Jm1}
\end{equation}
where the density is now that of the perturbed component (baryons).

Using the results of \S~\ref{evol_Vs} relative to the sound speed, we find 
for the evolution of the adiabatic Jeans length and mass in the case $\Omega_{\rm b} h^2 > 0.026$
\begin{equation}
\lambda_{\rm J}\propto\left\{\begin{array}{l}
 a^2\,\\ 
 a^{3/2}\,\\ 
 a\,\\
 a^{1/2} \,
 \\\end{array}\right.  \label{aJl}
\hspace{12mm}M_{\rm J}\propto{{\lambda _{\rm J}^3} \over {a^3}}\propto\left\{\begin{array}{l}
 a^3\hspace{20mm}a<a_{\rm eq}\,\\ 
 a^{3/2}\hspace{17mm}a_{\rm eq}<a<a_{\rm b\gamma }\,\\ 
 const.\hspace{13mm}a_{\rm b\gamma }<a<a_{\rm dec}\,\\
 a^{-3/2}\hspace{15mm}a_{\rm dec}<a \,
\end{array}\right.  \label{aJlm}
\end{equation}

In figures \ref{scalord4} and \ref{scalord5} we plot with the same horizontal scale the trends of the ordinary Jeans length and mass for $\Omega_{\rm b}h^2 = 0.08 $, a case in which the intermediate regime $ a_{\rm b\gamma } < a < a_{\rm dec} $ is present.

\begin{figure}[h]
  \begin{center}
    \leavevmode
    \epsfxsize = 9cm
    \epsffile{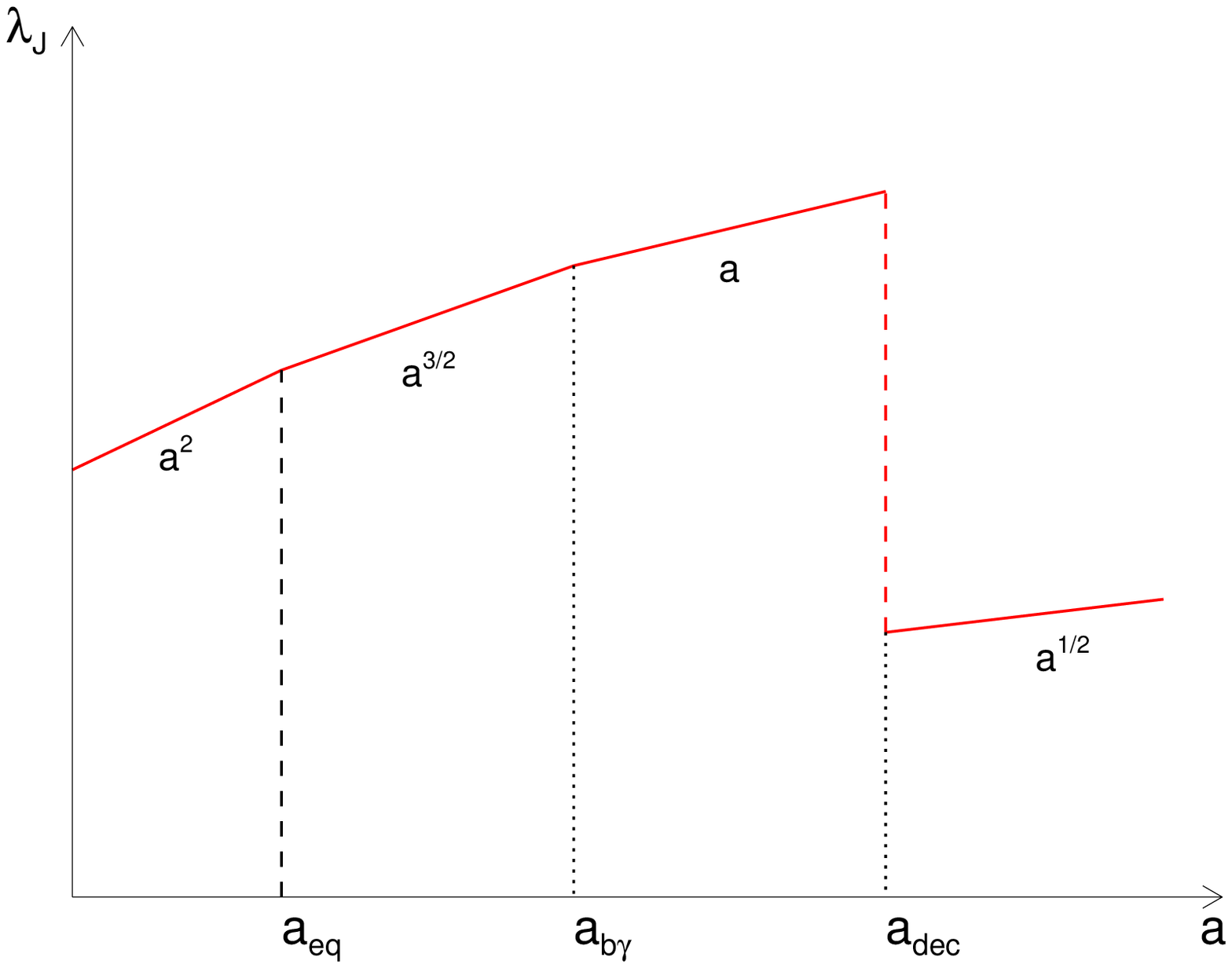}
  \end{center}
\caption{\small The trends of the ordinary Jeans length as discussed in the text for $\Omega_{\rm b}h^2 = 0.08 $, a case in which the intermediate regime $ a_{\rm b\gamma } < a < a_{\rm dec} $ is present. The horizontal scale is the same as in figure \ref{scalord1}.}
\label{scalord4}
\end{figure}

The greatest value of the Jeans mass is just before decoupling (see Padmanabhan (1993) \cite{padmbooksfu}), in the interval $a_{\rm b\gamma }<a<a_{\rm dec}$ , where
\be
M_{\rm J}(a \lsim a_{\rm dec}) = 1.47 \cdot  10^{14} M_\odot \left({\Omega_{\rm m}}\over{\Omega_{\rm b}}\right)^{1/2} \left(\Omega_{\rm m} h^2 \right)^{-2} \;,
\ee
that for $ \Omega_{\rm m} = \Omega_{\rm b} \simeq 0.1 h^{-2} $ is $ \sim 10^{16} M_\odot $. Just after decoupling we have
\be
M_{\rm J}(a \gsim a_{\rm dec}) = 2.5 \cdot  10^{3} M_\odot \left({\Omega_{\rm b}}\over{\Omega_{\rm m}}\right) \left(\Omega_{\rm m} h^2 \right)^{-1/2} \;,
\label{mj1after}
\ee
that for $ \Omega_{\rm m} = \Omega_{\rm b} \simeq 0.1 h^{-2} $ is $ \sim 10^{4} M_\odot $. This drop is very sudden and large, changing the Jeans mass by $F_1^{3/2} \simeq 1.7 \cdot 10^{-11} (\Omega _{\rm b} h^2)^{3/2}$. 

The maximum possible value of the Jeans mass is obtained for $ a_{\rm b\gamma } = a_{\rm dec} $, keeping constant $ a_{\rm eq} $ (i.e., substituting a fraction of baryons with the same amount of dark matter), and is given by
\be \label{mjmaxord}
M_{\rm J,max} = M_{\rm J}(a_{\rm b\gamma} = a_{\rm dec}) = 3.2 \cdot  10^{14} M_\odot \left({\Omega_{\rm m}}\over{\Omega_{\rm b}}\right)^{1/2} \left(\Omega_{\rm m} h^2 \right)^{-2} \left( a_{\rm dec} \over a_{\rm eq} \right)^{3/2} \;.
\ee

\begin{figure}[h]
  \begin{center}
    \leavevmode
    \epsfxsize = 9cm
    \epsffile{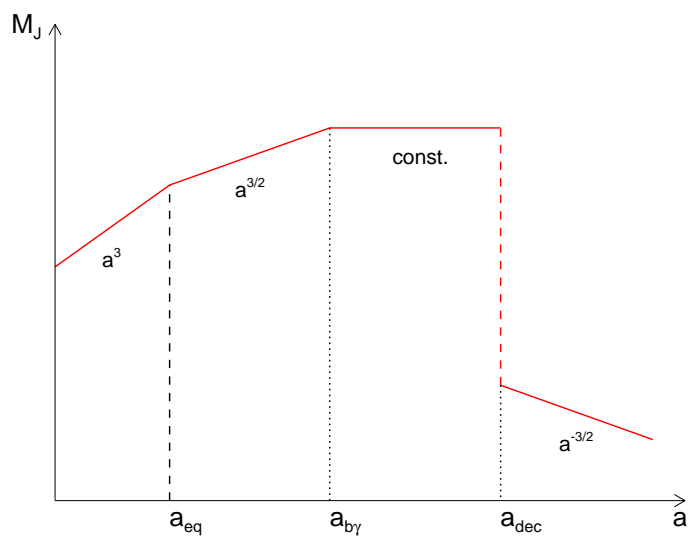}
  \end{center}
\caption{\small The trends of the ordinary Jeans mass as discussed in the text for $ \Omega_{\rm b}h^2 = 0.08 $, a case in which the intermediate regime $ a_{\rm b\gamma } < a < a_{\rm dec} $ is present. The horizontal scale is the same used for plotting $ \lambda_J $ in figure \ref{scalord4}.}
\label{scalord5}
\end{figure}

Otherwise, if $\Omega_{\rm b} h^2 \leq 0.026$, $a_{\rm b\gamma } > a_{\rm dec}$, there is no intermediate phase $a_{\rm b\gamma } < a < a_{\rm dec}$, and $M_{\rm J}(a \lsim a_{\rm dec})$ is larger
\be
M_{\rm J}(a \lsim a_{\rm dec}) \simeq 3.1 \cdot  10^{16} M_\odot \left({\Omega_{\rm b}}\over{\Omega_{\rm m}}\right) \left(\Omega_{\rm m} h^2 \right)^{-1/2} \;,
\ee
while after decoupling it takes the value in eq.~(\ref{mj1after}), so that the drop is larger, $F_2^{3/2} \simeq 8.3 \cdot 10^{-14}$.

We note that, with the assumption $\Omega_{\rm m}=\Omega_{\rm b}=1$, $M_{\rm J,max}$ (which is the first scale to become gravitationally unstable and collapse soon after decoupling) has the size of a supercluster of galaxies.


\subsubsection{Mirror sector}

In the mirror sector the Jeans length and mass are
\begin{equation}
\lambda_{\rm J}'  \simeq v_{\rm s}'\sqrt{{\pi} \over {G\rho_{\rm dom}}}
 \hspace{12mm}
 M_{\rm J}' = {4\over3}\pi\rho_{\rm b}'\left({\lambda_{\rm J}'} \over {2}\right)^3 \;.
\end{equation}
In this case it's no more sufficient to interchange $a_{\rm b\gamma }$ with $a_{\rm b\gamma }'$ and $a_{\rm dec}$ with $a_{\rm dec}'$, as made for the sound speed, because from relation (\ref{hiera1}) we note that in the mirror sector the photon-baryon equipartition happens before the matter-radiation equality (due to the fact that we are considering a mirror sector with more baryons and less photons than the ordinary one). It follows that, due to the shifts of the key epochs, the intervals of scale factor for the various trends are different. As usual, there are two different possibilities, $x > x_{\rm eq}$ and $x < x_{\rm eq}$ (which correspond roughly to $\Omega_{\rm b} h^2 > 0.026$ and $\Omega_{\rm b} h^2 < 0.026$ in an ordinary Universe), where, as discussed in \S ~\ref{subsec_mirror_vs}, for the second one the intermediate situation is absent.

\begin{figure}[h]
  \begin{center}
    \leavevmode
    \epsfxsize = 12cm
    \epsffile{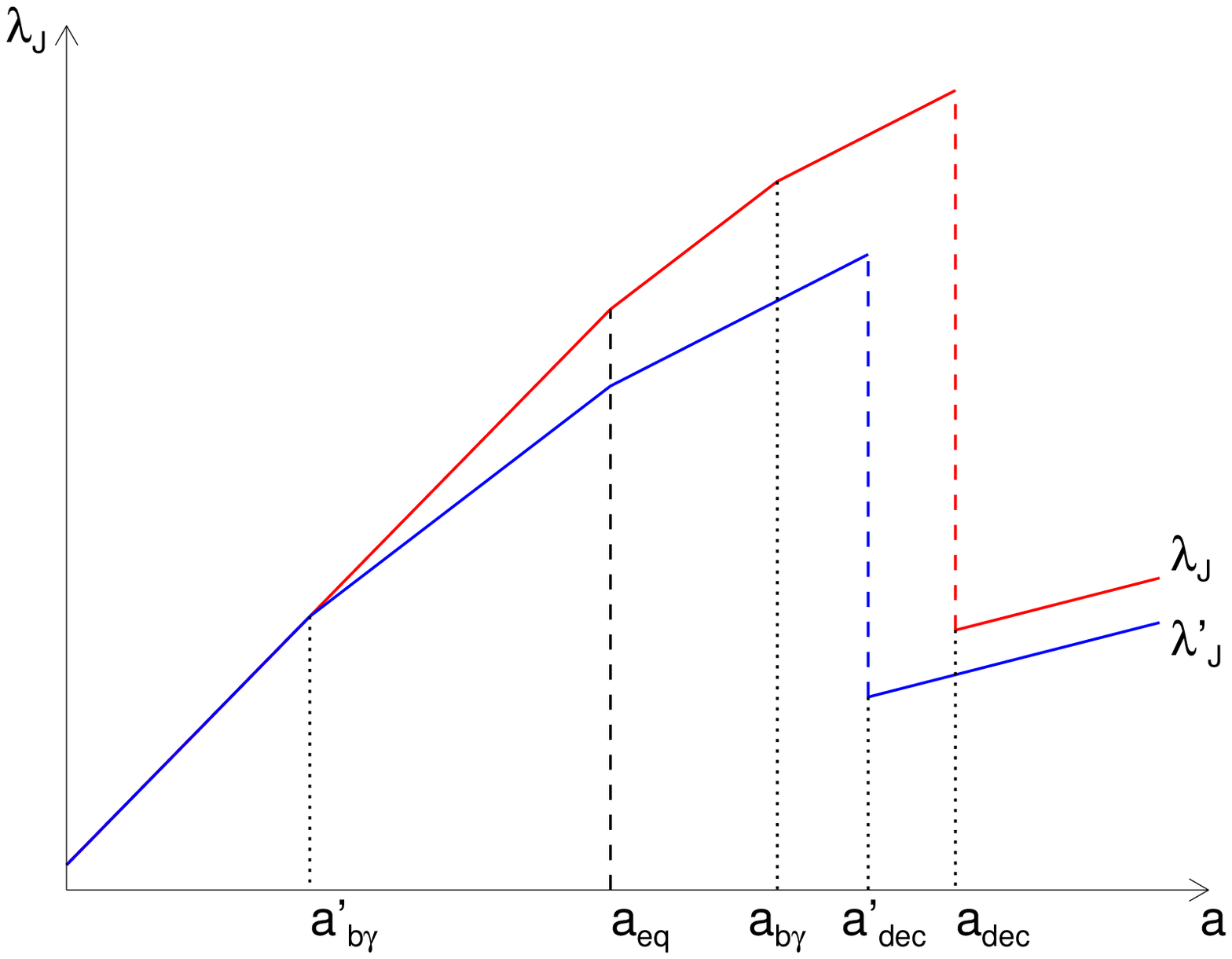}
  \end{center}
\caption{\small The trends of the mirror Jeans length (blue line) as a function of the scale factor, compared with the ordinary Jeans length (red line). The ordinary model has $\Omega_{\rm b}h^2 = 0.08 $ (as in figure \ref{scalord4}), while the mirror model has $ x = 0.6 $ and $ \beta = 2 $. The horizontal scale is the same as in figure \ref{scalord2}.
We note that the same behaviours reported for the ordinary sector in figure \ref{scalord4} are present in the mirror sector for different intervals of scale factor.}
\label{scalord6}
\end{figure}

Using the results of \S~\ref{evol_Vs} for the sound speed, we find 
the evolution of the adiabatic Jeans length and mass in the case $x > x_{\rm eq}$
\begin{equation}
\lambda_{\rm J}' \propto \left\{\begin{array}{l}
 a^2\,\\ 
 a^{3/2}\,\\ 
 a\,\\
 a^{1/2} \,
 \\\end{array}\right.
\hspace{12mm}M_{\rm J}' \propto {{(\lambda _{\rm J}')^3} \over {a^3}}\propto\left\{\begin{array}{l}
 a^3 \hspace{20mm} a < a_{\rm b\gamma }' \,\\ 
 a^{3/2} \hspace{17mm} a_{\rm b\gamma }' < a < a_{\rm eq} \,\\ 
 const. \hspace{13mm} a_{\rm eq} < a < a_{\rm dec}' \,\\
 a^{-3/2} \hspace{15mm} a_{\rm dec}' < a \,
\end{array}\right.
\end{equation}

We plot in figures \ref{scalord6} and \ref{scalord8} the trends of the mirror Jeans length and mass compared with those for the ordinary sector; the parameters of both mirror and ordinary models are the ones previously used, i.e. $\Omega_{\rm b}h^2 = 0.08 $, $ x = 0.6 > x_{\rm eq} $ and $ \beta = 2 $. If we remember eq.~(\ref {f1pf2p}), for the mirror model the drop in the Jeans mass at decoupling is $(F_1')^{3/2} = \beta^{3/2} x^{-9/2} (F_1)^{3/2} $, which, given our bounds on $ x $ and $ \beta $, is greater than $ (F_1)^{3/2} $. We give here some numerical example: for $x = 0.7$, $(F_1')^{3/2} \approx 5 \beta^{3/2} (F_1)^{3/2}$; for $x = 0.5$, $(F_1')^{3/2} \approx 23 \beta^{3/2} (F_1)^{3/2}$; for $x = 0.3$, $(F_1')^{3/2} \approx 225 \beta^{3/2} (F_1)^{3/2}$; for $x = 0.6$ and $ \beta = 2 $ (the case of figures \ref{scalord6} and \ref{scalord8}), $(F_1')^{3/2} \approx 28 (F_1)^{3/2}$.

\begin{figure}[h]
  \begin{center}
    \leavevmode
    \epsfxsize = 12cm
    \epsffile{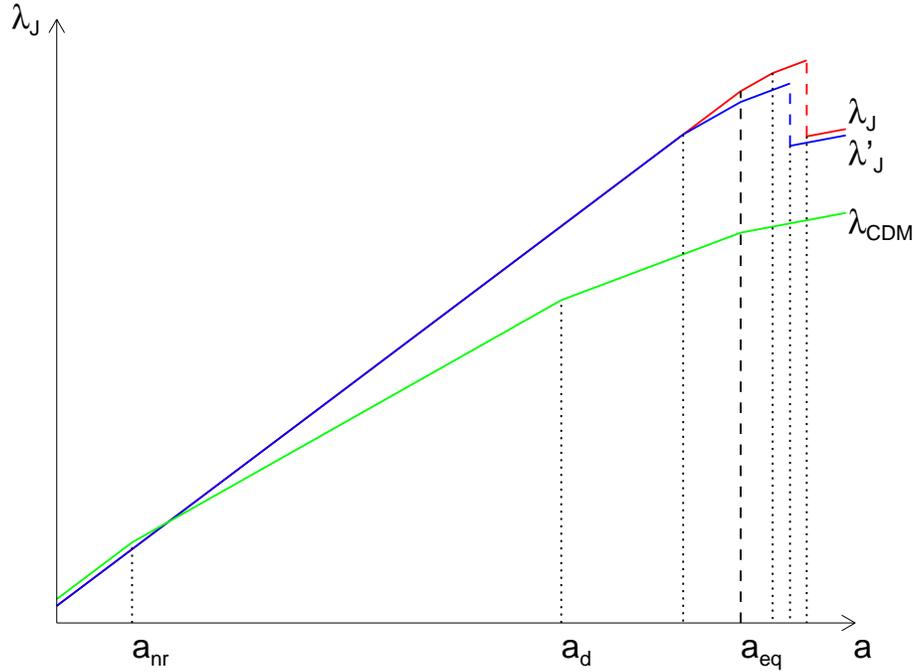}
  \end{center}
\caption{\small The trends of the mirror (blue line) and ordinary (red) Jeans length compared with that of a typical non baryonic cold dark matter candidate of mass $ \sim 1 \, {\rm GeV} $ (green); $ a_{\rm nr} $ and $ a_{\rm d} $ indicate the scale factors at which the dark matter particles become non relativistic or decouple, respectively. The models are the same as in figure \ref{scalord6}, but the horizontal scale is expanded by some decade to show the CDM Jeans length, as in figure \ref{scalord3}.}
\label{scalord7}
\end{figure}

\begin{figure}[h]
  \begin{center}
    \leavevmode
    \epsfxsize = 12cm
    \epsffile{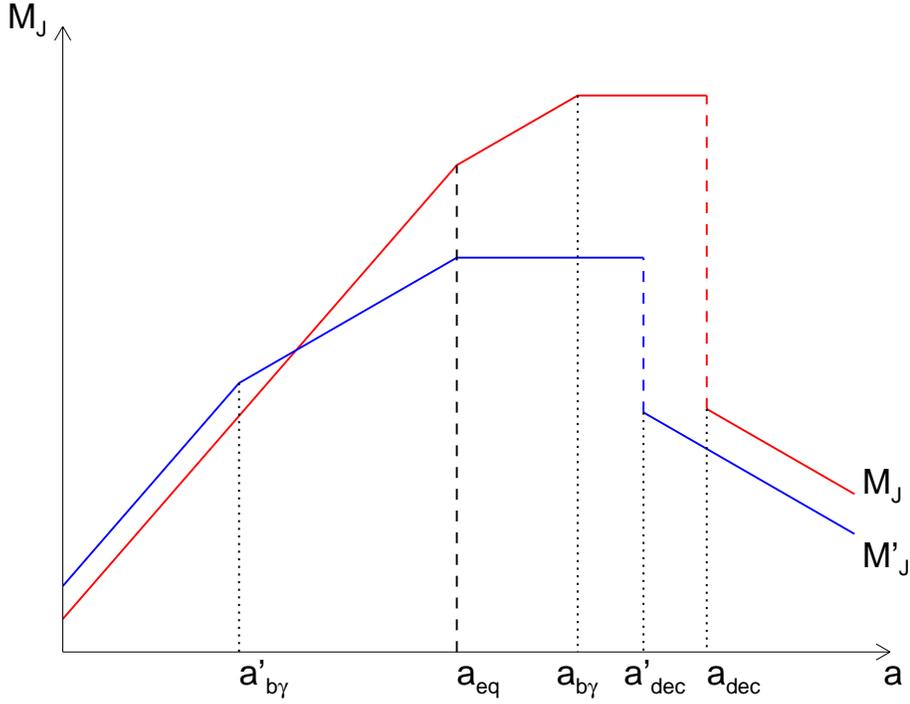}
  \end{center}
\caption{\small The trends of the mirror Jeans mass (blue line) as a function of the scale factor, compared with the ordinary Jeans mass (red line). The models and the horizontal scale are the same as in figure \ref{scalord6}.}
\label{scalord8}
\end{figure}

If we now consider the expression (\ref {shiftabg}), we have
\be
{ a_{\rm b\gamma }' \over a_{\rm eq} } = \left( { 1 + \beta } \over { \beta } \right) \left( { x^4 } \over { 1 + x^4 } \right) \;,
\ee
which can be used to express the value of the mirror Jeans mass in the interval $a_{\rm eq} < a < a_{\rm dec}'$ (where $ M_{\rm J}' $ takes the maximum value) in terms of the ordinary Jeans mass in the corresponding ordinary interval $a_{\rm b\gamma} < a < a_{\rm dec}$. We obtain
\be
M_{\rm J}'(a \lsim a_{\rm dec}') \approx 
  \beta^{-1/2} \left( { x^4 \over {1 + x^4} } \right)^{\rm 3/2} 
  \cdot M_{\rm J}(a \lsim a_{\rm dec}) \;,
\ee
which, for $\beta \geq 1$ and $x < 1$, means that the Jeans mass for the M baryons is lower than for the O ones over almost the entire ($\beta$-$x$) parameter space, with implications for the structure formation process. If, e.g., $ x = 0.6 $ and $ \beta = 2 $, then $ M_{\rm J}' \sim 0.03 \; M_{\rm J} $. We can also express the same quantity in terms of $ \Omega_{\rm b}$, $x$ and $\beta $ only, in the case that all the dark matter is in the form of mirror baryons, as
\be
M_{\rm J}'(a \lsim a_{\rm dec}') \approx 
  3.2 \cdot  10^{14} M_\odot \;
  \beta^{-1/2} ( 1 + \beta )^{-3/2} \left( x^4 \over {1+x^4} \right)^{3/2} 
  ( \Omega_{\rm b} h^2 )^{-2} \;.
\ee

It's important to stress that these quantities are strongly dependent on the values of the free parameters $ x $ and $ \beta $, which shift the key epochs and change their relative position. We can describe some case useful to understand the general behaviour, but if we want an accurate solution of a particular model, we must unambiguously identify the different regimes and solve in detail the appropriate equations.

\begin{figure}[h]
  \begin{center}
    \leavevmode
    \epsfxsize = 12cm
    \epsffile{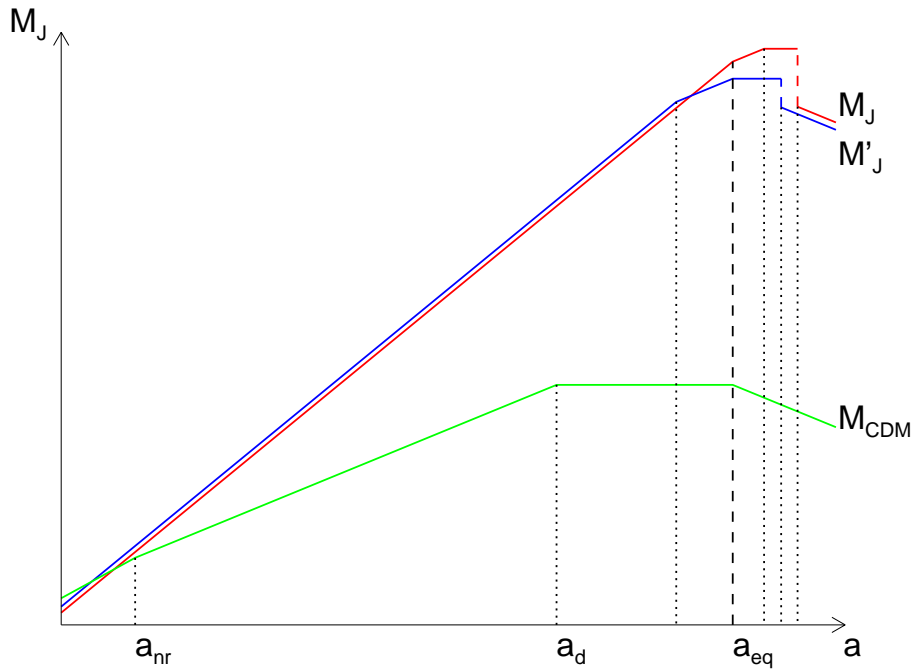}
  \end{center}
\caption{\small The trends of the mirror (blue line) and ordinary (red) Jeans mass compared with those of a typical non baryonic cold dark matter candidate of mass $ \sim 1 \, {\rm GeV} $ (green); $ a_{\rm nr} $ and $ a_{\rm d} $ indicate the scale factors at which the dark matter particles become non relativistic or decouple, respectively. The model parameters and the horizontal scale are the same as in figure \ref{scalord7}.}
\label{scalord9}
\end{figure}

In figures \ref{scalord7} and \ref{scalord9} we plot the trends of the mirror and ordinary Jeans length and mass compared with those of a typical non baryonic cold dark matter candidate of mass $ \sim $ 1GeV. Apart from the usual expansion of the horizontal scale, due to the much lower values of the CDM key epochs as compared to the baryonic ones, a comparison of the mirror scenario with the cold dark matter one shows that the maximal value of the CDM Jeans mass is $ 10^{15} $ times lower than that for mirror baryons, which is a very big value.
This implies that a very large range of mass scales, which in a mirror baryonic scenario oscillate before decoupling, in a cold dark matter scenario would grow unperturbed during all the time (for more details see \S~\ref{mirevolpert}).

For the case $ x < x_{\rm eq} $, both $ a_{\rm b\gamma}' $ and $ a_{\rm dec}' $ are smaller than the previous case $ x > x_{\rm eq} $, while the matter-radiation equality remains practically the same; as explained in \S~\ref{mirror_dm}, the mirror decoupling (with the related drop in the associated quantities) happens before the matter-radiation equality, and the trends of the mirror Jeans length and mass are the following
\begin{equation}
\lambda_{\rm J}' \propto \left\{\begin{array}{l}
 a^2\,\\ 
 a^{3/2}\,\\ 
 a\,\\
 a^{1/2} \,
 \\\end{array}\right.
\hspace{12mm}M_{\rm J}' \propto {{(\lambda _{\rm J}')^3} \over {a^3}}\propto\left\{\begin{array}{l}
 a^3 \hspace{20mm} a < a_{\rm b\gamma }' \,\\ 
 a^{3/2} \hspace{17mm} a_{\rm b\gamma }' < a < a_{\rm dec}' \,\\ 
 const. \hspace{13mm} a_{\rm dec}' < a < a_{\rm eq} \,\\
 a^{-3/2} \hspace{15mm} a_{\rm eq} < a \,
\end{array}\right.
\end{equation}

In figure \ref{scalord10} we plot the mirror Jeans mass for the three different possibilities: $ x < x_{\rm eq} $, $ x > x_{\rm eq} $ and $ x = x_{\rm eq} $ (the transition between the two regimes).

\begin{figure}[h]
  \begin{center}
    \leavevmode
    \epsfxsize = 12cm
    \epsffile{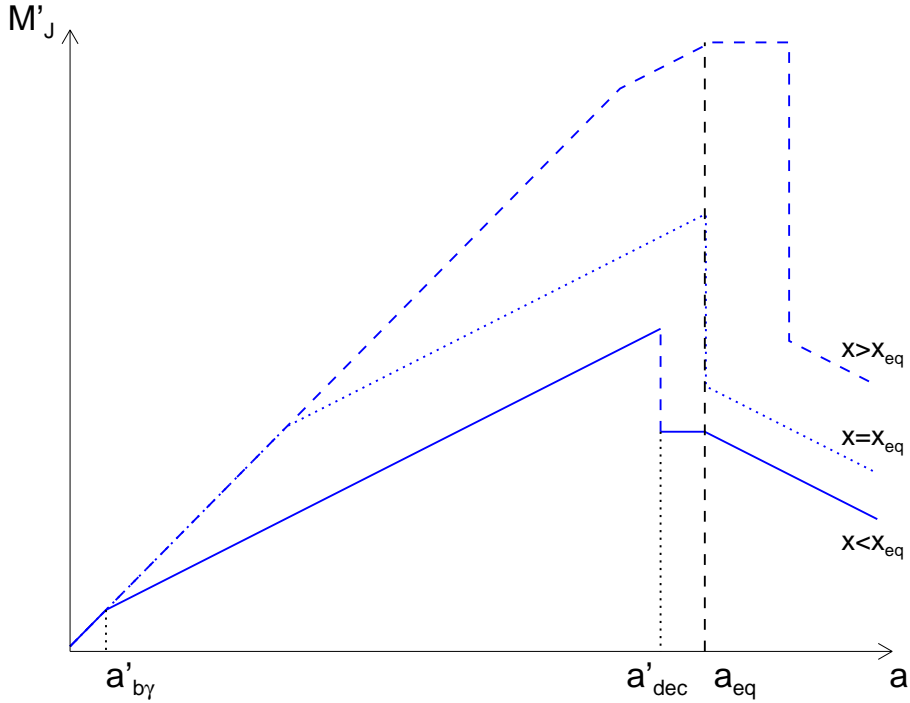}
  \end{center}
\caption{\small The trends of the mirror Jeans mass for the cases $ x < x_{\rm eq} $ (solid line), $ x = x_{\rm eq} $ (dotted) and $ x > x_{\rm eq} $ (dashed). The model with $ x > x_{\rm eq} $ is the same as in figure \ref{scalord8}, the others are obtained changing only the value of $ x $ and keeping constant all the other parameters. As clearly shown in the figure, the only key epoch which remains almost constant in the three models is the matter-radiation equality; the mirror baryon-photon equipartition and decoupling indicated are relative to the model with $ x < x_{\rm eq} $. It's also evident the change in the trends when $ x $ becomes lower than $ x_{\rm eq} $, due to the fact that $ a_{\rm dec}' $ becomes lower than $ a_{\rm eq} $.}
\label{scalord10}
\end{figure}


\subsection{Evolution of the Hubble mass}
\label{subsec_evol_MH2}

The trends of the Hubble length and mass for the baryonic component are the same as for for the non-baryonic ones (defined in \S ~\ref{subsec_evol_MH1})

\begin{equation}
\lambda_{\rm H} \propto \left\{\begin{array}{l}
 a^2\,\\ 
 a^{3/2}\,
 \\\end{array}\right.
\hspace{12mm}M_{\rm H} \propto {{(\lambda _{\rm H})^3} \over {a^3}}\propto\left\{\begin{array}{l}
 a^3 \hspace{20mm} a < a_{\rm eq} \,\\ 
 a^{3/2} \hspace{17mm} a > a_{\rm eq} \,
\end{array}\right.
\end{equation}

In this case, it should be emphasized that 
during the period of domination of photons ($a<a_{\rm b\gamma }$) the baryonic Jeans mass is of the same order of the Hubble mass. In fact, following definitions (\ref{nbJl}), (\ref{nbJm}) and (\ref{nbHm}), given $H^2 = (8\pi /3) G \rho _{\rm \gamma } \sim 1/{\lambda _{\rm H}^2}$, we find
\be
{{M_{\rm J}}\over{M_{\rm H}}} = {{\lambda _{\rm J}^3}\over{\lambda _{\rm H}^3}} = \left( {{\pi \sqrt{8} v_{\rm s} } \over {\sqrt{3} }} \right)^3 \simeq 26 \;.
\ee

We plot the trends of the Hubble mass in figures \ref {mirmasssca1} and \ref{mirmasssca2} together with other fundamental mass scales.


\subsection{Dissipative effects: collisional damping}
\label{disseff-colldamp}

We now turn our attention to physical dissipative processes that can modify the purely gravitational evolution of perturbations. In baryonic models the most important physical phenomenon is the interaction between baryons and photons in the pre-recombination era, and the consequent dissipation due to viscosity and heat conduction.


\subsubsection{Ordinary baryons}

Adiabatic perturbations in the photon-baryon plasma suffer from collisional damping around the time of recombination because the perfect fluid approximation breaks down. As we approach decoupling, the photon mean free path increases and photons can diffuse from the overdense into the underdense regions, thereby smoothing out any inhomogeneities in the primordial plasma. The effect is known as ``collisional dissipation'' or ``Silk damping'' (Silk (1967)~\cite{silk215}). 
To obtain an estimate of the effect, we consider the physical (proper) distance associated with the photon mean free path
\begin{equation}
\lambda _{\gamma} = {{1} \over {X_{\rm e}{\rm n}_{\rm e}\sigma_{\rm T}}} \simeq 10^{29}a^3X_{\rm e}^{-1}\left(\Omega_{\rm b}h^2\right)^{-1}\,{\rm cm} \;,  \label{pmp}
\end{equation}
where $X_{\rm e}$ is the electron ionization factor, ${\rm n}_{\rm e}\propto a^{-3}$ is the number density of the free electrons and $\sigma_{\rm T}$ is the cross section for Thomson scattering. Clearly, all baryonic perturbations with wavelengths smaller than $\lambda _{\gamma}$ will be smoothed out by photon free streaming. The perfect fluid assumption breaks down completely when $\lambda\ll\lambda _{\gamma}$. Damping, however, occurs on scales much larger than $\lambda _{\gamma}$ as the photons slowly diffuse from the overdense into the underdense regions, dragging along the still tightly coupled baryons. 
Integrating up to recombination time we obtain the total physical distance traveled by a typical photon (see Kolb \& Turner (1990)~\cite{kolbbookeu})
\begin{equation}
\lambda _{\rm S}= 
\sqrt{{{3\over5}}(\lambda _{\gamma})_{\rm rec}t_{\rm
rec}}\,\,\simeq\,\,3.5\left(\Omega_b h^2\right)^{-3/4}\,{\rm Mpc} \;,
\label{lS}
\end{equation}
and the associated mass scale, which is known as the ``Silk mass'', is given by
\begin{equation}
M_{\rm S} = {4\over3}\pi\rho_{\rm b}\left({{\lambda _{\rm S}} \over {2}}\right)^3\simeq6.2\times10^{12}\left(\Omega_b h^2\right)^{-5/4}\,{\rm M}_{\odot} \;,\label{Sm}
\end{equation}
which, assuming $ \Omega_b h^2 \simeq 0.02 $ (as actually estimated, see \S~\ref{sec-now-cosm}), gives $ M_{\rm S} \simeq 8 \times10^{14}~{\rm M}_{\odot} $.
The dissipative process we considered above causes 
that fluctuations on scales below the Silk mass are completely obliterated by the time of recombination and no structure can form on these scales. Alternatively, one might say that adiabatic perturbations have very little power left on small scales.


\subsubsection{Mirror baryons}

In the mirror sector too, obviously, the photon diffusion from the overdense to underdense regions induces a dragging of charged particles and washes out the perturbations at scales smaller than the mirror Silk scale 
\be
\lambda'_S \simeq 3 \times f(x)(\Omega_m h^2)^{-3/4} \;{\rm Mpc} \;, 
\label{mirsilklamb1}
\ee
where $f(x)=x^{5/4}$ for $x > x_{\rm eq}$ and $f(x) = (x/x_{\rm eq})^{3/2} x_{\rm eq}^{5/4}$ for $x < x_{\rm eq}$, and we considered the initial hypothesis $\Omega _m \simeq \Omega _b'$. 

Thus, the density perturbation scales which can run the linear growth after the matter-radiation equality epoch are limited by the length $\lambda'_S$. 
The smallest perturbations that survive the Silk damping will have the mass 
\be
M'_S \sim [f(x) / 2]^3 (\Omega_m h^2)^{-5/4} 10^{12}~ M_\odot \;,
\ee
which should be less than $2\times 10^{12} ~ M_\odot$ in view of the BBN bound $x <0.64$.  Interestingly, for $x\sim x_{\rm eq}$ we obtain
\be
M'_S (x = x_{\rm eq}) \sim 10^{7}~ (\Omega_m h^2)^{-5} ~M_\odot \;,
\label{mirsilkm2}
\ee
which, for the current estimate of $ \Omega_m h^2 $ (see \S~\ref{sec-now-cosm}), gives $ M'_S \sim 3 \times 10^{10} ~M_\odot $, a typical galaxy mass.

At this point it is very interesting a comparison between different damping scales, collisional (ordinary and mirror baryons) and collisionless (non-baryonic dark matter). Recalling what stated in \S~\ref{diseffsec}, we have that for hot dark matter (as a neutrino of mass $\sim$10 eV) $M_{FS}^\nu \sim 10^{15} ~ M_\odot$, while for a typical warm dark matter candidate of mass $\sim$1 keV, $M_{FS}^{WDM} \sim 10^{9} - 10^{10} ~ M_\odot$. From eq.~(\ref{mirsilkm2}) it is evident that the dissipative scale for mirror Silk damping is analogous to that for WDM free streaming. Consequently, the cutoff effects on the corresponding large scale structure power spectra are similar, though with important differences due to the presence of oscillatory features in the mirror baryons spectra, which makes them distinguishable one from the other (for a detailed presentation of the mirror power spectra see next chapter). In figure \ref{mirsilksca1} we show this comparison together with the trend of the mirror Silk mass over a cosmologically interesting range of $ x $.

\begin{figure}[h]
  \begin{center}
    \leavevmode
    \epsfxsize = 12cm
    \epsffile{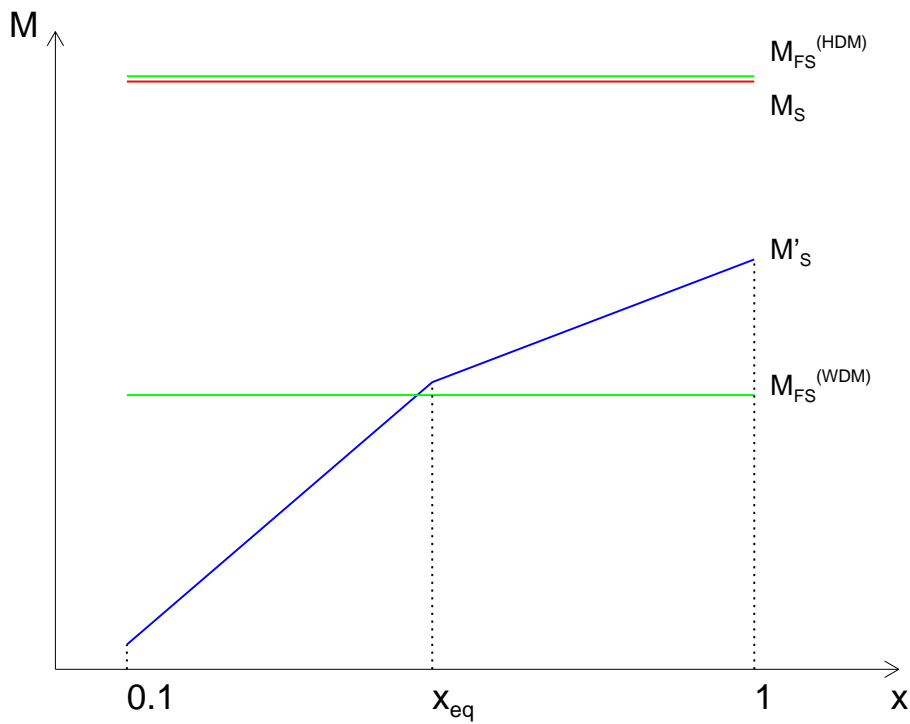}
  \end{center}
\caption{\small The trend of the mirror Silk mass (blue line) over a cosmologically interesting range of $ x $, which contains $ x_{eq} $ (we considered $ \Omega_{\rm m} h^2 \simeq 0.15 $, so $ x_{\rm eq} \simeq 0.3 $). The axis are both logarithmic. We show for comparison also the values of the ordinary Silk mass (red) and of the free streaming mass (green) for typical HDM and WDM candidates.}
\label{mirsilksca1}
\end{figure}


\passo
\def \mirror_bar_struc{Scenarios}
\section{\mirror_bar_struc}
\label{mirror_bar_struc}
\markboth{Chapter \ref{chap-mirror_univ_2}. ~ \mir_univ_2}
                    {\S \ref{mirror_bar_struc} ~ \mirror_bar_struc}

After the description of the fundamental scales for structure formation, let us now put together all the informations and discuss the mirror scenarios. They are essentially two, according to the value of $ x $, which can be higher or lower than $ x_{\rm eq} $, and are shown respectively in figures \ref {mirmasssca1} and \ref {mirmasssca2}, which will be our references during the present section.

Typically, adiabatic perturbations with sizes larger than the maximum value of the Jeans mass, which is $ M_{\rm J}'(a_{\rm eq}) $ for $ x > x_{\rm eq} $ and $ M_{\rm J}'(a_{\rm dec}') $ for $ x < x_{\rm eq} $, experience uninterrupted growth. In particular, as discussed in appendix \ref{app-strucform} and summarized in table \ref{tb2}, they grow as $\delta_{\rm b}\propto a^2$ before matter-radiation equality and as $\delta_{\rm b}\propto a$ after equality. Fluctuations on scales in the mass interval $ M_{\rm S}' < M < M_{\rm J,max} $ grow as $\delta_{\rm b}\propto a^2$ while they are still outside the Hubble radius. After entering the horizon and until recombination these modes oscillate like acoustic waves. The amplitude of the oscillation is constant before equilibrium but decreases as $a^{-1/4}$ between equipartition and recombination. After decoupling the modes become unstable again and grow as $\delta_{\rm b}\propto a$. Finally all perturbations on scales smaller than the value of the Silk mass 
are dissipated by photon diffusion. 

\begin{figure}[h]
  \begin{center}
    \leavevmode
    \epsfxsize = 12cm
    \epsffile{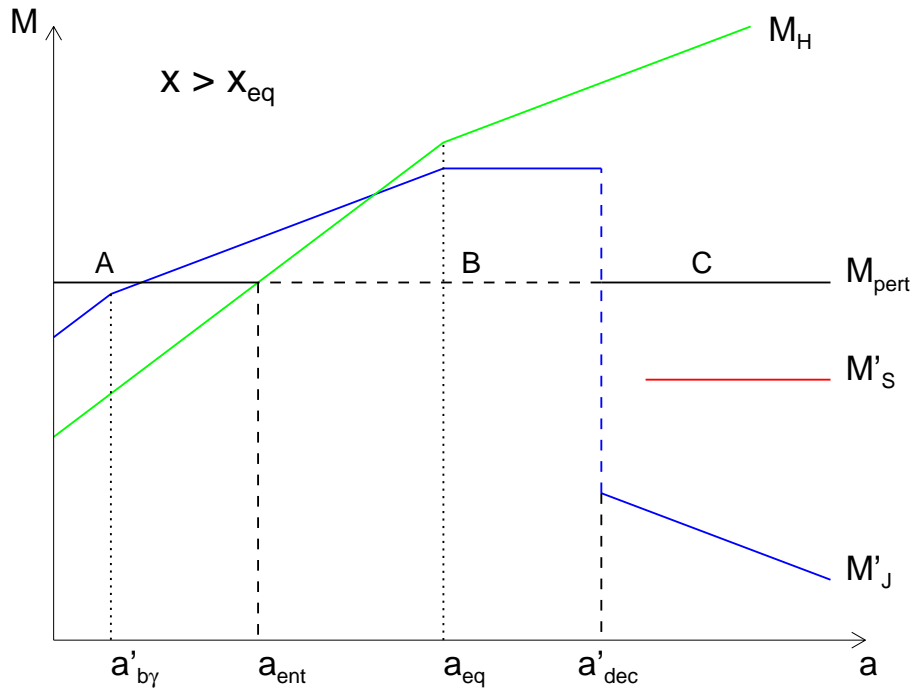}
  \end{center}
\caption{\small Typical evolution of a perturbed scale $ M_{\rm pert} $ (black line) in adiabatic mirror baryonic dark matter scenario with $ x > x_{\rm eq} $. 
The figure shows the Jeans mass $ M_{\rm J}' $ (blue), the Silk mass $ M_{\rm S}' $ (red) and the Hubble mass $ M_{\rm H} $ (green). The time of horizon crossing of the perturbation is indicated by $ a_{\rm ent} $. Are also indicated the three evolutionary stages: during stage A ($ a < a_{\rm ent} < a_{\rm eq} $) the mode grows as $\delta_{\rm b} \propto a^2$; throughout stage B ($ a_{\rm ent} < a < a_{\rm dec}' $) the perturbation oscillates; finally, in stage C ($ a > a_{\rm dec}' $) the mode becomes unstable again and grows as $ \delta_{\rm b} \propto a $. Note that fluctuations with size smaller than $ M_{\rm S}' $ are wiped out by photon diffusion.}
\label{mirmasssca1}
\end{figure}

Given this general behaviour, the schematic evolution of an adiabatic mode with a reference mass scale $ M_{\rm pert} $, with $ M_{\rm S}' < M_{\rm pert} < M_{\rm J}'(a_{\rm eq}) $, is depicted in figure \ref{mirmasssca1} for $ x > x_{\rm eq} $.
We distinguish between three evolutionary stages, called A, B and C, depending on the size of the perturbation and 
on the cosmological parameters $ \Omega_{\rm b} h^2 $, $ x $ and $ \beta $, which determine the behaviour of the mass scales, and in particular the key moments (time of horizon crossing and decoupling) and the dissipative Silk scale. 
During stage A, i.e. before the horizon crossing ($ a < a_{\rm ent} < a_{\rm eq} $), the mode grows as $\delta_{\rm b} \propto a^2$; throughout stage B ($ a_{\rm ent} < a < a_{\rm dec}' $) the perturbation enters the horizon, baryons and photons feel each other, and it oscillates; finally, in stage C ($ a > a_{\rm dec}' $), the photons and baryons decouple and the mode becomes unstable again and grows as $ \delta_{\rm b} \propto a $. 
Note that fluctuations with size greater than $ M_{\rm J}'(a_{\rm eq}) $ grow uninterruptedly (because after horizon crossing the photon pressure cannot balance the gravity), changing the trend from $ a^2 $ before MRE to $ a $ after it, while those with sizes smaller than $ M_{\rm S}' $ are completely washed out by photon diffusion.

After decoupling all surviving perturbations (those with $ M_{\rm pert} > M'_{\rm S} $) grow steadily until their amplitude becomes of order unity or larger. At that point the linear theory breaks down and one needs to employ a different type of analysis. 

\begin{figure}[h]
  \begin{center}
    \leavevmode
    \epsfxsize = 12cm
    \epsffile{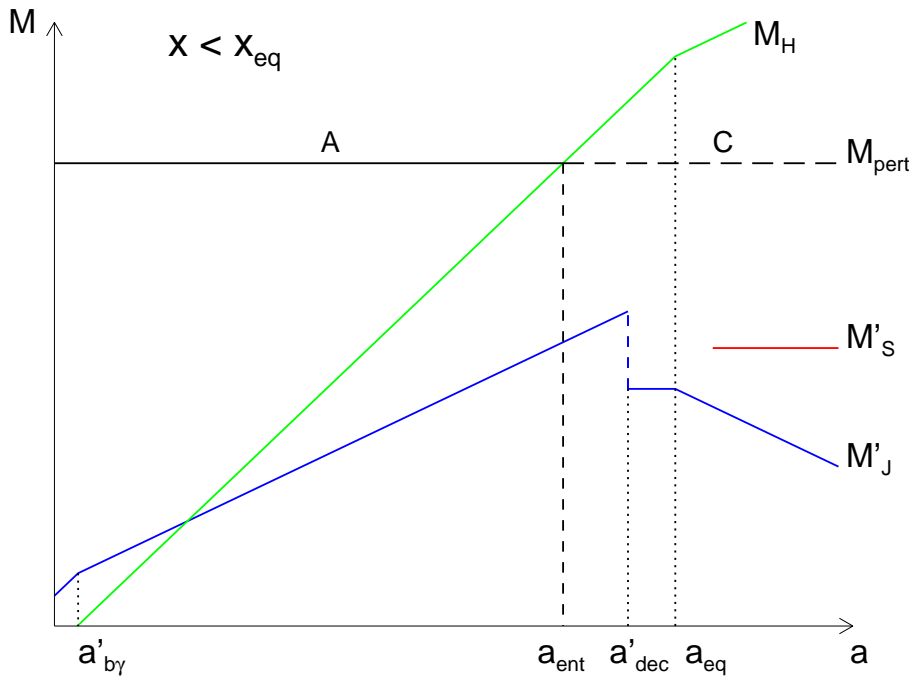}
  \end{center}
\caption{\small Typical evolution of a perturbed scale $ M_{\rm pert} $ (black line) in adiabatic mirror baryonic dark matter scenario with $ x < x_{\rm eq} $. 
The value of $ M_{\rm pert} $ is the same as in figure \ref{mirmasssca1}. The time of horizon crossing of the perturbation is indicated by $ a_{\rm ent} $. The figure shows the Jeans mass $ M_{\rm J}' $ (blue), the Silk mass $ M_{\rm S}' $ (red) and the Hubble mass $ M_{\rm H} $ (green). Unlike the case $ x > x_{\rm eq} $ (shown in the previous figure), now there are only the two evolutionary stages A ($ a < a_{\rm ent} $) and C ($ a > a_{\rm ent} $). Fluctuations with size smaller than $ M_{\rm S}' $ are wiped out by photon diffusion, but now the Silk mass is near to the maximum of the Jeans mass.}
\label{mirmasssca2}
\end{figure}

If we look, instead, at the schematic evolution of an adiabatic mode with the same reference mass scale $ M_{\rm pert} $ but for $ x < x_{\rm eq} $, as reported in figure \ref{mirmasssca2}, we immediately notice the lower values of the maximum Jeans mass and the Silk mass, which are now similar. So, for the plotted perturbative scale there are now only the two stages A and C. In general, depending on its size, the perturbation mass can be higher or lower than the Silk mass (and approximatively also than the maximum Jeans mass), so modes with $ M_{\rm pert} > M'_{\rm S} $ grow continuously before and after their horizon entry, while modes with $ M_{\rm pert} < M'_{\rm S} $ are completely washed out.

\begin{figure}[h]
  \begin{center}
    \leavevmode
    \epsfxsize = 12cm
    \epsffile{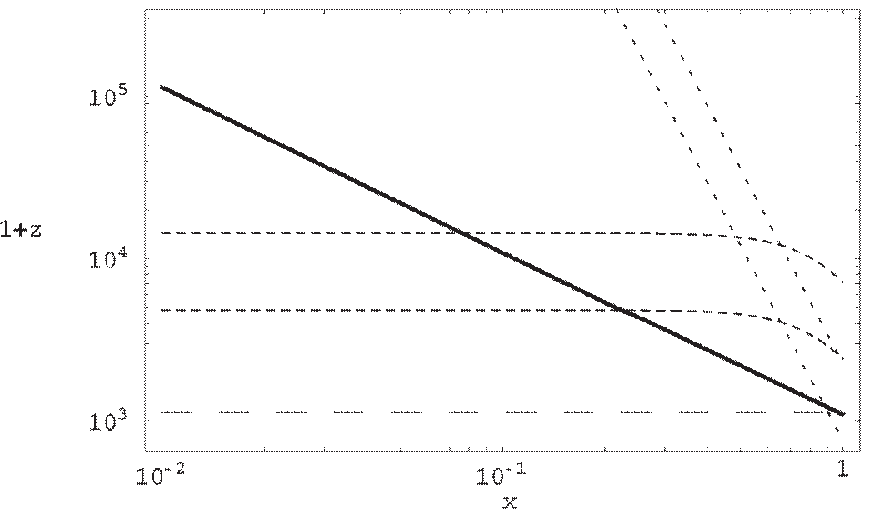}
  \end{center}
\caption{\small The M photon decoupling redshift $1+z_{dec}'$ as a function of $x$ (solid line).  The long-dash line marks the ordinary decoupling redshift $1+z_{\rm dec} = 1100$. We also show the matter-radiation equality redshift $1+z_{\rm eq}$ and the mirror Jeans-horizon mass equality redshift $1+z'_c$, for the cases $\Omega_m h^2 =0.2$ (respectively lower dash and lower dot) and $\Omega_m h^2=0.6$ (upper dash and dot). }
\label{figzz}
\end{figure}

Notice that $ M'_{\rm J} $ becomes smaller than the Hubble horizon mass $ M_{\rm H} $ starting from a redshift 
\be 
z_c= 3750 \: x^{-4} \: (\Omega_m h^2) \;, 
\ee
which is about $ z_{\rm eq} $ for $ x=0.64 $, but it sharply increases for smaller values of $ x $, as shown in figure \ref{figzz}. We can recognize this behaviour also watching at the intersections of the lines for $ M'_{\rm J} $ and $ M_{\rm H} $ in figures \ref{mirmasssca1} and \ref{mirmasssca2}. Thus, density perturbation scales which enter horizon at $ z \sim z_{\rm eq} $ have mass larger than $ M'_J $ and thus undergo uninterrupted linear growth immediately after $ t_{\rm eq} $. Smaller scales for which $ M'_{\rm J} > M_{\rm H} $ would instead first oscillate.  Therefore, the large scale structure formation is not delayed even if the mirror decoupling did not occur yet, i.e. even if $ x> x_{\rm eq} $. 

When compared with the non baryonic dark matter scenarios (see \S~\ref{non_bar_struct_form}), the main feature of the mirror baryonic dark matter scenario is that the M baryon density fluctuations should undergo the strong collisional damping around the time of M recombination, which washes out the perturbations at scales smaller than the mirror Silk scale. It follows that density perturbation scales which undergo the linear growth after the MRE epoch are limited by the length $\lambda'_S$. This could help in avoiding the excess of small scales (of few Mpc) in the CDM power spectrum mentioned in \S~\ref{cdmscen} without tilting the spectral index.
To some extent, the cutoff effect is analogous to the free streaming damping in the case of warm dark matter (WDM), but there are important differences. The point is that, alike usual baryons, the MBDM shows acoustic oscillations with an impact on the large scale power spectrum. In particular, it is tempting to imagine that the M baryon oscillation effects are related to the anomalous features observed in large scale structure power spectra (see next chapter for a complete discussion).  

In addition, the MBDM oscillations transmitted via gravity to the ordinary baryons, could cause observable anomalies in the CMB angular power spectrum for $l$'s larger than 200. This effect can be observed only if the M baryon Jeans scale $\lambda'_J$ is larger than the Silk scale of ordinary baryons, 
which sets a principal cutoff for CMB oscillations around $l\sim 1200$. As we will see in the next chapter, this would require enough large values of $x$, near the upper bound fixed by the BBN constraints, 
and, together with the possible effects on the large scale power spectrum, it can provide a direct test for the MBDM (verifiable by the higher sensitivity of next CMB and LSS experiments).

{\em Clearly, for small $x$ the M matter recombines before the MRE moment, and thus it behaves
as the CDM as far as the large scale structure is concerned.} However, there still can be crucial differences at smaller scales which already went non-linear, like galaxies. In our scenario, dark matter in galaxies and clusters can contain both CDM and MBDM components, or can be even constituted entirely by the mirror baryons. 

One can question whether the MBDM distribution in halos can be different from that of the CDM. Simulations show that the CDM forms triaxial halos with a density profile too clumped towards the center, and overproduces the small substructures within the halo. 
Since MBDM constitutes a kind of collisional dark matter, it may potentially avoid these problems, at least the one related with the excess of small substructures. 

It's also worth noting that, throughout the above discussion, we have assumed that the matter density of the Universe is close to unity. If, instead, the matter density is small and a vacuum density contribution is present, there is an additional complication due to the fact that the Universe may become curvature dominated starting from some redshift $ z_{curv} $. Given the current estimate $ \Omega_{\lambda} \simeq 0.7 $ (see \S~\ref{sec-now-cosm}), this transition has yet occurred and the growth of perturbations has stopped around $ z_{curv} $, when the expansion became too rapid for it.

At the end, we spend few words to mention that the main difficulty with the ordinary baryonic adiabatic scenario is the excess of angular fluctuations in the CMB temperature respect to the observational limits. More specifically, 
one needs $\delta_{\rm b}\simeq10^{-3}$ at recombination, 
but in the adiabatic picture matter inhomogeneities are accompanied by perturbations in the radiation field, and this will inevitably lead to temperature fluctuations of order $\delta{\rm T}/{\rm T}\simeq\delta_{\gamma}\simeq10^{-3}$ at decoupling, which is in direct disagreement with observations. 
In this sense the mirror baryonic adiabatic scenario has various ways to overcome this problem, first of all the fact that mirror recombination takes place $ x^{-1} $ times before the ordinary one (see \S~\ref{mirror_dm}), and structures have more time to grow after decoupling, in a way compatible with the CMB observations of $\delta{\rm T}/{\rm T} \simeq 10^{-5}$.


\passo
\def \mirevolpert{Evolution of perturbations}
\section{\mirevolpert}
\label{mirevolpert}
\markboth{Chapter \ref{chap-mirror_univ_2}. ~ \mir_univ_2}
                    {\S \ref{mirevolpert} ~ \mirevolpert}

As a result of the studies done in previous sections, in this section we finally consider the temporal evolution of perturbations, as function of the scale factor $ a $. All the plots are the results of numerical computations obtained making use of a Fortran code originally written for the ordinary Universe and modified to account for the mirror sector (for more details see the beginning of next chapter).

We used the synchronous gauge, described in \S~\ref{cmb6} and appendix \ref{mabert3}. The difference in the use of other gauges is limited to the gauge-dependent behaviour of the density fluctuations on scales larger than the horizon. The fluctuations can appear as growing modes in one coordinate system and as constant mode in another, that is exactly what occurs in the synchronous and the conformal Newtonian gauges.

In the figures we plot the evolution of the components of a mirror Universe, namely the cold dark matter\footnote{As non baryonic dark matter we consider only the cold dark matter, which is now the standard choice in cosmology.}, the ordinary baryons and photons, and the mirror baryons and photons, changing some parameter to evaluate their influence. Note that in figures are plotted the density contrasts {\em not} normalized to the average density.

First of all, we comment figure \ref{evol-x06-b2-k0510}b, which is the most useful to recognize the general features of the evolution of perturbations. Starting from the smallest scale factor, we see that all three matter components and the two radiative components grow with the same trend (as $ a^2 $), but the radiative ones have a slightly higher density contrast (with a constant rate until they are tightly coupled); this is simply the consequence of considering adiabatic perturbations, which are linked in their matter and radiation components by the adiabatic condition (\ref{adcond}). This is the situation when the perturbation is out of horizon, but, when it crosses the horizon, around $ a \sim 10^{-4} $, things drastically change. Baryons and photons, in each sector separately, become causally connected, feel each other, and begin to oscillate for the competitive effects of gravity and pressure. Meanwhile, the CDM density perturbation continues to grow uninterruptedly, at first reducing his rate from $ a^2 $ to $ \ln a $ (due to the rapid expansion during the radiation era), and later, as soon as MRE occurs (at $ a \sim 3 \times 10^{-3} $ for the considered model), increasing proportionally to $ a $. The oscillations of baryons and photons continue until their decoupling, which in the mirror sector occurs before than in the ordinary one (scaled by the factor $ x $). This moment is marked in the plot as the point where the lines for the two components move away one from the other. From this point, the photons continue the oscillations until they are completely damped, while the baryons rapidly fall into the potential wells created by the dark matter and begin to grow as $ a $. 
Note that it's important the way in which the oscillation reaches the decoupling; if it is compressing, first it slows down (as if it continues to oscillate, but disconnected from the photons), and then it expands driven by the gravity; if, otherwise, it is expanding, it directly continues the expansion and we see in the plot that it immediately stops to oscillate. In this figure we have the first behaviour in the mirror sector, the second one in the ordinary sector.

In figures \ref{evol-x06-b2-k0510}, \ref{evol-x06-b2-k1520} and \ref{evol-x06-b2-k2530} we compare the behaviour of different scales for the same model. The scales are given by $ \log k = -0.5, -1.0, -1.5, -2.0, -2.5, -3.0 $, where $ k = 2 \pi / \lambda $ is the wave number and is measured in ${\rm Mpc}^{-1}$. 
The effect of early entering the horizon of small scales (those with higher $ k $) is evident. 
Going toward bigger scales the superhorizon growth continue for a longer time, delaying more and more the beginning of acoustic oscillations, until it occurs out of the coordinate box for the bigger plotted scale ($ \log k = -3.0 $). Starting from the scale $ \log k = -1.5 $, the mirror decoupling occurs before the horizon entry of the perturbations, and the evolution of the mirror baryons density is similar to that of the CDM. The same happens to the ordinary baryons too, but for $ \log k \lsim -2.0 $ (since they decouple later), while the evolution of mirror baryons is yet indistinguishable from that of the CDM. For the bigger scales ($ \log k \lsim -2.5 $) the evolution of all three matter components is identical.

As previously seen, the decoupling is a crucial point for structure formation, and it assumes a fundamental role specially in the mirror sector, where it occurs before than in the ordinary one: mirror baryons can begin before to grow perturbations in their density distribution. For this reason it's important to analyze the effect of changing the mirror decoupling time, obtained changing the value of $ x $ and leaving unchanged all other parameters, as it is possible to do using figures \ref{evol-x06-b2-k0510}b, \ref{evol-x0504-b2-k10} and \ref{evol-x0302-b2-k10} for $ x = 0.6, 0.5, 0.4, 0.3, 0.2 $ and the same scale $ \log k = -1.0 $. It is evident the shift of the mirror decoupling toward lower values of $ a $ when reducing $ x $, according to the law (\ref{z'_dec}), which states a direct proportionality between the two. In particular, for $ x \lsim 0.3 $ mirror decoupling occurs before the horizon crossing of the perturbation, and mirror baryons mimic more and more the CDM, so that for $ x \simeq 0.2 $ the perturbations in the two components are indistinguishable. For the ordinary sector apparently there are no changes, but at a more careful inspection we note some difference due to the different amount of relativistic mirror species (proportional to $ x^4 $), which slightly shifts the matter-radiation equality. This effect is more clear in figure \ref{evol-x0206-b2-k10}, where we plot only the CDM and the ordinary baryons for the cases $ x = 0.2 $ and $ 0.6 $: for the lower value of $ x $ there are less mirror photons, the MRE occurs before and the perturbation in the collisionless component starts growing proportionally to the scale factor before; thus, when the baryons decouple, their perturbation rapidly grows to equalize that in the CDM, which meanwhile has raised more for the lower $ x $.

Obviously, these are cases where the CDM continues to be the dominant form of dark matter, and drives the growth of perturbations, given its continuous increase. In any case, if the dominant form of dark matter is made by mirror baryons the situation is practically the same, as visible comparing figures \ref{evol-x0504-b2-k10}b and \ref{evol-x04-b4nocdm-k10}a (where we see only slight differences on the CDM and mirror baryons behaviours in the central region of the plots), since mirror baryons decouple before than ordinary ones and fall into the potential wells of the CDM, reinforcing them.

Finally, in the interesting case where mirror baryons constitute {\em all} the dark matter, they drive the evolution of perturbations. In fact, in figure \ref{evol-x04-b4nocdm-k10}b we clearly see that the density fluctuations start growing in the mirror matter and the visible baryons are involved later, after being recombined, when they rewrite the spectrum of already developed mirror structures. This is another effect of a mirror decoupling occurring earlier than the ordinary one: the mirror matter can drive the growth of perturbations in ordinary matter and provide the rapid growth soon after recombination necessary to take into account of the evolved structures that we see today.

Given all the considerations made in this chapter, it is evident that the case of mirror baryons is very interesting for structure formation, because they are collisional between themselves but collisionless for the ordinary sector, or, in other words, they are self-collisional. In this situation baryons and photons in the mirror sector are tightly coupled until decoupling, and structures cannot grow before this time, but the mirror decoupling happens before the ordinary one, thus structures have enough time to grow according to the limits imposed by CMB and LSS (something not possible in a purely ordinary baryonic scenario). Another important feature of the mirror dark matter scenario is that, if we consider small values of $x$, the perturbation evolution is very similar to the CDM case, but with a fundamental difference: there exist a cutoff scale due to the mirror Silk damping, which kills the small scales, overcoming the problems of the CDM scenario with the excessive number of small satellites. These are important motivations to go further in the work and investigate the effects of the mirror sector on the CMB and LSS power spectra, as we will do in the next chapter.

\begin{figure}[h]
\begin{center}
\leavevmode
{\hbox 
{\epsfxsize = 12.5cm \epsfysize = 9.5cm \epsffile{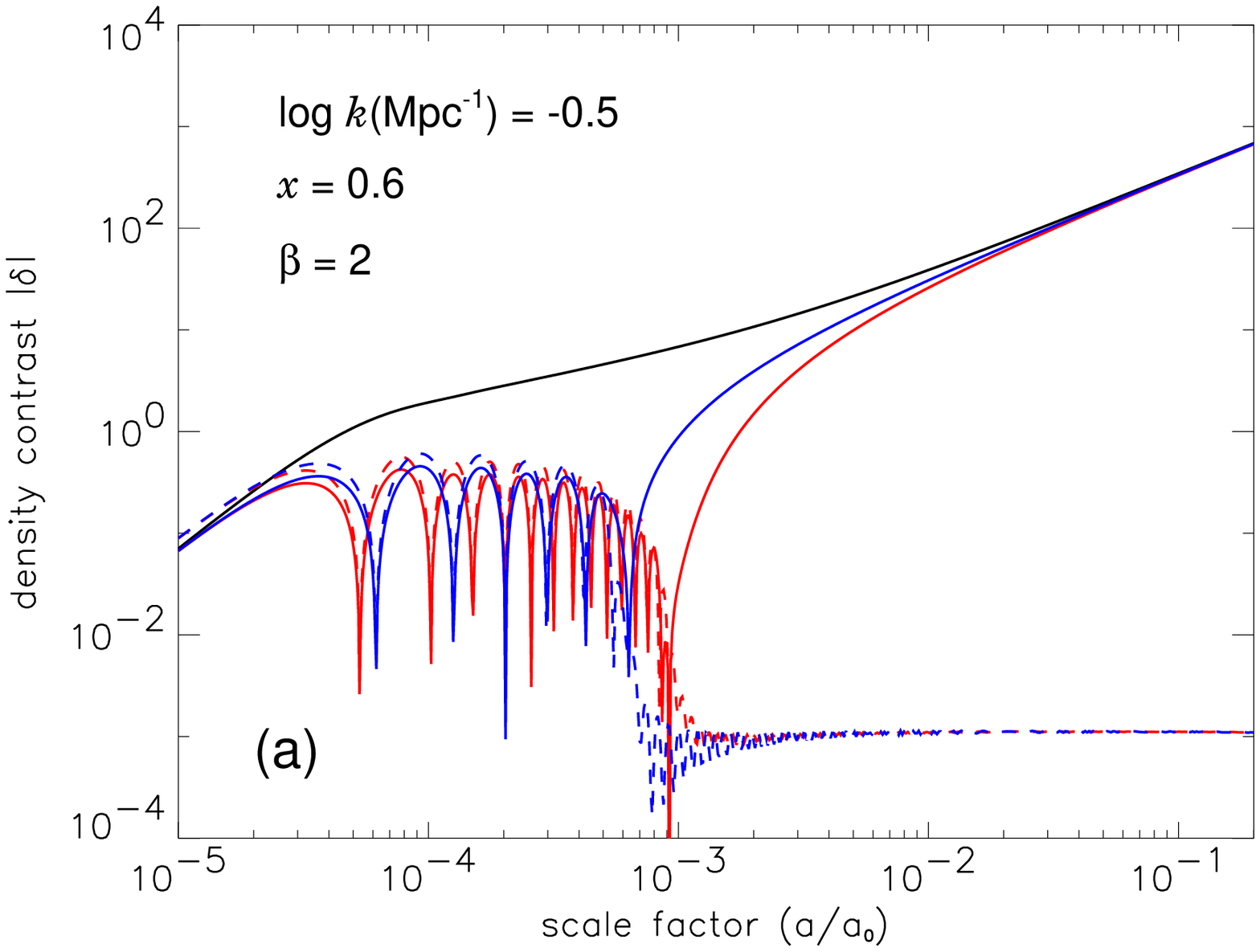} }
{\epsfxsize = 12.5cm \epsfysize = 9.5cm \epsffile{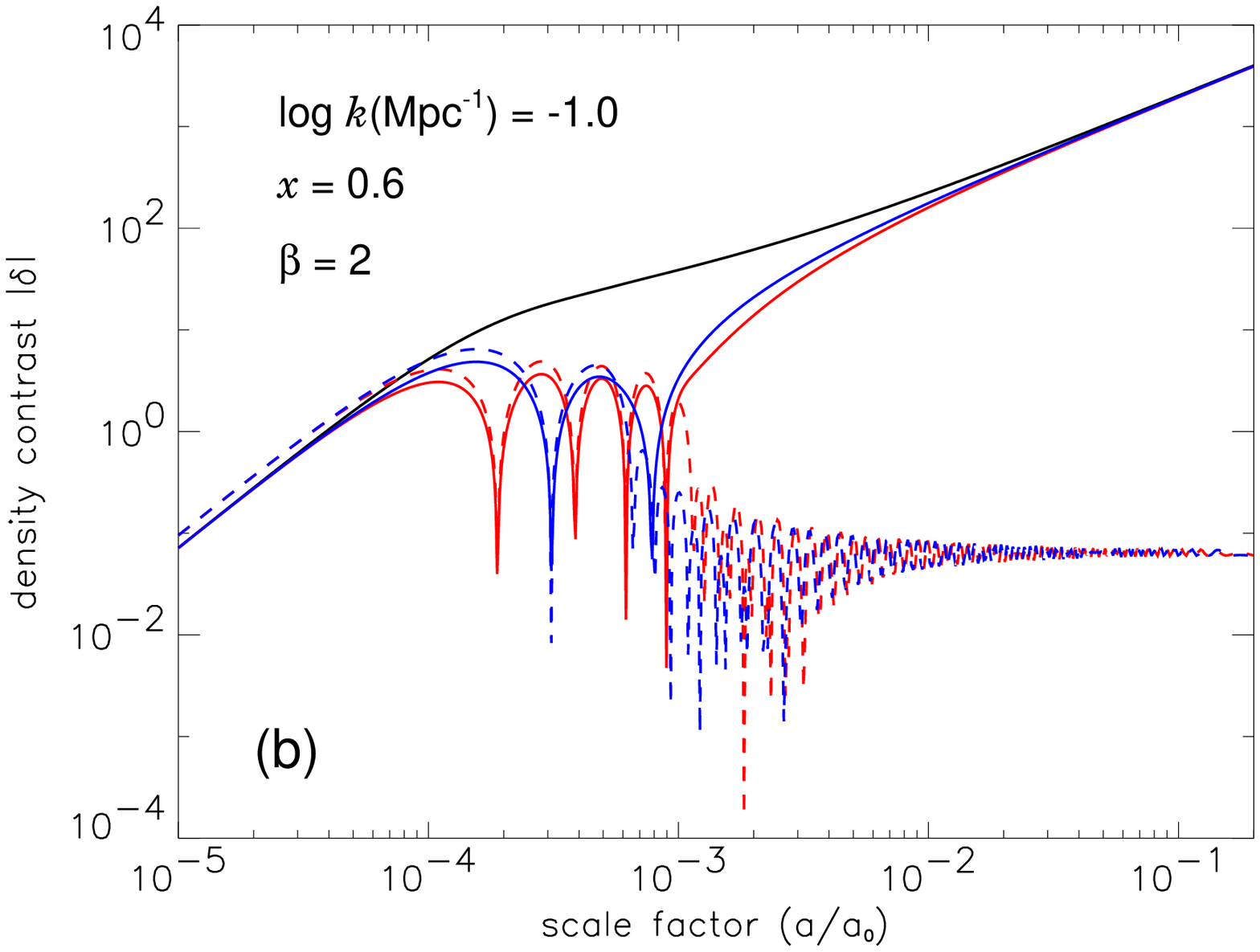} } }
\end{center}
\vspace{-.5cm}
\caption{\small Evolution of perturbations for the components of a Mirror Universe: cold dark matter (dark line), ordinary baryons and photons (solid and dotted red lines) and mirror baryons and photons (solid and dotted blue lines). The model is a flat Universe with $ \Omega_{\rm m} = 0.3 $, $ \Omega_{\rm b} h^2 = 0.02 $, $ \Omega'_{\rm b} h^2 = 0.04 $ ($ \beta = 2 $), $ h = 0.7 $, $ x = 0.6 $, and plotted scales are $ \log k ({\rm Mpc}^{-1}) = -0.5 $  ($ a $) and $ -1.0 $ ($ b $).}
\label{evol-x06-b2-k0510}
\end{figure}

\begin{figure}[h]
\begin{center}
\leavevmode
{\hbox 
{\epsfxsize = 12.5cm \epsfysize = 9.5cm \epsffile{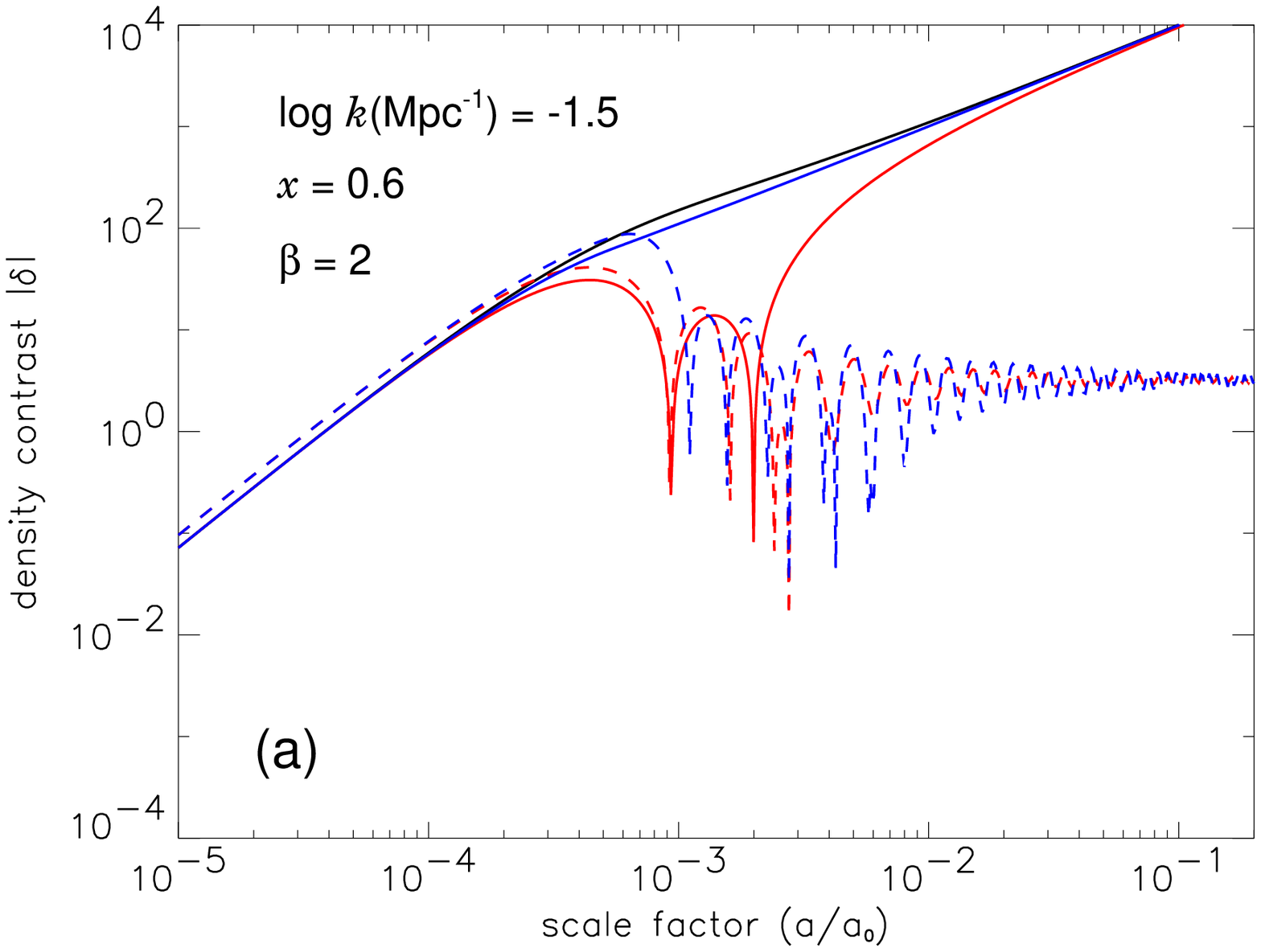} }
{\epsfxsize = 12.5cm \epsfysize = 9.5cm \epsffile{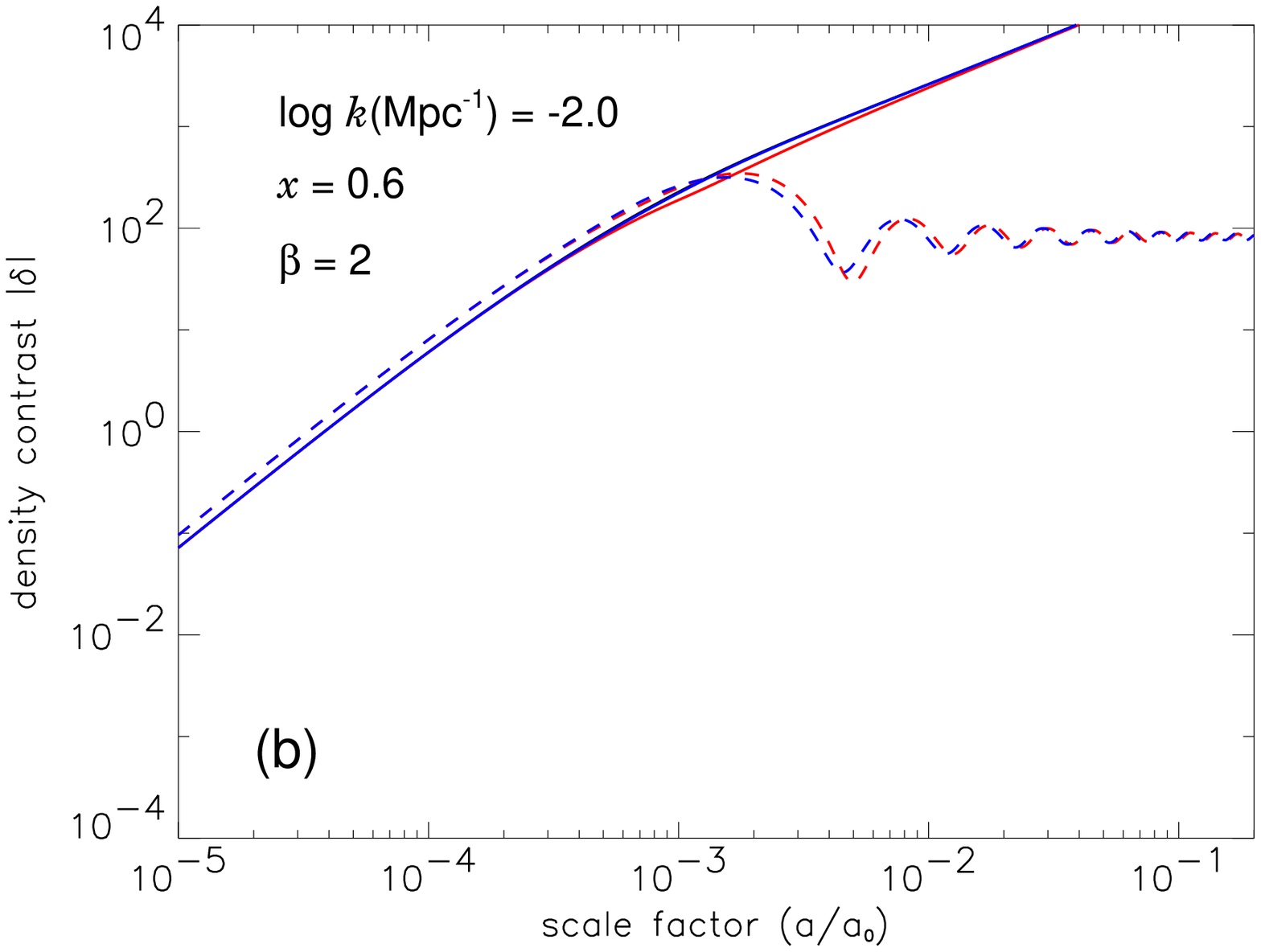} } }
\end{center}
\caption{\small The same as in figure \ref{evol-x06-b2-k0510}, but for scales $ \log k ({\rm Mpc}^{-1}) = -1.5 $  ($ a $) and $ -2.0 $ ($ b $).}
\label{evol-x06-b2-k1520}
\end{figure}

\begin{figure}[h]
\begin{center}
\leavevmode
{\hbox 
{\epsfxsize = 12.5cm \epsfysize = 9.5cm \epsffile{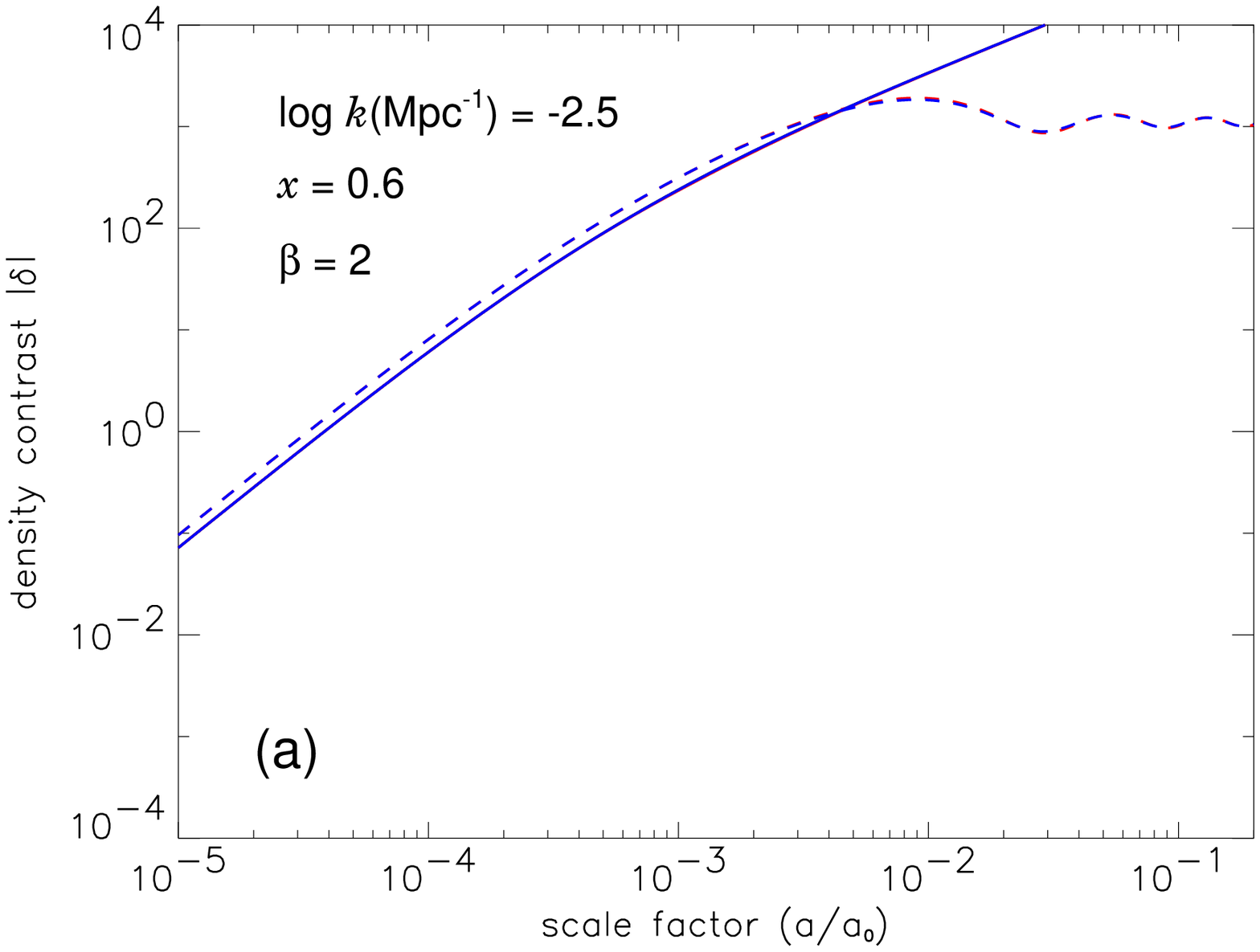} }
{\epsfxsize = 12.5cm \epsfysize = 9.5cm \epsffile{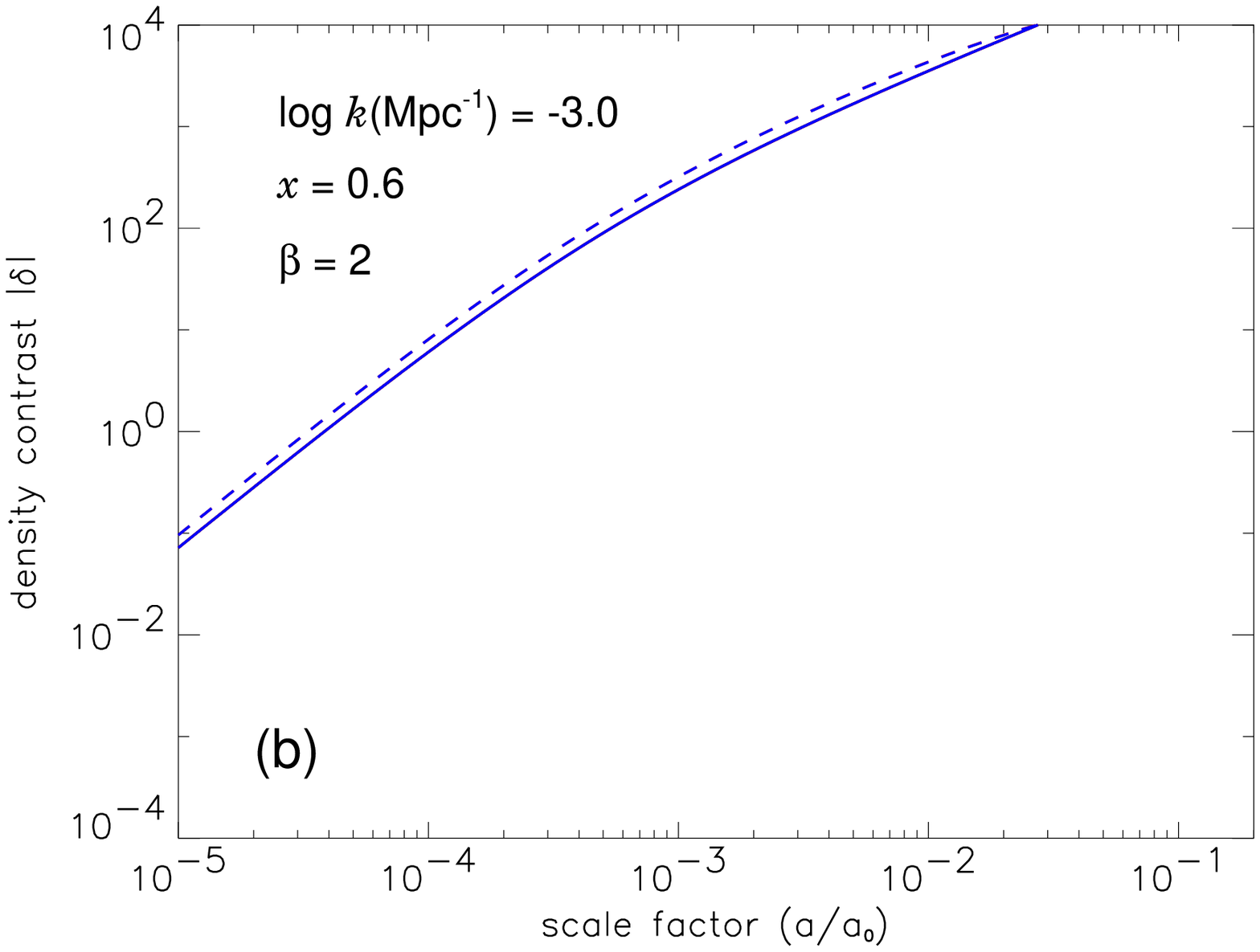} } }
\end{center}
\caption{\small The same as in figure \ref{evol-x06-b2-k0510}, but for scales $ \log k ({\rm Mpc}^{-1}) = -2.5 $  ($ a $) and $ -3.0 $ ($ b $).}
\label{evol-x06-b2-k2530}
\end{figure}

\begin{figure}[h]
\begin{center}
\leavevmode
{\hbox 
{\epsfxsize = 12.5cm \epsfysize = 9.5cm \epsffile{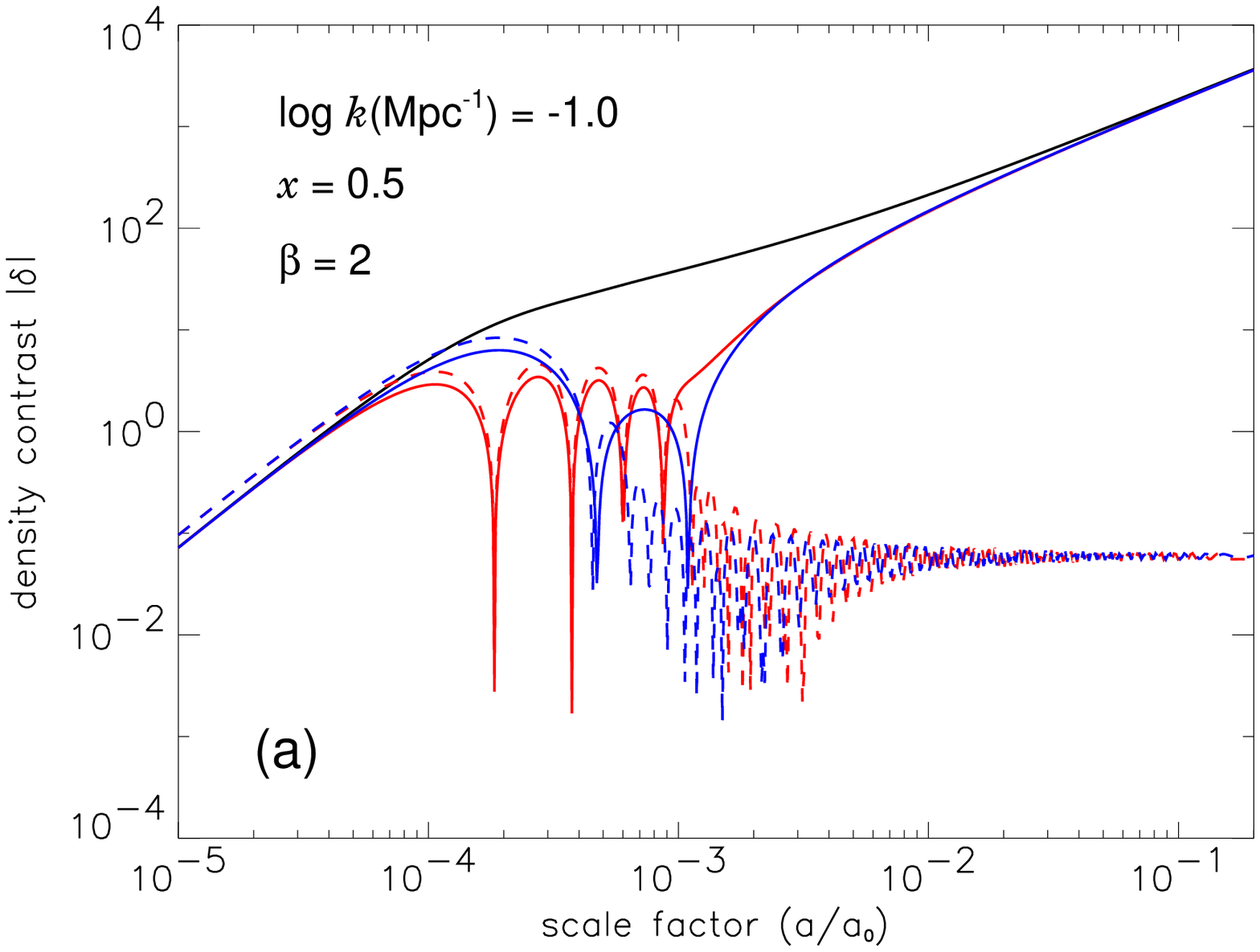} }
{\epsfxsize = 12.5cm \epsfysize = 9.5cm \epsffile{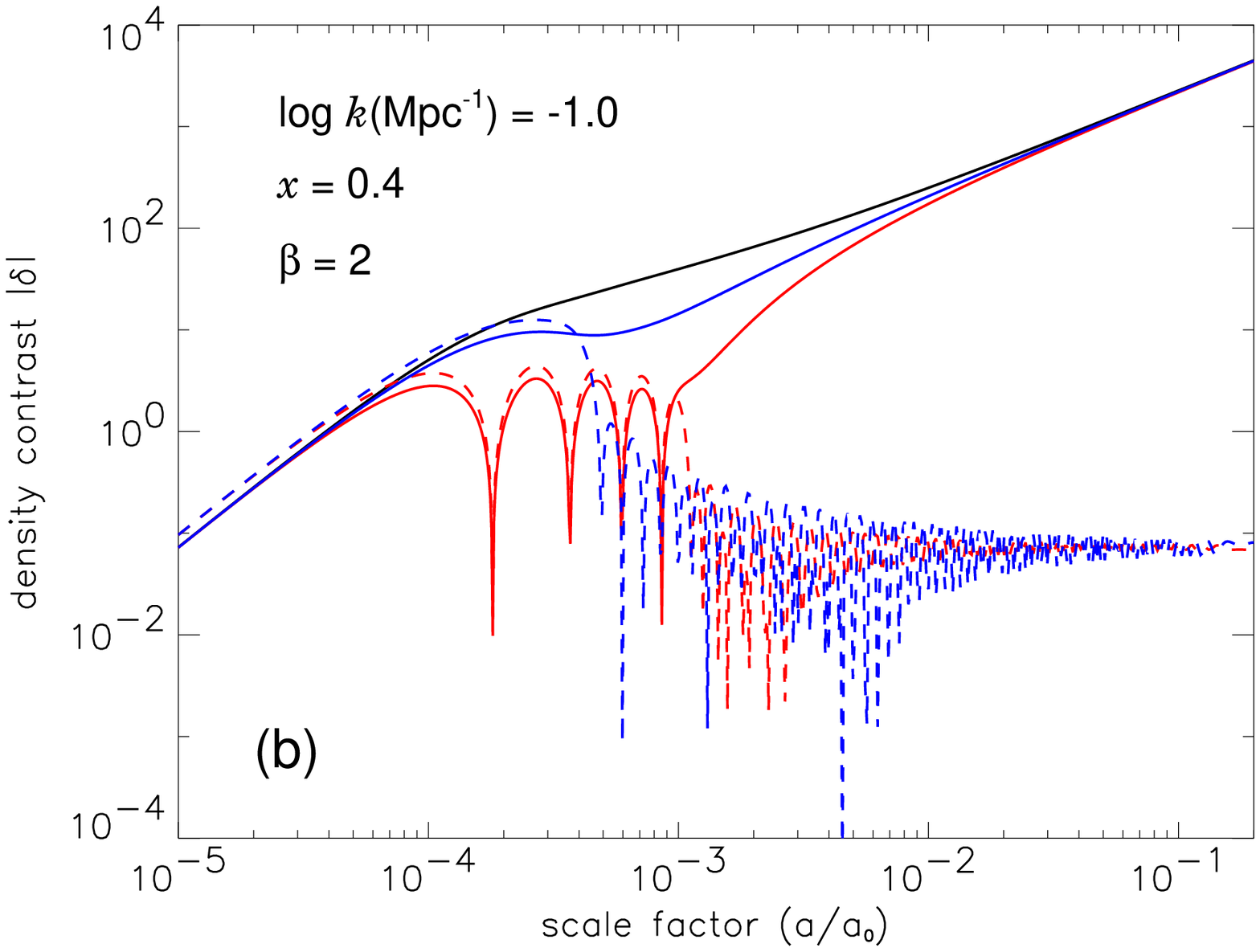} } }
\end{center}
\caption{\small Evolution of perturbations for the components of a Mirror Universe: cold dark matter (dark line), ordinary baryons and photons (solid and dotted red lines) and mirror baryons and photons (solid and dotted blue lines). The model is a flat Universe with $ \Omega_{\rm m} = 0.3 $, $ \Omega_{\rm b} h^2 = 0.02 $, $ \Omega'_{\rm b} h^2 = 0.04 $ ($ \beta = 2 $), $ h = 0.7 $, $ x = 0.5 $ ($ a $) or $ 0.4 $ ($ b $), and plotted scale is $ \log k ({\rm Mpc}^{-1}) = -1.0 $.}
\label{evol-x0504-b2-k10}
\end{figure}

\begin{figure}[h]
\begin{center}
\leavevmode
{\hbox 
{\epsfxsize = 12.5cm \epsfysize = 9.5cm \epsffile{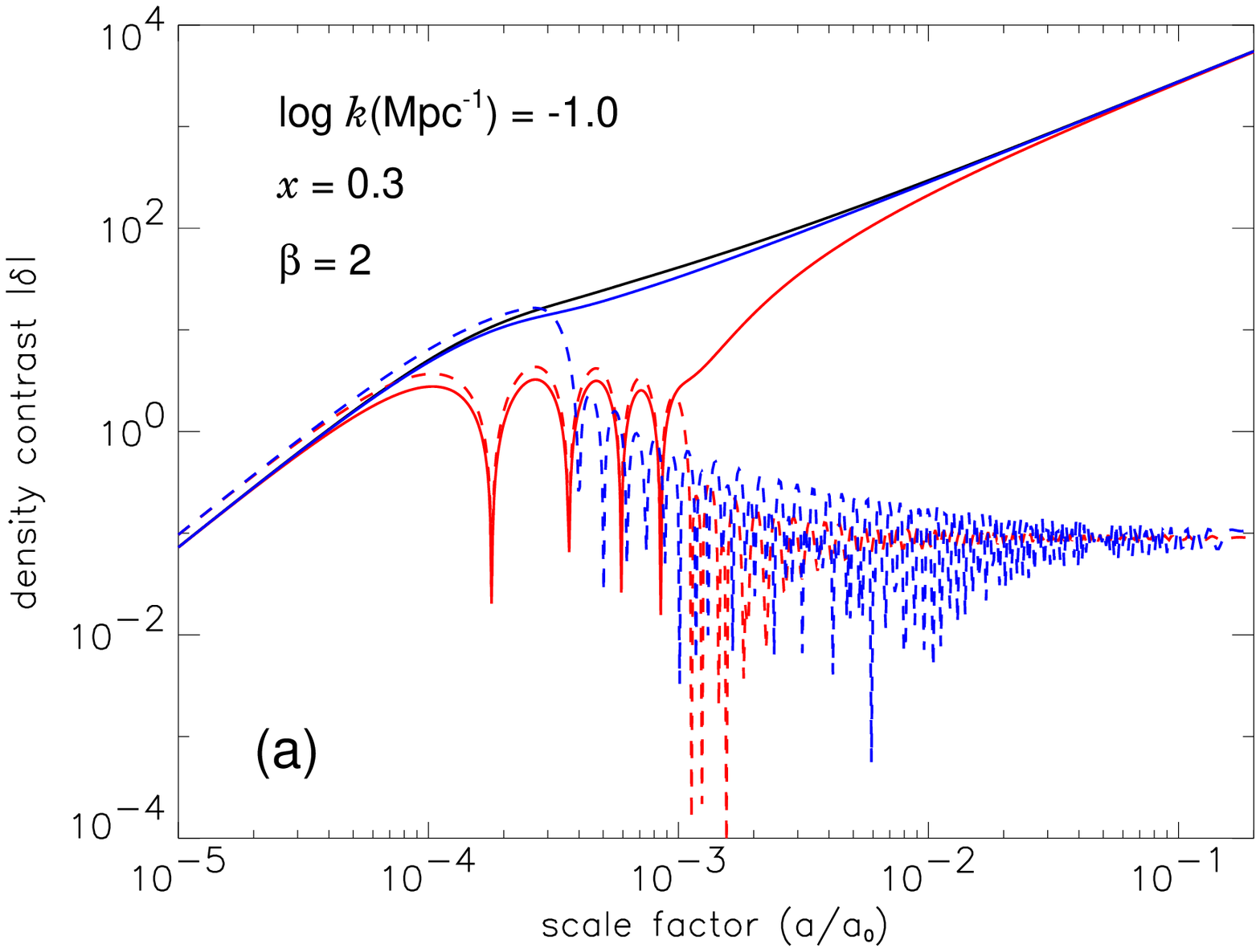} }
{\epsfxsize = 12.5cm \epsfysize = 9.5cm \epsffile{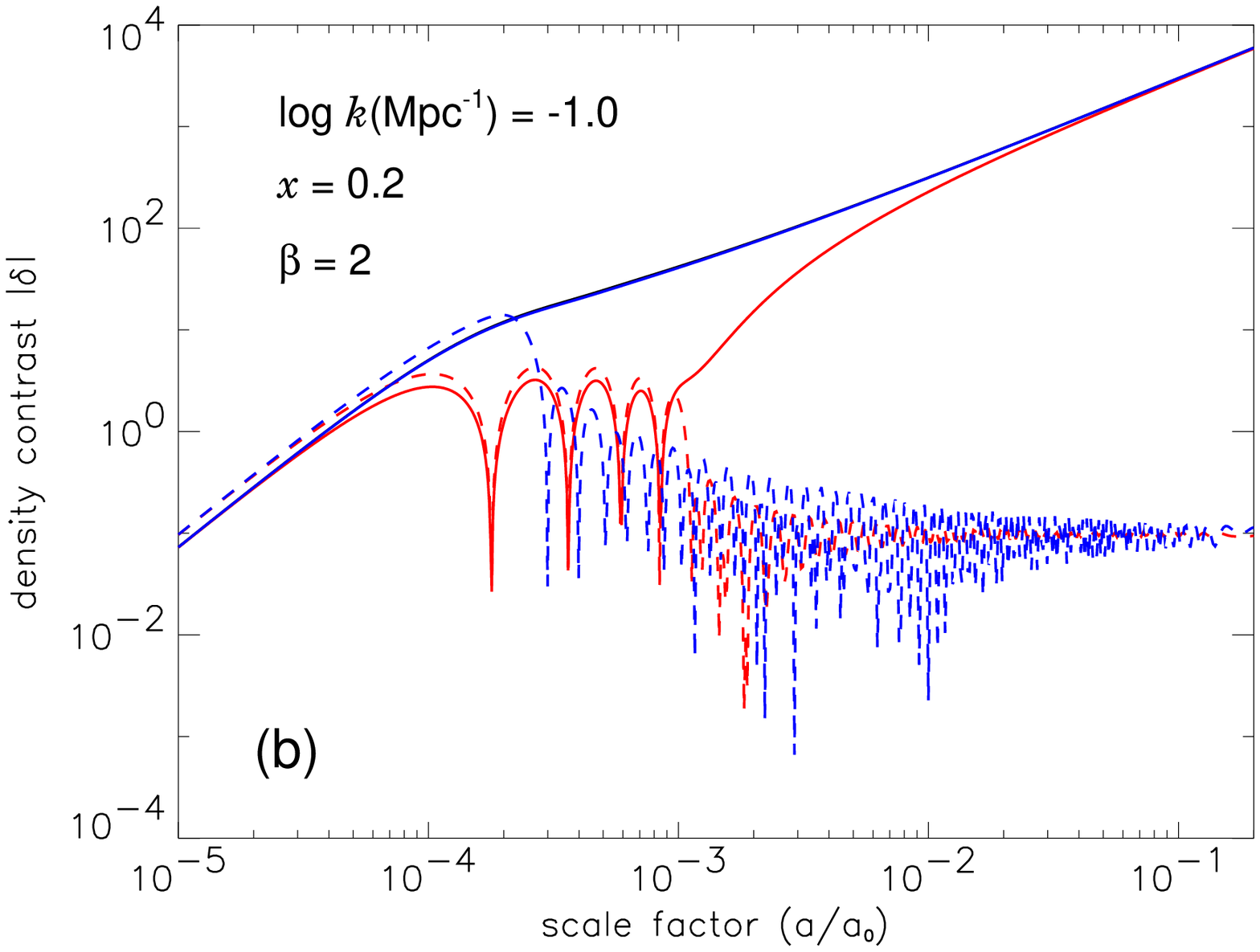} } }
\end{center}
\caption{\small The same as in figure \ref{evol-x0504-b2-k10}, but for  $ x = 0.3 $ ($ a $) and $ 0.2 $ ($ b $).}
\label{evol-x0302-b2-k10}
\end{figure}

\begin{figure}[h]
  \begin{center}
    \leavevmode
    \epsfxsize = 12.5cm
    \epsfysize = 9.5cm
    \epsffile{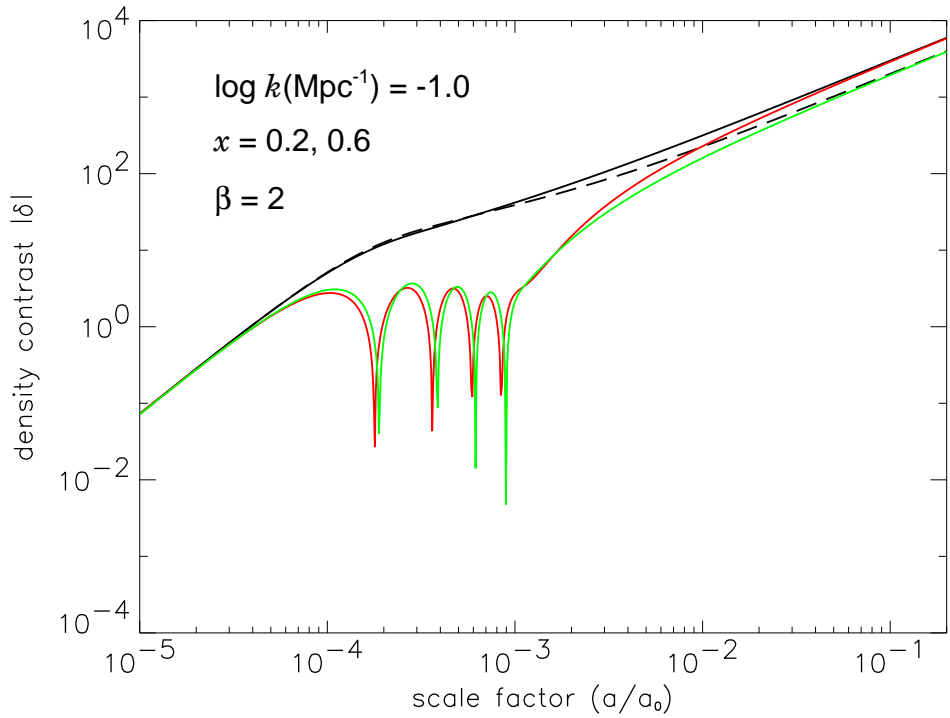}
  \end{center}
\caption{\small Evolution of perturbations in a Mirror Universe for cold dark matter (dark lines) and ordinary baryons (red and green lines). The models are flat, with $ \Omega_{\rm m} = 0.3 $, $ \Omega_{\rm b} h^2 = 0.02 $, $ \Omega'_{\rm b} h^2 = 0.04 $ ($ \beta = 2 $), $ h = 0.7 $, $ x = 0.2 $ (solid black and red lines) and $ 0.6 $ (dashed black and green lines), and $ \log k ({\rm Mpc}^{-1}) = -1.0 $.}
\label{evol-x0206-b2-k10}
\end{figure}

\begin{figure}[h]
  \begin{center}
    \leavevmode
{\hbox 
{\epsfxsize = 12.5cm \epsfysize = 9.5cm \epsffile{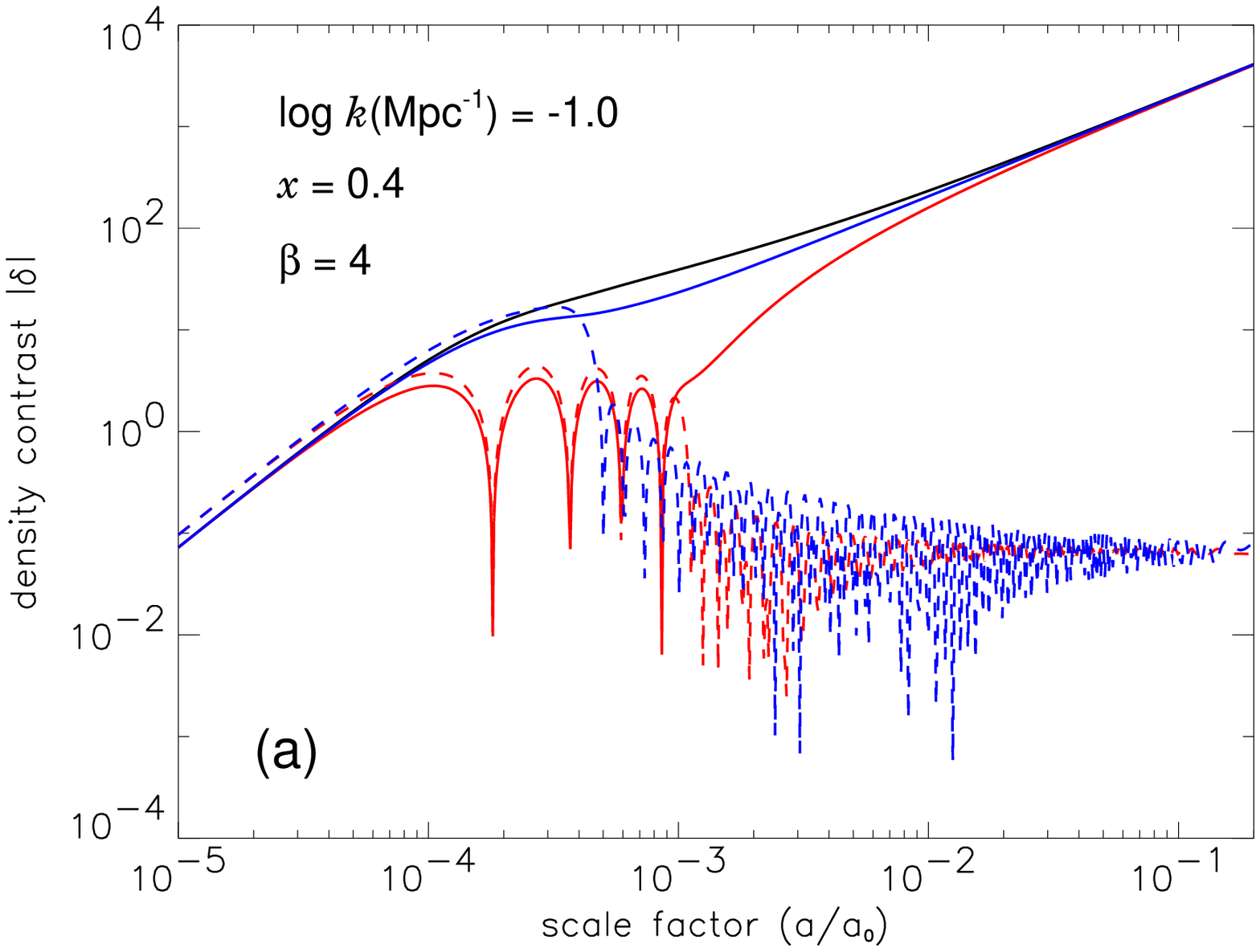} }
{\epsfxsize = 12.5cm \epsfysize = 9.5cm \epsffile{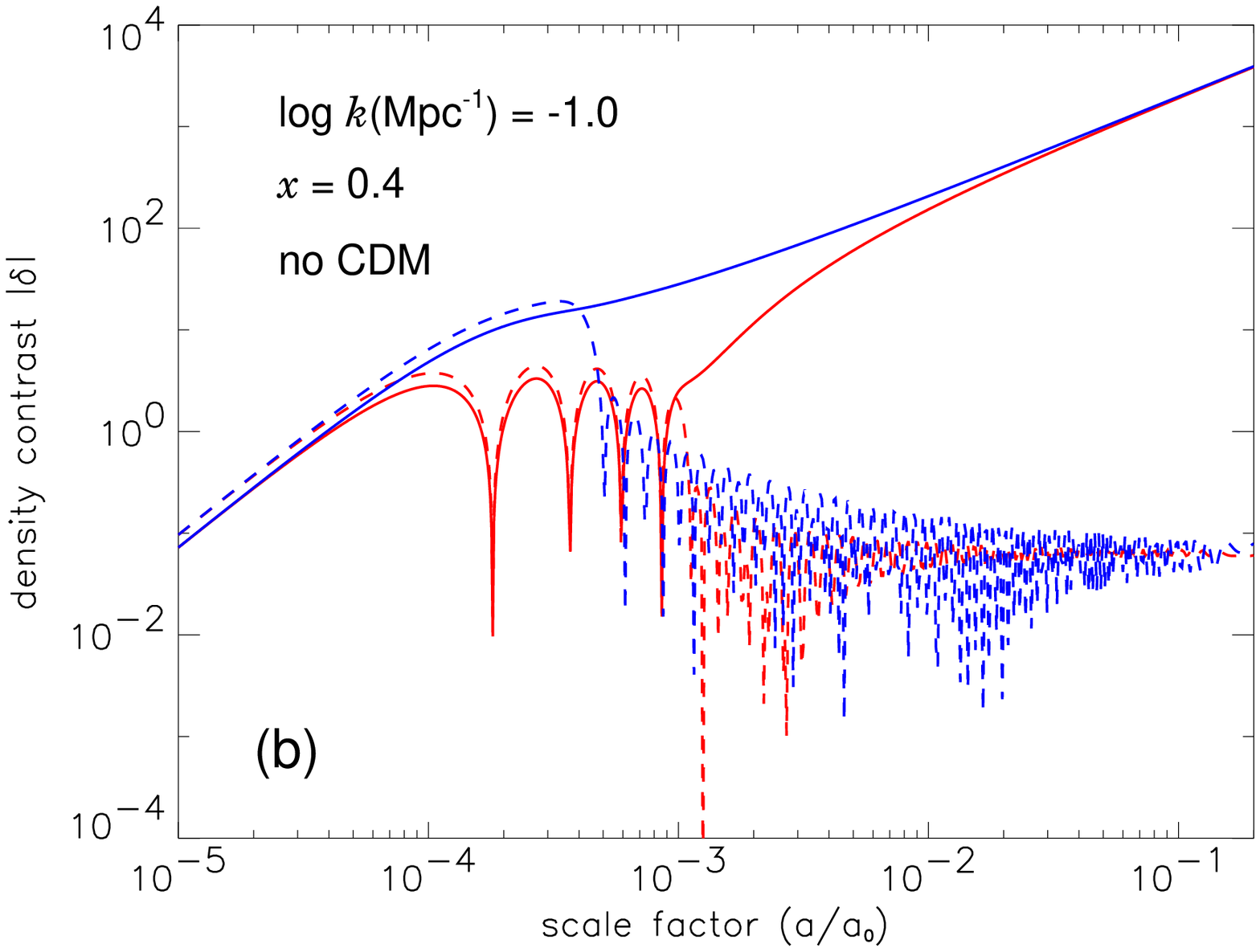} } }
\end{center}
\caption{\small Evolution of perturbations for the components of a Mirror Universe: cold dark matter (dark line), ordinary baryons and photons (solid and dotted red lines) and mirror baryons and photons (solid and dotted blue lines). The models are flat with $ \Omega_{\rm m} = 0.3 $, $ \Omega_{\rm b} h^2 = 0.02 $, $ \Omega'_{\rm b} = (2/3) (\Omega_{\rm m} - \Omega_{\rm b}) $ ($ \beta = 4 $) ($ a $) or $ (\Omega_{\rm m} - \Omega_{\rm b}) $ (no CDM) ($ b $), $ h = 0.7 $, $ x = 0.4 $, and $ \log k ({\rm Mpc}^{-1}) = -1.0 $.}
\label{evol-x04-b4nocdm-k10}
\end{figure}



\chapter{Cosmic microwave background and large scale structure for a Mirror Universe}
\label{chap-mirror_univ_3}
\def \mir_univ_3{CMB and LSS for a Mirror Universe}
\markboth{Chapter \ref{chap-mirror_univ_3}. ~ \mir_univ_3}
                    {Chapter \ref{chap-mirror_univ_3}. ~ \mir_univ_3}


\passo
\def \intro_mirror_3{Introduction}
\section{\intro_mirror_3}
\label{intro_mirror_3}
\markboth{Chapter \ref{chap-mirror_univ_3}. ~ \mir_univ_3}
                    {\S \ref{intro_mirror_3} ~ \intro_mirror_3}

Let us now consider the consequences of the existence of a mirror sector in terms of signatures on the Cosmic Microwave Background (CMB) and the Large Scale Structure (LSS). In the last decade their study is providing a great amount of observational data, and the continuous improvement is such that we are in the condition to verify different models of Universe, with the possibility to fit the cosmological parameters (for recent works see, e.g., Wang et al. (2002) \cite{wang} or Efstathiou et al. (2002) \cite{efstmnras330}). These powerful cosmological instruments could help us to understand the nature of the dark matter of the Universe analyzing the implications of different types of dark matter on their cosmological observables, and comparing them with the available experimental results. In this view we must study the consequences of the Mirror Universe on the CMB and LSS in order to compare our models with data, study their compatibility and possibly reduce the available parameter space.

To do this we need to numerically compute mirror models and their CMB and LSS power spectra (see \S~\ref{cmb4} and \S~\ref{lss1}). We used the evolutionary equations written by Ma and Bertschinger in 1995 \cite{mabert1} and described in \S~\ref{cmb6}, inserted in a Fortran code developed by the group of the prof. N. Vittorio of the University of Rome ``Tor Vergata'' and kindly provided to us. Obviously, this program was written for a standard Universe made only by the ordinary sector, and the first step of our work was to modify the code in order to simulate a Mirror Universe (which, as we know now, has two sectors instead). This required to modify and add several subroutines to the code. In fact, according to what said in previous chapters, we have to treat two self-interacting sectors, and thus double all the equations governing the evolution of the considered components, relativistic (photons and massless neutrinos) and non relativistic (baryons and cold dark matter). In addition, the two sectors communicate via gravity, so that they are coupled and influence each other through this interaction. Therefore, there are more regimes respect to the standard case, because now, instead of being simply the ordinary ones, they are made up of the couplings between the different ordinary and mirror regimes, the latter being time-shifted from the former according to the laws exposed in \S~\ref{mirror_dm} and \S~\ref{evol_Vs}.


\passo
\def \mirror_mod{The mirror models}
\section{\mirror_mod}
\label{mirror_mod}
\markboth{Chapter \ref{chap-mirror_univ_3}. ~ \mir_univ_3}
                    {\S \ref{mirror_mod} ~ \mirror_mod}

We computed many models for Mirror Universe, assuming adiabatic scalar perturbations, a flat space-time geometry, and different mixtures of ordinary and mirror baryons, photons and massless neutrinos, cold dark matter, and cosmological constant.

In Fourier space, all the {\bf k} modes in the linearized Einstein, Boltzmann, and fluid equations evolve independently; thus the equations can be solved for one value of {\bf k} at a time. Moreover, all modes with the same $ k $ (the magnitude of the comoving wavevector,) obey the same evolutionary equations.
We integrated the equations of motion numerically over the range $ -5.0 \leq \log k \leq -0.5 $ (where $ k $ is measured in $ {\rm Mpc}^{-1} $) using points evenly spaced with an interval of $ \Delta \log k = 0.01 $. The full integration was carried to $ z = 0 $.

As shown in \S~\ref{cmb6}, the Boltzmann equation for massless neutrinos (\ref{bolmn}) has been transformed into an infinite hierarchy of moment equations that must be truncated at some maximum multipole order $l_{\rm max}$.  One simple but inaccurate method is to set $F_{\nu l}=0$ for $l>l_{\rm max}$.\footnote{The problem with this scheme is that the coupling of multipoles in equations 
leads to the propagation of errors from $l_{\rm max}$ to smaller $l$.  Indeed, these errors can propagate to $l=0$ in a time $\tau\approx l_{\rm max}/k$ and then reflect back to increasing $l$, leading to amplification of errors in the interval
$0\le l\le l_{\rm max}$.} According to what suggested by Ma and Bertschinger (1995) \cite{mabert1} instead, an improved truncation scheme is based on extrapolating the behaviour of $F_{\nu l}$ to $l=l_{\rm max}+1$ as
\begin{equation}
\label{truncnu}
     F_{\nu\,(l_{\rm max}+1)}\approx{(2l_{\rm max}+1)\over k\tau}\,F_{\nu
       \,l_{\rm max}}-F_{\nu\,(l_{\rm max}-1)}\ .
\end{equation}
However, time-variations of the potentials during the radiation-dominated era make even equation (\ref{truncnu}) a poor approximation if $l_{\max}$ is chosen too small.

Thus, in order to have a relative accuracy better than $ 10 ^{-3} $ in our final results, in the computation of the potential and the density fields the photons and the massless neutrinos phase space distributions for both the ordinary and mirror sectors were expanded in Legendre series 
truncating the Boltzmann hierarchies at $ l_{\rm max} = 2000 $, using the truncation schemes given by equations (\ref{truncnu}) and (\ref{truncphot}).

At last, we normalize our results to the COBE data, following the procedure described by Bunn and White in 1997 \cite{bunnwhite480}.

\begin{table}[h]
\begin{center}
\begin{tabular}{|c|c|c|} \hline \hline
  Parameter & min. value & max. value \\
  \hline \hline
  $ \Omega_{\rm m} $ & 0.1 & 1.0 \\
  \hline
  $ \omega_{\rm b} $ & 0.010 & 0.030 \\
  \hline
  $ \omega'_{\rm b} $ & 0.0 & $ \omega_{\rm m} - \omega_{\rm b} $ \\
  \hline
  $ x $ & 0.1 & 1.0 \\
  \hline
  $ h $ & 0.5 & 0.9 \\
  \hline
  $ n $ & 0.90 & 1.10 \\
  \hline \hline
\end{tabular}
\end{center}
\caption{\small Parameters and their ranges used in mirror models. The values are not evenly spaced, but arbitrarily chosen in the parameter space. Not listed there are the total and vacuum densities, but, being flat models, they are $ \Omega_{\rm 0} = 1 $ and $ \Omega_{\Lambda} = 1 - \Omega_{\rm m} $.}
\label{parrange}
 \end{table}

We considered different values of the cosmological parameters, where now we add to the usual ones two new mirror parameters: the ratio of the temperatures in the two sectors $ x $ (see \S~\ref{term_mir_univ}) and the mirror baryons density $ \Omega'_{\rm b} $ (also expressed via the ratio of the baryonic densities in the two sectors $ \beta $, defined in \S~\ref{baryogen}). Starting from an ordinary reference model (we choose the so-called ``concordance model'' of Wang et al. (2002) \cite{wang}), we study the influence of the mirror sector varying the two parameters that describe it for a given ordinary sector. Furthermore, we evaluate the influence of all the parameters for a Mirror Universe, changing all of them. The values used for the parameters are not on a regular grid, but arbitrarily chosen for the only purpose to better understand the CMB and LSS for a Mirror Universe. At the moment we are not able to make a grid thin enough to perform a fit of all free parameters (even if we fix some of them) for two reasons: first, we have two further parameters, which lengthen a lot the computational time (by a factor $ 10^2 $), and second our program is much slower than others commonly used for a standard Universe, because our models are more complicated in terms of calculus and for our choice to privilege precision instead of performance (at least in this moment\footnote{One of the future steps of the work will be just to write a new program (probably based on the commonly used Seljak and Zaldarriaga's CMBFAST code \cite{selzal469,zalsel129}) much faster than the one now used, which should allow us to fit the parameters and compare the results with the other cosmological models.}). We list the parameters used and their value ranges in table \ref{parrange} (remember the comment on the grid choice);  the total and vacuum densities (not listed in the table) are fixed by our choice of a flat geometry: $ \Omega_{\rm 0} = 1 $ and $ \Omega_{\Lambda} = 1 - \Omega_{\rm m} $. In addition, in order to study the parameter dependence, we did computations also for different numbers of extra-neutrino species ($ \Delta N_\nu =0.5, 1.0, 1.5 $).


\passo
\def \cmb_2{The cosmic microwave background for a Mirror Universe}
\section{\cmb_2}
\label{cmb_2}
\markboth{Chapter \ref{chap-mirror_univ_3}. ~ \mir_univ_3}
                    {\S \ref{cmb_2} ~ \cmb_2}

\begin{figure}[p]
  \begin{center}
    \leavevmode
    \epsfxsize =12cm
    \epsffile{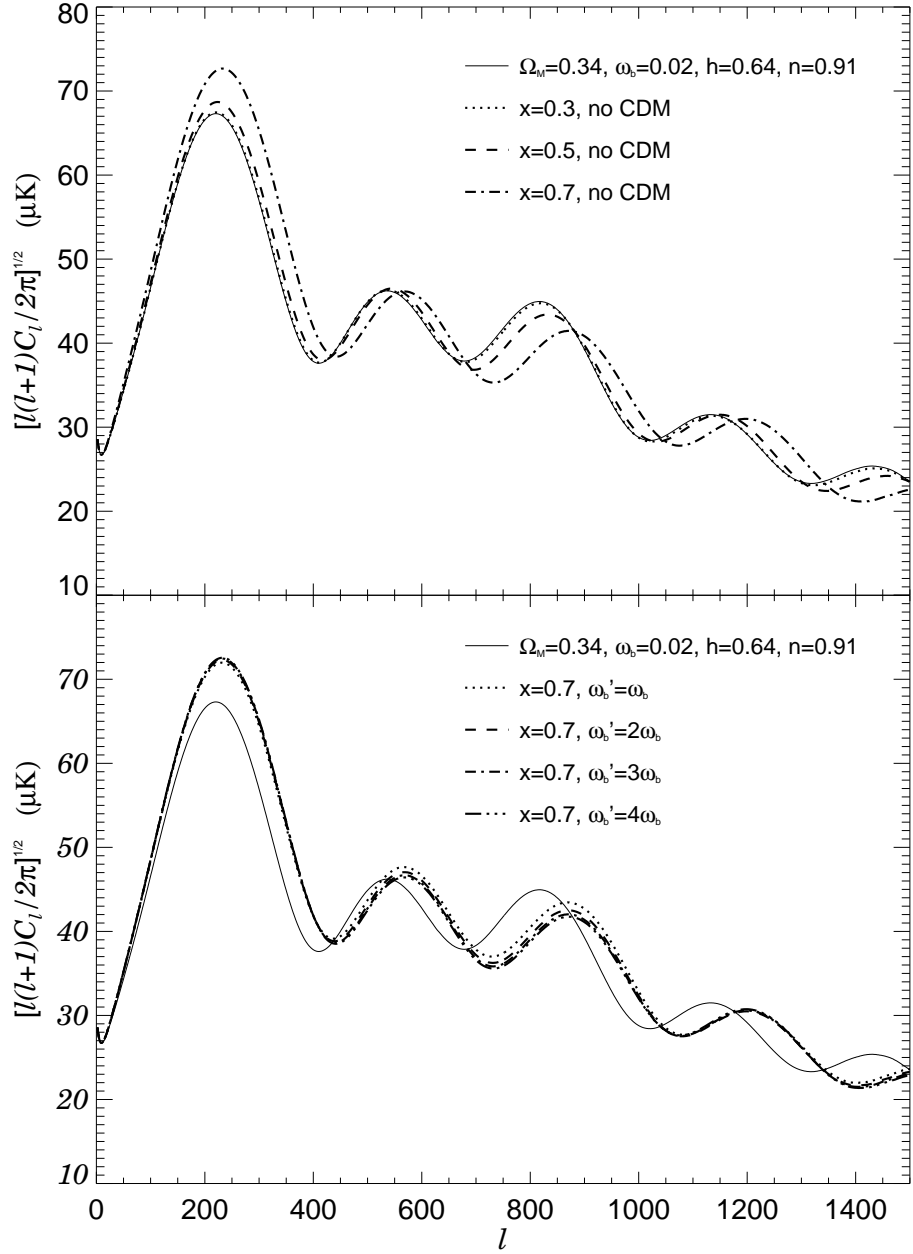}
  \end{center}
\vspace{-.4cm}
\caption{\small CMB angular power spectrum for different values of $ x $ and $ \omega_{\rm b}' = \Omega_{\rm b}' h^2 $, compared with a standard model (solid line). 
{\sl Top panel.} Mirror models with the same parameters as the ordinary one, and with $ x = 0.3, 0.5, 0.7 $, and $ \omega_{\rm b}' = \Omega_{\rm m} h^2 - \omega_{\rm b} $ (no CDM) for all models.
{\sl Bottom panel.} Mirror models with same parameters as the ordinary one, and with $ x = 0.7 $ and $ \omega_{\rm b}' = \omega_{\rm b}, 2 \omega_{\rm b}, 3 \omega_{\rm b}, 4 \omega_{\rm b} $.}
\label{cmblssfig1}
\end{figure}

As anticipated when we studied the structure formation, we expect that the existence of a mirror sector influences the cosmic microwave background radiation observable today; now we want to evaluate this effect.

We choose a starting standard model and add a mirror sector simply removing cold dark matter and adding mirror baryons. At present, we think that a good reference model is the so-called ``concordance model'' of Wang et al. (2002) \cite{wang} (in this case they performed an 11-parameters fit of all the available experimental data). We are aware that for the reader this model will be surely obsolete, given the current rate of new experimental data. However, this is not a shortcoming, because here we want only to put in evidence the differences respect to a representative reference model, and this is a good model for this purpose, unless new (very unlikely) revolutionary observations will change the scenario. The parameter values for this reference model are: $ \Omega_0 = 1 $, $ \Omega_{\rm m} = 0.34 $, $ \Omega_{\rm \Lambda} = 0.66 $, $ \omega_{\rm b} = \Omega_{\rm b} h^2 = 0.02 $, $ n_{\rm s} = 0.91 $, $ h = 0.64 $, with only cold dark matter (no massive neutrinos) and scalar adiabatic perturbations.

From this starting point, we first substitute all the cold dark matter with mirror baryonic dark matter (MBDM) and evaluate the CMB angular power spectrum varying $ x $ from 0.3 to 0.7. This is shown in top panel of figure \ref{cmblssfig1}, where mirror models are plotted together with the concordance model. The first evidence is that the deviation from the standard model is not linear in $ x $: it grows more for bigger $ x $ and for $ x \lsim 0.3 $ the power spectra are practically coincident. This is important, because it means that a Universe where all the dark matter is made of mirror baryons could be indistinguishable from a CDM model if we analyse the CMB only. We see the greatest separation from the standard model for $ x = 0.7 $, but it will increase for larger values of $ x $. 
The height of the first acoustic peak grows for $ x \gsim 0.3 $, while the position remains nearly constant. For the second peak occurs the opposite, i.e. the height remains practically constant, while the position shifts toward higher $ l $; for the third peak, instead, we have a shift both in height and position (the absolute shifts are similar to the ones for the first two peaks, but the height now decreases instead of increasing).
Observing also the other peaks, we recognize a general pattern: except for the first one, odd peaks change both height and location, even ones change location only.

In bottom panel of figure \ref{cmblssfig1} we show the intermediate case of a mixture of CDM and MBDM. We consider  $ x = 0.7$, a high value which permits us to see well the differences, and change $ \omega_{\rm b}' $ from  $ \omega_{\rm b} $ (20\% of $ \Omega_{\rm dm} $) to $ 4 \omega_{\rm b} $ (80\% of $ \Omega_{\rm dm} $). The dependence on the amount of mirror baryons is lower than on the ratio of temperatures $ x $. In fact, the position of the first peak is nearly stable for all the mirror models (except for a very low increase of height for growing $ \omega_{\rm b}' $), while differences appear for the other peaks. In the second peak the position is shifted as in the case without CDM independently from $ \omega_{\rm b}' $, while the height is inversely proportional to $ \omega_{\rm b}' $ with a separation appreciable for $ \omega_{\rm b}' \lsim 3 \omega_{\rm b} $. For the third peak the behaviour is the same as for the case without CDM, with a slightly stronger dependence on $ \omega_{\rm b}' $, while for the other peaks there is a weaker dependence on $ \omega_{\rm b}' $. A common feature is that the heights of the peaks are not linearly dependent on $ \omega_{\rm b}' $, while their positions are practically insensitive to $ \omega_{\rm b}' $ but depend only on $ x $.

We will analyse in more detail the $ x $ and $ \omega_{\rm b}' $ dependence of the peaks, together with other parameters, in \S~\ref{para_trends}.


\def \cmb_3{The mirror cosmic microwave background radiation}
\subsection{\cmb_3}
\label{cmb_3}
\markboth{Chapter \ref{chap-mirror_univ_3}. ~ \mir_univ_3}
                    {\S \ref{cmb_3} ~ \cmb_3}

\begin{figure}[p]
  \begin{center}
    \leavevmode
    \epsfxsize = 12cm 
    \epsffile{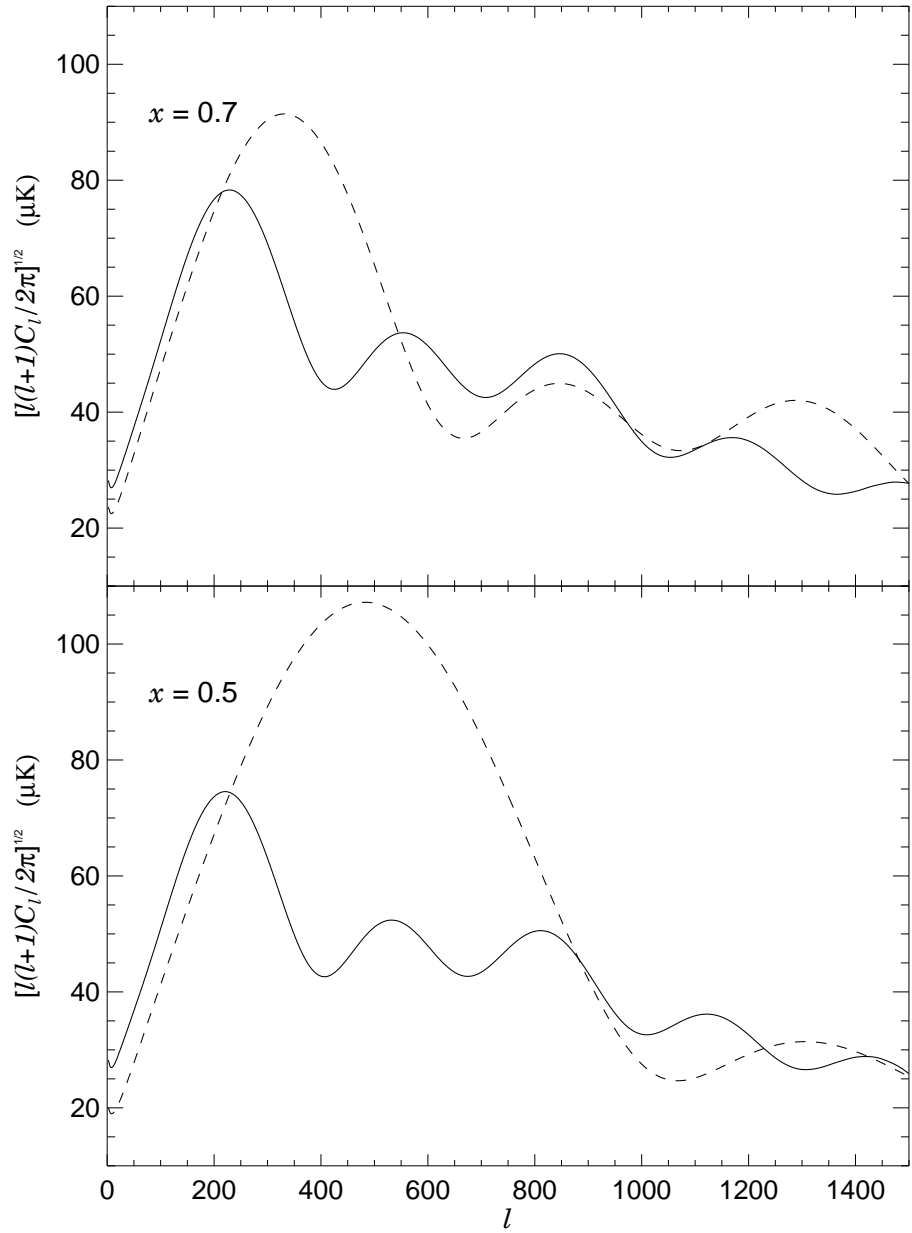}
  \end{center}
\caption{\small Angular power spectra for ordinary (solid line) and mirror (dashed line) CMB photons. The models have $ \Omega_0 = 1 $, $ \Omega_{\rm m} = 0.3 $, $ \omega_{\rm b} = \omega_{\rm b}' = 0.02 $, $ h = 0.7 $, $ n = 1.0 $, and $ x = 0.7 $ (top panel) and $ x = 0.5 $ (bottom panel).}
\label{cmblssfig2}
\end{figure}

In the same way as ordinary photons at decoupling from baryons formed the CMB we observe today, also mirror photons at their decoupling formed a mirror cosmic microwave background radiation, which, on the contrary, we cannot observe because they don't couple with the ordinary baryons of which we are made\footnote{Indeed, there is in principle the possibility of an influence of the mirror CMB on our photons in case of existence of a photon-mirror photon mixing (as supposed, for example, by Foot (2002) \cite{foot0207175} and references therein), but its detection is not possible with present and probably even future experiments.} (it would be possible for an hypothetical mirror observer, instead). Nevertheless, its study is not only speculative, being a way to better understand the cosmology of the Mirror Universe and our observable CMB.

We computed two models of mirror CMB, in order to have elements to compare with the corresponding observable CMBs. The chosen parameter values are those currently estimated (see \S~\ref{sec-now-cosm}), and the amount of mirror baryons is the same as the ordinary ones, while $ x $ is taken as 0.7 or 0.5 to explore different scenarios. Thus the parameters of the models are: $ \Omega_0 = 1 $, $ \Omega_{\rm m} = 0.3 $, $ \omega_{\rm b} = \omega_{\rm b}' = 0.02 $, $ x = 0.7$ or 0.5, $ h = 0.7 $, $ n = 1.0 $.
In figure \ref{cmblssfig2} we plot 
the ordinary and mirror CMB spectra corresponding to the same Mirror Universe. 

The first evidence is that, being scaled by the factor $ x $ the temperatures in the two sectors, also their temperature fluctuations will be scaled by the same amount, as evident if we look at the lowest $ \ell $ values (the fluctuations seeds are the same for both sectors). Starting from the top figure, we see that the first mirror CMB peak is much higher and shifted to higher multipoles than the ordinary one, while other peaks are both lower and at higher $ \ell $ values, with a shift growing with the order of the peak. 

Observing the bottom plot, we note the effect of a change of the parameter $ x $ on the mirror CMB: (i) for lower $ x $-values the first peak is higher (for $ x = 0.5 $ it is nearly one and half the ordinary one); (ii) the position shifts to much higher multipoles (with the same horizontal scale we can no more see some peaks). The reason is that a change of $ x $ corresponds to a change of the mirror decoupling time. The mirror photons, which decouple before the ordinary ones, see a smaller sound horizon, scaled approximately by the factor $ x $; since the first peak occurs at a multipole $ \ell \propto ({\rm sound\;horizon}) ^{-1} $, we expect it to shifts to higher $ \ell $-values by a factor $ x^{-1} $, that is exactly what we see in the figure.

We have verified (even if not shown in the figures) that increasing $ x $ the mirror CMB is more and more similar to the ordinary one, until for $ x = 1 $ the two power spectra are perfectly coincident (as expected, since in this case the two sectors have exactly the same temperatures, the same particle contents, and then their photons power spectra are necessarily the same). 

If we were able to detect both the ordinary and mirror CMB photons, we had two snapshots of the Universe at two different epochs, which were a powerful cosmological instrument, but unfortunately this is impossible, because mirror photons are by definition completely invisible for us.


\newpage
\passo
\def \lss_2{The large scale structure for a Mirror Universe}
\section{\lss_2}
\label{lss_2}
\markboth{Chapter \ref{chap-mirror_univ_3}. ~ \mir_univ_3}
                    {\S \ref{lss_2} ~ \lss_2}

Given the oscillatory behaviour of the mirror baryons (different from the smooth one of cold dark matter), we expect that MBDM induces specific signatures also on the large scale structure power spectrum (see \S~\ref{lss1}).

In order to evaluate this effect, we computed LSS power spectra using the same reference and mirror models used in \S~\ref{cmb_2} for the CMB analysis. So the two panels of figure \ref{cmblssfig3} are the LSS corresponding to those of figure \ref{cmblssfig1}. In order to remove the dependences of units on the Hubble constant, we plot on the $ x $-axis the wave number in units of $ h $ and on the $ y $-axis the power spectrum in units of $ h^3 $. The minimum scale (the maximum $ k $) plotted depends on the limit of the linear regime, placed between $ k/h $ = 0.3 and 0.4 Mpc$^{-1}$, according to what described by Hamilton \& Tegmark (2000) \cite{hamteg330}.

In top panel of the figure we show the dependence on $ x $ for different mirror models without CDM; in this case, where all the dark matter is made of mirror baryons, the oscillatory effect is obviously maximum. The first evidence is the strong dependence on $ x $ of the beginning of oscillations: it goes to higher scales for higher $ x $ and, starting from $ x \simeq 0.3 $, the power spectrum for a Mirror Universe approaches more and more the CDM one. This behaviour is a consequence of the $ x $-dependence of the mirror Silk scale (see \S~\ref{disseff-colldamp} and figure \ref{mirsilksca1}): this dissipative scale induces a cutoff in the power spectrum, which is damped with an oscillatory behaviour (it will be more evident in figures \ref{cmblssfig4} and \ref{cmblssfig5}, where we extend our models to smaller scales inside the non linear region). Oscillations begin at the same time of the damping, and they are so deep (because there are many mirror baryons) to go outside the coordinate box. In any case the mirror spectra are always below the ordinary one for every value of $ x $.

The dependence on the amount of mirror baryons is instead shown in the bottom panel of the figure, where only a fraction of the dark matter is made of mirror baryons, while the rest is CDM. Contrary to the CMB case, the matter power spectrum strongly depends on $ \omega_{\rm b}' $. The oscillations are deeper for increasing mirror baryons densities and the spectrum goes more and more away from the pure CDM one. We note also that the damping begins always at the same scale, and thus it depends only on $ x $ and not on $ \omega_{\rm b}' $, as we know from expression (\ref{mirsilklamb1}). The same considerations are valid for the oscillation minima, which become much deeper for higher mirror baryon densities, but shift very slightly to lower scales, so that their positions remain practically constant.

\begin{figure}[p]
  \begin{center}
    \leavevmode
    \epsfxsize = 12cm
    \epsffile{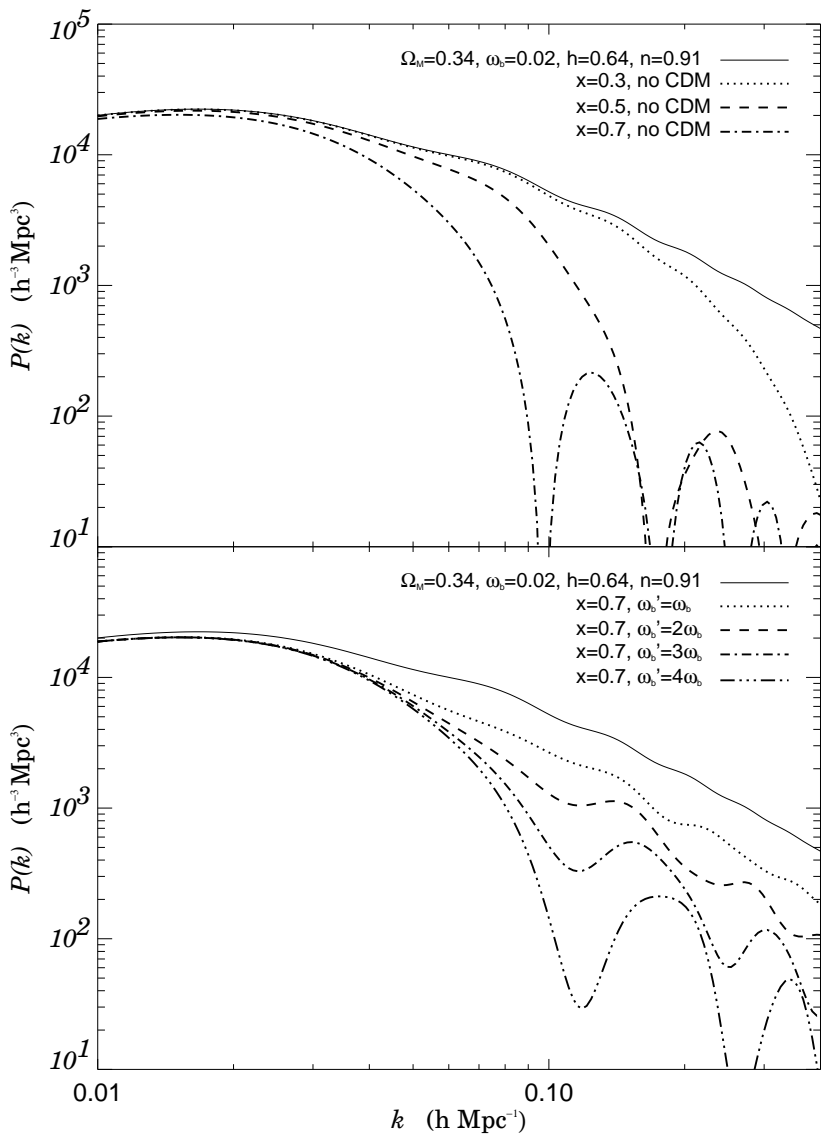}
  \end{center}
\addvspace{.8cm}
\caption{\small LSS power spectrum in the linear regime for different values of $ x $ and $ \omega_{\rm b}' = \Omega_{\rm b}' h^2 $,  compared with a standard model (solid line). In order to remove the dependences of units on the Hubble constant, we plot on the $ x $-axis the wave number in units of $ h $ and on the $ y $-axis the power spectrum in units of $ h^3 $.
{\sl Top panel.} Mirror models with the same parameters as the ordinary one, and with $ x = 0.3, 0.5, 0.7 $ and $ \omega_{\rm b}' = \Omega_{\rm m} h^2 - \omega_{\rm b} $ (no CDM) for all models.
{\sl Bottom panel.} Mirror models with the same parameters as the ordinary one, and with $ x = 0.7 $ and $ \omega_{\rm b}' = \omega_{\rm b}, 2 \omega_{\rm b}, 3 \omega_{\rm b}, 4 \omega_{\rm b} $.}
\label{cmblssfig3}
\end{figure}


\def \lss_non_lin{Extension to smaller (non linear) scales}
\subsection{\lss_non_lin}
\label{lss_non_lin}
\markboth{Chapter \ref{chap-mirror_univ_3}. ~ \mir_univ_3}
                    {\S \ref{lss_non_lin} ~ \lss_non_lin}

\begin{figure}[p]
  \begin{center}
    \leavevmode
    \epsfxsize = 12cm
    \epsffile{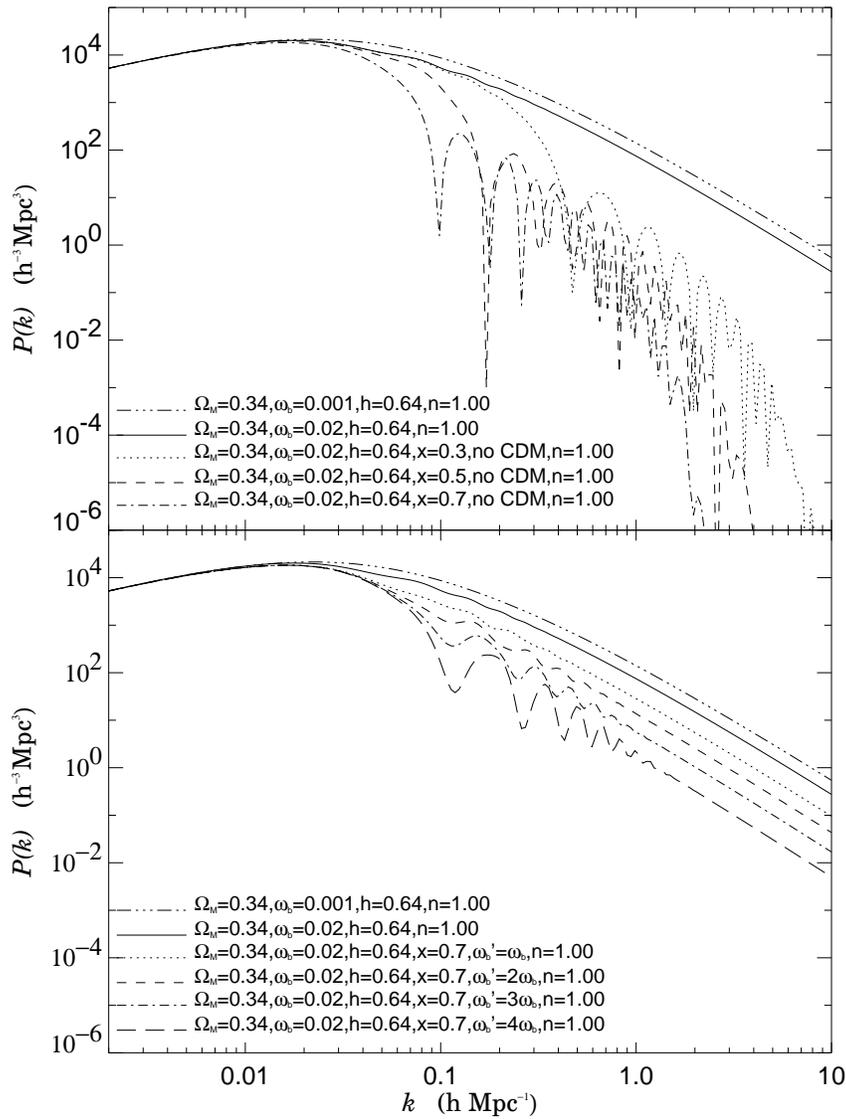}
  \end{center}
\addvspace{.8cm}
\caption{\small LSS power spectrum beyond the linear regime for different values of $ x $ and $ \omega_{\rm b}' = \Omega_{\rm b}' h^2 $,  compared with a standard model (solid line). The models have the same parameters as in figure \ref{cmblssfig3}, except for the spectral index, which is now set to 1.0. For comparison we also show a standard CDM model with a negligible amount of baryons ($ \Omega_{\rm b} \sim 0.2\% $).}
\label{cmblssfig4}
\end{figure}

Let us now extend the behaviour of the matter power spectrum to lower scales, which already became non linear. Obviously, since our treatment is based on the linear theory, it is no longer valid in non linear regime. Nevertheless, even if it is not useful for comparison with observations, the extension of our models to these scales is very useful to understand the behaviour of the power spectrum in a mirror baryonic dark matter scenario, in particular concerning the position of the cutoff (remember the discussion in \S~\ref{cdmscen} about the failure of the CDM scenario).

Therefore, in figures \ref{cmblssfig4} and \ref{cmblssfig5} we extend the power spectra up to $ k/h = 10 $ Mpc$^{-1}$ (corresponding to galactic scales), well beyond the limit of the linear regime, given approximately by $ k/h < 0.4 $ Mpc$^{-1}$.

In figure \ref{cmblssfig4} we plot in both panels the same models as in figure \ref{cmblssfig3}, except for the spectral index now set to 1.0, a value chosen to eliminate the effect of a power spectrum tilt on the cutoff.\footnote{Indeed, here we want to simply compare the standard and mirror power spectra, and a common tilt for both models does not influence our conclusions; nevertheless, we prefer to use a scale invariant spectrum.} For comparison we show also a standard model characterized by a matter density made almost completely of CDM, with only a small contamination of baryons ($ \Omega_{\rm b} \simeq 0.2\% $ instead of $ \simeq 4 \% $ of other models). In the top panel, the $ x $-dependence of the mirror power spectra is considered: the vertical scale extends to much lower values compared to figure \ref{cmblssfig3}, and we can clearly see the deep oscillations, but in particular it is evident the presence of the previously cited cutoff. For larger values of $ x $ oscillations begin earlier and cutoff moves to higher scales. Moreover, note that the model with almost all CDM has more power than the same standard model with baryons, which in turn has more power than all mirror models for any $ x $ and for all the scales. In the bottom panel we show the dependence on the baryon content. It is remarkable that all mirror models stop to oscillate at some low scale and then continue with a smooth CDM-like trend. This means that, after the cutoff due to mirror baryons, the dominant behaviour is the one characteristic of cold dark matter models (due to the lack of a cutoff for CDM). Clearly, for higher mirror baryon densities the oscillations continue down to smaller scales, but, contrary to the previous case, where all the dark matter was mirror baryonic, there will always be a scale below which the spectrum is CDM-like.

An interesting point of the mirror baryonic scenario is his capability to mimic a CDM scenario under certain circumstances and for certain measurements. To explain this point, in figure \ref{cmblssfig5} we show models with low $ x $-values (0.2 or 0.1) and all dark matter made of MBDM; we see that for $ x = 0.2 $ the standard and mirror power spectra are already practically coincident in the linear region. If we go down to $ x = 0.1 $ the coincidence is extended up to $ k/h \sim 1 $ Mpc$ ^{-1} $. In principle, we could still decrease $ x $ and lengthen this region of equivalence between the different CDM and MBDM models, but we have to remember that we are dealing with linear models extended to non linear scales, so neglecting all the non linear phenomena (such as merging or stellar feedback), that are very different for the CDM and the MBDM scenarios. In the same plot we also considered a model with $ x = 0.2 $ and dark matter composed equally by mirror baryons and by CDM. This model shows that in principle it's possible a tuning of the cutoff effect reducing the amount of mirror matter, in order to better reproduce the cutoff needed to explain, for example, the low number of small satellites in galaxies.

This work provided for us the linear transfer functions, which constitute the principal ingredient for the computation of the power spectrum at non linear scales. This calculation is out of the aim of this thesis, but is one of the next steps in the study of the Mirror Universe.

\begin{figure}[h]
  \begin{center}
    \leavevmode
    \epsfxsize = 12cm
    \epsffile{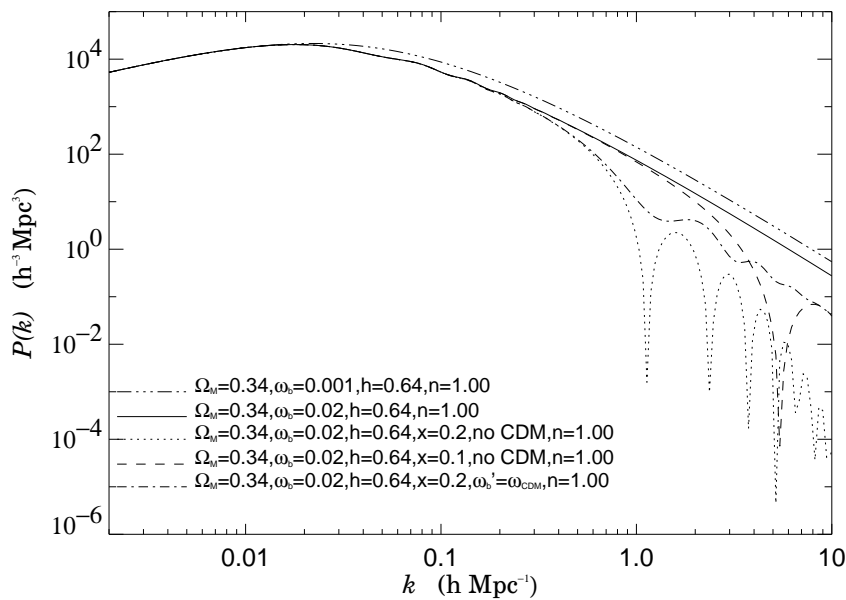}
  \end{center}
\addvspace{-6.5cm}
\caption{\small LSS power spectrum beyond the linear regime for two low values of $ x $ (0.1 and 0.2) and different amounts of mirror baryons ($ \Omega_{\rm b}' = \Omega_{\rm m} - \Omega_{\rm b} $ or $ \Omega_{\rm b}' = \Omega_{\rm CDM} $),  compared with a standard model (solid line). The other parameters are the same as in figure \ref{cmblssfig3}, except for the spectral index, which is now set to 1.0. For comparison we show also a standard CDM model with a negligible amount of baryons ($ \Omega_{\rm b} \sim 0.2\% $).}
\label{cmblssfig5}
\end{figure}


\newpage
\passo
\def \para_trends{Parameter trends}
\section{\para_trends}
\label{para_trends}
\markboth{Chapter \ref{chap-mirror_univ_3}. ~ \mir_univ_3}
                    {\S \ref{para_trends} ~ \para_trends}

The shapes, heights and locations of peaks 
and oscillations in the photons and matter power spectra are predicted by all models based on the inflationary scenario. Furthermore the details of the 
features in these power spectra depend critically on the chosen cosmological parameters, which in turn can be accurately determined by precise measurements of these patterns. In this section we briefly discuss the sensitivity of the $ C_\ell $'s and $ P(k) $'s on the values of some fundamental parameters in a mirror baryonic scenario.

In particular, the exact form of the CMB and LSS power spectra is greatly dependent on assumptions about the matter content of the Universe. Apart from the total density parameter $ \Omega_0 $ (our models are flat, so $ \Omega_0 $ is always 1), the composition of the Universe can be parametrized by its components $ \Omega_{\rm m} $ and $ \Omega_\Lambda $, and the components of the matter density $ \Omega_{\rm b}$, $ \Omega_{\rm b}' $, $ \Omega_{\rm CDM} $. Further parameters are the tilt of scalar fluctuations $ n $, the Hubble parameter {\it h}, and the ratio of temperatures in the two sectors $ x $. In addition, we consider also the dependence on the number of massless neutrino species $ N_\nu $, in order to compare it with the $ x $-dependence. This is important if we remember that the relativistic mirror particles can be parametrized in terms of effective number of extra-neutrino species (see \S~\ref{term_mir_univ}).

Starting from the reference model of parameters $ \Omega_{\rm m} = 0.3 $, $ \omega_{\rm b} = \omega_{\rm b}' = 0.02 $, $ x = 0.2 $, $ h = 0.7 $ and $ n = 1.0 $, we change one parameter each time, compute the respective models, and plot the CMB and LSS power spectra in order to show the dependence on it (figures \ref{cmblssfig8} - \ref{cmblssfig17}). Then, we compute the relative locations and heights of the first three acoustic peaks of the CMB angular power spectrum and plot them in figures \ref{cmblssfig6} and \ref{cmblssfig7} to compare their sensitivities to the parameters.

In the following we briefly analyse the dependences on every parameter, referring to the figure where the respective models are plotted.

\begin{itemize}

\item{\it Matter density} (fig.~\ref{cmblssfig8}): $ \Omega_{\rm m} $ varies from 0.1 to 0.5. In flat models a decrease in $ \Omega_{\rm m} $ implies two things: an increase in $ \Omega_\Lambda $ (with the consequent delay in matter-radiation equality) and a decrease in $ \Omega_{\rm CDM} $ (if we leave unchanged the O and M baryon densities). Both these things correspond to boosting and shifting effects on the acoustic peaks, while the matter power spectrum goes down, given the decreasing in the collisionless species (CDM) and the progressive relative growth of the baryon densities, which are responsible for the oscillatory features.

\item{\it O baryon density} (fig.~\ref{cmblssfig10}): $ \omega_{\rm b} = \omega_{\rm b}' $ varies from 0.01 to 0.03. An increase of the baryon fraction increases odd peaks (compression phase of the baryon-photon fluid) due to extra-gravity from baryons with respect to the even peaks (rarefaction phase of the fluid oscillation) in the CMB, and generate deeper oscillations in the LSS. In particular, the relative magnitudes of the first and second acoustic peaks are sensitive to $\omega_{\rm b}$, as we see in figure \ref{cmblssfig7}. These effects are completely due to O baryons. In fact, even if not shown, we have verified that an increase of $ \omega_{\rm b} $ with a constant $ \omega_{\rm b}' $ has exactly the same consequences for this value of $ x $, while to see the effect of M baryons we have to raise the temperature of the mirror sector.

\item {\it Hubble constant} (figs.~\ref{cmblssfig11} and \ref{cmblssfig12}): $ h $ varies from 0.50 to 0.90, but now we can leave constant either $ \Omega_{\rm b} = \Omega_{\rm b}' $ (fig.~\ref{cmblssfig11}) or $ \omega_{\rm b} = \omega_{\rm b}' $ (fig.~\ref{cmblssfig12}). In both cases a decrease in $ h $ corresponds to a delay in the epoch of matter-radiation equality and to a different expansion rate. This boosts the CMB peaks and slightly changes their location toward higher $ \ell $'s (similar to the effect of an increase in $ \Omega_\Lambda $), and induces a decrease in the LSS spectrum. There are slight differences between the two situations of $ \Omega $ or $ \omega $ constant, evident in particular on the first acoustic peak and on the matter oscillations. If fact, when we consider $ \omega_{\rm b,b'} $ constant, the baryon densities $ \Omega_{\rm b,b'} = \omega_{\rm b,b'} / h^2 $ grow for decreasing $ h $, then favouring the raise of the first peak in the CMB and the onset of oscillations in the LSS.

\item {\it Spectral index} (fig.~\ref{cmblssfig13}): $ n $ varies from 0.90 to 1.10. Increasing $ n $  will raise the power spectra at large $\ell$'s with respect to the low $\ell$'s and at large values of $ k $ with respect to low values. This is not so evident in figure (except before the first acoustic peak), where the curves seem nearly parallel as if they were simply vertically shifted; this means a low sensitivity to the spectral index in this range, as also evident in figures \ref{cmblssfig6} and \ref{cmblssfig7} for the CMB.

\item {\it Extra-neutrino species} (fig.~\ref{cmblssfig14}): $ \Delta N_\nu $ varies from 0.0 to 1.5. The effect of increasing the number of massless neutrino species is a slow raise of the first acoustic peak and a shift to higher $ \ell $ values for next peaks, together with a slight lowering of the matter power spectrum; all these changes are nearly proportional to $ \Delta N_\nu $, as shown also in figures \ref{cmblssfig6} and \ref{cmblssfig7}.

\item {\it Ratio of temperatures} (fig.~\ref{cmblssfig15}): $ x $ varies from 0.2 to 0.7. Concerning the CMB, the effect of raising $ x $ is qualitatively the same as an increase in $ \Delta N_\nu $, but more pronounced (for these ``cosmologically compatible'' ranges) and with a non-linear dependence. In the LSS spectrum, instead, the situation is different from the case of extra-neutrino species, as now a growth of $ x $ induces the onset of the oscillatory features at lower values of $ k $. This behaviour has been studied in more details in \S~\ref{cmb_2} and \S~\ref{lss_2}.

\item {\it M baryon density} (fig.~\ref{cmblssfig17}): $ \omega_{\rm b}' $ varies from 0.01 ($ \omega_{\rm b} $ / 2) to 0.08 ($ 4 \omega_{\rm b} $). The value of $ x $ is now raised to 0.7, because for 0.2 there aren't differences between models with different $ \omega_{\rm b}' $ values (we start observing small deviations only for the higher $ k $-values in the matter power spectrum). Also this behaviour has been studied in more details in \S~\ref{cmb_2} and \S~\ref{lss_2}; here we want to emphasize the low sensibility of the CMB on $ \omega_{\rm b}' $ (with a slightly stronger dependence starting from the third peak) and, on the contrary, the high sensitivity of the LSS. For the first one, an increase in $ \omega_{\rm b}' $ causes a very low increase of the height of the first peak and a progressive more pronounced decrease for the next peaks, while for the second one there is a fast deepening of the oscillations, slightly changing their locations.

\end{itemize}

In figures \ref{cmblssfig6} and \ref{cmblssfig7} we focus our attention on the CMB first three peaks, choosing some indicator which could quantify the sensitivity to the parameters previously discussed. In figure \ref{cmblssfig6} we analyse the locations of the peaks, plotting the differences of the locations between the various models and the reference model for the three peaks; in figure \ref{cmblssfig7} we plot the deviations of the differences between the heights of the peaks from the same quantities obtained for the reference model. In this way we obtain a clear picture of the trends of these indicators varying the parameters. These plots provide a useful reference in order to evaluate the influence of each parameter on the CMB and LSS power spectra, and they contain a number of informations; we can extract some of them particularly worth of noting.

Looking at the locations, we see a great sensitivity on the matter density and the Hubble constant, and a negligible one on the spectral index and the amount of mirror baryons. Concerning the extra-neutrino species and the temperature of the sectors, the sensitivities are comparable, but the trends are different: they are respectively a constant slope for $ \Delta N_\nu $ and an increasing one for $ x $.

As regards the peak temperatures, the most sensitive parameter, besides $ \Omega_{\rm m} $ and $ h $, is $ \omega_{\rm b} $; the dependence on $ n $ and $ \omega_{\rm b}' $ is a bit greater than what it is for the locations, and the differences between the trends with $ N_\nu $ and $ x $ are slightly more evident, specially for values $ x > 0.6 $.

\begin{figure}[p]
  \begin{center}
    \leavevmode
    \epsfxsize = 12cm
    \epsffile{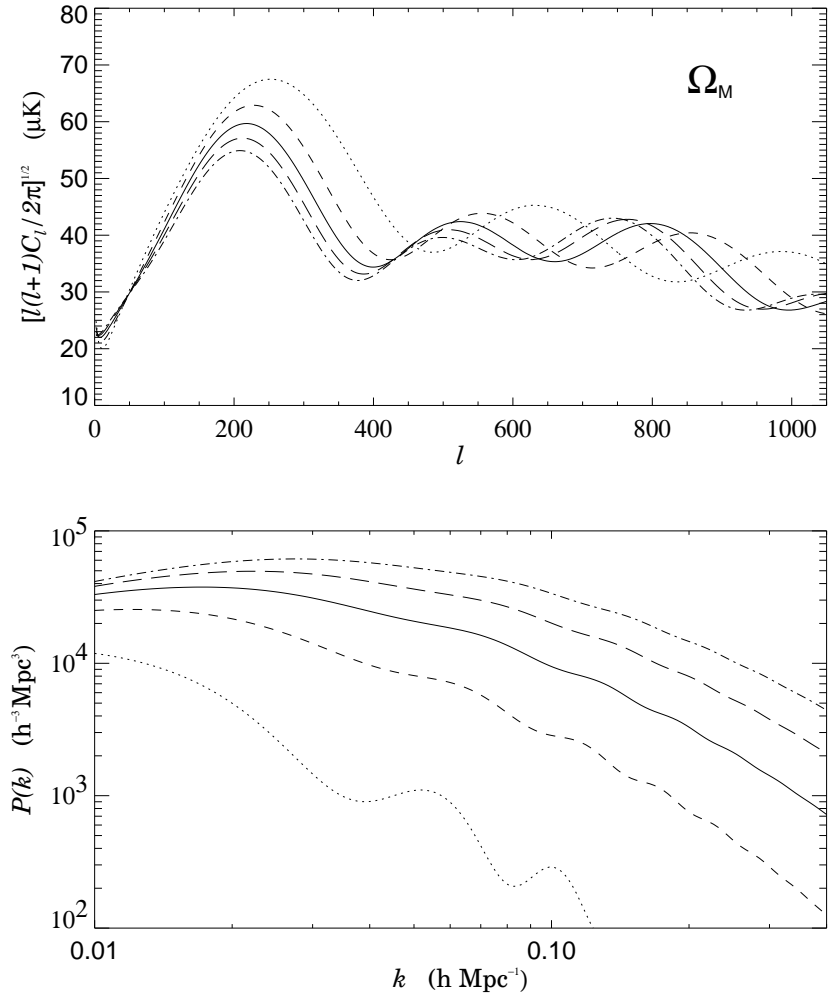}
  \end{center}
\caption{\small Dependence of the shape of photon and matter power spectra on the matter density $ \Omega_{\rm m} $. The reference model (solid line) has: $ \Omega_{\rm m} = 0.3 $, $ \omega_{\rm b} = \omega_{\rm b}' = 0.02 $, $ x = 0.2 $, $ h = 0.7 $ and $ n = 1.0 $. For other models, all the parameters are unchanged except for the one indicated: $ \Omega_{\rm m} = 0.1 $ (dot line), $ \Omega_{\rm m} = 0.2 $ (dash line), $ \Omega_{\rm m} = 0.4 $ (long dash line), $ \Omega_{\rm m} = 0.5 $ (dot-dash line).}
\label{cmblssfig8}
\end{figure}

\begin{figure}[p]
  \begin{center}
    \leavevmode
    \epsfxsize = 12cm
    \epsffile{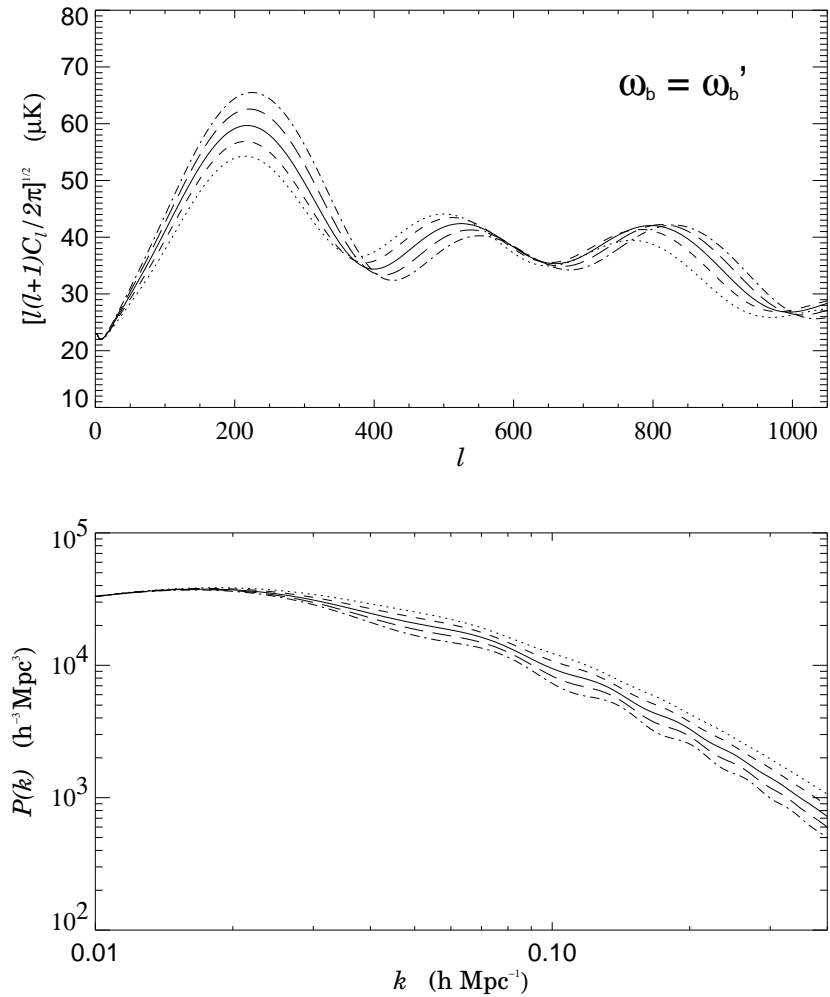}
  \end{center}
\caption{\small Dependence of the shape of photon and matter power spectra on the ordinary baryon density $ \omega_{\rm b} $ with $ \omega_{\rm b} = \omega_{\rm b}' $. The reference model (solid line) has: $ \Omega_{\rm m} = 0.3 $, $ \omega_{\rm b} = \omega_{\rm b}' = 0.02 $, $ x = 0.2 $, $ h = 0.7 $ and $ n = 1.0 $. For other models, all the parameters are unchanged except for the one indicated: $ \omega_{\rm b} = 0.010 $ (dot line), $ \omega_{\rm b} = 0.015 $ (dash line), $ \omega_{\rm b} = 0.025 $ (long dash line), $ \omega_{\rm b} = 0.03 $ (dot-dash line).}
\label{cmblssfig10}
\end{figure}

\begin{figure}[p]
  \begin{center}
    \leavevmode
    \epsfxsize = 12cm
    \epsffile{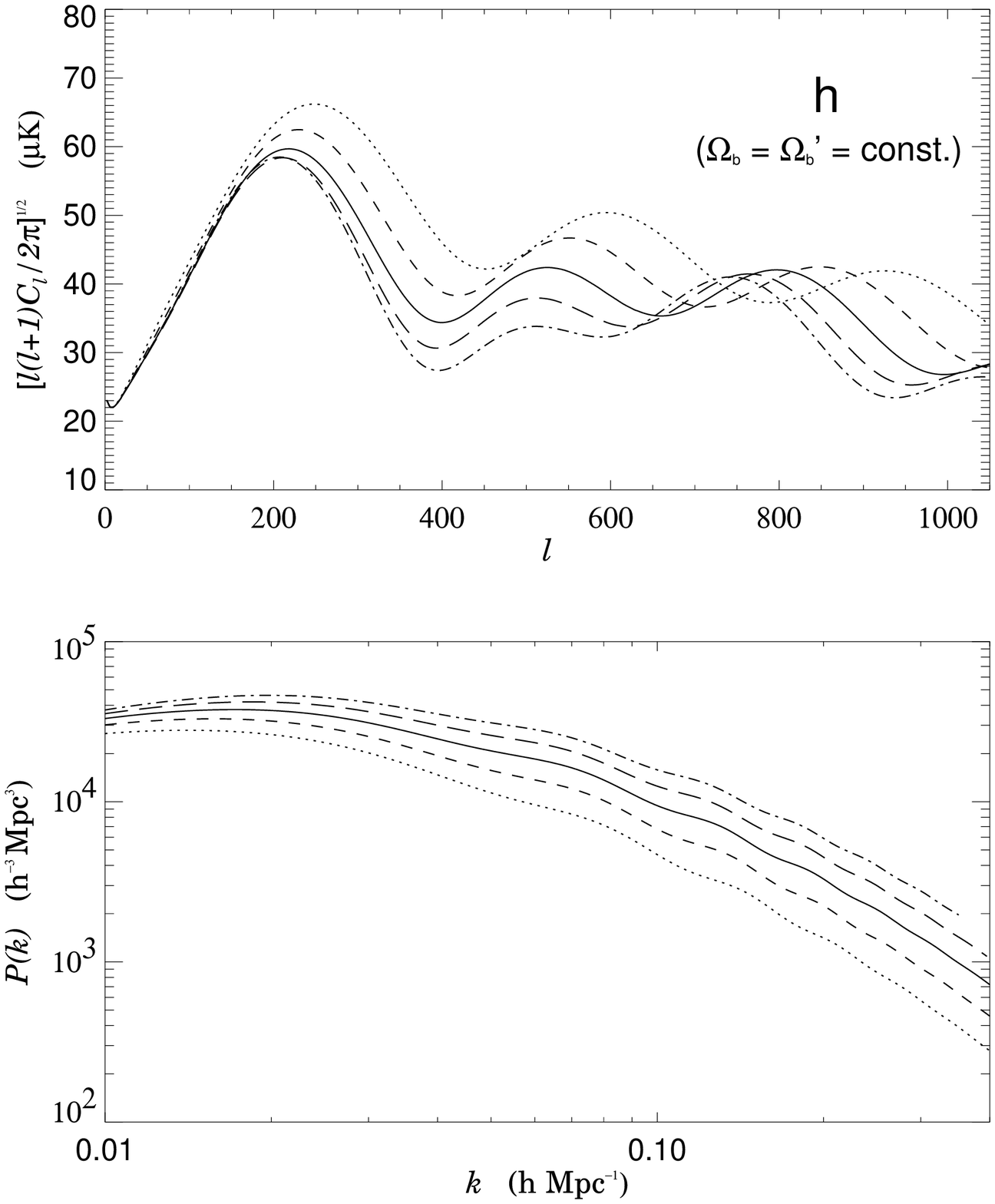}
  \end{center}
\caption{\small Dependence of the shape of photon and matter power spectra on the Hubble parameter $ h $ with $ \Omega_{\rm b} = \Omega_{\rm b}' = {\rm const} $. The reference model (solid line) has: $ \Omega_{\rm m} = 0.3 $, $ \Omega_{\rm b} = \Omega_{\rm b}' = 0.0408 $ (the value obtained for $ \omega_{\rm b} = 0.02 $ and $ h = 0.7 $), $ x = 0.2 $, $ h = 0.7 $ and $ n = 1.0 $. For other models, all the parameters are unchanged except for the one indicated: $ h = 0.5 $ (dot line), $ h = 0.6 $ (dash line), $ h = 0.8 $ (long dash line), $ h = 0.9 $ (dot-dash line).}
\label{cmblssfig11}
\end{figure}

\begin{figure}[p]
  \begin{center}
    \leavevmode
    \epsfxsize = 12cm
    \epsffile{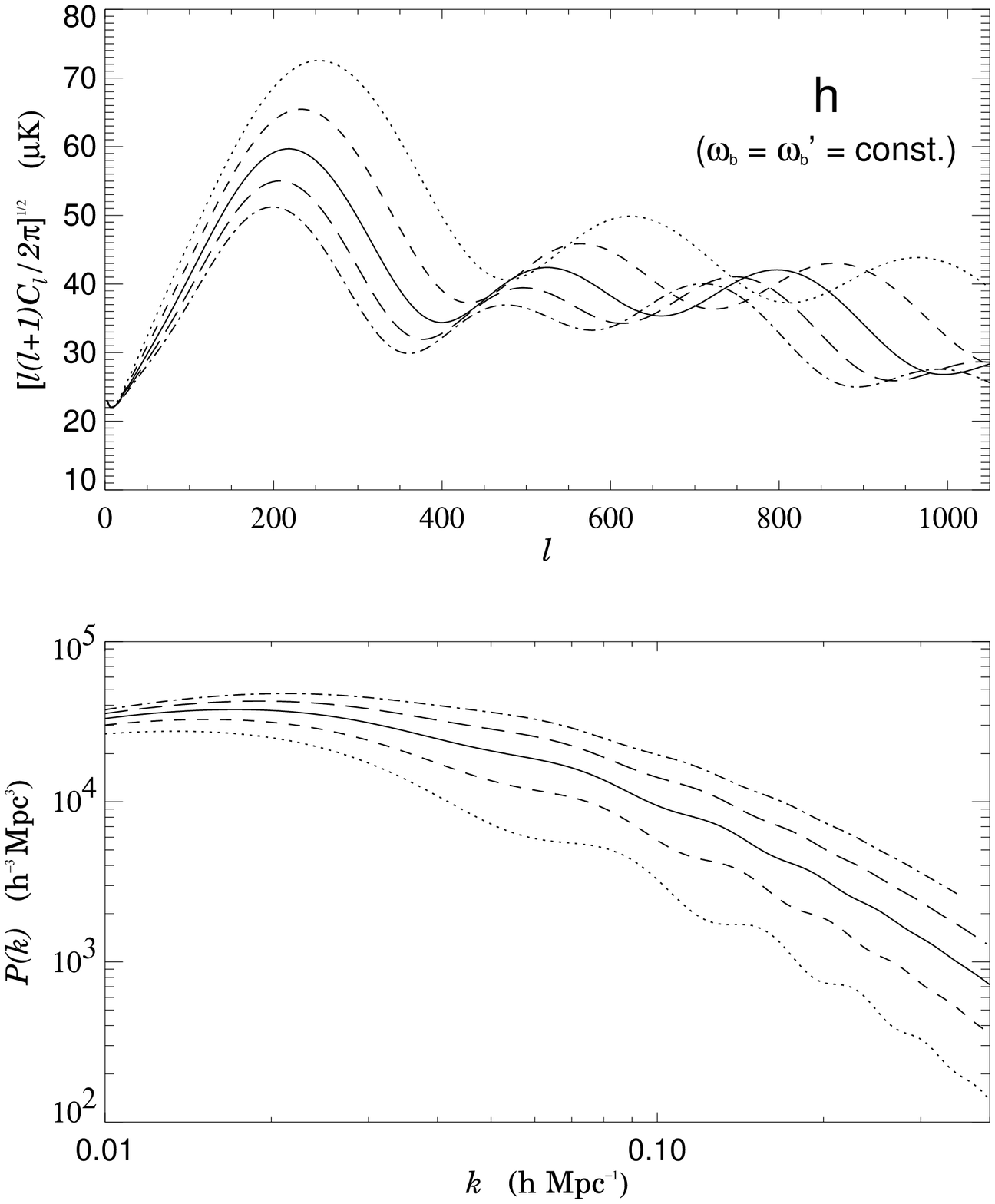}
  \end{center}
\caption{\small Dependence of the shape of photon and matter power spectra on the Hubble parameter $ h $ with $ \omega_{\rm b} = \omega_{\rm b}' = {\rm const} $. The reference model (solid line) has: $ \Omega_{\rm m} = 0.3 $, $ \omega_{\rm b} = \omega_{\rm b}' = 0.02 $, $ x = 0.2 $, $ h = 0.7 $ and $ n = 1.0 $. For other models, all the parameters are unchanged except for the one indicated: $ h = 0.5 $ (dot line), $ h = 0.6 $ (dash line), $ h = 0.8 $ (long dash line), $ h = 0.9 $ (dot-dash line).}
\label{cmblssfig12}
\end{figure}

\begin{figure}[p]
  \begin{center}
    \leavevmode
    \epsfxsize = 12cm
    \epsffile{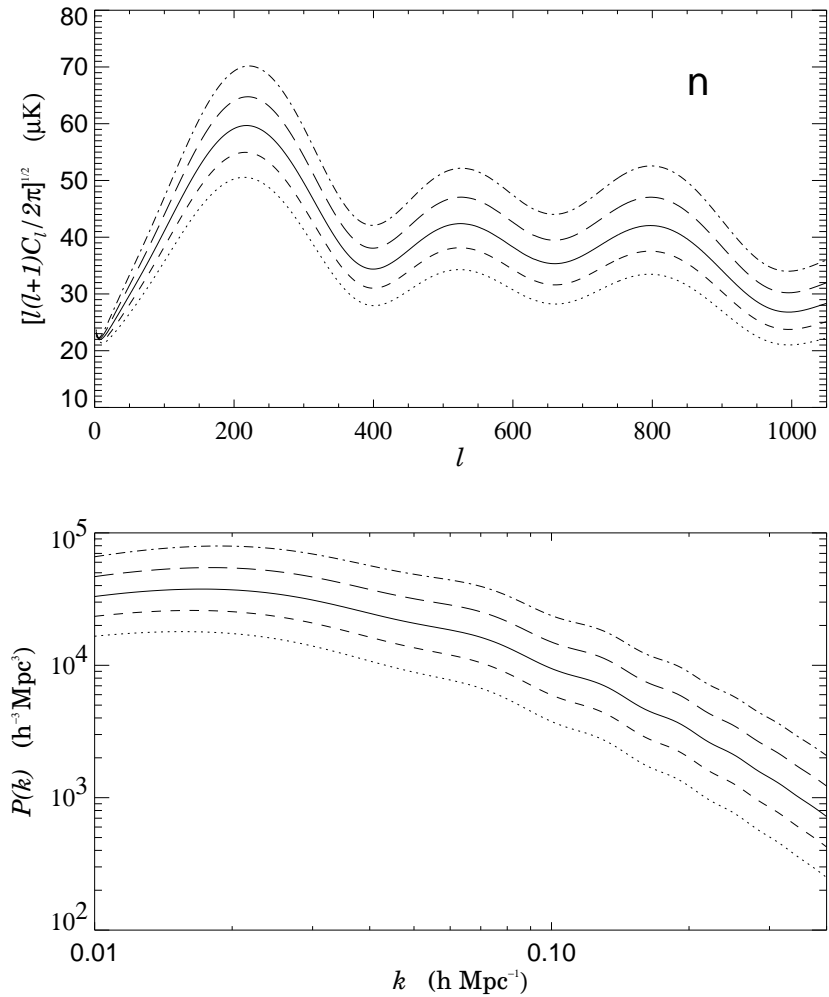}
  \end{center}
\caption{\small Dependence of the shape of photon and matter power spectra on the scalar spectral index $ n $. The reference model (solid line) has: $ \Omega_{\rm m} = 0.3 $, $ \omega_{\rm b} = \omega_{\rm b}' = 0.02 $, $ x = 0.2 $, $ h = 0.7 $ and $ n = 1.0 $. For other models, all the parameters are unchanged except for the one indicated: $ n = 0.90 $ (dot line), $ n = 0.95 $ (dash line), $ n = 1.05 $ (long dash line), $ n = 1.10 $ (dot-dash line).}
\label{cmblssfig13}
\end{figure}

\begin{figure}[p]
  \begin{center}
    \leavevmode
    \epsfxsize = 12cm
    \epsffile{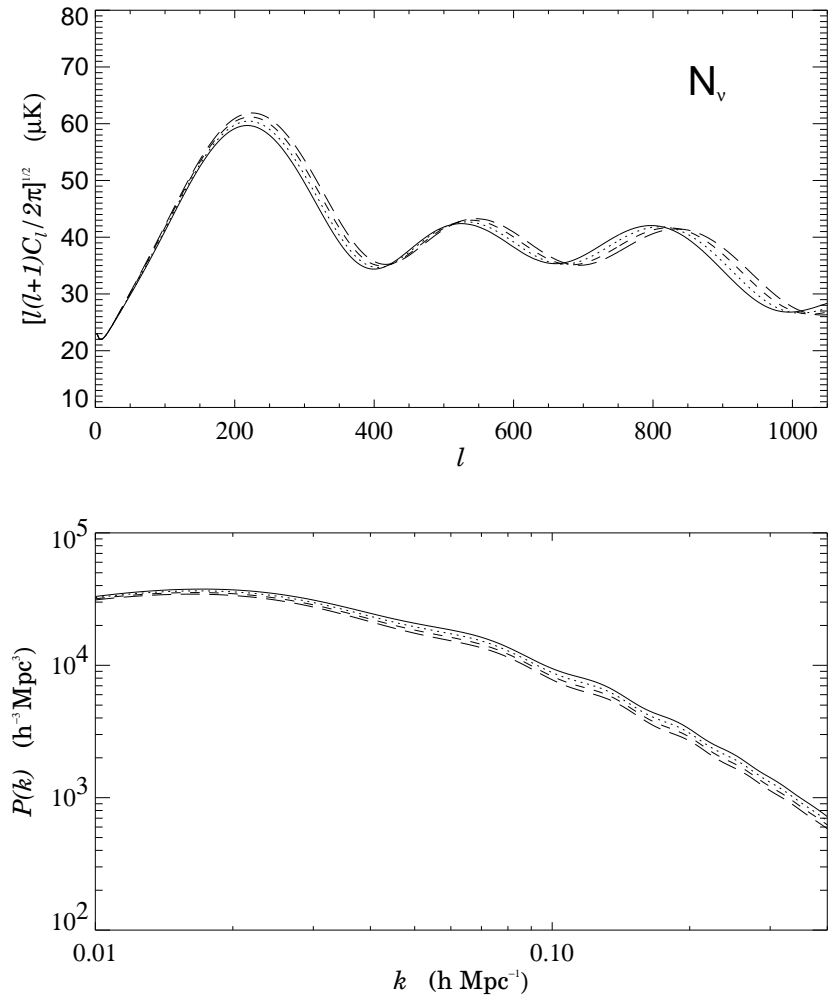}
  \end{center}
\caption{\small Dependence of the shape of photon and matter power spectra on the number of extra-neutrino species $ \Delta N_\nu $. The reference model (solid line) has: $ \Omega_{\rm m} = 0.3 $, $ \omega_{\rm b} = \omega_{\rm b}' = 0.02 $, $ x = 0.2 $, $ h = 0.7 $, $ n = 1.0 $, and $ \Delta N_\nu = 0 $. For other models, all the parameters are unchanged except for the one indicated: $ \Delta N_\nu = 0.5 $ (dot line), $ \Delta N_\nu = 1.0 $ (dash line), $ \Delta N_\nu = 1.5 $ (long dash line).}
\label{cmblssfig14}
\end{figure}

\begin{figure}[p]
  \begin{center}
    \leavevmode
    \epsfxsize = 12cm
    \epsffile{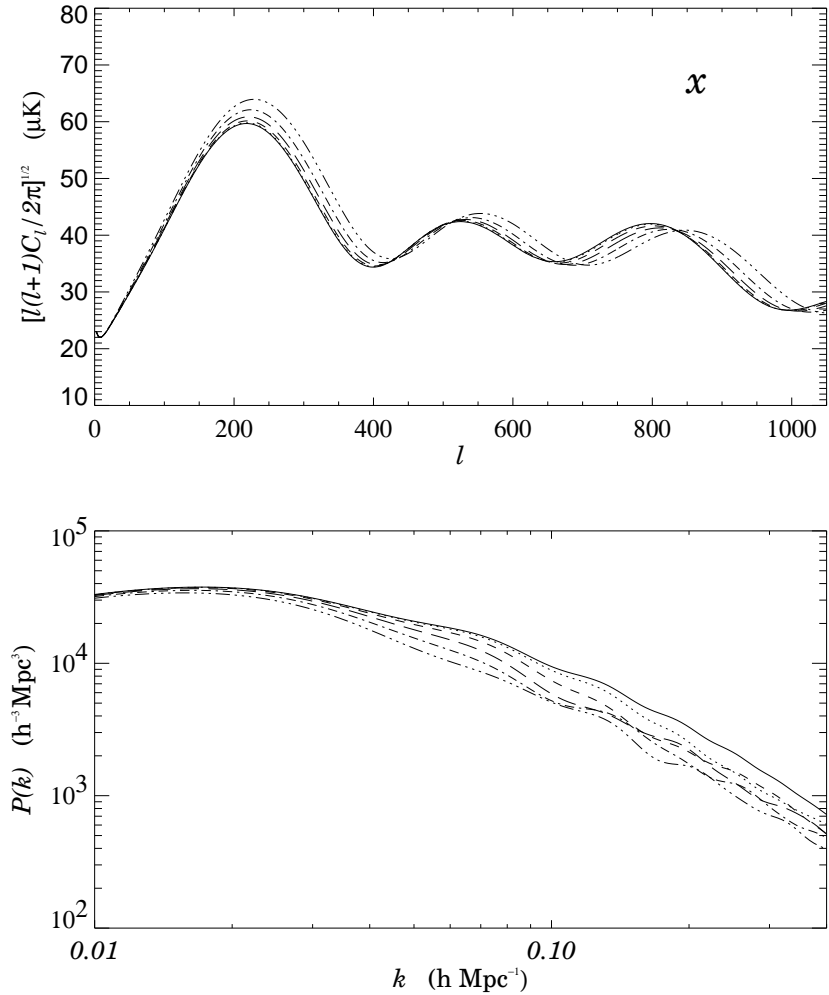}
  \end{center}
\caption{\small Dependence of the shape of photon and matter power spectra on the ratio of the temperatures of two sectors $ x $. The reference model (solid line) has: $ \Omega_{\rm m} = 0.3 $, $ \omega_{\rm b} = \omega_{\rm b}' = 0.02 $, $ x = 0.2 $, $ h = 0.7 $ and $ n = 1.0 $. For other models, all the parameters are unchanged except for the one indicated: $ x = 0.3 $ (dot line), $ x = 0.4 $ (dash line), $ x = 0.5 $ (long dash line), $ x = 0.6 $ (dot-dash line), $ x = 0.7 $ (dot-dot-dot-dash line).}
\label{cmblssfig15}
\end{figure}

\begin{figure}[p]
  \begin{center}
    \leavevmode
    \epsfxsize = 12cm
    \epsffile{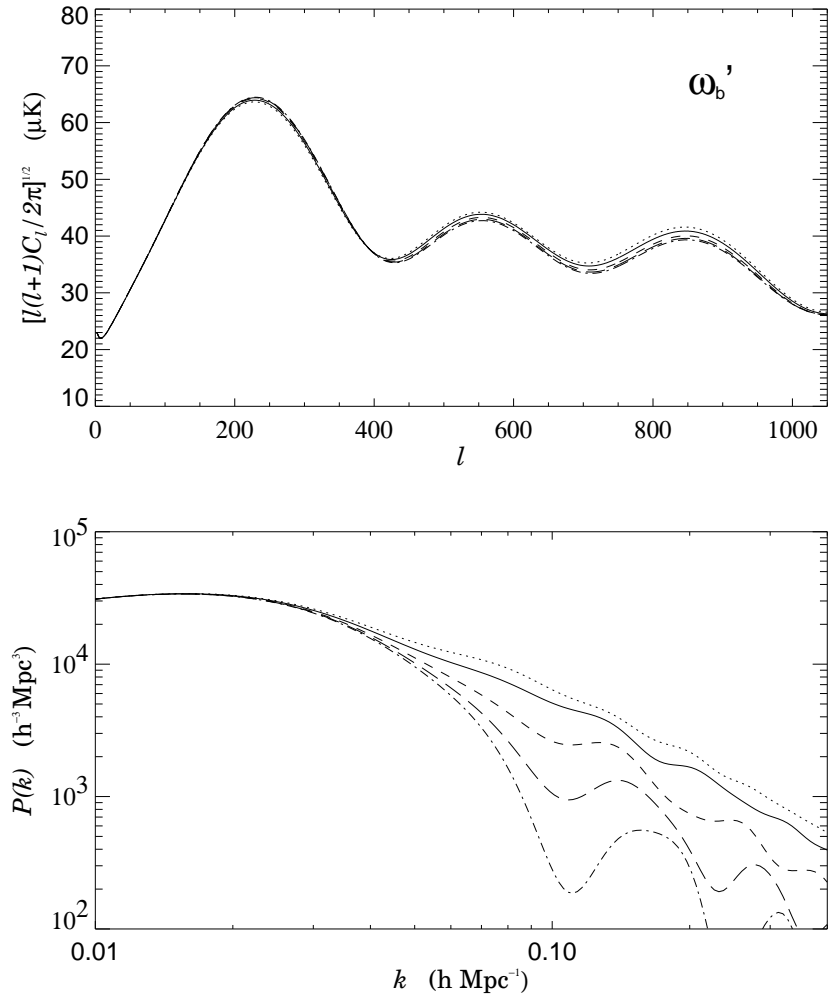}
  \end{center}
\caption{\small Dependence of the shape of photon and matter power spectra on the mirror baryon density $ \omega_{\rm b}' $ keeping constant $ \omega_{\rm b} $. The reference model (solid line) has: $ \Omega_{\rm m} = 0.3 $, $ \omega_{\rm b} = \omega_{\rm b}' = 0.02 $, $ x = 0.7 $ (not 0.2, as previous figures), $ h = 0.7 $ and $ n = 1.0 $. For other models, all the parameters are unchanged except for the one indicated: $ \omega_{\rm b}' = 0.01 = \omega_{\rm b} / 2 $ (dot line), $ \omega_{\rm b}' = 0.04 = 2  \omega_{\rm b} $ (dash line), $ \omega_{\rm b}' = 0.06 = 3 \omega_{\rm b} $ (long dash line), $ \omega_{\rm b}' = 0.08 = 4  \omega_{\rm b}$ (dot-dash line).}
\label{cmblssfig17}
\end{figure}

\begin{figure}[p]
  \begin{center}
    \leavevmode
    \epsfxsize = 15cm
    \epsffile{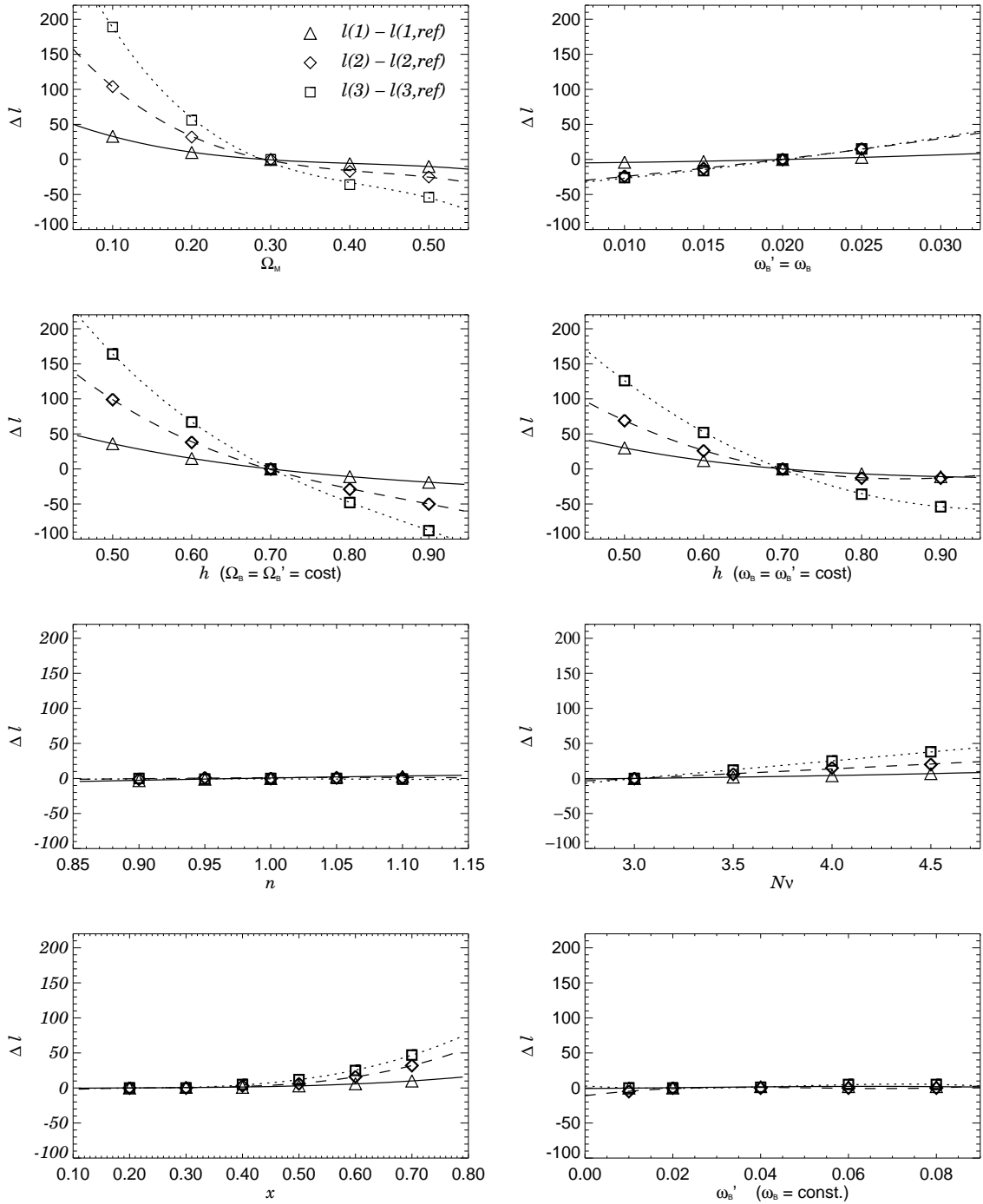}
  \end{center}
\caption{\small Dependences of the locations of the CMB acoustic peaks on the values of the cosmological parameters: $ \Omega_{\rm m} $, $ \omega_{\rm b} =  \omega_{\rm b}' $, $ h $ with $ \Omega_{\rm b} =  \Omega_{\rm b}' = {\rm const.} $, $ h $ with $ \omega_{\rm b} =  \omega_{\rm b}' = {\rm const.} $, $ n $, $ N_\nu $, $ x $, and $ \omega_{\rm b}' $ with $ \omega_{\rm b} $ constant and $ x = 0.7 $. The three indicators used here are the differences of the positions of the first three peaks of the models from the ones of the reference model. The reference model has: $ \Omega_{\rm m} = 0.3 $, $ \omega_{\rm b} = \omega_{\rm b}' = 0.02 $, $ x = 0.2 $, $ h = 0.7 $ and $ n = 1.0 $.}
\label{cmblssfig6}
\end{figure}

\begin{figure}[p]
  \begin{center}
    \leavevmode
    \epsfxsize = 15cm
    \epsffile{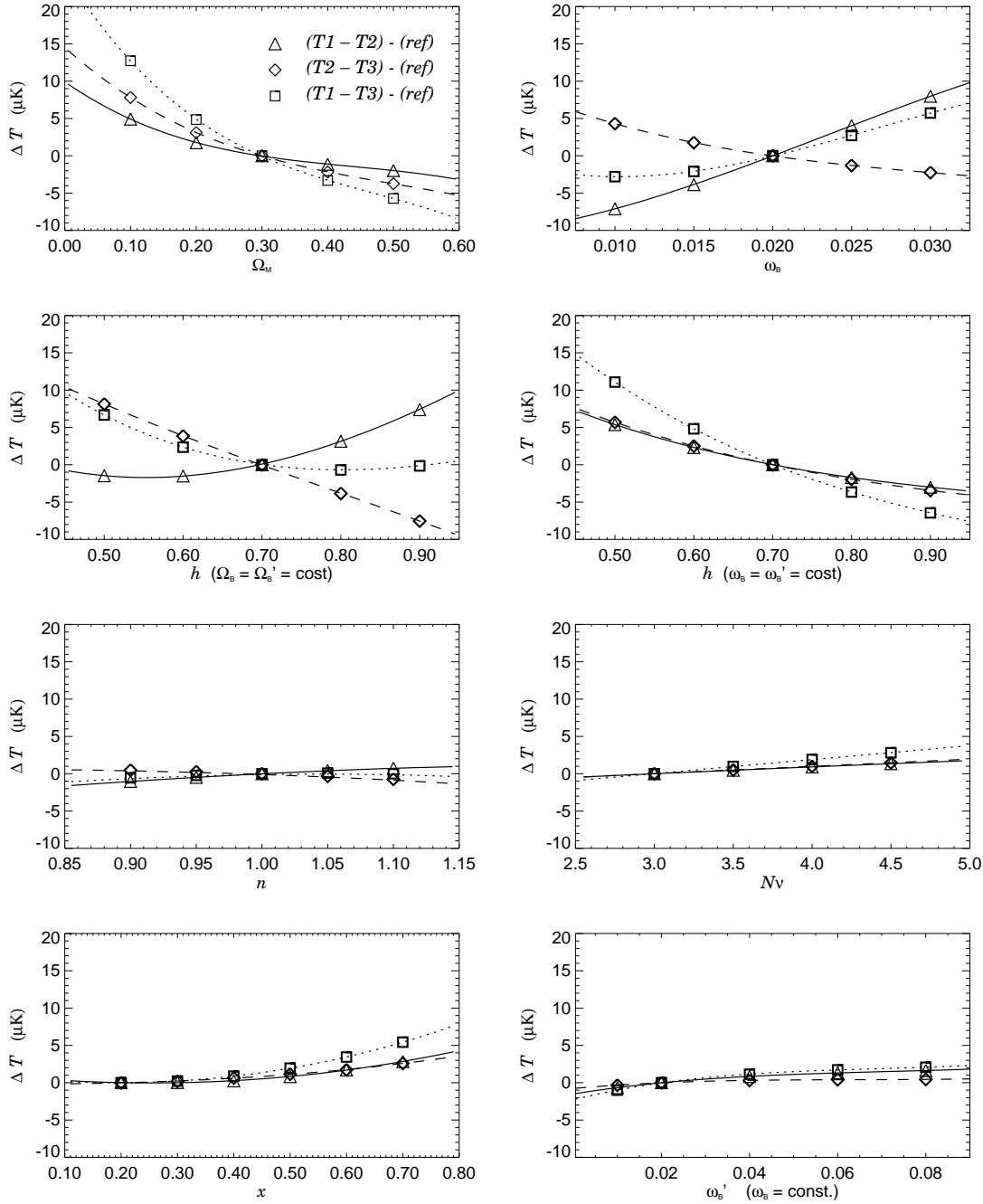}
  \end{center}
\caption{\small Dependences of the temperatures of the CMB acoustic peaks on the values of the cosmological parameters: $ \Omega_{\rm m} $, $ \omega_{\rm b} = \omega_{\rm b}' $, $ h $ with $ \Omega_{\rm b} =  \Omega_{\rm b}' = {\rm const.} $, $ h $ with $ \omega_{\rm b} =  \omega_{\rm b}' = {\rm const.} $, $ n $, $ N_\nu $, $ x $, and $ \omega_{\rm b}' $ with $ \omega_{\rm b} $ constant and $ x = 0.7 $. The three indicators used here are the deviations of the differences between the temperatures of the peaks from the same quantities obtained for a reference model. The reference model has: $ \Omega_{\rm m} = 0.3 $, $ \omega_{\rm b} = \omega_{\rm b}' = 0.02 $, $ x = 0.2 $, $ h = 0.7 $ and $ n = 1.0 $.}
\label{cmblssfig7}
\end{figure}


\newpage
\passo
\def \comp_obs{Comparison with observations}
\section{\comp_obs}
\label{comp_obs}
\markboth{Chapter \ref{chap-mirror_univ_3}. ~ \mir_univ_3}
                    {\S \ref{comp_obs} ~ \comp_obs}

So far we have studied the behaviour of the photon and matter power spectra varying many parameters with special attention to the two mirror parameters, i.e. the ratio of the temperatures of two sectors $ x $ and the amount of mirror baryons $ \omega_{\rm b}' $.

Now we want to compare these models with some experimental data, in order to evaluate the compatibility of the mirror scenario with observations and possibly restrict the parameter ranges.

As written in \S~\ref{mirror_mod}, we are not able to fit the parameters now. It is due to the slowness of our present version of the numerical code, but our game will be to choose some representative model and compare it with observations.

In the last decade the anisotropies observed in the CMB temperature became the most important source of information on the cosmological parameters: a lot of experiments (ground-based, balloon and satellite) were dedicated to its measurement. At the same time, many authors (see, e.g., Percival et al. \cite{perc327,percetal337} or Wang et al.\cite{wang}) proved that its joint analysis with the fluctuations in the matter distribution (they have both the same primordial origin) are a powerful instrument to determine the parameters of the Universe. As in \S~\ref{cmb_2} and \S~\ref{lss_2}, we analyse separately the variation of $ x $ and $ \omega_{\rm b}' $ in the mirror models, using now both the CMB and LSS informations at the same time.

In order to compare with observations, we use the best available data: for the CMB the COBE-DMR \cite{smoot92}, MAXIMA \cite{leea}, BOOMERANG \cite{nett}, DASI \cite{halv568,pryk568}
and CBI Mosaic \cite{pear0205388} data, and for the LSS the IRAS PSCz survey \cite{saun} (in particular the decorrelated power spectrum provided by Hamilton \& Tegmark (2002) \cite{hamteg330}). 

We start from figure \ref{cmblssfig18}, where we plot the models of the upper part of figures \ref{cmblssfig1} and \ref{cmblssfig3} together with the observations. This is useful to analyse models with different values of $ x $ and without CDM (i.e. all the dark matter is made of mirror baryons). In top panel, we see that with the accuracy of the current anisotropy measurements the CMB power spectra for mirror models are perfectly compatible with data. Indeed, the deviations from the standard model are weak, even in a Universe full of mirror baryons (see \S~\ref{cmb_2}). 
In lower panel, instead, the situation is completely different: oscillations due to mirror baryons are too deep to be in agreement with data, and only models with low values of $ x $ (namely $ x \lsim 0.3 $) are acceptable. This is an example of the great advantage of a joint analysis of CMB and LSS power spectra, being this conclusion impossible looking at the CMB only. 
Thus, we see the first strong constraint on 
the mirror parameters space: models with high mirror sector temperatures and all the dark matter made of mirror baryons have to be excluded.

In figure \ref{cmblssfig19} we show the models of the bottom panels of figures \ref{cmblssfig1} and \ref{cmblssfig3}, compared with the same observations, in order to analyse models with the same $ x $, but different mirror baryon contents. The above mentioned low sensitivity of the CMB power spectra on $ \omega_{\rm b}' $ doesn't give us indications for this parameter (even for high values of $ x $), but the LSS power spectrum helps us again, confirming a sensitivity to the mirror parameters greater than the CMB one. This plot tells us that also high values of $ x $ can be compatible with observations if we decrease the amount of mirror baryons in the Universe. It is a second useful indication: in case of high mirror temperatures we have to change the mirror baryon density in order to reproduce the oscillations present in the LSS data.

Therefore, after the comparison with experimental data, we are left with three possibilities for the Mirror Universe parameters:
\begin{itemize}
\item high $ x $ and low $ \omega_{\rm b}' $ (differences from the CDM in the CMB, and oscillations in the LSS with a depth modulated by the baryon density);
\item low $ x $ and high $ \omega_{\rm b}' $ (completely equivalent to the CDM for the CMB, and few differences for the LSS in the linear region);
\item low $ x $ and low $ \omega_{\rm b}' $ (completely equivalent to the CDM for the CMB, and nearly equivalent for the LSS in the linear region and beyond, according to the mirror baryon density).
\end{itemize}
Thus, with the current experimental accuracy, we can exclude only models with high $ x $ and high $ \omega_{\rm b}' $.

Observing the figures, we are tempted to do some (surely premature, given the low experimental accuracy and the lack of enough mirror models to fit the parameters) guesses. In fact, it's worth noting that in the CMB spectrum the mirror models seem to better fit the high first peak, while the mirror baryons could hopefully reproduce the oscillations present in the LSS power spectra. We are waiting for more accurate data (as expected for example from the MAP satellite) and preparing a faster numerical code, in order to fit the new data with a mirror model and establish if there are indications for a standard Universe or a Mirror Universe.

Our next step will be to consider some interesting mirror models and compute their power spectra. 

In figure \ref{cmblssfig20} we plot models with equal amount of ordinary and mirror baryons and a large range of temperatures. This is an interesting situation, because the case $ \Omega_{\rm b}' = \Omega_{\rm b} $ could be favoured in some baryogenesis scenario, as the one proposed by Bento and Berezhiani (2001) \cite{benber87}, which considered a lepto-baryogenesis mechanism that naturally lead to equal baryon number densities in both visible and hidden sectors. These models are even more interesting when we consider both their CMB and LSS power spectra. In top panel of figure we see that the temperature anisotropy spectra are fully compatible with observations, without large deviations from the standard case. In bottom panel, we have a similar situation for the matter power spectra, with some oscillations and a slightly greater slope, that could be useful to better fit the oscillations present in the data and to solve the discussed problem of the desired cutoff at low scales. Let us note that we are deliberately neglecting the biasing problem, given that an indication on its value can come only from a fit of the parameters; so, we have in fact a small freedom to vertically shift the curves in order to better fit the experimental data.

Models of a Mirror Universe where the dark matter is composed in equal parts by CDM and mirror baryons are plotted in figure \ref{cmblssfig21}. Now we concentrate on $ x $-values lower than the previous figure, because the greater mirror baryonic density would generate too many oscillations in the linear region of the matter power spectrum. In top panel we show that, apart from little deviations for the model with higher $ x $, all other models are practically the same. In bottom panel, instead, deviations are big, and we can still use LSS as a test for models. Indeed, models with $ x \gsim 0.4 $ are probably to exclude, even taking into account a possible bias. Models with lower $ x $ are all consistent with observations.


\def \stra_mod{Extremal models}
\subsubsection{\stra_mod}

At the end of this chapter we want to present some so-called ``extremal'' model. This name is due to the characteristic that some of their parameters are probably outside the present estimated limits, but nevertheless they show interesting features.

Let we consider a flat Mirror Universe where all the energy density is due to matter without cosmological constant contribution: $ \Omega_0 = \Omega_{\rm m} = 1 $. We can immediately place two objections to this scenario. The first is that supernovae measurements \cite{perl517} seem to indicate an accelerated expansion of the Universe, something easily understandable as a vacuum energy density effect. The second is related to the constraints on the matter density arising from the baryon fraction in galaxy clusters (combined with primordial nucleosynthesis data) \cite{erdo0202357} and from dynamical measurements of $ \Omega_{\rm m} $ \cite{borgetal561}: they both predict low matter density values. Against these arguments, we can say that there are still doubts on the reliability of the SNIa measurements; the accelerated expansion could be due to some other mechanism (see, e.g., Schwarz (2002) \cite{sch0209584}); cosmological constant has been always opposed by scientific community and there aren't commonly accepted theories on it; the cited estimates on the matter densities could not be valid in a Mirror Universe, given the differences between the mirror baryonic and the cold dark matter structure formation scenarios. 

Our main point here is to show 
that a flat Mirror Universe all made of matter could be compatible with observations of CMB and LSS power spectra alone, forgetting the possible constraints coming from other measurements.
To do this we have two possible kinds of models: those with low $ h $-values and those with high $ x $-values. The first models have the problem that the value of $ h $ is at the limit of the $ 2\sigma $ confidence interval estimated by the HST Key Project \cite{free} (but well inside other estimations cited in \S~\ref{sec-now-cosm}); in addition, many CMB and LSS joint analyses (included that containing the concordance model of Wang et al. (2002) \cite{wang}) obtain low values of $ h $, if they don't impose priors on its value. The second ones have the problem that high values of $ x $ are in contrast with BBN limits ($ x \lsim 0.64 $), as discussed in \S~\ref{term_mir_univ}. However, not very plausible though, one could consider the case with larger $x $ as well, up to $ x = 1 $. This could be achieved, without contradiction to the BBN limits, if after the BBN epoch (but before the matter-radiation equality) there is some additional entropy production in the mirror sector (or both in the O and M sectors), due to decay of some light metastable particles in both O and M photons. 

Just for demonstration, one can imagine some axion-like boson $g$ with mass $ M_g \sim 100 $ KeV which was in equilibrium sometimes in the early Universe, but has decoupled already before electroweak epoch, $ T \sim 100 $ GeV. In this case their temperature $ T_g $ at the nucleosynthesis epoch is related to the temperature of the rest of the thermal bath, $ T \sim 1 $ MeV, as $ T_g / T < [ \gs(1\:{\rm MeV}) / \gs(100\:{\rm GeV}) ]^{1/3} \approx 0.5 $, where $ \gs (1\:{\rm MeV}) = 10.75 $ and $ \gs(100\:{\rm GeV}) = 106.5 $ stand for the total number of the particle degrees of freedom in the standard model (see appendix \ref{usefulrel1-app}). Therefore their contribution in the nucleosynthesis is negligible, namely it corresponds to about $ \Delta N_\nu \approx 0.03 $. However, when the temperature drops down below 100 keV, the contribution of these particles in the energy density of the Universe becomes more relevant, and at some moment it can approach the contribution of the photon energy density (because after these particles become non relativistic, at $ T_g < M_g $, their energy density rescales as $ \propto M_g T_g^3 $, whereas the photon one as $ \propto T^4 $). If at this moment these particles decay into both ordinary and mirror photons, clearly with the same rate due to mirror symmetry, this would heat up both photon sectors.\footnote{One has to remember that this decay should take place before $ t \sim 10^6 {\rm s} $, otherwise it would affect the thermal Planck spectrum of the CMB photons.} Therefore, if before the decay of $ g $ the temperature of the M photons was smaller than the O ones, $ T' < 0.64\: T $ (as required by the BBN limits) so that energy densities were related as $ \rho'_\gamma / \rho_\gamma < (T' / T)^4 \approx 0.17 $, after decay the same amount of energy will be deposited to both sectors, $ \Delta \rho'_\gamma = \Delta \rho_\gamma $. Thus, e.g. if the energy density of the ordinary photons is nearly doubled, i.e. $ \Delta \rho_\gamma \approx \rho_\gamma $, then the O and M photons energy densities after decay should be related as $ \rho'_\gamma / \rho_\gamma \approx 0.5 $, which corresponds now to their temperature ratio $ x = T' / T \approx 0.5^{1/4} \approx 0.85 $. In other words, in the epoch relevant for CMB formation, due to additional entropy production, the temperature of mirror sector relative to the ordinary one can be much larger than at the nucleosynthesis epoch.\footnote{Of course, the ratio of the baryon to photon number densities, $ \eta = n_b / n_\gamma $, would be different before and after decay of $ g $, however one has to recall that the value of $ \eta $ determined by the BBN is compatible within a factor of two with the one provided by the CMB analysis.}

In figure \ref{cmblssfig22} we plot all these extremal models comparing them with the concordance model. For low $ h $-values, the mirror model is coincident with the concordance one for the first two peaks in the temperature anisotropy spectrum and higher for other peaks (still in agreement with observations). In the matter power spectrum, apart from the change in the slope at higher scales, there are some deviations at lower scales (still compatible with experimental data). Note that the ordinary model for the same $ h $ and $ \Omega_{\rm m} $ values is absolutely incompatible with observations, specially those for the LSS. 
As regards models with high $ x $, they are in good agreement not only with the CMB data, but also with the LSS ones, with some possible problem for the latter only at high scales.

We want to stress again that these models have the unique purpose of being speculative exercises or curiosities, without any ambition at this stage.

\begin{figure}[p]
  \begin{center}
    \leavevmode
    \epsfxsize = 13cm
    \epsffile{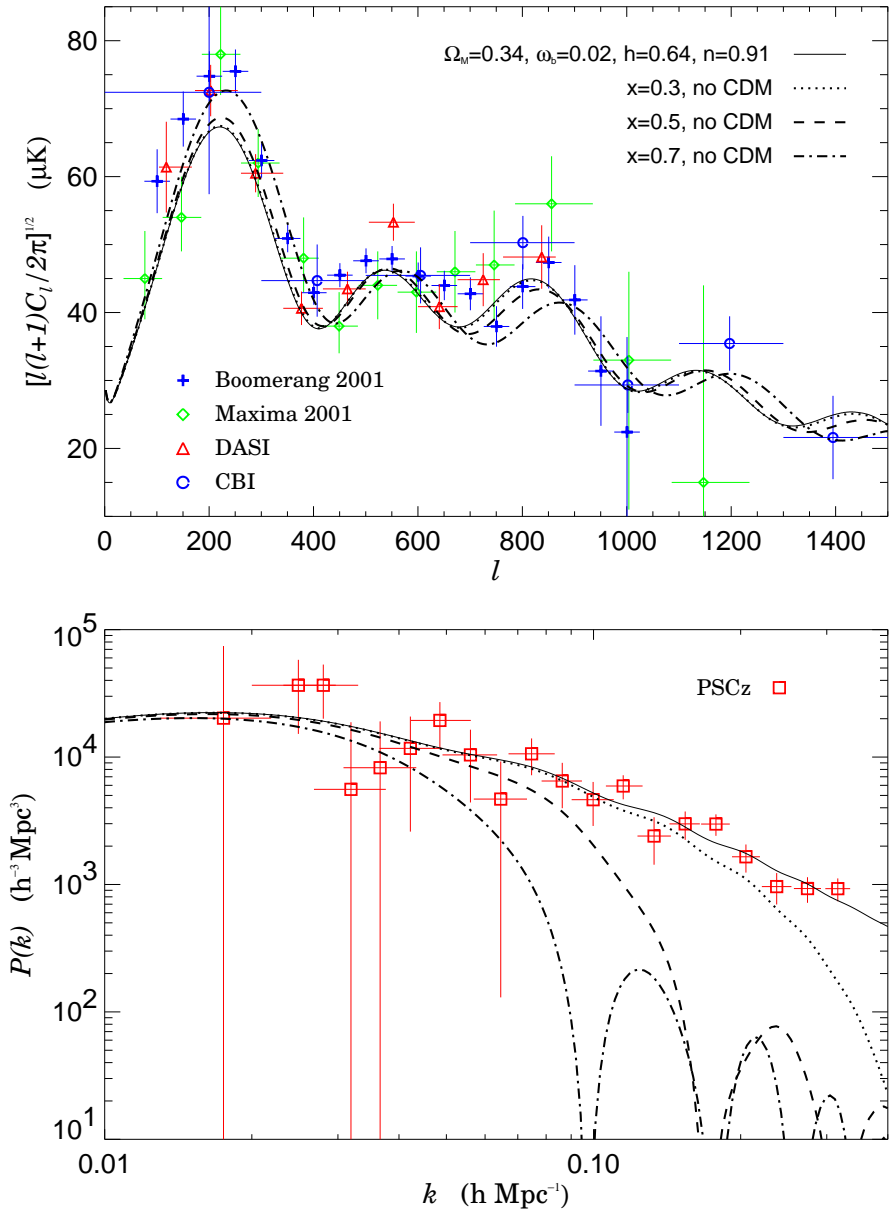}
  \end{center}
\addvspace{1.2cm}
\caption{\small CMB and LSS power spectra for various mirror models with different values of $ x $, compared with observations and with the concordance model of Wang et al. (2002) \cite{wang} (solid line) of parameters $ \Omega_0 = 1 $, $ \Omega_{\rm m} = 0.34 $, $ \Omega_{\rm \Lambda} = 0.66 $, $ \omega_{\rm b} = \Omega_{\rm b} h^2 = 0.02 $, $ n_{\rm s} = 0.91 $, $ h = 0.64 $. The mirror models have the same parameters as the standard one, but with $ x = 0.3, 0.5, 0.7 $ and $ \omega_{\rm b}' = \Omega_{\rm m} h^2 - \omega_{\rm b} $ (no CDM) for all models. {\sl Top panel.} Comparison of the photon power spectrum with the MAXIMA, BOOMERANG, DASI and CBI Mosaic data. {\sl Bottom panel.} Comparison of the matter power spectrum with the IRAS PSCz data.}
\label{cmblssfig18}
\end{figure}

\begin{figure}[p]
  \begin{center}
    \leavevmode
    \epsfxsize = 13cm
    \epsffile{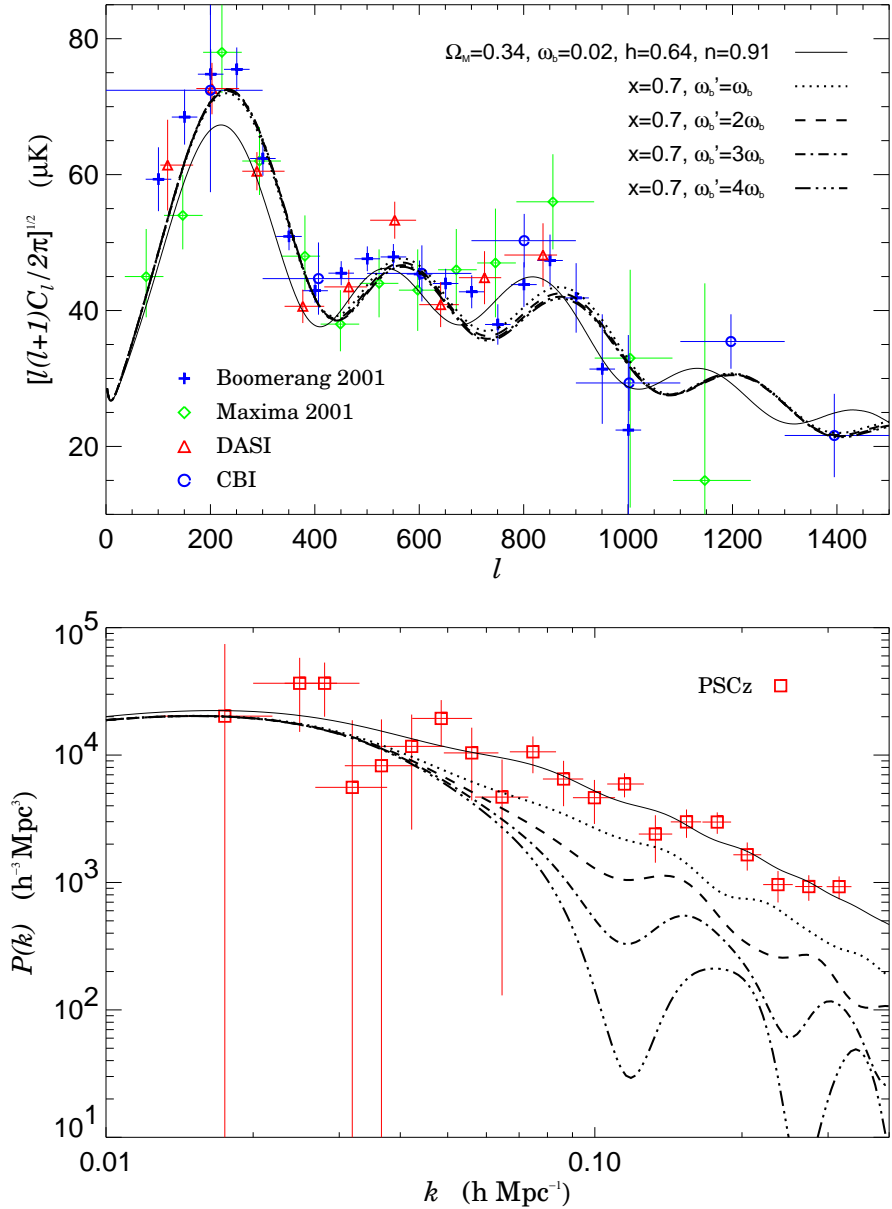}
  \end{center}
\addvspace{1.2cm}
\caption{\small CMB and LSS power spectra for various mirror models with different values of mirror baryon density, compared with observations and with the concordance model of Wang et al. (2002) \cite{wang} (solid line) of parameters $ \Omega_0 = 1 $, $ \Omega_{\rm m} = 0.34 $, $ \Omega_{\rm \Lambda} = 0.66 $, $ \omega_{\rm b} = \Omega_{\rm b} h^2 = 0.02 $, $ n_{\rm s} = 0.91 $, $ h = 0.64 $. The mirror models have the same parameters as the standard one, but with $ x = 0.7 $ and for $ \omega_{\rm b}' = \omega_{\rm b}, 2 \omega_{\rm b}, 3 \omega_{\rm b}, 4 \omega_{\rm b} $. {\sl Top panel.} Comparison of the photon power spectrum with the MAXIMA, BOOMERANG, DASI and CBI Mosaic data. {\sl Bottom panel.} Comparison of the matter power spectrum with the IRAS PSCz data.}
\label{cmblssfig19}
\end{figure}

\begin{figure}[p]
  \begin{center}
    \leavevmode
    \epsfxsize = 13cm
    \epsffile{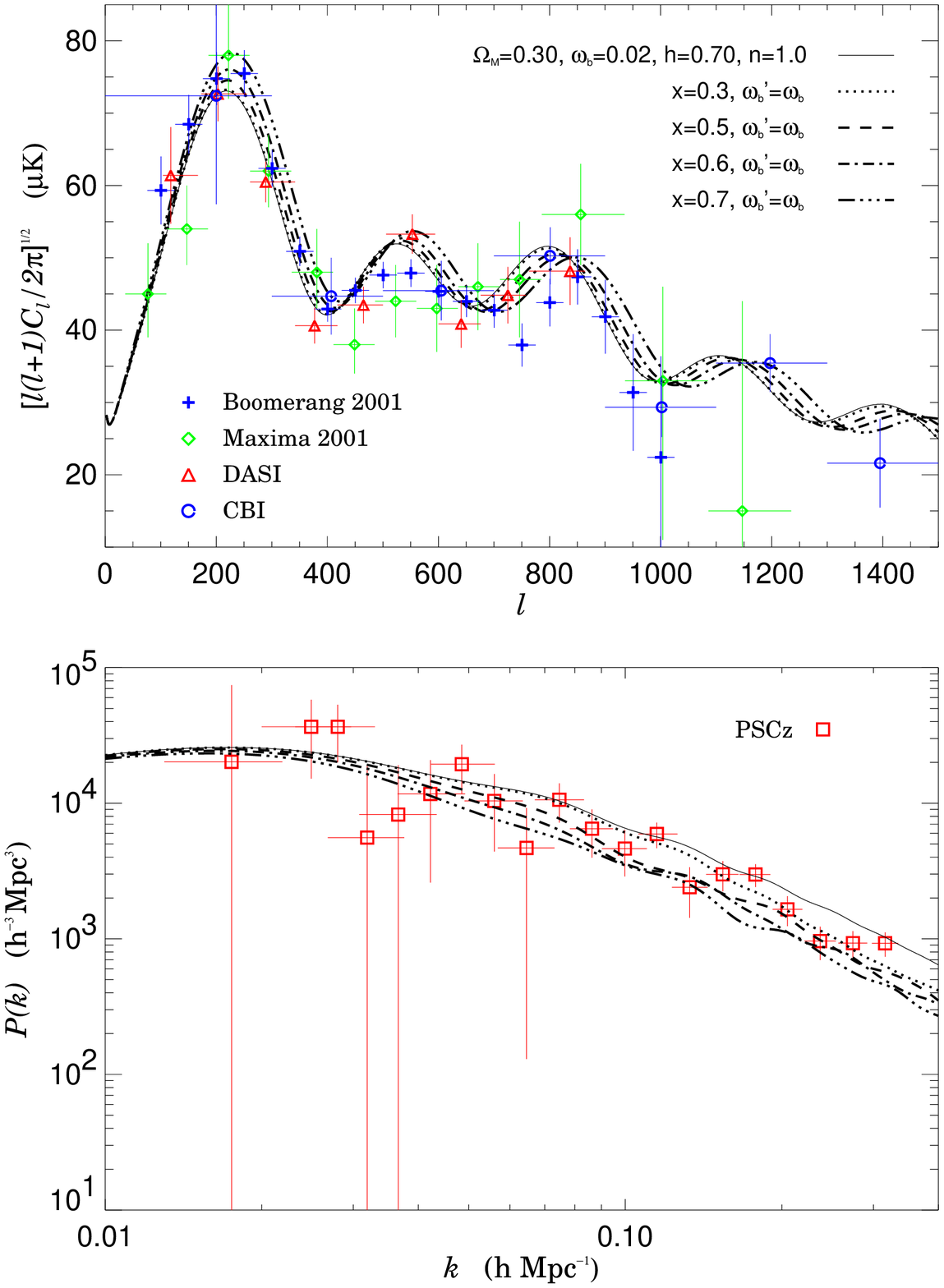}
  \end{center}
\addvspace{1.2cm}
\caption{\small CMB and LSS power spectra for various mirror models with different values of $ x $ and equal amounts of ordinary and mirror baryons, compared with observations and with a standard reference model (solid line) of parameters $ \Omega_0 = 1 $, $ \Omega_{\rm m} = 0.30 $, $ \Omega_{\rm \Lambda} = 0.70 $, $ \omega_{\rm b} = \Omega_{\rm b} h^2 = 0.02 $, $ n_{\rm s} = 1.0 $, $ h = 0.70 $. The mirror models have the same parameters as the standard one, but with $ \omega_{\rm b}' = \omega_{\rm b} $ and $ x = 0.3, 0.5, 0.6, 0.7 $. {\sl Top panel.} Comparison of the photon power spectrum with the MAXIMA, BOOMERANG, DASI and CBI Mosaic data. {\sl Bottom panel.} Comparison of the matter power spectrum with the IRAS PSCz data.}
\label{cmblssfig20}
\end{figure}

\begin{figure}[p]
  \begin{center}
    \leavevmode
    \epsfxsize = 13cm
    \epsffile{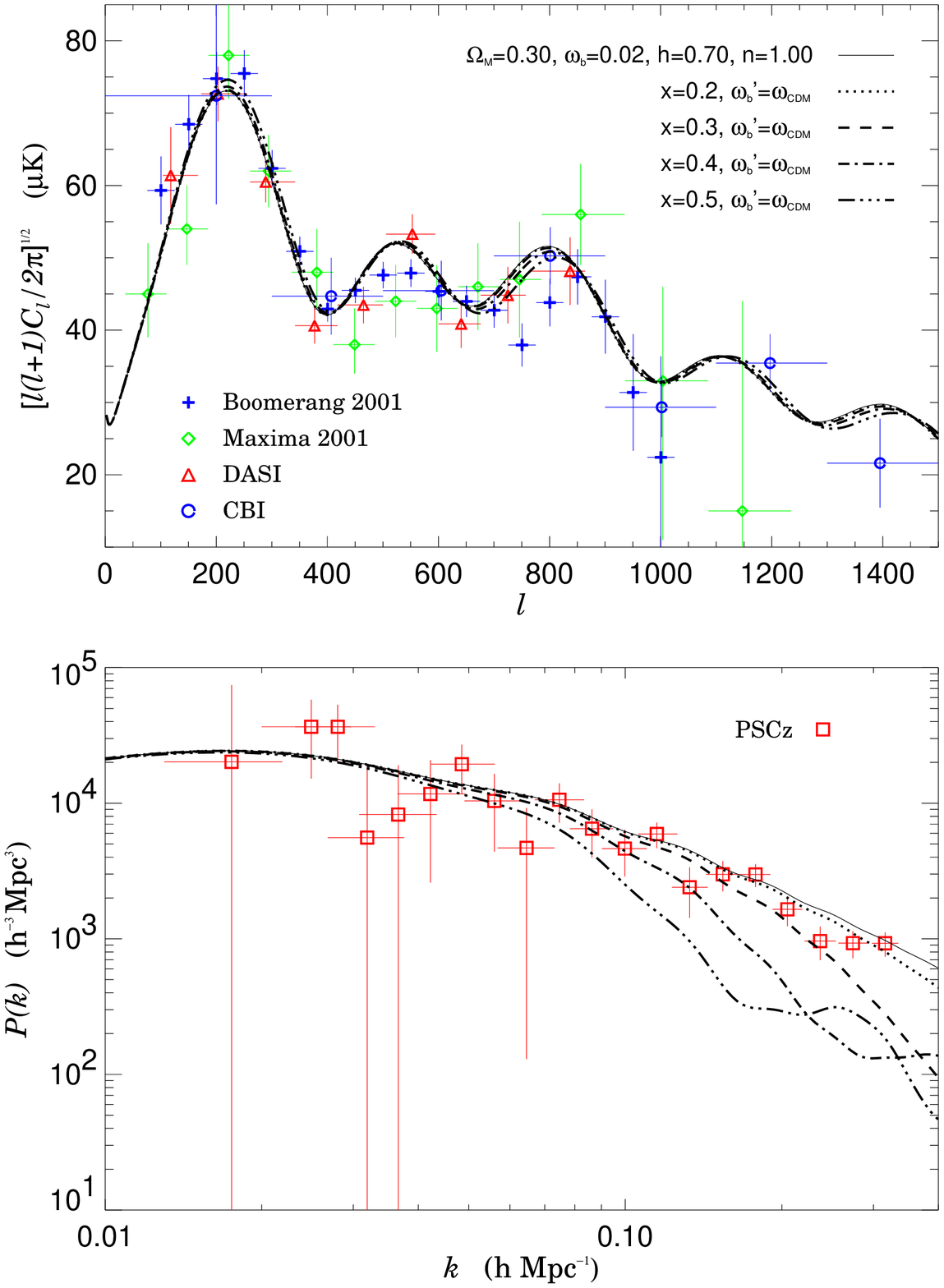}
  \end{center}
\addvspace{1.2cm}
\caption{\small CMB and LSS power spectra for various mirror models with different values of $ x $ and equal amounts of CDM and mirror baryons, compared with observations and with a standard reference model (solid line) of parameters $ \Omega_0 = 1 $, $ \Omega_{\rm m} = 0.30 $, $ \Omega_{\rm \Lambda} = 0.70 $, $ \omega_{\rm b} = \Omega_{\rm b} h^2 = 0.02 $, $ n_{\rm s} = 1.0 $, $ h = 0.70 $. The mirror models have the same parameters as the standard one, but with $ \omega_{\rm b}' = \omega_{\rm CDM} $ and $ x = 0.2, 0.3, 0.4, 0.5 $. {\sl Top panel.} Comparison of the photon power spectrum with the MAXIMA, BOOMERANG, DASI and CBI Mosaic data. {\sl Bottom panel.} Comparison of the matter power spectrum with the IRAS PSCz data.}
\label{cmblssfig21}
\end{figure}

\begin{figure}[p]
  \begin{center}
    \leavevmode
    \epsfxsize = 13cm
    \epsffile{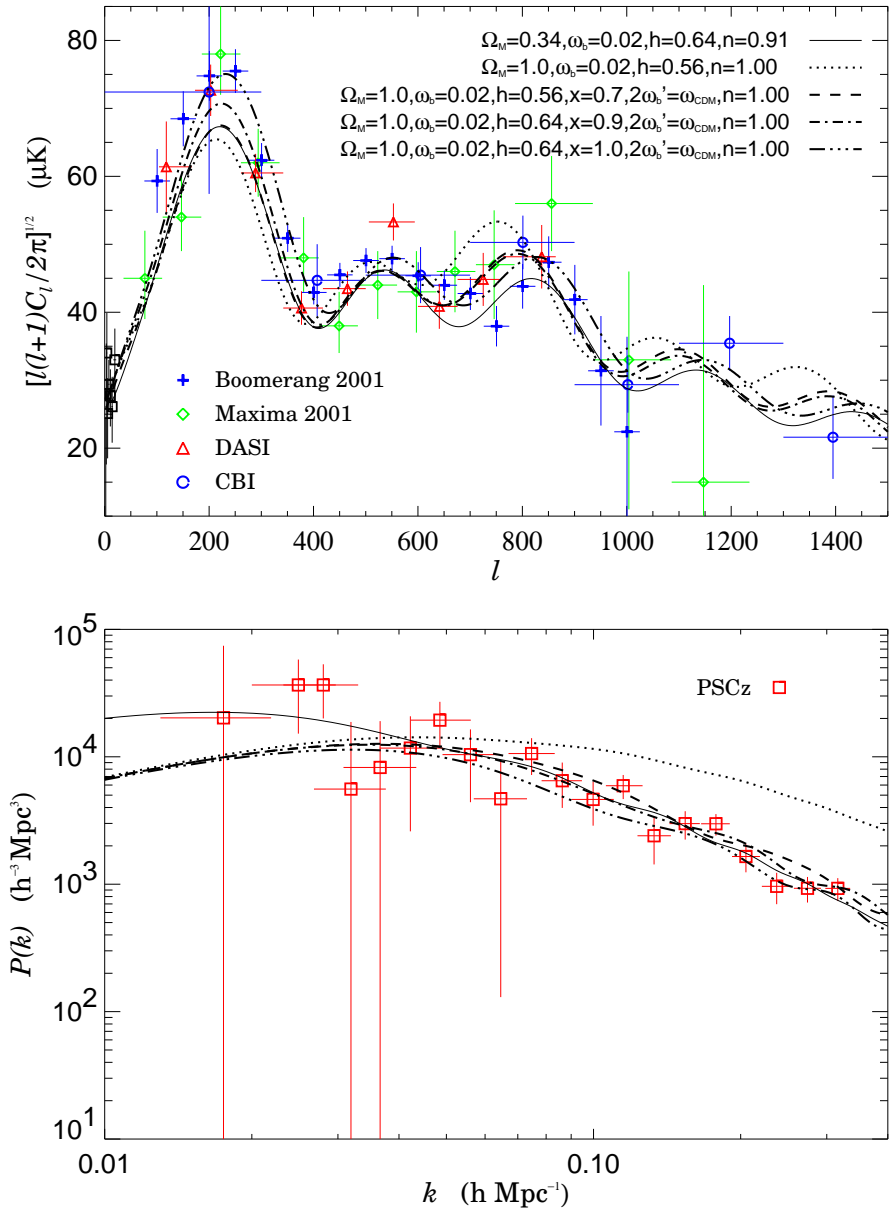}
  \end{center}
\addvspace{1.2cm}
\caption{\small CMB and LSS power spectra for standard and mirror models with $ \Omega_{\rm m} = 1 $ compared with observations and with the concordance model of Wang et al. (2002) \cite{wang} (solid line) of parameters $ \Omega_0 = 1 $, $ \Omega_{\rm m} = 0.34 $, $ \Omega_{\rm \Lambda} = 0.66 $, $ \omega_{\rm b} = \Omega_{\rm b} h^2 = 0.02 $, $ n_{\rm s} = 0.91 $, $ h = 0.64 $. The models can be divided into two groups: those with low $ h $-values  (dotted and dashed lines), and those with high $ x $-values (dot-dash and dot-dot-dot-dash).
{\sl Top panel.} Comparison of the photon power spectrum with the MAXIMA, BOOMERANG, DASI and CBI Mosaic data. {\sl Bottom panel.} Comparison of the matter power spectrum with the IRAS PSCz data.}
\label{cmblssfig22}
\end{figure}



\chapter{Mirror stars and other mirror astrophysical consequences}
\label{chap-mirror_univ_4}
\def \mir_univ_4{Mirror stars and other mirror astrophysical consequences}
\markboth{Chapter \ref{chap-mirror_univ_4}. ~ \mir_univ_4}
                    {Chapter \ref{chap-mirror_univ_4}. ~ \mir_univ_4}


\passo
\def \mirror_astr_pict{The mirror astrophysical picture inside the galaxy}
\section{\mirror_astr_pict}
\label{mirror_astr_pict}
\markboth{Chapter \ref{chap-mirror_univ_4}. ~ \mir_univ_4}
                    {\S \ref{mirror_astr_pict} ~ \mirror_astr_pict}

Until now we studied the so-called linear structure formation for a Mirror Universe, reaching the conviction that mirror matter could exist in the Universe, since it is compatible with all the available observations (BBN, CMB, LSS); in addition, we obtained some useful constraints on the parameters describing the mirror sector. Thus, we are pushed to continue our analysis of the Mirror Universe, and study also other astrophysical consequences of the mirror matter. It is well known that some topics (as the galaxy and star formations) present many difficulties and obscure points also for a standard Universe, where we have many direct observables, so it is even more difficult to treat them in a mirror scenario. Nevertheless, we can expose our ideas on various arguments, and go into more details where we well know the physics responsible for the processes, as for the mirror stellar evolution.

In previous chapters we described the evolution of a Mirror Universe from the inflation to the structure formation in linear scales, i.e., until the rich galaxy clusters. We also argued that for these scales the mirror matter should manifest as the CDM for some parameter choices ($ x < 0.3$-$0.2 $). Thus, it is crucial to extend our study to smaller scales which already went non-linear, like galaxies, and beyond until the smallest astrophysical structures, in order to understand the expected crucial differences between the mirror baryonic dark matter (MBDM) and CDM scenarios.

\subsubsection{The galaxy}

According to our picture, dark matter in galaxies and clusters is made of a mixture of mirror baryonic dark matter and cold dark matter, or can be even constituted entirely by the mirror baryons (only for low values of $ x $, given the bounds obtained in previous chapter). The presence of mirror matter has obviously consequences on the distribution of matter in galactic halos, given the different physics of mirror baryons and CDM: the first one is a self-collisional dark matter with exactly the same physical laws of our visible sector, while the second one is a collisionless matter component. In particular, simulations show that the CDM forms triaxial halos with a density profile too clumped towards the center, and overproduces the small substructures within the halo. In principle, MBDM could avoid both these problems. In chapter \ref{chap-mirror_univ_3} the study of the large scale structure demonstrated that the power spectrum for a Mirror Universe presents a cutoff at scales which depend on the temperature of the mirror sector and its baryon content, so that small substructures have much less power than in CDM scenarios. Regarding the halo profiles, the self-collisionality of mirror baryons could avoid the high central concentration typical of CDM simulations. Clearly we need a numerical study of galaxy formation (based on N-body simulations) to test these suggestions, and this will be just one of the next works after this thesis.

Meanwhile, given the complexity of the physics of galaxy formation (actually this process in still to be understood), we can introduce some general consideration on it. At a stage during the process of gravitational collapse of the protogalaxy, the opacity of the system becomes so high that the gas prefers to fragment into protostars. This complex phenomenon lead a part of the protogalactic gas to form the first stars (probably very massive, with $ M \sim 10^{2} $-$ 10^{3} M_\odot $). The details of this process are clearly dependent on the boundary conditions, which are different for the two sectors (as explained in chapter \ref{chap-mirror_univ_1}): (i) the temperature of the mirror sector is lower than that of the ordinary one by a factor $ x $; (ii) the mirror baryonic density is higher or equal than the ordinary one; (iii) the baryonic chemical composition is very different in the two sectors, since we know (see \S~\ref{nucleosyn}) that mirror sector is a helium-dominated world (with a concentration $ Y' $ dependent on $ x $ according to the formula (\ref{m_helium}), and in any case greater than $ Y $). 

Anyway, whatever the details of the scenario dependent on the exact composition of matter in the Mirror Universe, one has to take into account the occurrence that during the galaxy evolution the bulk of the M baryons could fastly fragment into stars. A difficult question to address here is related to the star formation in the M sector, also taking into account that its temperature/density conditions and chemical contents are much different from the ordinary ones. In any case, the fast star formation would extinguish the mirror gas and thus could avoid the M baryons to form disk galaxies as ordinary baryons do. The M protogalaxy, which at certain moment before disk formation essentially becomes the collisionless system of the mirror stars, could maintain a typical elliptical structure.\footnote{In other words, we speculate on the possibility that the mirror baryons form mainly the elliptical galaxies. For a comparison, in the ordinary world the observed bright galaxies are mainly spiral while the elliptical galaxies account about $20 ~\%$ of them. Remarkably, the latter contains old stars, very little dust and shows no sign of active star formation.} Certainly, in this consideration also the galaxy merging process should be taken into account.  As for the O matter, within the dark M matter halo it should typically show up as an observable elliptic or spiral galaxy, but some anomalous cases can be also possible, like certain types of irregular galaxies or even dark galaxies dominantly made out of M baryons. The central part of halo can nevertheless contain a large amount of ionized mirror gas and it is not excluded that it can have a quasi-spherical form. Even if the stellar formation is very efficient, the massive mirror stars in the dense central region fastly evolve (see \S~\ref{mirror_star_evol}) and explode as supernovae, leaving behind compact objects like neutron stars or black holes, and reproducing the mirror gas and dust. 

We note that, although the hydrogen cross section $\sigma_H$ is large, it does not necessarily implies that the galaxy core will collapse within a dynamical time, since the inner halo should be opaque for M particles. They undergo many scatterings and escape from the system via diffusion, so the energy drain can be small enough and the instability time can substantially exceed the age of the Universe \cite{hannestad}. 

\subsubsection{Mirror stars and MACHOs}

In order to understand the details of the process of galaxy formation and evolution, it is crucial to study the mirror star formation (beginning and speed) and evolution. Stars play an important role: the fraction of baryonic gas involved in their formation becomes collisionless on galactic scales, and supernovae explosions enrich the galaxy of processed collisional gas (stellar feedback). By this reason the following sections are devoted to the study of the mirror stellar evolution.

The existence of mirror stars is guaranteed by the existence of ordinary stars: given that two sectors have the same microphysics, stars necessarily form in both of them \cite{blin60}. Clearly, being different the boundary conditions, stars have some differences (see \S~\ref{mirror_star_evol}).
The fact that dark matter made of mirror baryons has the property of clumping into compact bodies such as mirror stars leads naturally to an explanation
for the mysterious Massive Astrophysical Compact Halo Objects (or MACHOs). 

These objects are revealed only by their gravitational effects. If we look at a star in the sky and at a certain time a compact invisible object passes near our line of sight, it acts as a {\sl gravitational microlens} and doubles the image we are seeing. If the resolution of the telescope is not large enough to resolve the two star images, we see only an enhancement of its brilliance, producing a typical symmetric, achromatic and unique light curve, which depends on the mass of the invisible object and on its distance from the line of sight (impact parameter). In figure \ref{macho1} we show an explanatory scheme of this phenomenon and typical light curves of microlensing events.

\begin{figure}[h]
  \begin{center}
    \leavevmode
{\hbox 
{\epsfxsize = 6cm \epsffile{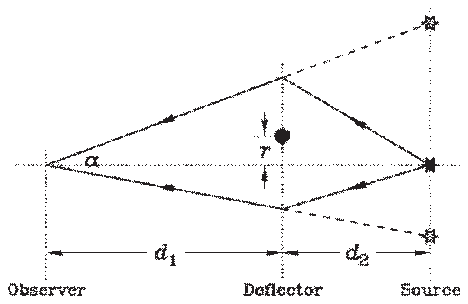} }
{\epsfxsize = 4cm \epsffile{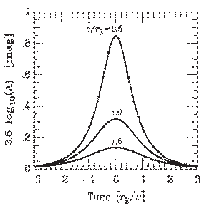} }  }
  \end{center}
\caption{\small {\sl Left panel.} Scheme of a microlensing in presence of a MACHO. {\sl Right panel.} Typical light curves of a microlensing event as a function of the impact parameter.}
\label{macho1}
\end{figure}

In the galactic halo (provided that it is the elliptical mirror galaxy) the mirror stars should be observed as MACHOs in gravitational microlensing  \cite{bere27,blin9801015,foot452,moha478,moha462}. The MACHO collaboration \cite{macho542} has been studying the nature of halo dark matter by using the gravitational microlensing technique. This 
experiment has collected 5.7 years of data and provided statistically strong evidence for dark matter in the form of invisible star sized objects, which is what you would expect if there was a significant amount of mirror matter in our galaxy. The MACHO collaboration has done a maximum likelihood analysis which implies a MACHO halo fraction of $20\%$ for a typical halo model with a $95\%$ confidence interval of $8\%$ to $50\%$. Their most likely MACHO mass is between $0.15 M_{\odot}$ and $0.9M_{\odot}$ (depending on the halo model), with an average around $M\simeq 0.5 ~M_\odot$, which is difficult to explain in terms of the brown dwarves with masses below the hydrogen ignition limit $M < 0.1 M_{\odot}$ or other baryonic objects \cite{Freese}. These observations are consistent with a mirror matter halo because the entire halo would not be expected to be in the form of mirror stars. Mirror gas and dust would also be expected because they are a necessary consequence of stellar evolution and should therefore significantly populate the halo. Thus, leaving aside the difficult question of the initial stellar mass function, one can remark that once the mirror stars could be very old and evolve faster than the ordinary ones, it is suggestive to think that most of massive ones, with mass above the Chandrasekhar limit $M_{\rm Ch} \simeq 1.5 ~ M_\odot$ have already ended up as supernovae, so that only the lighter ones 
remain as the microlensing objects. Perhaps, this is the first observational evidence of the mirror matter? 

\subsubsection{Other mirror astrophysical objects}

\noindent {\bf Mirror globular clusters.}~~It is also plausible that in the galactic halo some fraction of mirror stars exists in the form of compact substructures like globular or open clusters, in the same way as it happens for ordinary stars. In this case, for a significant statistics, one could observe interesting time and angular correlations between the microlensing events. 

\noindent {\bf Supernovae and gamma ray bursts.}~~Given that M baryons form the stars, some of them must also explode as M supernovae, with a rate in principle predictable after the study of the mirror star formation and evolution. Their explosion in our galaxy cannot be directly seen by ordinary observer, however it could be observed in terms of gravitational waves, and possibly revealed by their next generation detectors. In addition, if the M and O neutrinos are mixed \cite{akhm69,bere52,foota7,footd52,volk58}, it can lead the observable neutrino signal, which could be also accompanied by the weak gamma ray burst \cite{blin9902305,volk13}.

\noindent {\bf Supermassive black holes.}~~Another tempting issue is whether the M matter itself could help in producing big central black holes, with masses $\sim 10^7~ M_\odot$, which are thought to be main engines of the active galactic nuclei. 

\noindent{\bf Substellar scales.}~~If mirror matter exists in our galaxy, then binary systems consisting of ordinary and mirror matter should also exist. While systems containing approximately equal amounts of ordinary and mirror matter are unlikely due to e.g. differing rates of collapse for ordinary and mirror matter (due to different initial conditions: chemical composition, temperature distribution, etc.), systems containing predominately ordinary matter with a small amount of mirror matter (and viceversa) should exist. Remarkably, there is interesting evidence for the existence of such systems coming from extra-solar planet astronomy \cite{foot505,footsila0104251}.


\passo
\def \mirror_star_mod{The mirror star models}
\section{\mirror_star_mod}
\label{mirror_star_mod}
\markboth{Chapter \ref{chap-mirror_univ_4}. ~ \mir_univ_4}
                    {\S \ref{mirror_star_mod} ~ \mirror_star_mod}

Given all the above considerations on the importance of the study of mirror stars in order to explain the MACHOs, and the galaxy evolution and formation, we now turn to the study of the mirror star evolution.

As we know (see \S~\ref{intro_mirror}), the microphysics of the mirror sector is exactly the same as the visible one, the only changes are due to the boundary conditions. This is a very favourable condition for the study of M stars, because the necessary knowledge is the same than for the O ones, that we know very well. This means that M stars follow the same evolutionary stages than visible ones. A very brief review of stellar evolution will be given at the beginning of the next section.

The same physics for both O and M sectors means that the equations governing the mirror stellar evolution and the physical ingredients to put inside them (namely the equation of state, the opacity tables, and the nuclear reactions) are the same than for visible stars. The only change regards the composition of the M star. In fact, while the typical helium abundance for O stars is $ Y \simeq 0.24$, for the M stars we have $ Y' = 0.40$-$0.80 $. This interval is obtained considering that its lower limit is given by the primordial helium abundance coming from the mirror Big Bang nucleosynthesis studied in \S~\ref{nucleosyn}. In next section we will evaluate its impact on the evolution of M stars. 

If we consider a single isolated star\footnote{We are practically neglecting the interactions existing in systems of two or three stars.}, its evolutionary and structural properties depend only on the mass and the chemical composition. In particular, the latter is expressed by the abundances of hydrogen ($ X $), helium ($ Y $), and the so-called heavy elements or metals ($ Z $), i.e. all the elements heavier than H and He. \footnote{In this section we use the prime ($'$) to indicate mirror quantities only if they appear together with ordinary ones; otherwise we don't use it, taking in mind that high $ Y $-values refer to mirror stars and low $ Y $-values to the ordinary ones.}

We computed mirror star models using the evolutionary code FRANEC ({\sl Frascati RAphson Newton Evolutionary Code}), a numerical tool to solve the equations of stellar structures. As inputs for this code we chose the opacity tables of Alexander \& Ferguson (1994) \cite{alexfer94} for temperatures lower than 10000 $ K $ and those obtained in the Livermore laboratories by Rogers \& Iglesias (1996) \cite{rogigl96} for higher temperatures, the equation of state of Saumon, Chabrier and Van Horn (1995)  \cite{scvh95}, and the grey approximation for the integration of stellar atmospheres\footnote{The ``grey atmosphere'' approximation assumes local thermodynamical equilibrium with opacity independent of frequency. In this case the temperature in the stellar atmosphere is given by
\vspace{-.2cm}
\bea
T^4 (\tau)= {3 \over 4} T_e^4 \left( \tau + {2 \over 3} \right) \;, \nonumber
\eea
where $ T(\tau) $ is the temperature of an atmospheric layer located at the optical depth $ \tau $, and $ T_e $ is the effective temperature of the star.}. These inputs are valid over the entire ranges of temperatures and densities reached by our models.

We had no need to modify the code because, as explained before, mirror stars are evolutionary equivalent to ordinary stars with a high helium abundance. Thus, we computed stellar models for large ranges of masses and helium contents, and for a low metallicity $ Z $. A low $ Z $-value means that we are treating a so-called stellar population II, i.e.~an old stellar population, coming soon after the first one (the population III, without metals).

We summarize the values used for stellar parameters in table \ref{table1_7}, noting that the model with $ M = 0.8 M_\odot $ has been computed also for the not listed intermediate values $ Y = 0.30, 0.40, 0.60, 0.80 $, in order to better investigate the dependence of evolutionary properties on the helium content for a mass interesting as MACHO candidate.

\begin{table}[h]
\begin{center}
\begin{tabular}{|c|c|} \hline \hline
  parameter & values \\
  \hline \hline
  $ Z $ & $ 10^{-4} $ \\
  \hline
  $ Y $ & 0.24 - 0.50 - 0.70 \\
  \hline
  $ M / M_\odot $ & 0.5 - 0.6 - 0.8 - 1.0 - 1.5 - 2.0 - 3.0 - 4.0 - 5.0 - 7.0 - 10 \\
  \hline \hline
\end{tabular}
\end{center}
\caption{\small Parameters and their values for mirror star models. The models are for all the combinations of the parameter values. In addition the model with $ M = 0.8 M_\odot $ has been computed also for the not listed intermediate values $ Y = 0.30, 0.40, 0.60, 0.80 $.}
\label{table1_7}
\end{table}

We note that those which we are now calling mirror stars can perfectly be ordinary stars in an advanced evolutionary stage of the Universe, when the galactic matter is enriched by many elements processed by the previous stellar populations, thus forming stars with more helium in their structures.


\passo
\def \mirror_star_evol{Evolution of the mirror stars}
\section{\mirror_star_evol}
\label{mirror_star_evol}
\markboth{Chapter \ref{chap-mirror_univ_4}. ~ \mir_univ_4}
                    {\S \ref{mirror_star_evol} ~ \mirror_star_evol}

First of all we remember that in stellar astrophysics the evolution of a star is studied in the so-called {\sl H-R (Hertzsprung-Russell) diagram}, where we plot the luminosity $ L $ and the effective temperature $ T_e $ of the star\footnote{The {\sl effective temperature} $ T_e $ of a star is defined by 
\bea
L = 4 \pi R^2 \sigma T_e^4 \;, \nonumber
\eea
where $ \sigma $ is the Stephan-Boltzmann constant, $ L $ is the luminosity, and $ R $ is the radius at the height of the photosphere. Thus, $ T_e $ is the characteristic temperature of the stellar surface if it emits as a black body.}. In order to understand the evolutionary differences between ordinary and mirror stars, it is also necessary to give a very brief review of basic stellar evolutionary theory.

During the first fast phase of gravitational contraction at nearly constant effective temperature and decreasing luminosity, the star goes down along his Hayashi track\footnote{{\sl Hayashi track} is the evolutionary track of a totally convective stellar model. It is the coldest possible track for a star of a given mass, and it is located at the extreme right of the H-R diagram.}, negligibly slowing down its contraction only while the structure is fastly burning the few light elements (D, Li, Be, B) present. Contraction increases the central temperature $ T_c $, until stars with masses $ M \gsim 0.1 ~M_\odot $ \footnote{The exact value of the {\sl hydrogen burning minimum mass} $ M_{hbmm} $ is dependent on the metallicity. For our models $ Z = 10^{-4} $ and $ M_{hbmm} \simeq 0.1 ~M_\odot $, while for solar metallicity $ Z = 0.02 $ and $ M_{hbmm} \simeq 0.08 ~M_\odot $.} ignite the hydrogen burning\footnote{There are two possible ways of burning hydrogen: the first one, called PP or proton-proton chain, becomes efficient at $ T_c \sim 6 \times 10^6 ~{\rm K} $, while the second one, called CNO chain, at $ T_c \sim 15 \times 10^6 ~{\rm K} $. Since their efficiencies are dominant at different temperatures, the PP chain provides energy for smaller stellar masses, and the CNO does it for bigger ones.} in their cores at a temperature $ T_c \sim 6 \times 10^6 ~{\rm K} $, while the ones with lower masses do not ignite hydrogen and die as {\sl brown dwarfs}. After depletion of hydrogen in the core, the burning passes to a shell at the boundary of the core, which is now made of He. At this stage the star starts decreasing its effective temperature ({\sl turn-off}). Meanwhile the He core contracts until masses greater than $ \sim 0.5 ~M_\odot $ reach a central temperature $ T_c \sim 10^8 ~{\rm K} $ and start He-burning in C and O; stars with lower masses die as {\sl white dwarfs} and start their {\sl cooling sequences}. The nucleosynthesis process continues burning elements into heavier and heavier nuclei, but only the heaviest stars ($ M > 6-8 \; M_\odot $) complete the advanced evolutionary stages and die exploding as type II {\sl supernovae} and leaving in their place a {\sl neutron star} or a {\sl black hole}. An evolutionary feature common to all burning phases is that shell burnings induce convective phenomena which push the star toward his Hayashi track, namely toward lower effective temperatures, while central burnings push it toward higher $ T_e $. 

A key point is the evaluation of evolutionary times. The most lasting phase of life of a star is surely that corresponding to the central hydrogen burning (the so-called {\sl main sequence}), with time-scales of $ 10^{10} $ yr for masses near the solar mass. Thus, we can approximate the lifetime of a star with its main sequence time. Since both luminosity and effective temperature depend on the mass and chemical composition, clearly the lifetime too depends on them. We use now the proportionality relations \cite{claybook}
\be
L \propto \mu^{7.5} M^{5.5}
\label{lmum}
\ee
and
\be
T_e^4 \propto \mu^{7.5} \;,
\label{teffmu}
\ee
where $ \mu $ is the mean molecular weight. From eq.~(\ref{lmum}) we obtain that bigger masses need higher luminosities, so that they use all the available hydrogen earlier than the lighter ones. From both eqs.~(\ref{lmum}) and (\ref{teffmu}) we know that an increase of helium abundance corresponds to an increase of the mean molecular weight and consequently in both luminosity and effective temperature. 
The increase in luminosity means that the star needs more fuel to produce it, but at the same time its amount is lower, because higher $ Y $-values necessarily imply lower $ X $-values. Both these events act to shorten the lifetime of a mirror star, which has a high He content. This can be formalized in the following relation \cite{claybook}
\be
{\rm t_{MS}} \propto { X \over \mu^{1.4} } \;,
\label{txmu}
\ee
where $ t_{MS} $ is the main sequence lifetime.

After these predictions on evolutionary properties of He-rich stars, we analyse the quantitative results of our models. They can be divided into two groups. The first one is made of models with mass $ M = 0.8 M_\odot $ and many different $ Y $-values. The second one is made instead of models with only three different He contents and a large range of masses.

We start from figure \ref{mirstar_fig_tesi_1}, where we plot the models of $ 0.8 M_\odot $ in the H-R diagram. The models are followed until the He-burning ignition, i.e. along their main sequence, turn-off and red giant\footnote{A red giant is a cold giant star in the phase of H-burning in shell before the He-ignition.} phases, which practically occupy all their lifetimes. 
Our qualitative predictions are indeed confirmed. Models with more helium are more luminous and hot; for example the main sequence luminosity ratio of the model with $ Y = 0.80 $ to the one with $ Y = 0.24 $ is $ \sim 10^2 $. Other consequences of an He increase are a longer (in the diagram, not in time) phase of decreasing temperature at nearly constant luminosity, and a shorter red giant branch. 

From these models we computed the evolutionary times until the He-ignition, i.e. for the entire plotted tracks, and we summarize them in table \ref{table2_7}. As expected, the ages decrease for growing $ Y $, but we see now how much high is this correlation. For $ Y = 0.40 $ the lifetime is already about one third compared to a visible ($ Y = 0.24 $) star, while for the highest value, $ Y = 0.80 $, it is roughly $ 10^2 $ times lower. 
We can approximately say that an increase of $ 10 \% $ in helium abundance roughly divides by two the stellar lifetime. In figure \ref{mirstar_fig_tesi_3} we plot the evolutionary times listed in the table. We see that, using a logarithmic scale for the stellar age, we obtain a quasi-linear relation between it and the helium content for this range of parameters.

\begin{figure}[h]
  \begin{center}
    \leavevmode
    \epsfxsize = 13cm
    \epsffile{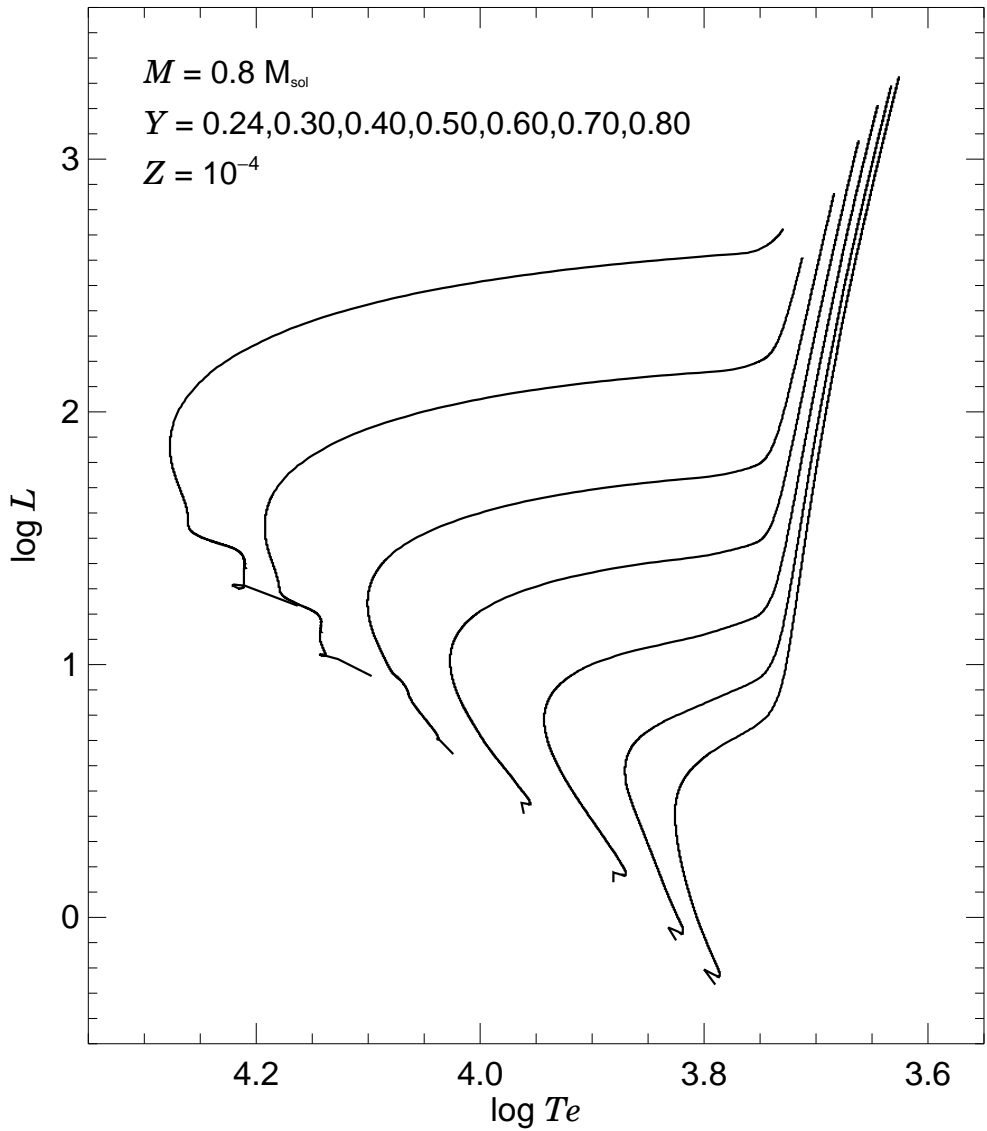}
  \end{center}
\caption{\small Evolutionary tracks in the H-R diagram of stars with $ M = 0.8 M_\odot $, $ Z = 10^{-4} $ and different helium contents $ Y = 0.24, 0.30, 0.40, 0.50, 0.60, 0.70, 0.80 $.}
\label{mirstar_fig_tesi_1}
\end{figure}

\begin{table}[h]
\begin{center}
\begin{tabular}{|c||c|c|c|c|c|c|c|c|} \hline \hline
  $ Y $ & 0.24 & 0.30 & 0.40 & 0.50 & 0.60 & 0.70 & 0.80 \\
  \hline
  age ($ 10^9 yr $) & 13.2 & 8.53 & 4.50 & 2.17 & 1.01 & 0.417 & 0.169 \\
  \hline \hline
\end{tabular}
\end{center}
\caption{\small Ages computed for stars of mass $ M = 0.8 M_\odot $ and the indicated helium contents.}
\label{table2_7}
\end{table}

\begin{figure}[h]
  \begin{center}
    \leavevmode
    \epsfxsize = 10cm
    \epsffile{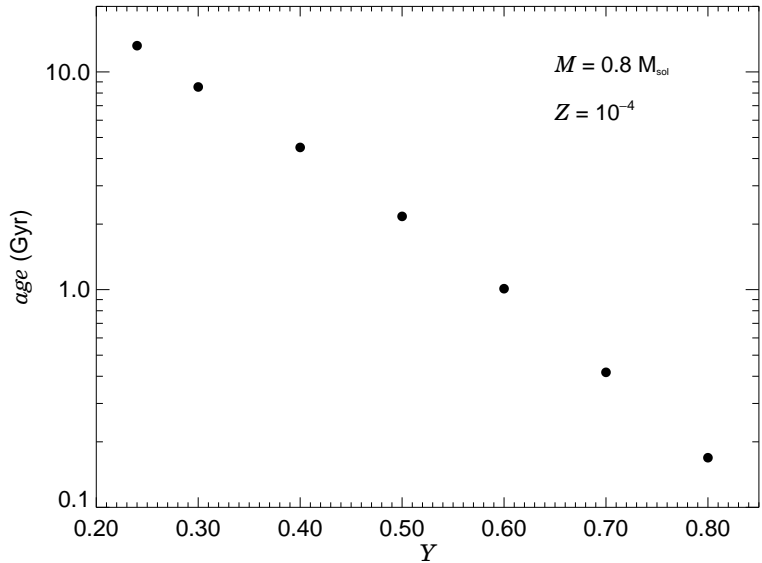}
  \end{center}
\caption{\small Evolutionary times (listed in table \ref{table2_7}) for models with $ M = 0.8 \; M_\odot $, $ Z = 10^{-4} $ and different helium contents $ Y = 0.24, 0.30, 0.40, 0.50, 0.60, 0.70, 0.80 $.}
\label{mirstar_fig_tesi_3}
\end{figure}

Let us now extend the analysis to models covering a large range of masses, from 0.5 $ M_\odot $ to 10 $ M_\odot $. Since the dependence on the helium content has been already studied for the 0.8 $ M_\odot $ case, we concentrate on only three $ Y $-values. Figure \ref{mirstar_fig_tesi_2} shows the evolutionary tracks for all the masses and only two helium contents, and again the models are followed up to the He-ignition. For every mass the $ Y $-dependence is the same as for the above discussed 0.8 $ M_\odot $ model. For models with masses $ M \gsim 2 ~M_\odot $ the growth in $ Y $ causes a considerable increase of the He-ignition effective temperature, together with the disappearance of the red giant branch.

In table \ref{table3_7} we list the lifetimes for all masses and the three indicated helium contents. The ratio of an ordinary star ($ Y = 0.24 $) evolutionary time to the high He-content mirror one ($ Y = 0.70 $) is between $ \sim 30 $ for 0.5 $ M_\odot $ and $ \sim 10 $ for 10 $ M_\odot $. These data are plotted in figure \ref{mirstar_fig_tesi_4}, where we see that the same dependence on the star masses holds for every helium content, with a shift toward lower ages for higher $ Y $-values.

This is an evidence that, under large mass ranges and different boundary conditions (in terms of temperatures of the mirror sector, and thus stellar helium content), the lifetimes of mirror stars are roughly an order of magnitude greater than the ones of visible stars. 

This means that, compared to O stars, M stars evolve faster and enrich earlier the galaxy of processed mirror gas, with implications for galaxy evolution. From the detailed study of this evolution together with the necessary information of the initial mirror stellar mass function, we could predict the expected population of mirror stars, in order to compare it with current MACHO observations. In addition, we could evaluate the amount of gravitational waves expected from mirror supernovae. These are just some interesting future applications of the present study.

\begin{figure}[h]
  \begin{center}
    \leavevmode
    \epsfxsize = 14cm
    \epsffile{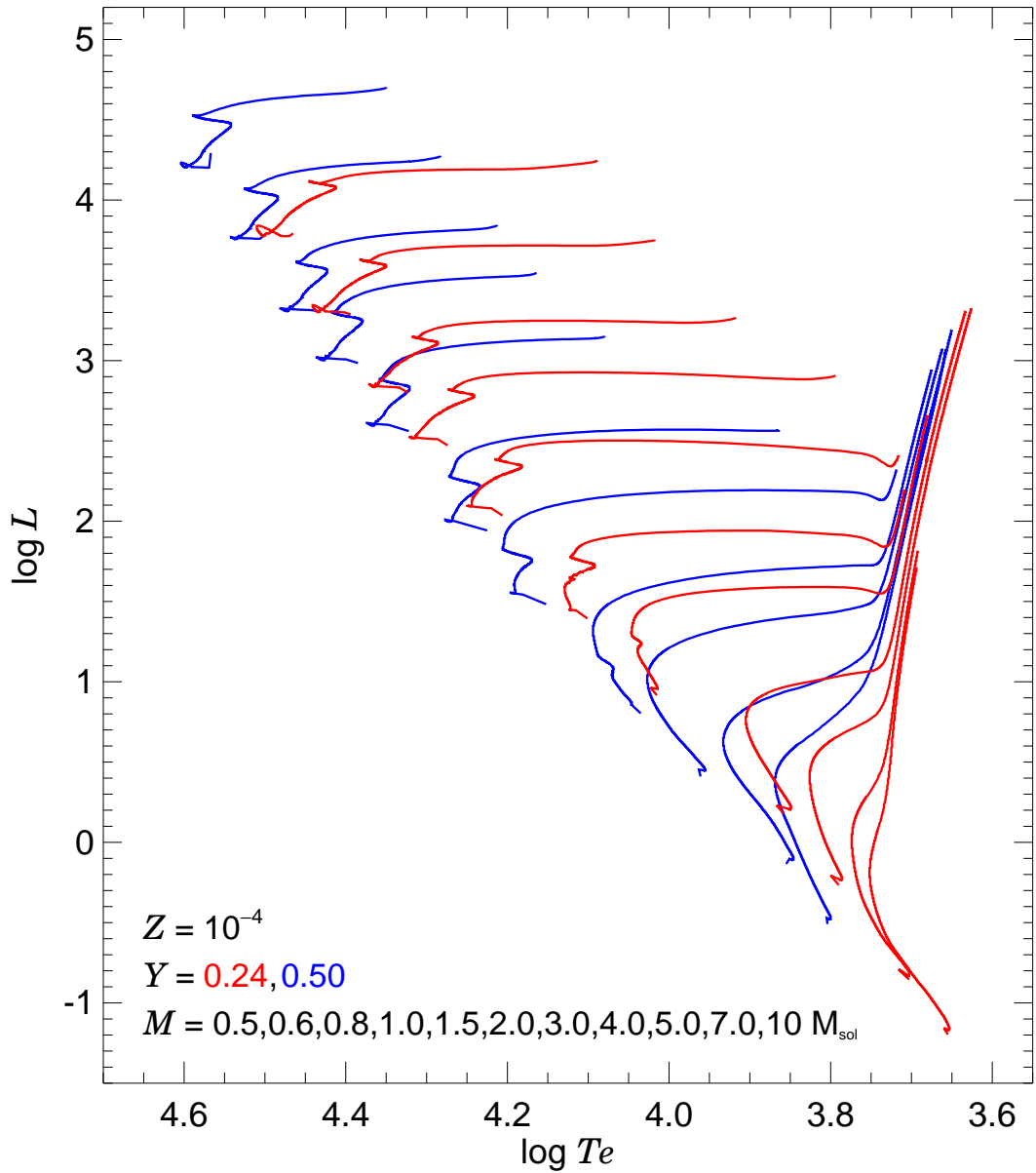}
  \end{center}
\caption{\small Evolutionary tracks in the H-R diagram of stars with different masses $ M = 0.5, 0.6, 0.8, 1.0, 1.5, 2.0, 3.0, 4.0, 5.0, 7.0, 10 \; M_\odot $, $ Z = 10^{-4} $ and two different helium contents $ Y = $ 0.24 (red line) and 0.50 (blue).}
\label{mirstar_fig_tesi_2}
\end{figure}

\begin{table}[h]
\begin{center}
\begin{tabular}{|c|c|c|c|} \hline \hline
  mass & age ($ yr $) & age ($ yr $) & age ($ yr $) \\
  ($ M/M_\odot $) & ($ Y = 0.24 $) & ($ Y = 0.50 $) & ($ Y = 0.70 $) \\
  \hline \hline
  0.5 & $ 7.06 \times 10^{10} $ & $ 1.12 \times 10^{10} $ & $ 1.92 \times 10^9 $ \\  \hline
  0.6 & $ 3.71 \times 10^{10} $ & $ 5.80 \times 10^9 $ & $ 1.04 \times 10^9 $ \\  \hline
  0.8 & $ 1.32 \times 10^{10} $ & $ 2.17 \times 10^9 $ & $ 4.17 \times 10^8 $ \\  \hline
  1.0 & $ 6.07 \times 10^9 $ & $ 1.05 \times 10^9 $ & $ 2.19 \times 10^8 $ \\  \hline
  1.5 & $ 1.60 \times 10^9 $ & $ 3.01 \times 10^8 $ & $ 9.02 \times 10^7 $ \\  \hline
  2.0 & $ 6.43 \times 10^8 $ & $ 1.40 \times 10^8 $ & $ 4.45 \times 10^7 $ \\  \hline
  3.0 & $ 2.14 \times 10^8 $ & $ 5.64 \times 10^7 $ & $ 1.78 \times 10^7 $ \\  \hline
  4.0 & $ 1.15 \times 10^8 $ & $ 3.13 \times 10^7 $ & $ 9.41 \times 10^6 $ \\  \hline
  5.0 & $ 7.26 \times 10^7 $ & $ 2.05 \times 10^7 $ & $ 6.56 \times 10^6 $ \\  \hline
  7.0 & $ 3.81 \times 10^7 $ & $ 1.17 \times 10^7 $ & $ 4.14 \times 10^6 $ \\  \hline
  10 & $ 2.09 \times 10^7 $ & $ 7.18 \times 10^6 $ & $ 2.78 \times 10^6 $ \\
  \hline \hline
\end{tabular}
\end{center}
\caption{\small Ages computed for stars of the indicated mass and helium content, with a metallicity $ Z = 10^{-4} $.}
\label{table3_7}
\end{table}

\begin{figure}[h]
  \begin{center}
    \leavevmode
    \epsfxsize = 11cm
    \epsffile{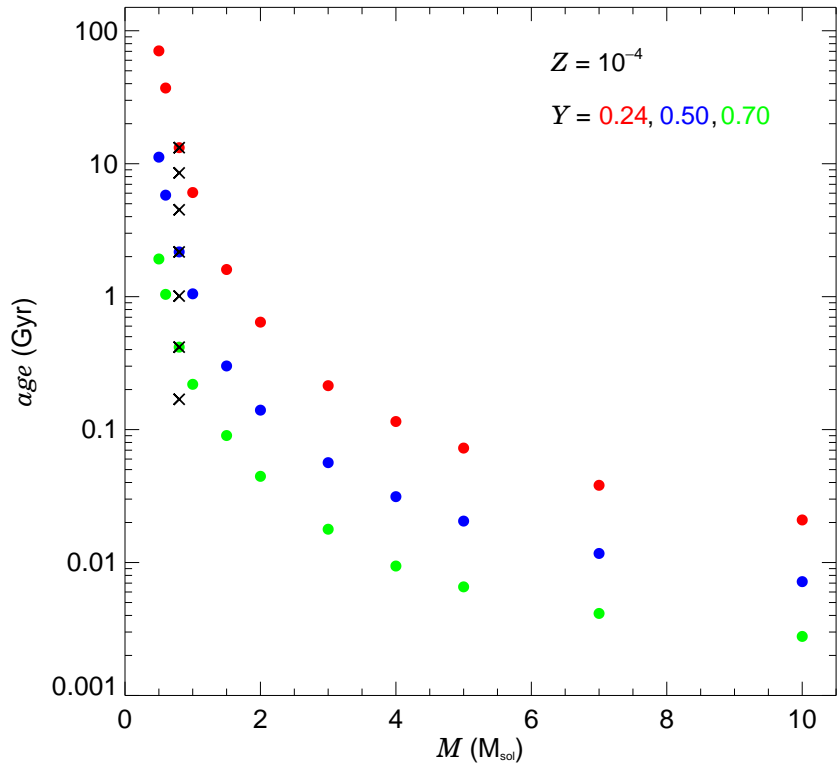}
  \end{center}
\caption{\small Evolutionary times (listed in table \ref{table3_7}) for models with $ M = $ 0.5, 0.6, 0.8, 1.0, 1.5, 2.0, 3.0, 4.0, 5.0, 7.0, 10$ \; M_\odot $, $ Z = 10^{-4} $ and three different helium contents $ Y = $ 0.24 (red dots), 0.50 (blue) and 0.70 (green). Are also indicated the models of figure \ref{mirstar_fig_tesi_3} (black).}
\label{mirstar_fig_tesi_4}
\end{figure}



\def \conclus{Conclusions}
\chapter{\conclus}
\label{conclus}
\markboth{Chapter \ref{conclus}. ~ \conclus}
                    {Chapter \ref{conclus}. ~ \conclus}


\passo

The aim of this thesis is contained in a question: ``is mirror matter a reliable dark matter candidate?'' Its emergence arises from the problems encountered by the standard candidate, which is now the cold dark matter, in some aspects, as for example the central galactic density profiles or the number of small satellites. In order to answer this question, we studied the cosmological implications of the parallel mirror world with the same microphysics as the ordinary one, but which couples the latter only gravitationally, and its consistence with present observational data, in particular the ones coming from the Big Bang nucleosynthesis, the cosmic microwave background radiation and the large scale structure.

The nucleosynthesis bounds on the effective number of extra light neutrinos demand that the mirror sector should have a smaller temperature than the ordinary one, $ T'< T $, with the limit $ x \simeq T'/T< 0.64 $ set by the constraint $ \Delta N_\nu < 1 $. By this reason its evolution should be substantially deviated from the standard cosmology as far as the crucial epochs like baryogenesis, nucleosynthesis, baryon-photon decoupling, etc. are concerned. 

Starting from an asymmetric inflationary scenario which could explain the different initial temperatures of the two sectors, 
in the context of both the GUT or electroweak baryogenesis scenarios the condition $T'<T$ yields that the mirror sector should produce a larger baryon asymmetry than the observable one, $\eta'_B>\eta_B$. 

Therefore, the temperature bound implies that the mirror sector contains less relativistic matter (photons and neutrinos) than the ordinary one, $\Omega'_r \ll \Omega_r$, so that in the relativistic expansion epoch the cosmological energy density is dominated by the ordinary component,  while the mirror one gives a negligible contribution.  However, for the non-relativistic epoch  the complementary situation can occur when  the mirror baryon matter density is bigger than the ordinary one, $\Omega'_B > \Omega_B$. Hence, the mirror baryonic dark matter can contribute the dark matter of the Universe along with the cold dark matter or even constitute a dominant dark matter component. We know also that mirror world must be a helium-dominated world, since in its sector the Big Bang nucleosynthesis epoch proceeds differently from the ordinary one, and it predicts the mirror helium abundance in the range $Y' =0.5$-$0.8$, considerably larger than the one for observable helium, $Y \simeq 0.24$. 
 
Since the existence of a mirror hidden sector changes the time of 
key epochs,
there are important consequences in the structure formation scenario for a Mirror Universe. We studied this scenario in presence of adiabatic scalar density perturbations, which are now the most probable kind of primordial fluctuations, in the context of the Jeans gravitational instability theory.

Given that the physics is the same in both sectors, key differences are the shifts of fundamental epochs, namely the matter-radiation equality occurs in a Mirror Universe before than in a standard one, $ (a_{\rm eq})_{\rm mir} < (a_{\rm eq})_{\rm ord} $, and the baryon-photon equipartition and the matter-radiation decoupling do the same: $ a_{\rm b\gamma}' < a_{\rm b\gamma} $; $ a_{\rm dec}' < a_{\rm dec} $. The first step is given by the study of the mirror sound speed and its comparison with the ordinary one and with the velocity dispersion of a typical cold dark matter candidate. From this study we obtain the mirror Jeans length and mass, again to be compared with the same quantities obtained for the ordinary sector and for the cold dark matter. There are two different possibilities, according to the value of $ x $, which can be higher or lower than $ x_{\rm eq} \approx 0.046 (\Omega_{\rm m} h^2)^{-1} $, the value for which mirror decoupling occurs at matter-radiation equality time. 

The values of the length and mass scales clearly depend on the mirror sector temperature and baryonic density, but we found that $ M_{\rm J}' $ is always smaller than $ M_{\rm J} $, with a typical ratio $ \sim 10 $ for $ x > x_{\rm eq} $, while for cold dark matter it is $ \sim 10^{15}$.

Another important quantity to describe the structure evolution is the dissipative scale, represented in the mirror sector by the mirror Silk scale. We found that it is much lower than the ordinary one, obtaining $ M_{\rm S}' \sim 10^{10} \:{\rm M}_\odot $ for $ x \simeq x_{\rm eq} $, a value similar to the free streaming scale for a typical warm dark matter candidate, and much higher than the one for cold dark matter.

We put together all these informations to build two different mirror scenarios, for $ x > x_{\rm eq} $ and $ x < x_{\rm eq} $. In the latter case we obtain a maximum mirror Jeans scale similar to the Silk mass, so that practically all perturbations with masses greater than the Silk mass grow uninterruptedly, just as in a cold dark matter scenario.

After this, we modified a numerical code existing for the standard Universe in order to take into account a hidden mirror sector, and computed the evolution of perturbations in the linear regime for all the components present in a Mirror Universe, namely the ordinary and mirror baryons and photons, and the cold dark matter. We did this for various mirror temperatures and baryon densities, and for different perturbative scales, finding all the features predicted by our structure formation study (as for example the mirror decoupling and the CDM-like behaviour for low $ x $-values).

Using the same numerical code we were able to predict the expected power spectra of cosmic microwave background and large scale structure for a flat Mirror Universe. We analysed the dependence of power spectra on mirror parameters $ x $ and $ \Omega_{\rm b}' $, and also on other cosmological parameters, as $ \Omega_{\rm m} $, $ \Omega_{\rm b} $, $ h $, $ n $, $ N_\nu $.

In CMB spectra we found various differences from a so-called standard concordance model for $ x \gsim 0.3 $ and a dependence not linear in $ x $, specially evident in the first and third peaks. The dependence on the mirror baryon density is instead very low. We computed also the power spectrum of mirror CMB photons, even if this study is now only an academic exercise, because by definition we can't reveal them.

Turning to the LSS power spectra, we showed an influence of the mirror sector bigger than for the CMB, with a great dependence on both mirror temperature and baryonic density. Both of them cause oscillations in power spectrum, but the first one influences the scale at which they start, while the second one their depth. In this case we see the mirror sector effects also for low $ x $-values, if we don't take a too small value for $ \Omega_{\rm b}' $.

We extended the models also to smaller (non linear) scales, in order to show the cutoff present in the mirror scenario. We demonstrated the existence of this cutoff, specially dependent on $ x $-value, but modulated also by $ \Omega_{\rm b}' $. This is an important feature of the mirror structure formation scenario, because it could explain the observed small number of satellites which is a problem for cold dark matter, and at the same time this is valid also for low $ x $-values, that give mirror baryons equivalent to CDM for the CMB and LSS at linear scales.

The next step was the comparison with observations. In this phase it is important the joint analysis of both CMB and LSS data, which gave us an important limit on the mirror parameter space. In fact, we obtained the conclusion that mirror models with high $ x $ and high $ \Omega_{\rm b}' $ are excluded by LSS observations, because they generate too deep oscillations in power spectra. This is an important bound, which limits to the three following possibilities to have a mirror sector:
\begin{itemize}
\item high $ x $ and low $ \Omega_{\rm b}' $ (differences from the CDM in the CMB, and oscillations in the LSS with a depth modulated by the baryon density);
\item low $ x $ and high $ \Omega_{\rm b}' $ (completely equivalent to the CDM for the CMB, and few differences for the LSS in the linear region);
\item low $ x $ and low $ \Omega_{\rm b}' $ (completely equivalent to the CDM for the CMB, and nearly equivalent for the LSS in the linear region and beyond, according to the mirror baryon density).
\end{itemize}
Thus, with the current experimental accuracy, we can exclude only models with high $ x $ and high $ \Omega_{\rm b}' $, but with the 
soon available high precision data on CMB (Map, Planck) and LSS (2dF and SDSS) we will be able to choose between the CDM and mirror cosmological scenarios.

In last chapter we qualitatively discussed the implications of the mirror baryons representing a kind of self-interacting dark matter for the galactic halo structure and MACHOs. Both of them can be explained as mirror consequences, the first one as mirror galaxy, the second ones as mirror stars. Finally, we computed the evolutionary properties of mirror stars using a numerical code. Stars in mirror sector are different only for their helium content ($ Y' = $ 0.4-0.8) from the visible ones ($Y \simeq $ 0.24). This difference implies much faster evolutionary times, dependent on the exact He content and thus on the temperature of the mirror sector.

Therefore, at the end of this work of thesis, we reached a partial answer: {\sl on the light of current observations of BBN, CMB and LSS in linear scales the mirror baryonic dark matter is not only fully in agreement with observations, but in some case it could be even preferable to the CDM scenario}. In addition, we obtained some useful constraints on the parameter space, which can address our future efforts to understand other aspects of the Mirror Universe.


\passo
\def \suggest_fut{Suggestions for future works}
\subsubsection{\suggest_fut}
\label{suggest_fut}

Let us conclude by briefly describing some of the planned future developments of this work. They are all linked each other and help to complete the picture of a Mirror Universe.

\noindent {\bf Fit of all cosmological parameters.}~~Since the program used in this thesis for computing CMB and LSS power spectra privileges the precision to the performance, it is too slow to make us able to build a parameter grid thin enough to fit the ordinary and mirror cosmological parameters. Thus, we will adapt a numerical code faster than the one now used (as for example the popular CMBFAST), in order to drastically decrease the computational time.

\noindent {\bf Mirror star formation.}~~An important question that still needs to be resolved is the mirror star formation. The different primordial chemical content of the mirror sector causes differences in opacities and then in fragmentation of protostellar clouds. The initial mirror stellar mass function, together with the mirror evolutionary data (that we obtained in this thesis), are the necessary inputs to insert as stellar feedback in N-body simulations of galaxy formation and evolution.

\noindent {\bf Estimation of present mirror star population.}~~With the knowledge of mirror star formation and evolution we could in principle predict the present stellar population of mirror stars, in order to compare it with current MACHO observations.

\noindent {\bf Gravitational waves from the mirror sector.}~~After the study of the present mirror star population, we can evaluate the spectrum of gravitational waves expected from the mirror sector as produced by supernovae explosions or binary systems of compact objects.

\noindent {\bf Extension to smaller (non linear) scales and galaxy formation.}~~Using the transfer functions obtained here in the linear approximation, we can use N-body simulations to include non-linear effects, as merging and stellar feedback. In this way we can also simulate the galaxy formation in presence of a mirror sector, compute the density profiles, and verify if, as we expect, the mirror scenario solves the open problems placed by the cold dark matter scenario.

\noindent {\bf Supermassive black hole formation.}~~Another important issue worthy of exploration is the influence of the mirror sector on supermassive black holes formation in galaxy centers. Maybe mirror matter could help in producing them.



\appendix


\chapter{Useful formula for cosmology and thermodynamics}
\def \appen-cosm-mod{Useful formula for cosmology and thermodynamics}
\label{appen-cosm-mod}


\passo
\def \gen_rel{General Relativity}
\section{\gen_rel}
\label{gen_rel}
\markboth{Appendix \ref{appen-cosm-mod}. ~ \appen-cosm-mod}
                    {\S \ref{gen_rel} ~ \gen_rel}

The essence of Einstein's theory is to transform gravitation from being a force to being a property of space-time, which may be curved. The interval between two events 
\be
\label{metric}
ds^2=g_{\alpha \beta}(x)~dx^{\alpha} dx^{\beta} \;,
\ee
\noindent is fixed by the {\sl metric tensor} $g_{\alpha \beta }$ which describes the space-time geometry\footnote{Repeated indices imply summation and $ \alpha, \beta $ run from 0 to 3; $ x^0 = t $ is the time coordinate and $x^i$ ($i = 1,2,3$) are space coordinates.} ($ g^{\alpha \mu }g_{\mu \beta } = \delta^{\alpha}_{\beta} $). 
For the Riemannian spaces, the {\sl tensor of curvature} is
\be
\label{riemanntensor}
R^\mu _{\nu \alpha \beta} = {\partial \Gamma^\mu _{\nu \beta} \over \partial x^\alpha } - {\partial \Gamma^\mu _{\nu \alpha } \over \partial x^\beta } + \Gamma^\mu _{\sigma \alpha } \Gamma^\sigma _{\upsilon \beta } - \Gamma^\mu _{\sigma \beta } \Gamma^\sigma _{\nu \alpha } \;,
\ee
\noindent where the $\Gamma$'s are {\sl Christoffel symbols}
\be
\label{christoffel}
\Gamma^\mu _{\alpha \beta } = {1 \over 2} \; g^{\mu \sigma }\left[{\partial g_{\sigma \alpha } \over \partial x^\beta }+{\partial g_{\sigma \beta } \over \partial x^\alpha }-{\partial g_{\alpha \beta } \over \partial x^\sigma }\right] \;.
\ee
The equation of motion of a free particle is determined by the space-time metric
\be
\label{geodesic}
{d^2 x^\alpha  \over ds^2} + \Gamma^\alpha _{\beta \gamma } {d x^\beta  \over ds}{d x^\gamma  \over ds} = 0 \;.
\ee
so that free particle moves on a {\sl geodesic}.

On the other hand, the metric $g_{\alpha \beta }$ is itself determined by the distribution of matter, described by the {\sl energy-momentum tensor} $ T_{\alpha \beta } $, according to the {\em Einstein equations}
\be
\label{einsteinequation}
G_{\alpha \beta } \equiv  R_{\alpha \beta } - {1\over 2} R g_{\alpha \beta } = {8\pi G} T_{\alpha \beta } + \Lambda g_{\alpha \beta } \;,
\ee
where 
\be
\label{riccitensor}
R_{\alpha \beta } = R^\gamma _{\alpha \gamma \beta } ~,~~~~~~~~~~~ R = g^{\alpha \beta }R_{\alpha \beta } \;
\ee
are respectively the {\sl Ricci tensor} and the {\sl Ricci scalar}, and $\Lambda$ is the {\sl cosmological constant}. For the FRW metric, the non zero components of the Ricci tensor and the value of the Ricci scalar are
\bea 
\label{ricci}
& & R_{00} = -3 {\ddot a \over a } ~,~~~~~~~~~ R_{ij} = - \left[ {\ddot a \over a } + 2{\dot a^2 \over a^2 } + { 2k \over a^2 } \right] g_{ij} \;, \nonumber \\
& & R = -6 \left[ {\ddot a \over a } + {\dot a^2 \over a^2 } + { k \over a^2 } \right] \;.
\eea
For a perfect fluid the energy-momentum tensor has the form
\be
\label{energymomentum}
T_{\alpha \beta } = (p + \rho) u_\alpha  u_\beta  - p g_{\alpha \beta } \;,
\ee
where $\rho$ and $p$ are respectively the energy density and pressure of the fluid, $u_\alpha = g_{\alpha \beta } \, dx^\beta / ds $ is the fluid four-velocity. Considering the symmetries of the FRW metric (uniformity and isotropy), which demand that $u^0 = 1$ and $u^i = 0$ in the comoving coordinate system, we obtain $T_{\alpha\beta} = diag(\rho,-p,-p,-p)$. This is valid also in presence of the cosmological constant, if we substitute $ p $ and $ \rho $ as indicated in the following
\be
\label{p_rho_wl}
p\, \to p - p_\Lambda ~,~~~~~~~~~~ \rho\, \to \rho + \rho_\Lambda ~;~~~~~~~~~~ \rho_\Lambda = - p_\Lambda = {\Lambda \over 8\pi G} \;.
\ee
Therefore, the Einstein equations for a Universe described by the FRW metric are reduced to equations (\ref{friedmann1}) and (\ref{friedmann3}), where the relations between the energy and pressure densities are in general related as $ p = w \rho $. In particular, for the dominance of relativistic and non relativistic matter we have respectively $ w = 1/3 $ and $ w = 0 $, while for the vacuum energy dominance one has $ w = -1 $.


\passo
\def \app-flat{Flat models}
\section{\app-flat}
\label{app-flat}
\markboth{Appendix \ref{appen-cosm-mod}. ~ \appen-cosm-mod}
                    {\S \ref{app-flat} ~ \app-flat}

Here we report two special cases of the relationships (\ref{aflat})-(\ref{rhoflat}): {\sl dust} or {\sl matter dominated Universe} ($w = 0$)
\bea
\label{aflatMD}
a(t) & = & a_0 {\left( t \over t_0 \right)}^{2/3} ~ \\
\label{tflatMD}
t & = & t_0 {\left( 1+z \right)}^{-3/2} ~ \\
\label{HflatMD}
H & = & {2 \over 3t} = H_0 {\left( 1+z \right)}^{3/2} ~ \\
\label{qflatMD}
q_0 & = & {1 \over 2} ~ \\
\label{t0flatMD}
t_0 & = & {2 \over {3 H_0}} ~ \\
\label{rhoflatMD}
\rho_m & = & { 1 \over {6 \pi G t^2}} ~
\eea

\addvspace{0.8cm}
\noindent and {\sl radiation dominated Universe} ($w = 1/3$)
\bea
\label{aflatRD}
a(t) & = & a_0 {\left( t \over t_0 \right)}^{1/2} ~ \\
\label{tflatRD}
t & = & t_0 {\left( 1+z \right)}^{-2} ~ \\
\label{HflatRD}
H & = & {1 \over 2t} = H_0 {\left( 1+z \right)}^2 ~ \\
\label{qflatRD}
q_0 & = & 1 ~ \\
\label{t0flatRD}
t_0 & = & {1 \over 2 H_0} ~ \\
\label{rhoflatRD}
\rho_r & = & { 3 \over {32 \pi G t^2}} ~
\eea

\addvspace{0.6cm}


\passo
\def \mabert3{The synchronous gauge}
\section{\mabert3}
\label{mabert3}
\markboth{Appendix \ref{appen-cosm-mod}. ~ \appen-cosm-mod}
                    {\S \ref{mabert3} ~ \mabert3}

Since our interests lie in the physics in an expanding Universe, we use comoving coordinates $x^\alpha = (\tau,{\bf x})$, with the expansion factor $a(\tau)$ of the Universe factored out.  The comoving coordinates are related to the proper time and positions $t$ and ${\bf r}$ by $dx^0 =d\tau = dt/a(\tau)$, $d{\bf x}=d{\bf r}/a(\tau)$. Dots will denote derivatives with respect to $\tau$: $\dot a\equiv\partial a/\partial\tau$.

The components $g_{00}$ and $g_{0i}$ of the metric tensor in the synchronous gauge are by definition unperturbed.  The line element is given by
\begin{equation}
  ds^2 = a^2(\tau)\{-d\tau^2 + (\delta_{ij} + h_{ij})dx^i dx^j\} \;.
\end{equation}
The metric perturbation $h_{ij}$ can be decomposed into a trace part $h \equiv h_{ii}$ and a traceless part consisting of three pieces, $h^\parallel_{ij}, h^\perp_{ij}$, and $h^T_{ij}$, where $h_{ij}=h\delta_{ij}/3 + h^\parallel_{ij}+h^\perp_{ij}+h^T_{ij}$ ;
$h^\parallel_{ij}$ can be written in terms of some scalar field $\mu$ and $h^\perp_{ij}$ in terms of some divergenceless vector {\bf A} as
\begin{eqnarray}
        h^\parallel_{ij} &=& \left( \partial_i\partial_j -
	{1\over 3} \delta_{ij} \nabla^2 \right) \mu \;, \nonumber\\
        h^\perp_{ij} &=& \partial_i A_j + \partial_j A_i \;, \qquad
        \partial_i A_i = 0 \;.
\end{eqnarray}
The two scalar fields $h$ and $\mu$ (or $h^\parallel_{ij}$) characterize the scalar mode of the metric perturbations, while $A_i$ (or $h^\perp_{ij}$) and $h^T_{ij}$ represent the vector and the tensor modes, respectively.

We will be working in the Fourier space $k$. We introduce two fields $h({\bf k},\tau)$ and $\eta({\bf k},\tau)$ in $k$-space and write the scalar mode of $h_{ij}$ as a Fourier integral
\begin{equation}
\label{hijk}
	h_{ij}({\bf x},\tau) = \int d^3k e^{i{\bf k}\cdot{\bf x}}
	\left\{ \hat{k}_i\hat{k}_j h({\bf k},\tau) +
	(\hat{k}_i\hat{k}_j - {1 \over 3}\delta_{ij})\,
        6\eta({\bf k},\tau) \right\} \;,\quad {\bf k} = k\hat{k} \;.
\end{equation}
Note that $h$ is used to denote the trace of $h_{ij}$ in both the real space and the Fourier space.

In spite of its wide-spread use, there are disadvantages associated with the synchronous gauge: spurious gauge modes contained in the solutions to the equations for the density perturbations, due to the arbitrary choice of the initial hypersurface and its coordinate assignments; coordinate singularities arising when two observers' trajectories intersect each other (a point in spacetime will have two coordinate labels, since the coordinates are defined by freely falling observers), so that a different initial hypersurface of constant time has to be chosen to remove these singularities.


\subsubsection{Einstein equations and energy-momentum conservation}

We find it most convenient to solve the linearized Einstein equations in the Fourier space $k$, and a cosmological constant is allowed through its inclusion in $\rho$ and $p$ using the (\ref{p_rho_wl}). In the synchronous gauge, the scalar perturbations are characterized by $h({\bf k},\tau)$ and $\eta({\bf k},\tau)$ in equation (\ref{hijk}).  In terms of $h$ and $\eta$, the time-time, longitudinal time-space, trace space-space, and longitudinal traceless space-space parts of the Einstein equations give the following four equations to linear order in $k$-space:
\begin{eqnarray}
    k^2\eta - {1\over 2}{\dot{a}\over a} \dot{h}
        &=& 4\pi G a^2 \delta T^0{}_{\!0}
	\;,\label{ein-syna}\\
    k^2 \dot{\eta} &=& 4\pi Ga^2 ({\rho}+p)
	\theta	 \;,\label{ein-synb}\\
    \ddot{h} + 2{\dot{a}\over a} \dot{h} - 2k^2
	\eta &=& -8\pi G a^2 \delta T^i{}_{\!i}
	\;,\label{ein-sync}\\
    \ddot{h}+6\ddot{\eta} + 2{\dot{a}\over a}\left(\dot{h}+6\dot{\eta}
	\right) - 2k^2\eta &=& -24\pi G a^2 ({\rho}+p)\sigma	
	\;.\label{ein-synd}
\end{eqnarray}
\label{ein-syn}
The variables $\theta$ and $\sigma$ are defined as
\begin{equation}
\label{theta}
 	({\rho}+p)\theta \equiv i k^j \delta T^0{}_{\!j} ~,
	\qquad	({\rho}+p)\sigma \equiv -(\hat{k}_i\hat{k}_j
	- {1\over 3} \delta_{ij})\Sigma^i{}_{\!j} \;,
\end{equation}
and $\Sigma^i{}_{\!j} \equiv T^i{}_{\!j}-\delta^i{}_{\!j} T^k{}_{\!k}/3$ denotes the traceless component of $T^i{}_{\!j}$. When the different components of matter and radiation (i.e., CDM, HDM, baryons, photons, and massless neutrinos) are treated separately, $({\rho}+p)\theta = \sum_i ({\rho}_i+p_i)\theta_i$ and $({\rho}+p)\sigma = \sum_i ({\rho}_i+p_i)\sigma_i\,$, where the index $i$ runs over the particle species.

Now we derive the transformation relating $\delta T^\alpha{}_{\!\beta}$, making use of the definition of energy-momentum tensor (\ref{energymomentum}). For a fluid moving with a small coordinate velocity $v^i \equiv dx^i/d\tau$, $v^i$ can be treated as a perturbation of the same order as $\delta\rho=\rho-\bar{\rho}$, $\delta p=p-\bar{p}$, and the metric perturbations.  Then to linear order in the perturbations the energy-momentum tensor is given by
\begin{eqnarray}
	T^0{}_{\!0} &=& -(\bar{\rho} + \delta\rho) \;,\nonumber\\
	T^0{}_{\!i} &=& (\bar{\rho}+\bar{p}) v_i  = -T^i{}_{\!0}\;,\nonumber\\
	T^i{}_{\!j} &=& (\bar{p} + \delta p) \delta^i{}_{\!j}
		+ \Sigma^i{}_{\!j} \;,\qquad
		\Sigma^i{}_{\!i}=0 \;,
\end{eqnarray}
where we have allowed an anisotropic shear perturbation $\Sigma^i{}_{\!j}$ in $T^i{}_{\!j}$.  

The conservation of energy-momentum is a consequence of the Einstein
equations.  Let $w \equiv p / \rho$ describe the equation of state (see \S \ref{sec-RW-metric}). Then the perturbed part of energy-momentum conservation equations
\begin{equation}
\label{Econs}
	T^{\alpha\beta}{}_{\!;\alpha} = \partial_\alpha T^{\alpha\beta}
	+ \Gamma^\beta{}_{\!\mu\nu} T^{\mu\nu}
	+ \Gamma^\mu{}_{\!\mu\nu} T^{\beta\nu} = 0
\end{equation}
in $k$-space implies
\begin{eqnarray}
\label{fluid}
	\dot{\delta} &=& - (1+w) \left(\theta+{\dot{h}\over 2}\right)
	  - 3{\dot{a}\over a} \left({\delta p \over \delta\rho} - w
	  \right)\delta  \;, \nonumber\\
	\dot{\theta} &=& - {\dot{a}\over a} (1-3w)\theta - {\dot{w}\over
	     1+w}\theta + {\delta p/\delta\rho \over 1+w}\,k^2\delta
	     - k^2 \sigma \;.
\end{eqnarray}
These equations are valid for a single uncoupled fluid, or for the net (mass-averaged) $\delta$ and $\theta$ for all fluids.  They need to be modified for individual components if they interact with each other.  An example is the baryonic fluid in our model, which couples to the photons before recombination via Thomson scattering. An extra term representing momentum transfer between the two components needs to be added to the $\dot\delta$ equation for the baryons.

For the isentropic primordial perturbations considered in this thesis, the equations above simplify since $\delta p=c_s^2\delta\rho$, where $c_s^2= dp/d\rho=w+\rho dw/d\rho$ is the adiabatic sound speed squared.  For the photons and baryons (the only collisional fluid components with pressure), $w$ is a constant ($w=1/3$ for photons and $w\approx0$ for baryons, since they are non relativistic at the times of interest).  Thus, $\delta p/\delta\rho - w=0$. 


\passo
\def \equil_therm{Thermodynamics of the Universe}
\section{\equil_therm}
\label{equil_therm}
\markboth{Appendix \ref{appen-cosm-mod}. ~ \appen-cosm-mod}
                    {\S \ref{equil_therm} ~ \equil_therm}

In a gas of a given species with $g$ internal degrees of freedom and energy $E = \sqrt{|{\bvec p}|^2 + m^2}$, {\sl kinetic equilibrium} is established by sufficiently rapid elastic scattering processes; in this case, for an ideal gas, the equilibrium phase-space density is
\begin{equation}
\label{kinetic_equil}
f(p) \; = \; \left[ \exp\left( {E \, - \, \mu} \over {T} \right) \; {\pm} \; 1 \right]^{-1} \;,
\end{equation}
where +/$-$ refers to Fermi-Dirac/Bose-Einstein statistics and $\mu$ is the chemical potential. In general each species has its own equilibrium temperature $T$, and the entire Universe can be represented as a plasma with different temperatures. However, if several species strongly interact among them, they will reach a mutual equilibrium and a common temperature; this is indeed the situation at early times. As the Universe expands and cools down, some species may start interacting more and more weakly and eventually decouple.
We can consider the photon temperature $T_\gamma$ as the plasma reference temperature $T$ of the Universe. 

In {\it chemical equilibrium}, established by processes which can create and destroy particles (differently from kinetic equilibrium), the chemical potential is additively conserved. So it is zero for particles such as photons and $Z^0$, which can be emitted and absorbed in any number, and consequently opposite for a particle and its antiparticle, which can annihilate into such bosons.

The quantities of interest are the number density $n$, energy density $\rho$ and pressure $p$ of a given species, 
defined for a generic species of mass $m$ with chemical potential $\mu $ at temperature $T$ as
\bea
n & = & {g \over 2\pi ^2} \, \int_m^{\infty} { { {\left( E^2 - m^2 \right)^{1/2}} \over { {\exp \left[ \left( E \, - \, \mu \right) / {T} \right]} \; {\pm} \; 1} } \, E dE } \;, \\
\rho & = & {g \over 2\pi ^2} \, \int_m^{\infty} { { {\left( E^2 - m^2 \right)^{1/2}} \over { {\exp \left[ \left( E \, - \, \mu \right) / {T} \right]} \; {\pm} \; 1} } \, E^2 dE } \;,
\label{rho_equil2} \\
p & = & {g \over 6\pi ^2} \, \int_m^{\infty} { { {\left( E^2 - m^2 \right)^{3/2}} \over { {\exp \left[ \left( E \, - \, \mu \right) / {T} \right]} \; {\pm} \; 1} } \, dE } \;.
\label{p_equil2}
\eea
For non relativistic species ($T \ll m$) we have (for both Fermi-Dirac and Bose-Einstein statistics)
\bea
n & \simeq & g \, \left( {m T} \over {2 \pi} \right)^{3/2} \, \exp \left( - \, { {m - \mu} \over {T} } \right) \;,
\label{n_nonrel} \\
\rho & \simeq & n \, \left( m \, + \, {3 \over 2} \, T \right) \simeq n \, m \;,
\label{rho_nonrel} \\
p & \simeq & n \, T \; \ll \; \rho \;.
\eea
The average energy per particle $\langle E \rangle \equiv \rho / n$ is instead given by 
\be
\langle E \rangle \simeq m \; + \; {3 \over 2} \, T \;.
\ee
For relativistic species ($T \gg m$) we obtain in the non degenerate case ($T\gg\mu$)
\bea
n & \simeq & \left\{ \ba{llll} \left( 3/4 \right) & \left( \zeta(3) / {\pi^2} \right) \, g \,T^3 & \;\;\;\; ~~~~~~~~~~~ \, FD \\ & \\
 & \left( \zeta(3) / {\pi^2} \right) \, g \, T^3  & \;\;\;\; ~~~~~~~~~~~ \, BE
\ea \right. \label{n_rel_nondeg} \\ 
\rho & \simeq & \left\{ \ba{llll} \left( 7/8 \right) & \left( {\pi^2} / 30 \right) \, g \, T^4 & \, ~~~~~~~~~~~~~~~~~FD \\ & \\
 & \left( {\pi^2} / 30 \right) \, g \, T^4 & \, ~~~~~~~~~~~~~~~~~BE
\ea \right. \label{ro_rel_nondeg} \\
p & \simeq & \left( 1 / 3 \right) \, \rho  \\
\langle E \rangle & \simeq & \left\{ \ba{llll} \left( 7/6 \right) & \left( \pi^4 / 30 \right) \, \zeta(3) \, T  \; \simeq \; 3.15 \, T & \; FD \\ & \\ 
 & \left( \pi^4 / 30 \right) \, \zeta(3) \, T \; \simeq \; 2.70 \, T & \; BE 
\ea \right.
\eea
In the relations above $\zeta(x)$ is the Riemann zeta-function, and $\zeta(3) \simeq 1.2021$.

Another important quantity for the evolution of a thermodynamic system is its entropy $S$.
Assuming zero chemical potentials\footnote{This is a very good approximation, as all evidence indicates that $|\mu| \ll T$.}
the entropy per comoving volume is defined, up to an additive constant, by \footnote{With a chemical potential we find $S = a^3 (\rho + p - \mu n)/T$.}
\begin{equation}
  S \; = \; V \, {{p \, + \, \rho} \over {T}} \; = \; a^3 \, {{p \, + \, \rho} \over {T}} \;.
\end{equation}
Note that throughout most of the history of the Universe (in particular the early Universe) the reaction rates of particles in the thermal bath, $\Gamma$, were much greater than the expansion rate, $H$, and thermal equilibrium\footnote{This is not true during phase transitions, when entropy is not conserved, in general.} should have been maintained. In this case the entropy in a comoving volume is conserved
\begin{equation} \label{S_conserv}
  dS \; = \; 0 \;.
\end{equation}
It is also useful to define the {\sl entropy density} in the comoving volume,
\begin{equation}\label{entropy_dens}
  s \; \equiv \; {S \over V} \; = \;  {{p \, + \, \rho} \over {T}}  \; \propto \; a^{-3} \;.
\end{equation}

Note that if ${\cal S}=4\rho_r/3T_r$ is the radiation entropy, then ${\rm S}={\cal S}/k_{\rm B}n_{\rm b}$ is the entropy per baryon, where $T_r$ is the radiation temperature, $k_{\rm B}$ is the Boltzmann constant and $n_{\rm b}$ is the baryon number density. Given that $\rho_r \propto T_r^4$ and $\rho_{\rm b}\propto n_{\rm b}$, the entropy per baryon satisfies the relation 
\be \label{entrradbar}
{\rm S}\propto\rho_{\gamma}^{3/4}/\rho_{\rm b} ~. 
\ee


\passo
\def \usefulrel1-app{Particle numbers at key epochs}
\section{\usefulrel1-app}
\label{usefulrel1-app}
\markboth{Appendix \ref{appen-cosm-mod}. ~ \appen-cosm-mod}
                    {\S \ref{usefulrel1-app} ~ \usefulrel1-app}

From decoupling till present days, neutrinos remain relativistic and therefore continue to retain their equilibrium distribution; hence the degrees of freedom characterizing the present energy density and entropy ($\gamma$ + 3 $\nu$) are (using eqs.~(\ref{degfree}) and (\ref{degfrees}))
\bea
 \gs^0 & = & 2 \; +  \; {7 \over 8} \, \left( 3 {\times} 2 \right) \, \left( {4 \over 11} \right)^{{4 \over 3}} \; \simeq \; 3.36 \;, \\
 \gss^0 & = & 2 \; +  \; {7 \over 8} \, \left( 3 {\times} 2 \right) \, \left( {4 \over 11} \right) \; \simeq \; 3.91 \;.
\eea

At BBN epoch $ T \sim 1 $ MeV and the present particles ($\gamma$ + $e^{+}e^{-}$ + 3 $\nu$) give
\be
 \gs = \gss = 2 \; 
                     +  \; {7 \over 8} \cdot 4 \, 
                     +  \; {7 \over 8} \cdot \left( 3 {\times} 2 \right) \; 
                     = \; 10.75 \;.
\ee

At $ T > 100 $ GeV all SM particles -- 8 gluons, $W^\pm$, $Z$ and $ \gamma $, 3 families of quarks (3 colors per each) and charged leptons, 3 families of $ \nu $ and the Higgs doublet -- are relativistic, and we have
\be
g_\ast^{\rm SM} = \gss^{\rm SM} = 106.75 \;.
\ee
In a minimal supersymmetric SM (MSSM) this number should be nearly doubled
\be
g_\ast^{\rm MSSM} = \gss^{\rm MSSM} = 221.5 \;.
\ee

At $ T \gg 100 $ GeV there could be some other particle in the thermal bath, so that $ g_\ast \gsim g_\ast^{\rm SM} $ or $ g_\ast \gsim g_\ast^{\rm MSSM} $ in the supersymmetric case.



\chapter{Structure formation}
\def \app-strucform{Structure formation}
\label{app-strucform}


\passo
\def \newth-app{Linear Newtonian theory}
\section{\newth-app}
\label{newth-app}
\markboth {Appendix \ref{app-strucform}. ~ \app-strucform}
                    {\S \ref{newth-app} ~ \newth-app}

\noindent {\bf The general fluid equations.} 
We model the Universe as a fluid, so that all the relevant quantities are described by smoothly varying functions of position. 
After the first bound structures, such as galaxies, form, they are treated like particles along with the genuine particles that remain unbound. 

For an expanding fluid it is convenient to adopt a comoving reference frame, specified by space coordinates $x_{\alpha}$ and a universal time $t$; if we consider a fluid with density $\rho$ and pressure $p$, moving with ``peculiar" velocity $u_{\alpha}$ in a gravitational potential $\Phi$,
its evolution is governed by the standard Eulerian equations for a self gravitating medium 
\begin{eqnarray}
{{\partial\rho}\over{\partial t}}+ 3H\rho+
  {{1}\over{a}}\partial_{\alpha}(\rho u^{\alpha})&=&0 \;,  \label{Neqc1}\\
{{{\rm d}^2a}\over{{\rm d}t^2}}x_{\alpha}+ {{\partial u_{\alpha}}\over{\partial t}}+
  Hu_{\alpha} + {{1}\over{a}}u^{\beta}\partial_{\beta}u_{\alpha}+
  {{1}\over{a\rho}}\partial_{\alpha}p + {{1}\over{a}}\partial_{\alpha}\Phi&=&0 \;,  \label{Neqc2}\\
\partial^2\Phi-4\pi Ga^2\rho&=&0 \;.  \label{Neqc3}
\end{eqnarray}
Expressions (\ref{Neqc1})-(\ref{Neqc3}) are respectively known as {\sl the
continuity, the Euler and the Poisson equations}. They describe mass conservation, momentum conservation and the Newtonian gravitational potential. 

\noindent {\bf The unperturbed background.} 
The simplest non-static solution to the system (\ref{Neqc1})-(\ref{Neqc3}) describes a smoothly expanding ($u_{\alpha}=0$), homogeneous and isotropic fluid (i.e.~$\rho_0=\rho_0(t)\,,~p_0=p_0(t)$). In particular, the unperturbed background Universe is characterized by the system
\begin{eqnarray}
{{{\rm d}\rho_0}\over{dt}}+ 3H\rho_0&=&0 \;,  \label{Neqb1}\\
{{{\rm d}^2a}\over{dt^2}}x_{\alpha} + {{1}\over{a}}\partial_{\alpha}\Phi_0&=&0 \;,
\label{Neqb2}\\
\partial^2\Phi_0- 4\pi Ga^2\rho_0&=&0\;,  \label{Neqb3}
\end{eqnarray}
with solutions
\begin{equation}
\rho_0\propto
a^{-3}\;,\hspace{5mm}v^{\alpha}_0=Hr_{\alpha}\hspace{5mm}{\rm and}\hspace{5mm}\Phi_0={2 \over 3}\pi Ga^2\rho_0x^2 \;.
\label{Nbs}
\end{equation}

\noindent {\bf The linear regime.} 
Consider perturbations about the aforementioned background solution
\begin{equation}
\rho=\rho_0+\delta\rho\,, \hspace{5mm} p=p_0+\delta p\,,
\hspace{5mm} v^{\alpha}=v_0^{\alpha}+\delta v^{\alpha}\,,
\hspace{5mm} \Phi=\Phi_0+\delta\Phi\,.  \label{deltas}
\end{equation}
where $\delta\rho$, $\delta p$, $\delta v^{\alpha}$ and $\delta\Phi$ are the perturbed first order variables with spatial as well as temporal dependence, $\delta\rho=\delta\rho(t,x_{\alpha})$. In the linear regime the perturbed quantities are much smaller than their zero order counterparts, $\delta\rho\ll\rho_0$. During this period higher order terms, for example the product $\delta\rho\delta v_{\alpha}$, are negligible. This means that different perturbative modes evolve independently and therefore can be treated separately. Note that $\delta v_{\alpha}\equiv u_{\alpha}$ is simply the peculiar velocity describing deviations from the smooth Hubble expansion. Also, the fluid pressure is related to the density via the equation of state of the medium, which is $p=p(\rho)$ for ``barotropic'' fluids (see \S\ref{sec-RW-metric}).  

In perturbation analysis it is advantageous to employ dimensionless variables for first order quantities. Here, we will be using the dimensionless ``density contrast'' $\delta\equiv\delta\rho/\rho_0$. Throughout the linear regime $\delta\rho\ll\rho_0$ meaning that $\delta\ll1$. Substituting eq.~(\ref{deltas}) into (\ref{Neqc1}) and keeping up to first order terms only, we obtain the relation for the linear evolution of density fluctuations
\begin{equation}
{{\partial\delta}\over{\partial t}} + {{1} \over {a}}\partial_{\alpha}\delta v^{\alpha}=0\,.  \label{ldelta}
\end{equation}
Also, substituting eq.~(\ref{deltas}) into (\ref{Neqc2}) and linearizing we obtain the propagation equation for the velocity perturbation
\begin{eqnarray}
{{\partial\delta v_{\alpha}}\over{\partial t}}+ H\delta v_{\alpha}+
{{v_{\rm s}^2} \over {a\rho_0}}\partial_{\alpha}\delta+
{{1}\over{a}}\partial_{\alpha}\delta\Phi=0 \;,  
\label{ldeltavi}
\end{eqnarray}
where $v_{\rm s}$ is the {\sl adiabatic sound speed}, defined, according to eq.~(\ref {eos}), as
\be
\label{soundspeed}
v_s = {\left({\partial p} \over {\partial \rho} \right)}_S^{1/2} = w^{1/2} \;,
\ee
where $S$ denotes the entropy. Thus, using the considerations exposed in \S\ref{sec-RW-metric}, we find that in a dust fluid $v_s = 0$, while in a radiative fluid $v_s = {1 / {\sqrt 3}}$. Note that eq.~(\ref {soundspeed}) indicates that $w > 1$ is impossible, because it would imply $v_s > 1$. Otherwise, if $w < 0$, then it is no longer related to the sound speed, which would be imaginary.

Finally, inserting eq.~(\ref{deltas}) into the Poisson equation and linearizing we arrive at the relation
\begin{equation}
\partial^2\delta\Phi- 4\pi Ga^2\rho_0\delta=0 \;
\label{ldeltaPhi}
\end{equation}
for the evolution of the perturbed gravitational potential. Results (\ref{ldelta})-(\ref{ldeltaPhi}) determine the behaviour of the density perturbation completely. When combined they lead to a second order differential equation that describes the linear evolution of the density contrast. In particular, taking the time derivative of (\ref{ldelta}) and employing eqs.~(\ref{ldeltavi}) and (\ref{ldeltaPhi}), we find to linear order 
\be
{{\partial^2\delta} \over {\partial t^2}} - {{v_{\rm s}^2} \over {a^2}}\partial^2\delta = -2H{{\partial\delta} \over {\partial t}} + 4\pi G\rho_0\delta \;. \label{Ndel}
\ee

\noindent {\bf The Jeans Length.} 
Equation (\ref{Ndel}) is a wave-like equation with two extra terms in the right-hand side; one due to the expansion of the Universe and the other due to gravity. It is, therefore, natural to seek plane wave solutions of the form
\begin{equation}
\delta=\sum_{\rm k}\tilde{\delta}_{({\rm k})}{\rm e}^{i{\rm
k}_{\alpha}x^{\alpha}} \;, \label{Fourier}
\end{equation}
where $\tilde{\delta}_{({\rm k})}=\tilde{\delta}_{({\rm k})}(t)$ and ${\rm k}_{\alpha}$ is the comoving wavevector.
We obtain
\begin{equation}
{{{\rm d}^2\tilde{\delta}_{({\rm k})}} \over {{\rm d}t^2}} =
  -2H{{{\rm d}\tilde{\delta}_{({\rm k})}} \over {{\rm d}t}}+ \left(4\pi G\rho_0 
  - {{v_{\rm s}^2{\rm k}^2} \over {a^2}}\right)\tilde{\delta}_{({\rm k})}\;, \label{Ndelk}
\end{equation}
which determines the evolution of the ${\rm k}$-th perturbative mode. The first term in the right-hand side of the above is due to the expansion and always suppresses the growth of $\tilde{\delta}_{({\rm k})}$. The second reflects the conflict between pressure support\footnote{For a baryonic gas the pressure is the result of particle collisions, while for ``dark matter'' collisions are negligible and pressure comes from the readjustment of the orbits of the collisionless species. In both cases it is the velocity dispersion of the perturbed component which determines the Jeans length.}  and gravity. When $4\pi G\rho_0\gg v_{\rm
s}^2k^2/a^2$ gravity dominates. On the other hand, pressure support wins if $v_{\rm s}^2k^2/a^2\gg4\pi G\rho_0$. The threshold $4\pi G\rho_0=v_{\rm s}^2k^2/a^2$ defines the length scale
\begin{equation}
\lambda_{\rm J}=v_{\rm s}\sqrt{{\pi} \over {G\rho_0}} \;. \label{Jl}
\end{equation}
The physical scale $\lambda_{\rm J}$, known as the {\bf Jeans length}, constitutes a characteristic feature of the perturbation. It separates the gravitationally stable modes from the unstable ones. Fluctuations on scales well beyond $\lambda_{\rm J}$ grow via gravitational instability, while modes
with $\lambda\ll\lambda_{\rm J}$ are stabilized by pressure.
The Jeans length corresponds to the {\bf Jeans mass}, defined as the mass contained within a sphere of radius $\lambda_{\rm J}/2$
\begin{equation}
M_{\rm J}={4\over3}\pi\rho\left({\lambda_{\rm J}} \over {2}\right)^3 \;, \label{Jm}
\end{equation}
where $\rho$ is the density of the perturbed component.

\noindent {\bf Multi-component fluids.}
When dealing with a multi-component medium (e.g.~baryons, mirror baryons, photons, neutrinos or other exotic particles), the ${\rm k}$-th perturbative mode in the non-relativistic component evolves according to
\begin{equation}
{{{\rm d}^2\delta_{i}} \over {{\rm d}t^2}} = -2H{{{\rm d}\delta_{i}} \over {{\rm d}t}} + \left[4\pi G\rho_0\sum_{j}\epsilon_{j}\delta_{j} - {{\left(v_{\rm s}^2\right)_{i}{\rm k}^2} \over {a^2}}\delta_{i}\right]\,, \label{mcNdel}
\end{equation}
where the index $i$ refers to the component in question. The sum is over all species and $\epsilon_{i}=\rho_{i}/\rho_0$ provides a measure of each component's contribution to the total background density $\rho_0=\sum_{i}\rho_{i}$. 
To first approximation, $H$ is determined by the component that dominates gravitationally, while $v_{\rm s}$ is the velocity dispersion of the perturbed species which provide the pressure support.

\noindent {\bf Solutions.}
We will now look for solutions to eqs.~(\ref{Ndelk}) and (\ref{mcNdel}) in the following four different situations: (i)~perturbations in the dominant non-relativistic component (baryonic or not) for $t>t_{\rm eq}$; (ii)~fluctuations in the
non-baryonic matter for $t<t_{\rm eq}$; (iii)~baryonic perturbations in the presence of a dominant collisionless species. 

\noindent {\bf Perturbed Einstein-de Sitter Universe.}
Consider a dust dominated (i.e.~$p=0=v_{\rm s}^2$) FRW cosmology with flat spatial section. This model, also known as the Einstein-de Sitter Universe, is though to provide a good description of our Universe after recombination. To zero order $a\propto t^{2/3}$, $H=2/3t$ and $\rho_0=1/6\pi Gt^2$ (see \S\ref{app-flat}). Perturbing this background, we look at scales well below the Hubble radius, where the Newtonian treatment is applicable. Using definition (\ref{Jl}) and the relation $\lambda=2\pi a/{\rm k}$, eq.~(\ref{Ndelk}) becomes
\footnote{Note that from now on we will drop the tilde (${}^{\sim}$) and the wavenumber (${\rm k}$) for convenience.}
\begin{equation}
t^2{{{\rm d}^2\delta} \over {{\rm d}t^2}} + {4\over3}t{{{\rm d}\delta} \over {{\rm d}t}} - {2\over3}\left[1 -\left({\lambda_{\rm J}} \over {\lambda}\right)^2\right]\delta=0\,. \label{EdS}
\end{equation}
For modes well within the horizon but still much larger than the Jeans length ($\lambda_{\rm J}\ll\lambda\ll\lambda_H$), we find
\begin{equation}
\delta={\cal C}_1t^{2/3}+{\cal C}_2t^{-1} \label{EdS1}
\end{equation}
for the evolution of the density contrast. As expected, there are two solutions: one growing and one decaying. Any given perturbation is expressed as a linear combination of the two modes. At late times, however, only the growing mode is
important.
Therefore, after matter-radiation equality perturbations in the non-relativistic component grow proportionally to the scale factor (recall that $a\propto t^{2/3}$ for dust). Note that baryonic perturbations cannot grow until matter has decoupled from radiation at recombination (we always assume that $t_{\rm eq}<t_{\rm rec}$). Dark matter particles, on the other hand, are already collisionless and fluctuations in their density can grow immediately after equipartition. After recombination, perturbations in the baryons also grow proportionally to the scale factor.

On scales well below the Jeans length (i.e.~for $\lambda\ll \lambda_{\rm J}$), eq.~(\ref{EdS}) admits the solution
\begin{equation}
\delta={\cal C}t^{-1/6}{\rm e}^{\pm i\sqrt{2/3}(\lambda_{\rm
J}/\lambda){\rm ln}t} \;, \label{EdS2}
\end{equation}
which describes a damped oscillation. Thus, small-scale perturbations in the non-relativistic matter are suppressed by pressure. 
Also, before recombination baryons and photons are tightly coupled and $(v_{\rm s}^2)_{\rm b}\propto T_{\rm b}\simeq T_{\gamma}\propto a^{-1}$, 
after decoupling $(v_{\rm s}^2)_{\rm b}\propto T_{\rm b}\propto a^{-2}$.

\noindent {\bf Mixture of radiation and dark matter.}
Consider the radiation dominated regime when $a\propto t^{1/2}$ and $H=1/2t$. On scales much smaller than the Hubble radius we can employ the Newtonian theory to study perturbations in the non-relativistic matter. 
Applying eq.~(\ref{mcNdel}) to a mixture of radiation and collisionless particles ($v_{\rm s}=0$) prior to equipartition ($\rho_{\rm DM}\ll\rho_{\gamma}\simeq\rho_0$),
given that the small-scale photon distribution is smooth (i.e.~$\langle\delta_{\gamma}\rangle\simeq0$),
we have
\begin{equation}
t{{{\rm d}^2\delta_{\rm DM}} \over {{\rm d}t^2}} + {{{\rm d}\delta_{\rm DM}} \over {{\rm d}t}} = 0 \;, \label{r+dmNdel1}
\end{equation}
that admits the solution
\begin{equation}
\delta_{\rm DM}={\cal C}_1+ {\cal C}_2{\rm ln}t \;. \label{r+dm}
\end{equation}
Thus, in the radiation epoch small scale perturbations in the collisionless component experience a very slow logarithmic growth, even when $\lambda>\lambda_{\rm J}$.
The stagnation or freezing-in of matter perturbations prior to equilibrium is generic to models with a period of rapid expansion dominated by relativistic particles and is sometimes referred to as the ``Meszaros effect''~\cite{M}.

\noindent {\bf Mixture of dark matter and baryons.}
During the period from equilibrium to recombination perturbations in the dark component grow by a factor of $a_{\rm rec}/a_{\rm eq}={\rm T}_{\rm eq}/{\rm T}_{\rm rec} \simeq 21 \Omega_{\rm m} h^2$. At the same time baryonic fluctuations do not experience any growth because of the tight coupling between photons and baryons. After decoupling, perturbations in the ordinary matter will also start
growing driven by the gravitational potential of the collisionless species. 
Consider the post-recombination Universe
dominated by non-baryonic dark matter. 
For baryonic fluctuations on scales larger that $\lambda_{\rm J}$, eq.~(\ref{mcNdel}) gives
\begin{equation}  \label{dm+bNdel1}
a^{3/2}{{{\rm d}} \over {{\rm d}a}}\left(a^{-1/2}{{{\rm d}\delta_{\rm b}} \over {{\rm d}a}}\right)+ 2{{{\rm d}\delta_{\rm b}} \over {{\rm d}a}} = {3\over2}{\cal C} \;,  
\end{equation}
where ${\cal C}$ is a constant.
The initial conditions at recombination are $\delta_{\rm b}=0$, because of the tight coupling between the baryons and the smoothly distributed photons, and $\delta_{\rm DM}\neq0$ given that the dark matter particles are already collisionless. The solution
\begin{equation}
\delta_{\rm b}=\delta_{\rm DM}\left(1 - {{a_{\rm rec}} \over {a}}\right)  \label{dm+b}
\end{equation}
shows that $\delta_{\rm b}\rightarrow\delta_{\rm DM}$ as $a\gg a_{\rm rec}$. In other words, baryonic fluctuations quickly catch up with perturbations in the dark matter component after decoupling. Alternatively, one might say that the baryons fall into the ``potential wells'' created by the collisionless species.


\passo
\def \relth-app{Linear relativistic theory}
\section{\relth-app}
\label{relth-app}
\markboth {Appendix \ref{app-strucform}. ~ \app-strucform}
                    {\S \ref{relth-app} ~ \relth-app}

\noindent {\bf The relativistic equations.} 
The basis of the relativistic analysis is the Einstein field equations (\ref {einsteinequation}), describing the interaction between matter and spacetime geometry (see appendix \ref{gen_rel}).
The Einstein tensor $G_{\alpha \beta}$ has the extremely important property of having an identically vanishing divergence, that is $\nabla^\beta G_{\alpha \beta }=0$. When applied to eq.~(\ref{einsteinequation}) neglecting the cosmological constant, the latter yields the conservation law
\begin{equation}
\nabla^\beta T_{\alpha \beta }=0 \;.  \label{cl}
\end{equation}
For a perfect fluid the stress-energy tensor takes the simple form (\ref{energymomentum}).
Here, similarly to the Newtonian analysis, we assume a barotropic fluid with $p=p(\rho)$. 
The relativistic analogues of the continuity and Euler equations are obtained from the timelike and spacelike parts of the conservation law (\ref{cl}). In particular, 
we obtain the energy density conservation law
\begin{equation}
x\dot{\rho}+ 3H(\rho+p)=0 \;,  \label{GRce}
\end{equation}
and the momentum density conservation equation
\begin{equation}
(\rho+p)\dot{u}_a+ {\rm D}_ap=0 \;. \label{GRmc}
\end{equation}
Equations (\ref{GRce}) and (\ref{GRmc}) are supplemented by
\begin{equation}
R_{ab}u^au^b={1\over2}\kappa(\rho+3p) \;,  \label{EFE1}
\end{equation}
which is the relativistic analogue of the Poisson equation and relates the spacetime geometry to the matter sources. Note that the cosmological constant has been set to zero.

\noindent {\bf The linear regime.} 
Perturbing eq.~(\ref{GRce}) and defining $\delta=\delta\rho/\rho_0$, 
the perturbed continuity equation to first order gives
\begin{equation}
\delta'- 3wH_0{\delta}+ 3(1+w)\delta H=0 \;, \label{GRdotdel}
\end{equation}
where the dash indicates differentiation with respect to $t$ and $w=p_0/\rho_0$ determines the equation of state of the medium.  
Also, starting from eq.~(\ref{GRmc}) we have
\begin{equation}
\left(\delta H\right)'+ 2H_0\delta H + {4\over3}\pi
G\rho_0\delta + {{v_{\rm s}^2} \over {3(1+w)}}{\rm D}^2\delta=0
\label{GRdotvp}
\end{equation}
to first order, where $\delta H$ describes scalar deviations from the smooth background expansion. 
Using eqs.~(\ref{GRdotdel}) and (\ref{GRdotvp}), and Fourier decomposing, in terms of the scale factor $a$ we arrive at
\begin{equation}
a^2{{d^2\tilde{\delta}_{({\rm k})}} \over {da^2}} +
  {3\over2}(1-5w+2v_{\rm s}^2)a{{d\tilde{\delta}_{({\rm k})}} \over {da}} -
  {3\over2}\left[(1+8w-3w^2-6v_{\rm s}^2) - 
  {2\over3}{{{\rm k}^2v_{\rm s}^2} \over 
  {a^2H_0^2}}\right]\tilde{\delta}_{({\rm k})}=0
\label{GRddotdel1}
\end{equation} 
for the evolution of the ${\rm k}$-th perturbative mode.

\noindent {\bf Solutions.} 
We will now seek solutions to the relativistic perturbation equations to supplement the Newtonian results of the previous section. The cases to be considered are: (i)~super-horizon sized perturbations in the dominant non-relativistic component; (ii)~fluctuations in the relativistic matter before matter-radiation equality both inside and outside the Hubble radius.

\noindent {\bf Perturbed Einstein - de Sitter Universe.} 
On scales beyond $\lambda_{\rm H}$ one needs to engage general relativity even when dealing with non-relativistic matter. For pressureless dust $w=0=v_{\rm s}^2$ and for modes lying beyond the Hubble radius ($\lambda\gg\lambda_{\rm H}$ and ${\rm k}^2v_{\rm s}^2/a^2H_0^2\ll1$), eq.~(\ref{GRddotdel1}) becomes
\footnote{Again we have dropped the tilde and the wavenumber for simplicity.}
\begin{equation}
a^2{{{\rm d}^2\delta} \over {{\rm d}a^2}} +
{3\over2}a{{{\rm d}\delta} \over {{\rm d}a}} -
{3\over2}\delta=0 \;, \label{GRESddotdel1}
\end{equation}
with
\begin{equation}
\delta={\cal C}_1a+{\cal C}_2a^{-3/2} \;. \label{GRES}
\end{equation}
Thus, large-scale perturbations in the non-relativistic component grow as $\delta_{\rm b}\propto a\propto t^{2/3}$.

\noindent {\bf The radiation dominated era.} 
Before equality radiation dominates the energy density of the Universe and $w=1/3=v_{\rm s}^2$. During this period eq.~(\ref{GRddotdel1}) gives
\begin{equation}
a^2{{d^2\delta_{\gamma}} \over {da^2}} - 2\left(1
-{1\over6}{{{\rm k}^2} \over {a^2H_0^2}}\right)\delta_{\gamma}=0 \;. \label{GRrddotdel}
\end{equation}
On large scales, when $\lambda\gg\lambda_{\rm H}$ and ${\rm k}^2/a^2H_0^2\ll1$, the above reduces to
\begin{equation}
a^2{{d^2\delta_{\gamma}} \over {da^2}} - 2\delta_{\gamma}=0 \;,
\label{GRrlsddotdel}
\end{equation}
with a power law solution of the form
\begin{equation}
\delta_{\gamma}={\cal C}_1a^2+ {\cal C}_2a^{-1} \;.    \label{GRrls}
\end{equation}
Hence, before matter-radiation equality large-scale perturbations in the radiative fluid grow as $\delta_{\gamma}\propto a^{2}$. Note that eq.~(\ref{GRrlsddotdel}) also governs the evolution of the non-relativistic component (baryonic or not), since it does not incorporate any pressure effect. Therefore, solution (\ref{GRrls}) also applies to baryons and collisionless matter.

On sub-horizon scales, with $\lambda\ll\lambda_{\rm H}$ and ${\rm k}^2/a^2H_0^2\gg1$, eq.~(\ref{GRrddotdel}) becomes
\begin{equation}
a^2{{d^2\delta_{\gamma}} \over {da^2}} +
{1\over3}{{{\rm k}^2} \over {a^2H_0^2}}\delta_{\gamma}=0  \label{GRrssddotdel}
\end{equation}
and admits the oscillatory solution
\begin{equation}
\delta_{\gamma}={\cal C}{\rm e}^{\,i\lambda_{\rm H}/\lambda} \;. 
\label{GRrss}
\end{equation}
Thus, in the radiation era small-scale fluctuations in the relativistic component oscillate like sound waves. Given that $\lambda_{\rm H}/\lambda\gg1$, the oscillation frequency is very high. As a result, $\langle\delta_{\gamma}\rangle\simeq0$ on scales well below the Hubble radius. In other words, the radiative fluid is expected to have a smooth distribution on small scales.

Note that in the radiation era the transition from growing to oscillatory modes occurs at $\lambda\sim\lambda_{\rm H}$, which implies that before equality the role of the Jeans length is played by the Hubble radius.


\begin{thebibliography}{2}
\markboth{Bibliography}{Bibliography}
\addcontentsline{toc}{chapter}{Bibliography}
\footnotesize

\bibitem{akhm69}
E. Akhmedov, Z. Berezhiani, G. Senjanovi\'c, {\sl Phys. Rev. Lett.} {\bf 69}, 3013 (1992). 

\bibitem{macho542}
C. Alcock et al (Macho Collab.), {\sl Astrophys. J.} {\bf 542}, 281 (2000).

\bibitem{alexfer94}
D.R. Alexander \& J.W. Ferguson, {\sl Astrophys. J.} {\bf 437}, 879 (1994).

\bibitem{blasi1} 
R. Aloisio, P. Blasi, A.V. Olinto, {\sl astro-ph/0206036} (2002).

\bibitem{alpher48} 
R.A. Alpher \& R.C. Herman, {\sl Nature} {\bf 162}, 774 (1948).

\bibitem{Manyfold} 
N. Arkani-Hamed, S. Dimopoulos, G. Dvali, N. Kaloper, {\sl J. High Energy Phys.} {\bf 2000}, 10 (2000).

\bibitem{babul1} 
A. Babul, S.D.M. White, {\sl Mon. Not. R. Astr. Soc.} {\bf 253}, 31P (1991).

\bibitem{bacmatvit2002} 
C. Baccigalupi, A. Balbi, S. Matarrese, F. Perrotta, N. Vittorio, {\sl Phys. Rev.} {\bf D65}, 063520 (2002).

\bibitem{balb0005124} 
A. Balbi {\it et al.}, {\sl  Astrophys. J.} {\bf 545}, L1 (2000).

\bibitem{balmatvit2003} 
A. Balbi, C. Baccigalupi, F. Perrotta, S. Matarrese, N. Vittorio, {\sl astro-ph/0301192} (2003). 

\bibitem{bardeen1} 
J.M. Bardeen, {\sl Phys. Rev.} {\bf D22}, 1882 (1980).

\bibitem{bar} 
J.M. Bardeen, J.R. Bond, N. Kaiser, A.S. Szalay, {\sl Astrophys. J.} {\bf 304}, 15 (1986).

\bibitem{bell479}
N.F. Bell, {\sl Phys. Lett.} {\bf B479}, 257 (2000).

\bibitem{benber87} 
L. Bento \& Z. Berezhiani, {\sl Phys. Rev. Lett.} {\bf 87}, 231304 (2001).

\bibitem{benberciar} 
L. Bento, Z. Berezhiani \& P. Ciarcelluti, {\sl ``Mirror world and baryon asymmetry: unified origin of visible and dark matter''}, in preparation.

\bibitem{bere27} 
Z. Berezhiani, {\sl Acta Phys. Polon.} {\bf B27}, 1503 (1996).

\bibitem{bere417} 
Z. Berezhiani, {\sl Phys. Lett.} {\bf B417}, 287 (1998). 

\bibitem{bccv1} 
Z. Berezhiani, P. Ciarcelluti, D. Comelli, F. Villante, astro-ph/0312605.

\bibitem{BCT}
Z. Berezhiani, D. Comelli, N. Tetradis, {\sl Phys. Lett.} {\bf B431}, 286 (1998). 

\bibitem{bere503} 
Z. Berezhiani, D. Comelli, F. Villante, {\sl Phys. Lett.} {\bf B503}, 362 (2001).

\bibitem{bere375} 
Z. Berezhiani, A. Dolgov, R.N. Mohapatra, {\sl Phys. Lett.} {\bf B375}, 26 (1996).

\bibitem{bere0009290} 
Z. Berezhiani, L. Gianfagna, M. Giannotti, {\sl Phys. Lett.} {\bf B500}, 286 (2001).

\bibitem{bere52} 
Z. Berezhiani, R. Mohapatra, {\sl Phys. Rev.} {\bf D52}, 6607 (1995). 

\bibitem{Venya} 
V. Berezinsky, A. Vilenkin, {\sl Phys. Rev.} {\bf D62}, 083512 (2000).  

\bibitem{binney} 
J.J. Binney \& N.W. Evans, {\sl Mon. Not. R. Astr. Soc.} {\bf 327}, L27 (2001).

\bibitem{blin9801015}
S. Blinnikov, in {\sl Baryonic Matter in the Universe and Its Spectroscopic Studies} - Atami (Japan) - November 22 - 24, 1998, {\sl astro-ph/9801015}.  

\bibitem{blin9902305} 
S. Blinnikov, in {\sl XXVII ITEP Winter School} - Snegiri - February 16-24, 1999, {\sl astro-ph/9902305}.

\bibitem{blin36}
S. Blinnikov, M. Khlopov, {\sl Yad. Fiz.} {\bf 36}, 675 (1982).

\bibitem{blin60}
S. Blinnikov, M. Khlopov, {\sl Astron. Zh.} {\bf 60}, 632 (1983).

\bibitem{blu} 
G.R. Blumenthal, S.M. Faber, J.R. Primack, M.J. Rees, {\sl Nature} {\bf 311}, 517 (1984).

\bibitem{bode} 
P. Bode {\it at al.}, {\sl Astrophys. J.} {\bf 556}, 93 (2001).

\bibitem{bond84} 
J.R. Bond \& G. Efstathiou, {\sl Astrophys. J. Lett.} {\bf 285}, L45 (1984).

\bibitem{bondsza1} 
J.R. Bond \& A. Szalay , {\sl Astrophys. J.} {\bf 276}, 443 (1983).

\bibitem{borgetal561} 
S. Borgani {\it at al.}, {\sl Astrophys. J.} {\bf 561}, 13 (2001).

\bibitem{bower1} 
R.G. Bower, P. Coles, C.S. Frenk, S.D.M. White, {\sl Astrophys. J.} {\bf 405}, 403 (1993).

\bibitem{bullock00} 
J.S. Bullock, A. Dekel, T.S. Kolatt, A.V. Kravtsov, C. Porciani, J.R. Primack, {\sl Astrophys. J.} {\bf 555}, 240 (2000).

\bibitem{bunnwhite480} 
E.F. Bunn \& M. White, {\sl Astrophys. J.} {\bf 480}, 6 (1997).

\bibitem{burl} 
S. Burles, K. M. Nollett, M. S. Turner, {\sl Astrophys. J.} {\bf 552}, L1 (2001).

\bibitem{sze} 
J.E. Carlstrom {\it at al.}, in {\em Constructing the Universe with Clusters of Galaxies}, IAP conference, eds. F. Durret and G. Gerbal (2000), {\sl astro-ph/0103480}.

\bibitem{ioecassisi}
S. Cassisi, V. Castellani, P. Ciarcelluti, G. Piotto, and M. Zoccali, {\sl Mon. Not. R. Astr. Soc.} {\bf 315}, 679 (2000).

\bibitem{ciarc2001} 
S. Cassisi, V. Castellani, P. Ciarcelluti, {\sl Mem. S.A.It.} {\bf 72}, 743C (2001).

\bibitem{0409629} 
P. Ciarcelluti, astro-ph/0409629.

\bibitem{io1} 
P. Ciarcelluti, astro-ph/0409630.

\bibitem{io2} 
P. Ciarcelluti, astro-ph/0409633.

\bibitem{bcc1} 
P. Ciarcelluti, S. Cassisi, Z. Berezhiani, {\sl ``Evolutionary and structural properties of mirror dark stars (MACHOs)''}, in preparation.

\bibitem{claybook} 
D.D. Clayton, {\sl Principles of Stellar Evolution and Nucleosynthesis}, The University of Chicago Press (1983).

\bibitem{coles95} 
P. Coles \& F. Lucchin, {\sl Cosmology: The Origin and Evolution of Cosmic Structure}, Wiley (1995).

\bibitem{colin00} 
R. Colin, V. Avila-Reese, and O. Valenzuela, {\sl Astrophys. J.} {\bf  542}, 622 (2000).

\bibitem{dber} 
P. de Bernardis {\it at al.}, {\sl Nature} {\bf 404}, 955 (2000).

\bibitem{debattista} 
V.P. Debattista \& J.A. Sellwood, {\sl Astrophys. J. Lett.} {\bf 493}, L5 (1998).

\bibitem{dekels86} 
A. Dekel \& J. Silk, {\sl Astrophys. J.} {\bf 303}, 39 (1986).

\bibitem{popolo1} 
A. Del Popolo, M. Gambera, {\sl Astronomy \& Astrophysics} {\bf 337}, 96 (1998).

\bibitem{popolo2} 
A. Del Popolo, M. Gambera, in {\sl Proceedings of the VIII Conference on Theoretical Physics: General Relativity and Gravitation} - Bistritza (Romania) -  June 15-18, 1998.

\bibitem{Gia} 
G. Dvali, Q. Shafi, R. Schaefer, {\sl Phys. Rev. Lett.} {\bf 73} (1994) 1886. 

\bibitem{erdo0202357} 
P. Erdogdu, S. Ettori and O. Lahav, {\sl astro-ph/0202357} (2002).

\bibitem{efs} 
G. Efstathiou, in {\em The physics of the early Universe}, eds. A. Heavens, J. Peacock, A. Davies (SUSSP) (1990).

\bibitem{efstmnras330} 
G. Efstathiou {\it at al.}, {\sl Mon. Not. R. Astr. Soc.} {\bf 330L}, 29 (2002).

\bibitem{navstein01} 
V.R. Eke, J.F. Navarro, M. Steinmetz, {\sl Astrophys. J.} {\bf 554}, 114 (2001).

\bibitem{ellib118}
J. Ellis, A. Linde, D. Nanopoulos, {\sl Phys. Lett.} {\bf B118}, 59 (1982).

\bibitem{enqvd62} 
K. Enqvist, H. Kurki-Suonio, J. Valiviita, {\sl Phys. Rev.} {\bf D62}, 103003 (2000).

\bibitem{enqvd65} 
K. Enqvist, H. Kurki-Suonio, J. Valiviita, {\sl Phys. Rev.} {\bf D65}, 043002 (2002).

\bibitem{erdogdu1} 
P. Erdogdu, S. Ettori, O. Lahav, {\sl astro-ph/0202357} (2002).

\bibitem{evrard} 
A. E. Evrard, {\sl Mon. Not. R. Astr. Soc.} {\bf 292}, 289 (1997).

\bibitem{feld0002}
G. Felder, L. Kofman, A. Linde, {\sl J. High Energy Phys.} {\bf 0002}, 027 (2000).

\bibitem{fixsen96} 
D.J. Fixsen {\it at al.}, {\sl Astrophys. J.} {\bf 473}, 576 (1996).

\bibitem{flores} 
R.A. Flores \& J.R. Primack, {\sl Astrophys. J. Lett.} {\bf 427}, L1 (1994).
 
\bibitem{foot452}
R. Foot, {\sl Phys. Lett.} {\bf B452}, 83 (1999).  

\bibitem{foot505}
R. Foot, {\sl Phys. Lett.} {\bf B505}, 1 (2001).

\bibitem{foot0207175} 
R. Foot, {\sl astro-ph/0207175} (2002).

\bibitem{foot272}
R. Foot, H. Lew, R. Volkas, {\sl Phys. Lett.} {\bf B272}, 67 (1991).

\bibitem{foota7}
R. Foot, H. Lew, R. Volkas, {\sl Mod. Phys. Lett.} {\bf A7}, 2567 (1992).

\bibitem{foota9}
R. Foot, H. Lew, R. Volkas, {\sl Mod. Phys. Lett.} {\bf A9}, 169 (1994).

\bibitem{foot7}
R. Foot, R.  Volkas, {\sl Astro. Part. Phys.} {\bf 7}, 283 (1997).

\bibitem{foot61}
R. Foot, R.  Volkas, {\sl Phys. Rev.} {\bf D61}, 043507 (2000).

\bibitem{footsila0104251}
R. Foot, Z.K. Silagadze, {\sl Acta Phys. Polon.} {\bf B32}, 2271 (2001).

\bibitem{footd52}
R. Foot, R. Volkas, {\sl Phys. Rev.} {\bf D52}, 6595 (1995).

\bibitem{free} 
W. L. Freedman {\it at al.}, {\sl Astrophys. J.} {\bf 553}, 47 (2001).

\bibitem{Freese}
K. Freese, {\sl Phys. Rep.} {\bf 333}, 183 (2000).

\bibitem{fre} 
C.S. Frenk, S.D.M. White, G. Efstathiou, {\sl Astrophys. J.} {\bf 327}, 507 (1988).

\bibitem{Fukugita} 
M. Fukugita, {\sl hep-ph/0012214} (2001).

\bibitem{FY86} 
M. Fukugita, T. Yanagida, {\sl Phys. Lett.} {\bf B174}, 45 (1986).

\bibitem{FY90} 
M. Fukugita, T. Yanagida, {\sl Phys. Rev.} {\bf D42}, 1285 (1990).

\bibitem{gamow48} 
G. Gamow, {\sl Phys. Rev.} {\bf 74}, 506 (1948).

\bibitem{giud9911}
G.F. Giudice, A. Riotto, I. Tkackev, {\sl JHEP} {\bf 9911}, 036 (1999).

\bibitem{glas167}
S.L. Glashow, {\em ibid.} {\bf B167}, 35 (1986).

\bibitem{guth23} 
A.H. Guth, {\sl Phys. Rev.} {\bf D23}, 347 (1981).

\bibitem{halv568} 
N. W. Halverson {\it at al.}, {\sl Astrophys. J.} {\bf 568}, 38 (2002).

\bibitem{hamteg330} 
A.J.S. Hamilton \& M. Tegmark, {\sl Mon. Not. R. Astr. Soc.} {\bf 330}, 506 (2002).

\bibitem{hannestad} 
S. Hannestad, {\sl astro-ph/9912558} (1999).

\bibitem{hodg47}
H. Hodges, {\sl Phys. Rev.} {\bf D47}, 456 (1993).

\bibitem{hold166}
B. Holdom, {\sl Phys. Lett.} {\bf B166}, 196 (1985).

\bibitem{huwh97} 
W. Hu \& M. White, {\sl Astrophys. J.} {\bf 479}, 568 (1997).

\bibitem{huss} 
A. Huss, B. Jain, and M. Steinmetz, {\sl Astrophys. J.} {\bf517}, 64 (1999).

\bibitem{igna0005125} 
A. Ignatiev, R. Volkas, {\sl Phys. Rev.} {\bf D62}, 023508 (2000).

\bibitem{igna0005238}
A. Ignatiev, R. Volkas, {\sl Phys. Lett.} {\bf B487}, 294 (2000).

\bibitem{jean199} 
J. Jeans, {\sl Phil Trans.} {\bf 199A}, 49 (1902).

\bibitem{jeanbookac}
J. Jeans, {\sl Astronomy and Cosmology}, Cambridge University Press (1928).

\bibitem{Kawano}
L. Kawano, {\it Preprint} {\sl FERMILAB-Pub-92/04-A} (1992).

\bibitem{khlo68}
M. Khlopov et al., {\sl Astron. Zh.} {\bf 68}, 42 (1991).

\bibitem{Klypin:1999uc} 
A.A. Klypin, A.V. Kravtsov, O. Valenzuela, \& F. Prada, {\sl Astrophys. J.} {\bf 522}, 82 (1999).

\bibitem{kobz3}
Y. Kobzarev, L. Okun, I. Pomeranchuk, {\sl Yad. Fiz.} {\bf 3}, 1154 (1966).

\bibitem{kosa1} 
H. Kodama \& M. Sasaki, {\sl Prog. Theo. Phys. Suppl.} {\bf 78}, 1 (1984).

\bibitem{KST}
E. Kolb, D. Seckel, M. Turner, {\sl Nature} {\bf 514} (1985) 415.  

\bibitem{kolbbookeu} 
E.W. Kolb \& M.S. Turner, {\sl The Early Universe}, Addison-Wesley (1990).

\bibitem{kolb50}  
E.W. Kolb \& S.L. Vadas, {\sl Phys. Rev.} {\bf D50}, 2479 (1994).

\bibitem{kormendy}
J. Kormendy \& G. R. Knapp (eds.), {\em Dark Matter in the Universe: IAU Symposium No. 117\/} (Reidel, Dordrecht, 1987).

\bibitem{krauss2} 
L. M. Krauss, {\sl Astrophys. J.} {\bf 501}, 461 (1998).

\bibitem{KRS155} 
V.A. Kuzmin, V.A. Rubakov, M.E. Shaposhnikov, {\sl Phys. Lett.} {\bf B155}, 36 (1985). 

\bibitem{lang0005004} 
A. Lange {\it at al.}, {\sl Phys. Rev.} {\bf D63}, 042001 (2001).

\bibitem{lazvla56} 
G. Lazarides, N. Vlachos, {\sl Phys. Rev.} {\bf D56}, 4562 (1997).

\bibitem{leea} 
A. T. Lee {\it at al.}, {\sl Astrophys. J.} {\bf 561}, L1 (2001).

\bibitem{li104}
T.D. Li and C.N. Yang, {\sl Phys. Rev.} {\bf 104}, 254 (1956).

\bibitem{liddbookcilss} 
A.R. Liddle and D.H. Lyth, {\sl Cosmological Inflation and Large-Scale Structure}, Cambridge University Press (2000).

\bibitem{lidd291} 
A.R. Liddle \& D.H. Lyth, {\sl Phys. Lett.} {\bf B291}, 391 (1992). 

\bibitem{lifshitz1} 
E.M. Lifshitz, {\sl J. Phys. USSR} {\bf 10}, 116 (1946).

\bibitem{LSV}
E. Lisi, S. Sarkar, F.L. Villante, {\sl Phys. Rev.} {\bf D 59}, 123520 (1999). 

\bibitem{lynd136} 
D. Lynden-Bell, {\sl Mon. Not. R. Astr. Soc.} {\bf 136}, 101 (1967).

\bibitem{mabert1} 
C. Ma \& E. Bertschinger, {\sl Astrophys. J.} {\bf 455}, 7 (1995).

\bibitem{maso0205384} 
B. S. Mason {\it at al.}, {\sl astro-ph/0205384} (2002).

\bibitem{mather1} 
J. C. Mather {\it at al.}, {\sl Astrophys. J.} {\bf 512}, 511 (1999).

\bibitem{mcna277} 
S.J. Mcnally and J.A. Peacock, {\sl Mon. Not. R. Astr. Soc.} {\bf 277}, 143 (1995).

\bibitem{M} 
P. Meszaros, {\sl Astron. Astrophys.} {\bf 37}, 225 (1974).

\bibitem{moha478} 
R.N. Mohapatra, V. Teplitz, {\sl Astrophys. J.} {\bf 478}, 29 (1997).

\bibitem{moha462} 
R.N. Mohapatra, V. Teplitz, {\sl Phys. Lett.} {\bf B462}, 302 (1999).

\bibitem{moha0001362} 
R.N. Mohapatra, V. Teplitz, {\sl Phys. Rev.} {\bf D62}, 063506 (2000).

\bibitem{moore94} 
B. Moore, {\sl Nature} {\bf 370}, 620 (1994).

\bibitem{Moore:1999wf} 
B. Moore {\it et al}, {\sl Astrophys. J. Lett.} {\bf 524}, L19 (1999).

\bibitem{moore99} 
B. Moore {\it at al.}, {\sl Mon. Not. R. Astr. Soc.} {\bf 310}, 1147 (1999).

\bibitem{MooreSub} 
B. Moore,  S. Ghigna,  F. Governato, G. Lake, T. Quinn,  J. Stadel, P. Tozzi, {\sl Astrophys. J.} {\bf 524}, L19 (1999).

\bibitem{NFW} 
J.F. Navarro, C.S. Frenk, and S.D. White, {\sl Astrophys. J.} {\bf 462}, 563 (1996).
  
\bibitem{nett} 
C. B. Netterfield {\it at al.}, {\sl Astrophys. J.} {\bf 571}, 604 (2002).

\bibitem{padmbooksfu} 
T. Padmanabhan, {\sl Structure Formation in the Universe}, Cambridge University Press (1993).

\bibitem{paddybook} 
T. Padmanabhan, {\sl Cosmology and Astrophysics Through Problems} (Cambridge University Press) (1996).

\bibitem{pavsic}
M. Pavsic, {\sl Int. J. Theor. Phys.} {\bf 9}, 229 (1974) [hep-ph/0105344].

\bibitem{peacockd1} 
J.A. Peacock \& S.J. Dodds, {\sl Mon. Not. R. Astr. Soc.} {\bf 267}, 1020 (1994).

\bibitem{pear0205388} 
T. J. Pearson {\it at al.}, {\sl astro-ph/0205388} (2002).

\bibitem{peeb263} 
P.J.E. Peebles, {\sl  Astrophys. J. Lett.} {\bf 263}, L1 (1982).

\bibitem{peebles70} 
P.J.E. Peebles \& J.T. Yu, {\sl Astrophys. J.} {\bf 162}, 815 (1970).

\bibitem{penwil65} 
A. Penzias \& R. Wilson, {\sl Astrophys. J.} {\bf 142}, 419 (1965).

\bibitem{perc327} 
W. J. Percival {\it at al.}, {\sl Mon. Not. R. Astr. Soc.} {\bf 327}, 1297 (2001).

\bibitem{percetal337} 
W.J. Percival {\it at al.}, {\sl Mon. Not. R. Astr. Soc.} {\bf 337}, 1068 (2002).

\bibitem{perl391} 
S. Perlmutter et al, {\sl Nature} {\bf 391}, 51 (1998).

\bibitem{perl517} 
S. Perlmutter et al, {\sl Astrophys. J.} {\bf 517}, 565 (1999).

\bibitem{pres239} 
W.H. Press \& E.T. Vishniac, {\sl Astrophys. J.} {\bf 239}, 1 (1980).

\bibitem{pryk568} 
C. Pryke {\it at al.}, {\sl Astrophys. J.} {\bf 568}, 46 (2002).

\bibitem{ratra1} 
B. Ratra, {\sl Phys. Rev.} {\bf D38}, 2399 (1988).

\bibitem{rees85} 
M.J. Rees, {\sl Mon. Not. R. Astr. Soc.} {\bf 213}, 75P (1985).

\bibitem{horizon} 
W. Rindler, {\sl Mon. Not. R. Astr. Soc.} {\bf 116}, 663 (1956).

\bibitem{bau} 
A. Riotto, M. Trodden, {\sl Ann. Rev. Nucl. Part. Sci.} {\bf 49} (1999) 35.

\bibitem{rogigl96}
F.J. Rogers \& C.A. Iglesias, {\sl Astrophys. J.} {\bf 464}, 943 (1996).

\bibitem{row} 
M. Rowan-Robinson, in {\em Third International Conference on Identification of Dark Matter}, World Scientific (2001), astro-ph/0012026.

\bibitem{ruba65} 
V. Rubakov, {\sl JETP Lett.} {\bf 65}, 621 (1997). 

\bibitem{sakha67} 
A.D. Sakharov, {\sl Sov. Phys. JETP Lett.} {\bf 5}, 24 (1967).

\bibitem{scvh95} 
D. Saumon, G. Chabrier, H.M. Van Horn, {\sl Astrophys. J. Suppl.} {\bf 99}, 713 (1995).

\bibitem{saun} 
W. Saunders {\it at al.}, {\sl Mon. Not. R. Astr. Soc.} {\bf 317}, 55 (2000).

\bibitem{ss1985} 
R. Schaeffer \& J. Silk, {\sl Astrophys. J.} {\bf 292}, 319 (1985).

\bibitem{schm507} 
B.P. Schmidt et al, {\sl  Astrophys. J.} {\bf 507}, 46 (1998).

\bibitem{sch0209584} 
D.J. Schwarz, {\sl astro-ph/0209584} (2002).

\bibitem{scot} 
P. E. Scott {\it at al.}, {\sl astro-ph/0205380} (2002).

\bibitem{selzal469} 
U. Seljak \& M. Zaldarriaga, {\sl Astrophys. J.} {\bf 469}, 437 (1996).

\bibitem{shu225} 
F.H. Shu, {\sl  Astrophys. J.} {\bf 225}, 83 (1978).

\bibitem{siev} 
J. L. Sievers {\it at al.}, {\sl astro-ph/0205387} (2002).

\bibitem{sila60}
Z. Silagadze, {\sl Phys. At. Nucl.} {\bf 60}, 272 (1997).

\bibitem{silk215} 
J. Silk, {\sl Nature} {\bf 215}, 1155 (1967).

\bibitem{silk68} 
J. Silk, {\sl Astrophys. J.} {\bf 151}, 459 (1968).

\bibitem{silk85} 
J. Silk, {\sl Astrophys. J.} {\bf 297}, 1 (1985).

\bibitem{smoot92} 
G.F. Smoot {\it at al.}, {\sl Astrophys. J. Lett.} {\bf 396}, L1 (1992).

\bibitem{sldolgov} 
J. Sommer-Larsen and A. Dolgov, {\sl Astrophys. J.} {\bf551}, 608 (2001).

\bibitem{sidm} 
D.N. Spergel and P.J. Steinhardt, {\sl Phys. Rev. Lett.} {\bf 84}, 3760 (2000).

\bibitem{stew302} 
E.D. Stewart \& D.H. Lyth, {\sl Phys. Lett.} {\bf B302}, 171 (1993).

\bibitem{taylor1} 
J. E. Taylor \& J. Silk, {\sl astro-ph/0207299} (2002).

\bibitem{tegmark1} 
M. Tegmark, M. Zaldarriaga and A. J. S. Hamilton, {\sl Phys. Rev.} {\bf D63}, 043007 (2001).

\bibitem{tetra57} 
N. Tetradis, {\sl Phys. Rev.} {\bf D57}, 5997 (1998).  

\bibitem{TothOstriker} 
G. Toth, J.P. Ostriker, {\sl Astrophys. J.} {\bf 389}, 5 (1992).

\bibitem{albada1} 
T. S. van Albada \& R. Sancisi, {\sl Phil. Trans. R. Lond. A} {\bf 320}, 447 (1986).

\bibitem{vdbosch01} 
F. van den Bosch, {\sl Mon. Not. R. Astr. Soc.} {\bf 327}, 1334 (2001).

\bibitem{vittorio84} 
N. Vittorio \& J. Silk, {\sl Astrophys. J.} {\bf 285}, 39 (1984).

\bibitem{volk58}
R. Volkas, Y. Wong, {\sl Phys.Rev.} {\bf D58}, 113001 (1998).

\bibitem{volk13} 
R. Volkas, Y. Wong, {\sl Astropart. Phys.} {\bf 13}, 21 (2000).

\bibitem{wang} 
X. Wang, M. Tegmark \& M. Zaldarriaga, {\sl Phys. Rev.} {\bf D65}, 123001 (2002).

\bibitem{weinbookgc} 
S. Weinberg, {\sl Gravitation and Cosmology}, Wiley (1972).

\bibitem{Weinberg}
M. Weinberg, {\sl Mon. Not. R. Astr. Soc.} {\bf 299}, 499 (1998).

\bibitem{white} 
S. D. M. White {\it et al}, {\sl Mon. Not. R. Astr. Soc.} {\bf 262}, 1023 (1993).

\bibitem{whi} 
D.M. White, C.S. Frenk, M. Davis, G. Efstathiou, {\sl Astrophys. J.} {\bf 313}, 505 (1987).

\bibitem{whitess94} 
M. White, D. Scott, and J. Silk, {\sl Ann. Rev. Astron. \& Astrophys.} {\bf 32}, 329 (1994).

\bibitem{zalsel129} 
M. Zaldarriaga \& U. Seljak, {\sl Astrophys. J. Suppl.} {\bf 129}, 431 (2000).

\bibitem{Z} 
Y.B. Zeldovich, {\sl Soviet Phys. Usp.} {\bf 9}, 602 (1967).

\bibitem{Ze}
Y.B. Zeldovich, {\sl Astrofisika} {\bf 6}, 319 (1970).

\bibitem{zwicky}
F. Zwicky, {\sl Helv. Phys. Acta} {\bf 6}, 110 (1933).

\end{thebibliography}
\end{document}